\documentclass[twoside,final,11pt]{entics} 
\usepackage{enticsmacro}
\usepackage{main,graphicx} 
\usepackage[T1]{fontenc}
\usepackage[utf8]{inputenc}
\usepackage{microtype}
\DisableLigatures[-]{encoding = T1, family = tt* }

\usepackage[conor]{agda}
\usepackage{agda-unicode}
\usepackage{xstring}
\usepackage{agda-custom}

\def\conf{MFPS 2026} 	
\volume{6}			
\def\volu{6}			
\def\papno{13}			
\def\mydoi{19668}
\def\lastname{Mosses, Cockx, Reus}
\def\runtitle{Denotational Semantics in Agda} 
\def\copynames{P. Mosses, J. Cockx, B. Reus}    

\begin{document}

\begin{frontmatter}
  \title{Mechanising Denotational Semantics in Agda}
  \author{Peter D. Mosses\thanksref{delft}\thanksref{swansea}\thanksref{mosses}}	
  \author{Jesper Cockx\thanksref{delft}\thanksref{cockx}}		
  \author{Bernhard Reus\thanksref{sussex}\thanksref{reus}}		
  \address[delft]{Department of Software Technology\\ Delft University of Technology\\	
    Delft, Netherlands}  							
  \address[swansea]{Faculty of Science and Engineering\\ Swansea University\\
    Swansea, United Kingdom}  							
  \thanks[mosses]{Email: \href{mailto:p.d.mosses@tudelft.nl} {\texttt{\normalshape
      p.d.mosses@tudelft.nl}}} 
  \thanks[cockx]{Email: \href{mailto:j.g.h.cockx@tudelft.nl} {\texttt{\normalshape
      j.g.h.cockx@tudelft.nl}}} 
  \address[sussex]{School of Engineering and Informatics\\ University of Sussex\\
    Brighton, United Kingdom} 
  \thanks[reus]{Email:  \href{mailto:bernhard@sussex.ac.uk} {\texttt{\normalshape
        bernhard@sussex.ac.uk}}}

\begin{abstract} 
  Mechanisation of a mathematical definition (also referred to as formalisation)
  has many benefits. Here, we focus on mechanising denotational semantic definitions
  of programming languages by embedding them in the Agda language. The Agda
  type-checker detects and reports any issues with the wellformedness and type
  correctness of the embedded definitions.
  
  To minimise the effort required, and to facilitate correlation of the original
  definition with its Agda embedding, mechanisation should not involve significant
  reformulation or extension. Here, we show how  to embed conventional Scott–Strachey
  denotational definitions in Agda with almost no changes to $\lambda$-notation or
  domain equations.

  Agda notation for definitions of types and functions corresponds closely
  to the conventional meta-notation of denotational semantics. We have developed
  a collection of Agda modules with postulated types for commonly used domain
  constructors and their associated operations. 

  We illustrate our approach with mechanisations of three denotational definitions:
  a model of the untyped $\lambda$-calculus, Plotkin’s denotational
  semantics of PCF, and a semantics of a sublanguage of Scheme. In previous work,
  similar mechanisations in Agda have revealed several unsuspected wellformedness
  issues in published denotational definitions.

  Some of the postulated properties of our definitions are inconsistent with a
  classical set-theoretic interpretation of Agda. We conjecture that they would be
  consistent with an interpretation of Agda in a higher-order intuititionistic logic,
  such as that used by Simpson in his work on synthetic domain theory. After recalling
  previous work on mechanising domain theory, we consider the possibility of
  mechanising synthetic domain theory in Agda.
\end{abstract}


\end{frontmatter}

\section{Introduction}\label{introduction}

The Scott–Strachey style of denotational semantics is based on Christopher
Strachey’s idea of defining the semantics of high-level programming languages
by mapping phrases to pure higher-order functions \cite{Strachey1966TFS},
combined with the rigorous foundations provided by Dana Scott’s
development of domain theory \cite{Scott1970OMT,Scott1972CL,Scott1971TMS}.
It has been widely adopted in theoretical studies of programming language semantics,
and presented in many textbooks.

A significant drawback for potential users of denotational semantics is the current lack of
an IDE to support development and testing of denotational definitions.
It is well known that manual development and review of large formal definitions are error-prone:
mechanical checks are essential to ensure wellformedness \cite{Klein2012RYR}.

We have developed a shallow embedding of Scott–Strachey style denotational definitions in Agda,
to support their formulation and checking wellformedness.
Crucially, definitions of denotations in $\lambda$-notation are embedded almost verbatim,
with only minor lexical changes (such as inserting spaces between adjacent variables).
If the embedding of a definition is rejected by Agda,
the location of the error in the Agda code generally makes it easy to find the
source of the wellformedness issue in the original definition.
Moreover, the directness of the Agda embedding minimises the risk of introducing
significant deviations.

The original motivation for developing an Agda embedding of denotational
semantics was to check the steps in proofs about a semantics of inheritance \cite{Cook1989DSI,Cook1994DSI}.
However, type-checking the embedding revealed some unsuspected wellformedness issues 
in the semantic definition \cite{Mosses2024TVD}.
Using a subsequent version of our embedding,
we have detected previously unreported wellformedness issues in the denotational semantics
given in the Scheme reports \cite{Mosses2025CDS}.
This has confirmed the usefulness and practical applicability of the embedding.

Our current embedding of denotational semantics in Agda is based on declarations
of postulated types for domain constructors and their associated operations.
We abstract from the mathematical structure of domains, as it does not affect
how denotational semantic definitions are formulated.
In future work, we aim to \emph{define} the postulated domain constructors and
operations based on postulated axioms for \emph{synthetic domain theory} (SDT).
As Alex Simpson hinted in \cite[sect.~3]{Simpson2004CAR}:
``\emph{it seems likely that, with appropriate reformulations, the development [of SDT]
could be carried out in the (predicative) context of Martin-L\"of’s Type Theory}''.

\paragraph{Contributions}

We start by briefly recalling some features of denotational semantics and Agda
(Section~\ref{Background-background}),
although readers should preferably already be familiar with the basic concepts of denotational
semantics (e.g., in Scott–Strachey style \cite{Scott1971TMS,Stoy1977DSS,Tennent1976DSP})
and with dependently-typed languages such as Agda \cite{Agda-Language,Agda-Wikipedia}.

We then present the Agda code that supports our embedding of denotational definitions,
and give examples to illustrate its use:
\begin{itemize}
  \item
  We introduce our postulated types for domain constructors and their associated operations
  (Section~\ref{Notation-postulated-domain-notation}).
  \item
  We give an Agda embedding of a model of the untyped $\lambda$-calculus
  (Section~\ref{Examples/LC-untyped-lambda-calculus}).
  When defining the denotations of abstraction and application,
  Agda requires explicit use of a bijection between a domain and
  the domain of endofunctions on it,
  but otherwise the embedding is as simple as possible.
  \item
  Our Agda embedding of Plotkin’s denotational semantics of the PCF language
  (Section~\ref{Examples/PCF-pcf-a-programming-language-for-computable-functions})
  illustrates the use of dependent types in formalising the semantics of
  an intrinsically-typed programming language.
  This example exploits Agda’s support for Unicode symbols to minimise notational differences
  from Plotkin’s definition.
  \item
  The last illustrative example that we present here is an embedding of the denotational semantics
  of a simple sublanguage of \emph{Scheme} (Section~\ref{Examples/Scm-scm-a-sublanguage-of-scheme}).
  It uses all the domain constructors introduced in Section~\ref{Notation-postulated-domain-notation}.
  \item
  We postulate basic properties of operations associated with some domain constructors
  (Section~\ref{Properties-postulated-properties}),
  and use these properties as rewrite rules to support testing the denotations of terms
  (Appendix~\ref{Tests-illustrative-tests}).
\end{itemize} 
Finally, we give an overview of mechanised domain theory and synthetic domain theory (SDT),
and consider the possibility of mechanising SDT in Agda (Section~\ref{mechanising}).

The Agda code shown in this paper is a highlighted rendering of our Agda source files,
but elides some details.
See the accompanying repository for the complete code and a generated website
with hyperlinked, highlighted code listings \cite{MFPS2026-Agda}.
The \LaTeX\ sources for sections~2–5 and the appendix of this paper were generated from the repository.
In the PDF, each declared or referenced Agda module name
is hyperlinked to the declaration of that module in the website,
to facilitate browsing the code while reading the paper.

\section{Background}\label{Background-background}

This section first recalls the main features of Scott--Strachey
denotational semantics, and summarises the notation conventionally used
in denotational descriptions of programming languages. It then mentions
significant features of the Agda language, and briefly explains some
Agda syntax used in the subsequent sections.
\begin{AgdaSuppressSpace}
\begin{code}[hide]%
\>[0]\AgdaKeyword{module}\AgdaSpace{}%
%
\AgdaModule{Background}%
\AgdaSpace{}%
\AgdaKeyword{where}\<%
\\
\>[0][@{}l@{\AgdaIndent{0}}]%
\>[2]\AgdaKeyword{import}\AgdaSpace{}%
%
\AgdaModule{Background.Denotational-Semantics}%
\<%
\\
\>[2]\AgdaKeyword{import}\AgdaSpace{}%
%
\AgdaModule{Background.Agda}%
\<%
\end{code}
\end{AgdaSuppressSpace}

\subsection{Denotational Semantics}\label{Background/Denotational-Semantics-denotational-semantics}

\subsubsection{Features}\label{Background/Denotational-Semantics-features}

A denotational semantic description of a programming language specifies
\emph{abstract syntax}, \emph{semantic domains}, and \emph{semantic
functions}.

\begin{description}
\tightlist
\item[Abstract syntax.]
An abstract syntax defines semantically-relevant sets of phrases (e.g.,
commands, declarations, expressions) and their compositional structure.
Phrases are regarded as \emph{abstract syntax trees} (ASTs), rather than
as strings of symbols.

Abstract syntax is specified by a \emph{context-free grammar}. The
nonterminal symbols of the grammar correspond to named sets of ASTs.
BNF-like productions determine notation for AST node constructors.
\item[Semantic domains.]
The \emph{denotation} of a phrase models the contribution of that phrase
to the semantics of all programs in which it occurs. Semantic domains
are sets of abstract mathematical values (e.g., functions, tuples, truth
values) used to construct denotations.

Semantic domains are specified by \emph{domain equations} that define
named semantic domains in terms of standard \emph{domain constructors}.
The definitions can be (mutually) recursive.
\item[Semantic functions.]
These functions map abstract phrases to their denotations. They are
required to be \emph{compositional}: the denotation of a phrase
generally depends on the denotations of its sub-phrases, but it cannot
depend on their syntax.

Semantic functions are specified inductively by \emph{semantic
equations}. Their AST arguments are enclosed by
\(\llbracket\dots\rrbracket\), avoiding potential confusion between
notation for AST constructors and elements of domains.
\end{description}

\subsubsection{Domain Notation}\label{Background/Denotational-Semantics-domain-notation}

Domains were originally required to be \emph{continuous lattices}
\cite{Scott1970OMT,Scott1972CL,Stoy1977DSS,Tennent1976DSP}.
Various more general notions of domain have been suggested (e.g., see
\cite{Abramsky1995DT})
but here it is only important that endofunctions between domains have
\emph{fixed points}, recursive \emph{domain equations} have well-defined
solutions (up to isomorphism), and each domain comes with a \emph{bottom
element}. The conventional notation for domain constructors and their
accompanying operations, summarised below, is independent of the exact
notion of domain.

\begin{description}
\tightlist
\item[Domains.]
Every domain \(D\) has a \emph{bottom} element \(\bot_D\) that
represents absence of information (e.g., due to an error or
nontermination of a computation), usually written just \(\bot\).
\item[Function domains.]
A function \(\phi : D \to E\) is (Scott-)\emph{continuous} if it is
monotone and preserves limits of directed subsets
of~\(D\)~\cite{Abramsky1995DT}.
The \emph{function domain} \(F = D \to E\) consists of all continuous
(total) functions from \(D\) to \(E\). The set of all functions from a
set~\(A\) to a domain also forms a domain, ordered pointwise.

Functions between domains are always continuous when they are defined
only using abstraction (written \(\lambda \delta.\epsilon\)),
application (written \(\phi\,\delta\)), the operation \(\textit{fix}\)
that maps each endofunction \(\phi : D \to D\) to its least fixed point,
and the continuous operations associated with other domain constructors.
\item[Flat domains.]
For any set \(A\) the \emph{flat domain} \(A_\bot\) is formed by adding
\(\bot\) as a fresh element. Functions on sets are implicitly extended
to (continuous) functions on flat domains, returning~\(\bot\) when any
argument is~\(\bot\).

Elements~\(\tau\) of the domain of \emph{truth-values}
\(\textbf{T} = \{ \textit{true}, \textit{false} \}_\bot\) are used in
\emph{conditionals} written \(\tau \to \delta_1, \delta_2\,\), where
\(\delta_1, \delta_2 : D\), \(\textit{true} \to \delta_1, \delta_2\) is
\(\delta_1\,\), \(\textit{false} \to \delta_1, \delta_2\) is
\(\delta_2\,\), and \(\bot_{\textbf{T}} \to \delta_1, \delta_2\) is
\(\bot_D\,\).
\item[Sum domains.]
The \emph{separated sum domain} \(X = \ldots + Y + \ldots\) consists of
injected elements written `\(\upsilon \textsf{ in } X\)' (where
\(\upsilon : Y\) for some summand \(Y\)) together with \(\bot_X\).

The \(\textbf{T}\)-valued operation \(\chi \mathbin{\textsf{E}} Y\)
(written \(\chi \in Y\) in
\cite{Scheme}) tests whether
\(\chi : X\) is the injection of some \(\upsilon : Y\); if so,
\(\chi \mid Y\) projects \(\chi\) to \(\upsilon\), otherwise to
\(\bot_X\). (This notation has the pragmatic advantage of being
independent of the order of the summands.)
\item[Product domains.]
The \emph{product domain} \(P = D \times E\) consists of pairs written
\(\langle \delta, \epsilon \rangle\), where \(\delta : D\),
\(\epsilon : E\), and \(\bot_P = \langle \bot_D, \bot_E \rangle\). When
\(\pi : P\), the operations \(\pi \downarrow 1\) and
\(\pi \downarrow 2\) select its components.

The domain of \(n\)-\emph{tuples}
\(\langle \delta_1, \ldots, \delta_n \rangle\) of elements
\(\delta_i : D\) is written~\(D^n\), and the domain of all finite
\emph{sequences} is written~\(D^*\). Further operations on~\(D^*\)
include the empty sequence~\(\langle \rangle\),
concatenation~\(\delta^*_1 \mathbin{\S} \delta^*_2\),
length~\(\textit{\#}\,\delta^*\), \(n\)th~component
\(\delta^* \downarrow n\), and \(n\)th~tail
\(\delta^* \mathbin{\dagger} n\) (\(n \geq 1\)).
\item[Recursive domains.]
In a collection of domain equations \(D_i = E_i\), the \(D_i\) are
distinct domain names, and the \(E_i\) are domain terms formed from
standard domain constructors and domain names, allowing unrestricted
recursion. However, the solution of the equations is only up to an
isomorphism (written \(\textit{unfold} : D_i \to E_i\)~,
\(\textit{fold} : E_i \to D_i\) in
\cite{Abramsky1995DT},
and \(\phi : D_i \to E_i\)~, \(\psi : E_i \to D_i\) in
\cite{Reynolds1998TPL}, but
usually left implicit).
\item[Updates.]
The \emph{update} \(\phi[\epsilon / \delta]\) of a function
\(\phi : D \to E\) maps (non-\(\bot\)) \(\delta : D\) to
\(\epsilon : E\), and all other elements \(\delta' : D\) to
\(\phi(\delta')\). The domain \(D\) has to be flat (with a continuous
equality); the same notation is used when \(D\) is a set.
\end{description}
\begin{AgdaSuppressSpace}
\begin{code}[hide]%
\>[0]\AgdaKeyword{module}\AgdaSpace{}%
%
\AgdaModule{Background.Denotational-Semantics}%
\AgdaSpace{}%
\AgdaKeyword{where}\<%
\end{code}
\end{AgdaSuppressSpace}

\subsection{Agda}\label{Background/Agda-agda}

\subsubsection{Features}\label{Background/Agda-features}

Agda
\cite{Agda-Language,Agda-Wikipedia}
is both a strongly typed functional \emph{programming language} with
support for first-class \emph{dependent types} and a \emph{proof
assistant} based on the Curry--Howard correspondence between
propositions and types. A number of features set Agda apart from a
typical functional programming language:

\begin{itemize}
\item
  Types in Agda are first-class values of a \emph{universe}
  \(\textsf{Set}_i\) where \(\textsf{Set} = \textsf{Set}_0\), and each
  universe \(\textsf{Set}_i\) belongs to the next universe
  \(\textsf{Set}_{i+1}\).
\item
  Agda has \emph{dependent function types} \AgdaRef{(x\ :\ A)\ →\ B} or
  \AgdaRef{∀\ x\ →\ B} where the type \AgdaRef{B} can depend on the value
  \AgdaRef{x}.
\item
  All functions in Agda are \emph{total}: evaluating a function call is
  guaranteed to return a value of the correct type in finite time. To
  ensure totality, Agda's type checker includes a termination check for
  recursive definitions, a positivity check for inductive datatypes, and
  a consistency check for universe levels.
\end{itemize}

Thanks to its dependent types and totality, Agda's type system can be
used as a higher-order logic for writing mathematical statements and
proofs, where types correspond to propositions and type checking
corresponds to checking the validity of the proof.

\subsubsection{Agda Syntax}\label{Background/Agda-agda-syntax}

The following syntactic constructs are used in our Agda code.

\begin{description}
\tightlist
\item[Comments.]
End-of-line comments are initiated by \AgdaRef{-\/-}, while multi-line
comments are between \AgdaRef{\{-} and \AgdaRef{-\}}.
\item[Naming.]
Declared symbols, variables, and type constructors all share a single
namespace. Names can be any sequence of non-whitespace ASCII and Unicode
characters except \AgdaRef{.;\{\}()@"}, excluding reserved symbols and
keywords such as \AgdaRef{→} or \AgdaRef{where}. Underscores in names play
a special role for mixfix notation.
\item[Mixfix notation.]
Agda supports \emph{mixfix notation} for defining operators, with
underscores in the name of a symbol indicating argument positions. For
example, if we declare \AgdaRef{\_+\_\ :\ Nat\ →\ Nat\ →\ Nat}, we can
use it as \AgdaRef{1\ +\ 1} (the spaces are required, since \AgdaRef{1+1}
is a valid Agda name!).
\item[Anonymous functions.]
Lambda abstractions use the syntax \AgdaRef{λ\ x\ →\ u} instead of
\AgdaRef{λ\ x.\ u}.
\item[Implicit arguments.]
Arguments marked by single curly braces \AgdaRef{\{…\}} in the type of a
symbol are considered to be \emph{implicit}. These arguments may be
omitted, and are then inferred by Agda's type checker.
\item[Type classes.]
Agda has no direct support for type classes, but they can be simulated
using Agda's \emph{instance arguments} to resolve type class instances.
Instance arguments are marked by double curly braces \AgdaRef{\{\{…\}\}}
and are resolved automatically by using definitions marked as
\AgdaRef{instance}.
\item[Rewrite rules.]
Agda has support for \emph{rewrite rules}
\cite{Cockx2020TTU}
that are applied automatically during type checking. Rewrite rules are
declared by marking an equality proof \AgdaRef{eq\ :\ a\ ≡\ b} with a
\AgdaRef{\{-\#\ REWRITE\ eq\ \#-\}} pragma. Agda can optionally check
confluence of rewrite rules, but currently does not check their
termination. Since only \emph{proven} (or postulated) equalities can be
added as rewrite rules, non-confluence and non-termination cannot affect
the soundness of Agda's type checker, only its
completeness~\cite{Cockx2021TRT}.
\end{description}
\begin{AgdaSuppressSpace}
\begin{code}[hide]%
\>[0]\AgdaKeyword{module}\AgdaSpace{}%
%
\AgdaModule{Background.Agda}%
\AgdaSpace{}%
\AgdaKeyword{where}\<%
\end{code}
\end{AgdaSuppressSpace}

\section{Postulated Domain Notation}\label{Notation-postulated-domain-notation}

This section postulates Agda notation for the domain constructors and
associated functions used in the
illustrative
examples (Section~\ref{Examples-illustrative-examples}).

 See the accompanying repository and the generated website
\cite{MFPS2026-Agda} for
the complete Agda code with the details elided here (including most
module imports, fixity declarations, and declarations of the types of
meta-variables). 

The Agda code that declares our embedding of domain notation is divided
into the modules imported below:
\begin{AgdaSuppressSpace}
\begin{code}[hide]%
\>[0]\AgdaSymbol{\{-\#}\AgdaSpace{}%
\AgdaKeyword{OPTIONS}\AgdaSpace{}%
\AgdaPragma{--rewriting}\AgdaSpace{}%
\AgdaPragma{--confluence-check}\AgdaSpace{}%
\AgdaPragma{--lossy-unification}\AgdaSpace{}%
\AgdaSymbol{\#-\}}\<%
\\
\>[0]\<%
\end{code}
\begin{code}%
\>[0]\AgdaKeyword{module}\AgdaSpace{}%
%
\AgdaModule{Notation}%
\AgdaSpace{}%
\AgdaKeyword{where}\<%
\\
\>[0][@{}l@{\AgdaIndent{0}}]%
\>[2]\AgdaKeyword{import}\AgdaSpace{}%
%
\AgdaModule{Notation.Domains}%
\<%
\\
\>[2]\AgdaKeyword{import}\AgdaSpace{}%
%
\AgdaModule{Notation.Functions}%
\<%
\\
\>[2]\AgdaKeyword{import}\AgdaSpace{}%
%
\AgdaModule{Notation.Flat}%
\<%
\\
\>[2]\AgdaKeyword{import}\AgdaSpace{}%
%
\AgdaModule{Notation.Flat.Booleans}%
\<%
\\
\>[2]\AgdaKeyword{import}\AgdaSpace{}%
%
\AgdaModule{Notation.Flat.Naturals}%
\<%
\\
\>[2]\AgdaKeyword{import}\AgdaSpace{}%
%
\AgdaModule{Notation.Sums}%
\<%
\\
\>[2]\AgdaKeyword{import}\AgdaSpace{}%
%
\AgdaModule{Notation.Products}%
\<%
\\
\>[2]\AgdaKeyword{import}\AgdaSpace{}%
%
\AgdaModule{Notation.Products.Tuples}%
\<%
\\
\>[2]\AgdaKeyword{import}\AgdaSpace{}%
%
\AgdaModule{Notation.Products.Sequences}%
\<%
\\
\>[2]\AgdaKeyword{import}\AgdaSpace{}%
%
\AgdaModule{Notation.Recursion}%
\<%
\\
\>[2]\AgdaKeyword{import}\AgdaSpace{}%
%
\AgdaModule{Notation.Updates}%
\<%
\end{code}
\end{AgdaSuppressSpace}
 In the PDF of this paper, Agda code blocks are highlighted, but
in general, names are \emph{not} hyperlinked to their declarations.
However, each declared or referenced \emph{module name} is hyperlinked
to the declaration of that module in the website, to facilitate browsing
the Agda code when online.

\begin{AgdaAlign}
\subsection{Domains}\label{Notation/Domains-domains}

Domains are embedded in Agda as elements of the type \AgdaRef{Domain}. A
domain \AgdaRef{D} is not itself a type, but it has a \emph{carrier} type
\AgdaRef{⟪\ D\ ⟫\ :\ Set}, which always contains an element
\AgdaRef{⊥\{D\}} (written \AgdaRef{⊥} when Agda can infer \AgdaRef{D}).
\begin{AgdaSuppressSpace}
\begin{code}%
\>[0]\AgdaKeyword{module}\AgdaSpace{}%
%
\AgdaModule{Notation.Domains}%
\AgdaSpace{}%
\AgdaKeyword{where}\<%
\\
\>[0][@{}l@{\AgdaIndent{0}}]%
\>[2]\AgdaKeyword{postulate}\<%
\\
\>[2][@{}l@{\AgdaIndent{0}}]%
\>[4]%
\AgdaPostulate{Domain}%
\AgdaSpace{}%
\AgdaSymbol{:}\AgdaSpace{}%
\AgdaPrimitive{Set}%
\>[30]\AgdaComment{--\ Domain\ is\ the\ type\ of\ all\ domains}\<%
\\
\>[4]\AgdaOperator{%
\AgdaPostulate{⟪\AgdaUnderscore{}%
⟫}}\AgdaSpace{}%
\AgdaSymbol{:}\AgdaSpace{}%
%
\AgdaPostulate{Domain}%
\AgdaSpace{}%
\AgdaSymbol{→}\AgdaSpace{}%
\AgdaPrimitive{Set}%
\>[30]\AgdaComment{--\ ⟪\ D\ ⟫\ is\ the\ carrier\ type\ of\ D}\<%
\\
\>[4]%
\AgdaPostulate{⊥}%
\AgdaSpace{}%
\AgdaSymbol{:}\AgdaSpace{}%
\AgdaSymbol{\{}\AgdaBound{D}\AgdaSpace{}%
\AgdaSymbol{:}\AgdaSpace{}%
%
\AgdaPostulate{Domain}%
\AgdaSymbol{\}}\AgdaSpace{}%
\AgdaSymbol{→}\AgdaSpace{}%
\AgdaOperator{%
\AgdaPostulate{⟪}%
}\AgdaSpace{}%
\AgdaBound{D}\AgdaSpace{}%
\AgdaOperator{%
\AgdaPostulate{⟫}%
}%
\>[30]\AgdaComment{--\ ⊥\{D\}\ is\ the\ 'bottom'\ element\ of\ D}\<%
\\
\>[4]%
\AgdaPostulate{𝟙}%
\AgdaSpace{}%
\AgdaSymbol{:}\AgdaSpace{}%
%
\AgdaPostulate{Domain}%
%
\>[30]\AgdaComment{--\ 𝟙\ is\ a\ unit\ domain}\<%
\\
\>[2]\AgdaKeyword{variable}\AgdaSpace{}%
\AgdaGeneralizable{A}\AgdaSpace{}%
\AgdaGeneralizable{B}\AgdaSpace{}%
\AgdaGeneralizable{C}\AgdaSpace{}%
\AgdaSymbol{:}\AgdaSpace{}%
\AgdaPrimitive{Set}\AgdaSymbol{;}\AgdaSpace{}%
\AgdaGeneralizable{D}\AgdaSpace{}%
\AgdaGeneralizable{E}\AgdaSpace{}%
\AgdaGeneralizable{F}\AgdaSpace{}%
\AgdaSymbol{:}\AgdaSpace{}%
%
\AgdaPostulate{Domain}%
\<%
\end{code}
\end{AgdaSuppressSpace}
Some previous papers on embedding denotational semantics in Agda
\cite{Mosses2025CDS,Mosses2025CSE,Mosses2025LAF}
defined domains to be types: \AgdaRef{Domain\ =\ Set}. However,
postulating \AgdaRef{⊥\ :\ D} for all \AgdaRef{D\ :\ Domain} was then
\emph{inconsistent} with the existence of an empty type in Agda.
Postulating \AgdaRef{Domain\ :\ Set} avoids that inconsistency.

The notation for domains postulated here supports \emph{type-checking}
embeddings of denotational semantics in Agda such as those in the
illustrative examples. It does \emph{not} define or constrain the
\emph{mathematical structure} of domains, nor the algebraic and
universal properties of the associated functions.

\subsection{Function Domains}\label{Notation/Functions-function-domains}

The conventional notation in denotational definitions for the domain of
continuous functions from \(D\) to \(E\) is \(D \to E\) or
\([D \to E]\). However, Agda reserves the notation \AgdaRef{D\ →\ E} for
the type of \emph{all} (total) functions from type \AgdaRef{D} to type
\AgdaRef{E}; here, we use the notation \AgdaRef{D\ →ᶜ\ E} for embedding
continuous function domains:
\begin{AgdaSuppressSpace}
\begin{code}[hide]%
\>[0]\AgdaSymbol{\{-\#}\AgdaSpace{}%
\AgdaKeyword{OPTIONS}\AgdaSpace{}%
\AgdaPragma{--rewriting}\AgdaSpace{}%
\AgdaPragma{--confluence-check}\AgdaSpace{}%
\AgdaSymbol{\#-\}}\<%
\end{code}
\begin{code}%
\>[0]\<%
\\
\>[0]\AgdaKeyword{module}\AgdaSpace{}%
%
\AgdaModule{Notation.Functions}%
\AgdaSpace{}%
\AgdaKeyword{where}\<%
\end{code}
\begin{code}[hide]%
\>[0][@{}l@{\AgdaIndent{1}}]%
\>[2]\AgdaKeyword{open}\AgdaSpace{}%
\AgdaKeyword{import}\AgdaSpace{}%
%
\AgdaModule{Notation.Domains}%
\<%
\\
\>[2]\AgdaKeyword{open}\AgdaSpace{}%
\AgdaKeyword{import}\AgdaSpace{}%
%
\AgdaModule{Agda.Builtin.Equality}%
\AgdaSpace{}%
\AgdaKeyword{public}\AgdaSpace{}%
\AgdaKeyword{using}\AgdaSpace{}%
\AgdaSymbol{(}\AgdaOperator{%
\AgdaDatatype{\AgdaUnderscore{}%
≡\AgdaUnderscore{}}}\AgdaSymbol{)}\<%
\\
\>[2]\AgdaKeyword{open}\AgdaSpace{}%
\AgdaKeyword{import}\AgdaSpace{}%
%
\AgdaModule{Agda.Builtin.Equality.Rewrite}%
\AgdaSpace{}%
\AgdaKeyword{using}\AgdaSpace{}%
\AgdaSymbol{()}\<%
\end{code}
\begin{code}%
\>[2]\AgdaKeyword{postulate}\AgdaSpace{}%
\AgdaOperator{%
\AgdaPostulate{\AgdaUnderscore{}%
→ᶜ\AgdaUnderscore{}}}\AgdaSpace{}%
\AgdaSymbol{:}\AgdaSpace{}%
%
\AgdaPostulate{Domain}%
\AgdaSpace{}%
\AgdaSymbol{→}\AgdaSpace{}%
%
\AgdaPostulate{Domain}%
\AgdaSpace{}%
\AgdaSymbol{→}\AgdaSpace{}%
%
\AgdaPostulate{Domain}%
\<%
\end{code}
\begin{code}[hide]%
\>[2]\AgdaComment{--\ D\ →ᶜ\ E\ is\ the\ domain\ of\ continuous\ functions\ from\ D\ to\ E}\<%
\\
\>[2]\AgdaKeyword{infixr}\AgdaSpace{}%
\AgdaNumber{0}\AgdaSpace{}%
\AgdaOperator{%
\AgdaPostulate{\AgdaUnderscore{}%
→ᶜ\AgdaUnderscore{}}}\<%
\end{code}
\end{AgdaSuppressSpace}
Both λ-abstraction and application preserve continuity. In conventional
denotational semantics, functions between domains are defined using
λ-abstraction and application from primitive continuous functions
associated with specific domain constructors, so they are
\emph{automatically} continuous. This motivates treating the carrier
\AgdaRef{⟪\ D\ →ᶜ\ E\ ⟫} of the embedding of a function domain as a type
of continuous functions. (\emph{Proving} functions defined in λ-notation
to be continuous in Agda requires pairing each λ-abstraction with an
explicit proof of its continuity, which is quite impractical --
especially when embedding denotations defined in continuation-passing
style.)

However, to support type-checking the \emph{direct} embedding of
λ-notation from conventional denotational definitions in Agda, it
appears to be necessary to \emph{rewrite} the carrier types of function
domains to ordinary function types:
\begin{AgdaSuppressSpace}
\begin{code}%
\>[2]\AgdaKeyword{postulate}\AgdaSpace{}%
%
\AgdaPostulate{dom-cts}%
\AgdaSpace{}%
\AgdaSymbol{:}\AgdaSpace{}%
\AgdaOperator{%
\AgdaPostulate{⟪}%
}\AgdaSpace{}%
\AgdaGeneralizable{D}\AgdaSpace{}%
\AgdaOperator{%
\AgdaPostulate{→ᶜ}%
}\AgdaSpace{}%
\AgdaGeneralizable{E}\AgdaSpace{}%
\AgdaOperator{%
\AgdaPostulate{⟫}%
}\AgdaSpace{}%
\AgdaOperator{%
\AgdaDatatype{≡}%
}\AgdaSpace{}%
\AgdaSymbol{(}\AgdaOperator{%
\AgdaPostulate{⟪}%
}\AgdaSpace{}%
\AgdaGeneralizable{D}\AgdaSpace{}%
\AgdaOperator{%
\AgdaPostulate{⟫}%
}\AgdaSpace{}%
\AgdaSymbol{→}\AgdaSpace{}%
\AgdaOperator{%
\AgdaPostulate{⟪}%
}\AgdaSpace{}%
\AgdaGeneralizable{E}\AgdaSpace{}%
\AgdaOperator{%
\AgdaPostulate{⟫}%
}\AgdaSymbol{)}\<%
\\
\>[2]\AgdaSymbol{\{-\#}\AgdaSpace{}%
\AgdaKeyword{REWRITE}\AgdaSpace{}%
%
\AgdaPostulate{dom-cts}%
\AgdaSpace{}%
\AgdaSymbol{\#-\}}\<%
\end{code}
\end{AgdaSuppressSpace}
Similarly, the notation \AgdaRef{A\ →ˢ\ D} is the embedding of the domain
of all functions from an ordinary type \AgdaRef{A} to a domain \AgdaRef{D}
(which are trivially continuous, ordered pointwise):
\begin{AgdaSuppressSpace}
\begin{code}%
\>[2]\AgdaKeyword{postulate}\AgdaSpace{}%
\AgdaOperator{%
\AgdaPostulate{\AgdaUnderscore{}%
→ˢ\AgdaUnderscore{}}}\AgdaSpace{}%
\AgdaSymbol{:}\AgdaSpace{}%
\AgdaPrimitive{Set}\AgdaSpace{}%
\AgdaSymbol{→}\AgdaSpace{}%
%
\AgdaPostulate{Domain}%
\AgdaSpace{}%
\AgdaSymbol{→}\AgdaSpace{}%
%
\AgdaPostulate{Domain}%
\<%
\end{code}
\begin{code}[hide]%
\>[2]\AgdaComment{--\ A\ →ˢ\ D\ is\ the\ domain\ of\ all\ functions\ from\ A\ to\ D}\<%
\\
\>[2]\AgdaKeyword{infixr}\AgdaSpace{}%
\AgdaNumber{0}\AgdaSpace{}%
\AgdaOperator{%
\AgdaPostulate{\AgdaUnderscore{}%
→ˢ\AgdaUnderscore{}}}\<%
\end{code}
\begin{code}%
\>[2]\AgdaKeyword{postulate}\AgdaSpace{}%
%
\AgdaPostulate{set-cts}%
%
\>[21]\AgdaSymbol{:}\AgdaSpace{}%
\AgdaOperator{%
\AgdaPostulate{⟪}%
}\AgdaSpace{}%
\AgdaGeneralizable{A}\AgdaSpace{}%
\AgdaOperator{%
\AgdaPostulate{→ˢ}%
}\AgdaSpace{}%
\AgdaGeneralizable{D}\AgdaSpace{}%
\AgdaOperator{%
\AgdaPostulate{⟫}%
}\AgdaSpace{}%
\AgdaOperator{%
\AgdaDatatype{≡}%
}\AgdaSpace{}%
\AgdaSymbol{(}\AgdaGeneralizable{A}\AgdaSpace{}%
\AgdaSymbol{→}\AgdaSpace{}%
\AgdaOperator{%
\AgdaPostulate{⟪}%
}\AgdaSpace{}%
\AgdaGeneralizable{D}\AgdaSpace{}%
\AgdaOperator{%
\AgdaPostulate{⟫}%
}\AgdaSymbol{)}\<%
\\
\>[2]\AgdaSymbol{\{-\#}\AgdaSpace{}%
\AgdaKeyword{REWRITE}\AgdaSpace{}%
%
\AgdaPostulate{set-cts}%
\AgdaSpace{}%
\AgdaSymbol{\#-\}}\<%
\end{code}
\end{AgdaSuppressSpace}
Embeddings of \emph{endofunctions} \AgdaRef{φ} on a domain \AgdaRef{D}
should always have fixed points \AgdaRef{fix\ φ}, with \AgdaRef{fix}
itself also being continuous:
\begin{AgdaSuppressSpace}
\begin{code}%
\>[2]\AgdaKeyword{postulate}\AgdaSpace{}%
%
\AgdaPostulate{fix}%
\AgdaSpace{}%
\AgdaSymbol{:}\AgdaSpace{}%
\AgdaOperator{%
\AgdaPostulate{⟪}%
}\AgdaSpace{}%
\AgdaSymbol{(}\AgdaGeneralizable{D}\AgdaSpace{}%
\AgdaOperator{%
\AgdaPostulate{→ᶜ}%
}\AgdaSpace{}%
\AgdaGeneralizable{D}\AgdaSymbol{)}\AgdaSpace{}%
\AgdaOperator{%
\AgdaPostulate{→ᶜ}%
}\AgdaSpace{}%
\AgdaGeneralizable{D}\AgdaSpace{}%
\AgdaOperator{%
\AgdaPostulate{⟫}%
}\<%
\end{code}
\end{AgdaSuppressSpace}

\subsection{Flat Domains}\label{Notation/Flat-flat-domains}

Lifting an ordinary set \(A\) by adding a \(\bot\) element gives a flat
domain, usually written \(A_\bot\). Our Agda embedding postulates a
corresponding domain constructor \AgdaRef{A\ +⊥}, together with a
function \AgdaRef{↑} for injecting elements of \AgdaRef{A} into
\AgdaRef{A\ +⊥}, and an operator \AgdaRef{f\ ♯} for extending functions on
\AgdaRef{A} to arguments in \AgdaRef{A\ +⊥}.
\begin{AgdaSuppressSpace}
\begin{code}[hide]%
\>[0]\AgdaSymbol{\{-\#}\AgdaSpace{}%
\AgdaKeyword{OPTIONS}\AgdaSpace{}%
\AgdaPragma{--rewriting}\AgdaSpace{}%
\AgdaPragma{--confluence-check}\AgdaSpace{}%
\AgdaSymbol{\#-\}}\<%
\\
\>[0]\<%
\end{code}
\begin{code}%
\>[0]\AgdaKeyword{module}\AgdaSpace{}%
%
\AgdaModule{Notation.Flat}%
\AgdaSpace{}%
\AgdaKeyword{where}\<%
\end{code}
\begin{code}[hide]%
\>[0]\<%
\\
\>[0][@{}l@{\AgdaIndent{1}}]%
\>[2]\AgdaKeyword{open}\AgdaSpace{}%
\AgdaKeyword{import}\AgdaSpace{}%
%
\AgdaModule{Notation.Domains}%
\<%
\\
\>[2]\AgdaKeyword{open}\AgdaSpace{}%
\AgdaKeyword{import}\AgdaSpace{}%
%
\AgdaModule{Notation.Functions}%
\<%
\\
\>[0]\<%
\end{code}
\begin{code}%
\>[0][@{}l@{\AgdaIndent{1}}]%
\>[2]\AgdaKeyword{postulate}\<%
\\
\>[2][@{}l@{\AgdaIndent{0}}]%
\>[4]\AgdaOperator{%
\AgdaPostulate{\AgdaUnderscore{}%
+⊥}}%
\>[9]\AgdaSymbol{:}\AgdaSpace{}%
\AgdaPrimitive{Set}\AgdaSpace{}%
\AgdaSymbol{→}\AgdaSpace{}%
%
\AgdaPostulate{Domain}%
%
\>[40]\AgdaComment{--\ A\ +⊥\ constructs\ a\ flat\ domain}\<%
\\
\>[4]%
\AgdaPostulate{↑}%
%
\>[9]\AgdaSymbol{:}\AgdaSpace{}%
\AgdaOperator{%
\AgdaPostulate{⟪}%
}\AgdaSpace{}%
\AgdaGeneralizable{A}\AgdaSpace{}%
\AgdaOperator{%
\AgdaPostulate{→ˢ}%
}\AgdaSpace{}%
\AgdaSymbol{(}\AgdaGeneralizable{A}\AgdaSpace{}%
\AgdaOperator{%
\AgdaPostulate{+⊥}%
}\AgdaSymbol{)}\AgdaSpace{}%
\AgdaOperator{%
\AgdaPostulate{⟫}%
}%
\>[40]\AgdaComment{--\ (↑\ a)\ injects\ a\ into\ A\ +⊥}\<%
\\
\>[4]\AgdaOperator{%
\AgdaPostulate{\AgdaUnderscore{}%
♯}}%
\>[9]\AgdaSymbol{:}\AgdaSpace{}%
\AgdaOperator{%
\AgdaPostulate{⟪}%
}\AgdaSpace{}%
\AgdaSymbol{(}\AgdaGeneralizable{A}\AgdaSpace{}%
\AgdaOperator{%
\AgdaPostulate{→ˢ}%
}\AgdaSpace{}%
\AgdaGeneralizable{D}\AgdaSymbol{)}\AgdaSpace{}%
\AgdaOperator{%
\AgdaPostulate{→ᶜ}%
}\AgdaSpace{}%
\AgdaSymbol{(}\AgdaGeneralizable{A}\AgdaSpace{}%
\AgdaOperator{%
\AgdaPostulate{+⊥}%
}\AgdaSymbol{)}\AgdaSpace{}%
\AgdaOperator{%
\AgdaPostulate{→ᶜ}%
}\AgdaSpace{}%
\AgdaGeneralizable{D}\AgdaSpace{}%
\AgdaOperator{%
\AgdaPostulate{⟫}%
}%
\>[40]\AgdaComment{--\ f\ ♯\ extends\ f\ to\ map\ ⊥\ to\ ⊥}\<%
\end{code}
\end{AgdaSuppressSpace}

\subsubsection{Booleans}\label{Notation/Flat/Booleans-booleans}

The McCarthy conditional operation \AgdaRef{β\ ⟶\ δ₁\ ,\ δ₂} extends the
usual ternary conditional choice to domains. It returns \AgdaRef{⊥}
whenever its first argument is \AgdaRef{⊥}.
\begin{AgdaSuppressSpace}
\begin{code}[hide]%
\>[0]\AgdaSymbol{\{-\#}\AgdaSpace{}%
\AgdaKeyword{OPTIONS}\AgdaSpace{}%
\AgdaPragma{--rewriting}\AgdaSpace{}%
\AgdaPragma{--confluence-check}\AgdaSpace{}%
\AgdaPragma{--lossy-unification}\AgdaSpace{}%
\AgdaSymbol{\#-\}}\<%
\\
\>[0]\<%
\end{code}
\begin{code}%
\>[0]\AgdaKeyword{module}\AgdaSpace{}%
%
\AgdaModule{Notation.Flat.Booleans}%
\AgdaSpace{}%
\AgdaKeyword{where}\<%
\end{code}
\begin{code}[hide]%
\>[0]\<%
\\
\>[0][@{}l@{\AgdaIndent{1}}]%
\>[2]\AgdaKeyword{open}\AgdaSpace{}%
\AgdaKeyword{import}\AgdaSpace{}%
%
\AgdaModule{Notation.Domains}%
\<%
\\
\>[2]\AgdaKeyword{open}\AgdaSpace{}%
\AgdaKeyword{import}\AgdaSpace{}%
%
\AgdaModule{Notation.Functions}%
\<%
\\
\>[2]\AgdaKeyword{open}\AgdaSpace{}%
\AgdaKeyword{import}\AgdaSpace{}%
%
\AgdaModule{Notation.Flat}%
\<%
\end{code}
\begin{code}%
\>[2]\AgdaKeyword{open}\AgdaSpace{}%
\AgdaKeyword{import}\AgdaSpace{}%
%
\AgdaModule{Data.Bool.Base}%
\AgdaSpace{}%
\AgdaKeyword{public}\AgdaSpace{}%
\AgdaKeyword{using}\AgdaSpace{}%
\AgdaSymbol{(}%
\AgdaDatatype{Bool}%
\AgdaSymbol{;}\AgdaSpace{}%
%
\AgdaInductiveConstructor{false}%
\AgdaSymbol{;}\AgdaSpace{}%
%
\AgdaInductiveConstructor{true}%
\AgdaSymbol{;}\AgdaSpace{}%
\AgdaOperator{%
\AgdaFunction{if\AgdaUnderscore{}%
then\AgdaUnderscore{}else\AgdaUnderscore{}}}\AgdaSymbol{;}\AgdaSpace{}%
%
\AgdaFunction{not}%
\AgdaSymbol{)}\<%
\\
\\[\AgdaEmptyExtraSkip]%
\>[2]%
\AgdaFunction{Bool⊥}%
\AgdaSpace{}%
\AgdaSymbol{=}\AgdaSpace{}%
%
\AgdaDatatype{Bool}%
\AgdaSpace{}%
\AgdaOperator{%
\AgdaPostulate{+⊥}%
}\<%
\\
\>[2]\AgdaOperator{%
\AgdaFunction{\AgdaUnderscore{}%
⟶\AgdaUnderscore{},\AgdaUnderscore{}}}\AgdaSpace{}%
\AgdaSymbol{:}\AgdaSpace{}%
\AgdaOperator{%
\AgdaPostulate{⟪}%
}\AgdaSpace{}%
%
\AgdaFunction{Bool⊥}%
\AgdaSpace{}%
\AgdaOperator{%
\AgdaPostulate{→ᶜ}%
}\AgdaSpace{}%
\AgdaGeneralizable{D}\AgdaSpace{}%
\AgdaOperator{%
\AgdaPostulate{→ᶜ}%
}\AgdaSpace{}%
\AgdaGeneralizable{D}\AgdaSpace{}%
\AgdaOperator{%
\AgdaPostulate{→ᶜ}%
}\AgdaSpace{}%
\AgdaGeneralizable{D}\AgdaSpace{}%
\AgdaOperator{%
\AgdaPostulate{⟫}%
}%
\>[38]\AgdaComment{--\ β\ ⟶\ δ₁\ ,\ δ₂\ is\ conditional\ choice}\<%
\\
\>[2]\AgdaOperator{%
\AgdaFunction{\AgdaUnderscore{}%
⟶\AgdaUnderscore{},\AgdaUnderscore{}}}\AgdaSpace{}%
\AgdaSymbol{=}\AgdaSpace{}%
\AgdaSymbol{(λ}\AgdaSpace{}%
\AgdaBound{b}\AgdaSpace{}%
\AgdaBound{δ₁}\AgdaSpace{}%
\AgdaBound{δ₂}\AgdaSpace{}%
\AgdaSymbol{→}\AgdaSpace{}%
\AgdaOperator{%
\AgdaFunction{if}%
}\AgdaSpace{}%
\AgdaBound{b}\AgdaSpace{}%
\AgdaOperator{%
\AgdaFunction{then}%
}\AgdaSpace{}%
\AgdaBound{δ₁}\AgdaSpace{}%
\AgdaOperator{%
\AgdaFunction{else}%
}\AgdaSpace{}%
\AgdaBound{δ₂}\AgdaSymbol{)}\AgdaSpace{}%
\AgdaOperator{%
\AgdaPostulate{♯}%
}\<%
\end{code}
\begin{code}[hide]%
\>[2]\AgdaKeyword{infixr}\AgdaSpace{}%
\AgdaNumber{20}\AgdaSpace{}%
\AgdaOperator{%
\AgdaFunction{\AgdaUnderscore{}%
⟶\AgdaUnderscore{},\AgdaUnderscore{}}}\<%
\end{code}
\end{AgdaSuppressSpace}
This module also defines \AgdaRef{Eq\ A} for use as an instance
parameter, restricting operation definitions to types \AgdaRef{A} such
that \AgdaRef{\_==\_\ :\ A\ →\ A\ →\ Bool}:
\begin{AgdaSuppressSpace}
\begin{code}%
\>[2]\AgdaKeyword{record}\AgdaSpace{}%
%
\AgdaRecord{Eq}%
\AgdaSpace{}%
\AgdaSymbol{(}\AgdaBound{A}\AgdaSpace{}%
\AgdaSymbol{:}\AgdaSpace{}%
\AgdaPrimitive{Set}\AgdaSymbol{)}\AgdaSpace{}%
\AgdaSymbol{:}\AgdaSpace{}%
\AgdaPrimitive{Set}\AgdaSpace{}%
\AgdaKeyword{where}\AgdaSpace{}%
\AgdaKeyword{field}\AgdaSpace{}%
\AgdaOperator{%
\AgdaField{\AgdaUnderscore{}%
==\AgdaUnderscore{}}}\AgdaSpace{}%
\AgdaSymbol{:}\AgdaSpace{}%
\AgdaBound{A}\AgdaSpace{}%
\AgdaSymbol{→}\AgdaSpace{}%
\AgdaBound{A}\AgdaSpace{}%
\AgdaSymbol{→}\AgdaSpace{}%
%
\AgdaDatatype{Bool}%
\<%
\\
\>[2]\AgdaKeyword{open}\AgdaSpace{}%
%
\AgdaRecord{Eq}%
\AgdaSpace{}%
\AgdaSymbol{\{\{...\}\}}\AgdaSpace{}%
\AgdaKeyword{public}\<%
\end{code}
\end{AgdaSuppressSpace}
The \AgdaRef{Bool⊥}-valued operation \AgdaRef{δ₁\ ==⊥\ δ₂} on
\AgdaRef{A\ +⊥} gives \AgdaRef{⊥} when either operand is \AgdaRef{⊥}:
\begin{AgdaSuppressSpace}
\begin{code}%
\>[2]\AgdaOperator{%
\AgdaFunction{\AgdaUnderscore{}%
==⊥\AgdaUnderscore{}}}\AgdaSpace{}%
\AgdaSymbol{:}\AgdaSpace{}%
\AgdaSymbol{\{\{}%
\AgdaRecord{Eq}%
\AgdaSpace{}%
\AgdaGeneralizable{A}\AgdaSymbol{\}\}}\AgdaSpace{}%
\AgdaSymbol{→}\AgdaSpace{}%
\AgdaOperator{%
\AgdaPostulate{⟪}%
}\AgdaSpace{}%
\AgdaSymbol{(}\AgdaGeneralizable{A}\AgdaSpace{}%
\AgdaOperator{%
\AgdaPostulate{+⊥}%
}\AgdaSymbol{)}\AgdaSpace{}%
\AgdaOperator{%
\AgdaPostulate{→ᶜ}%
}\AgdaSpace{}%
\AgdaSymbol{(}\AgdaGeneralizable{A}\AgdaSpace{}%
\AgdaOperator{%
\AgdaPostulate{+⊥}%
}\AgdaSymbol{)}\AgdaSpace{}%
\AgdaOperator{%
\AgdaPostulate{→ᶜ}%
}\AgdaSpace{}%
%
\AgdaFunction{Bool⊥}%
\AgdaSpace{}%
\AgdaOperator{%
\AgdaPostulate{⟫}%
}\<%
\\
\>[2]\AgdaBound{δ₁}\AgdaSpace{}%
\AgdaOperator{%
\AgdaFunction{==⊥}%
}\AgdaSpace{}%
\AgdaBound{δ₂}\AgdaSpace{}%
\AgdaSymbol{=}\AgdaSpace{}%
\AgdaSymbol{((λ}\AgdaSpace{}%
\AgdaBound{a₁}\AgdaSpace{}%
\AgdaSymbol{→}\AgdaSpace{}%
\AgdaSymbol{(((λ}\AgdaSpace{}%
\AgdaBound{a₂}\AgdaSpace{}%
\AgdaSymbol{→}\AgdaSpace{}%
%
\AgdaPostulate{↑}%
\AgdaSpace{}%
\AgdaSymbol{(}\AgdaBound{a₁}\AgdaSpace{}%
\AgdaOperator{%
\AgdaField{==}%
}\AgdaSpace{}%
\AgdaBound{a₂}\AgdaSymbol{))}\AgdaSpace{}%
\AgdaOperator{%
\AgdaPostulate{♯}%
}\AgdaSymbol{)}\AgdaSpace{}%
\AgdaBound{δ₂}\AgdaSymbol{))}\AgdaSpace{}%
\AgdaOperator{%
\AgdaPostulate{♯}%
}\AgdaSymbol{)}\AgdaSpace{}%
\AgdaBound{δ₁}\<%
\\
\\[\AgdaEmptyExtraSkip]%
\>[2]\AgdaKeyword{instance}\<%
\\
\>[2][@{}l@{\AgdaIndent{0}}]%
\>[4]%
\AgdaFunction{eqBool}%
\AgdaSpace{}%
\AgdaSymbol{:}\AgdaSpace{}%
%
\AgdaRecord{Eq}%
\AgdaSpace{}%
%
\AgdaDatatype{Bool}%
\<%
\\
\>[4]\AgdaOperator{%
\AgdaField{\AgdaUnderscore{}%
==\AgdaUnderscore{}}}\AgdaSpace{}%
\AgdaSymbol{\{\{}%
\AgdaFunction{eqBool}%
\AgdaSymbol{\}\}}\AgdaSpace{}%
\AgdaBound{b₁}\AgdaSpace{}%
\AgdaBound{b₂}\AgdaSpace{}%
\AgdaSymbol{=}\AgdaSpace{}%
\AgdaOperator{%
\AgdaFunction{if}%
}\AgdaSpace{}%
\AgdaBound{b₁}\AgdaSpace{}%
\AgdaOperator{%
\AgdaFunction{then}%
}\AgdaSpace{}%
\AgdaBound{b₂}\AgdaSpace{}%
\AgdaOperator{%
\AgdaFunction{else}%
}\AgdaSpace{}%
%
\AgdaFunction{not}%
\AgdaSpace{}%
\AgdaBound{b₂}\<%
\end{code}
\end{AgdaSuppressSpace}

\subsubsection{Naturals}\label{Notation/Flat/Naturals-naturals}

Agda allows decimal notation for natural numbers, as well as unary
notation using \AgdaRef{zero} and \AgdaRef{suc}.
\begin{AgdaSuppressSpace}
\begin{code}[hide]%
\>[0]\AgdaSymbol{\{-\#}\AgdaSpace{}%
\AgdaKeyword{OPTIONS}\AgdaSpace{}%
\AgdaPragma{--rewriting}\AgdaSpace{}%
\AgdaPragma{--confluence-check}\AgdaSpace{}%
\AgdaSymbol{\#-\}}\<%
\\
\>[0]\<%
\end{code}
\begin{code}%
\>[0]\AgdaKeyword{module}\AgdaSpace{}%
%
\AgdaModule{Notation.Flat.Naturals}%
\AgdaSpace{}%
\AgdaKeyword{where}\<%
\end{code}
\begin{code}[hide]%
\>[0]\<%
\\
\>[0][@{}l@{\AgdaIndent{1}}]%
\>[2]\AgdaKeyword{open}\AgdaSpace{}%
\AgdaKeyword{import}\AgdaSpace{}%
%
\AgdaModule{Notation.Flat}%
\<%
\\
\>[2]\AgdaKeyword{open}\AgdaSpace{}%
\AgdaKeyword{import}\AgdaSpace{}%
%
\AgdaModule{Notation.Flat.Booleans}%
\<%
\end{code}
\begin{code}%
\>[2]\AgdaKeyword{open}\AgdaSpace{}%
\AgdaKeyword{import}\AgdaSpace{}%
%
\AgdaModule{Agda.Builtin.Nat}%
\AgdaSpace{}%
\AgdaKeyword{public}%
\>[39]\AgdaKeyword{using}\AgdaSpace{}%
\AgdaSymbol{(}%
\AgdaDatatype{Nat}%
\AgdaSymbol{;}\AgdaSpace{}%
%
\AgdaInductiveConstructor{suc}%
\AgdaSymbol{;}\AgdaSpace{}%
\AgdaOperator{\AgdaPrimitive{\AgdaUnderscore{}+\AgdaUnderscore{}}}\AgdaSymbol{;}\AgdaSpace{}%
\AgdaOperator{\AgdaPrimitive{\AgdaUnderscore{}-\AgdaUnderscore{}}}\AgdaSymbol{)}\<%
\\
\>[39]\AgdaKeyword{renaming}\AgdaSpace{}%
\AgdaSymbol{(}\AgdaOperator{\AgdaPrimitive{\AgdaUnderscore{}==\AgdaUnderscore{}}}\AgdaSpace{}%
\AgdaSymbol{to}\AgdaSpace{}%
\AgdaOperator{\AgdaPrimitive{\AgdaUnderscore{}==ᴺ\AgdaUnderscore{}}}\AgdaSymbol{)}\<%
\\
\>[2]%
\AgdaFunction{Nat⊥}%
\AgdaSpace{}%
\AgdaSymbol{=}\AgdaSpace{}%
%
\AgdaDatatype{Nat}%
\AgdaSpace{}%
\AgdaOperator{%
\AgdaPostulate{+⊥}%
}\<%
\\
\>[2]\AgdaKeyword{instance}\<%
\\
\>[2][@{}l@{\AgdaIndent{0}}]%
\>[4]%
\AgdaFunction{eqNat}%
\AgdaSpace{}%
\AgdaSymbol{:}\AgdaSpace{}%
%
\AgdaRecord{Eq}%
\AgdaSpace{}%
%
\AgdaDatatype{Nat}%
\<%
\\
\>[4]\AgdaOperator{%
\AgdaField{\AgdaUnderscore{}%
==\AgdaUnderscore{}}}\AgdaSpace{}%
\AgdaSymbol{\{\{}%
\AgdaFunction{eqNat}%
\AgdaSymbol{\}\}}\AgdaSpace{}%
\AgdaSymbol{=}\AgdaSpace{}%
\AgdaOperator{\AgdaPrimitive{\AgdaUnderscore{}==ᴺ\AgdaUnderscore{}}}\<%
\end{code}
\end{AgdaSuppressSpace}

\subsection{Sum Domains}\label{Notation/Sums-sum-domains}

The separated sum \AgdaRef{D\ +\ E} of two domains corresponds to lifting
the disjoint union of their carrier sets. The following operations can
be used directly for binary sums, and iterated for domains with more
than two summands.
\begin{AgdaSuppressSpace}
\begin{code}[hide]%
\>[0]\AgdaSymbol{\{-\#}\AgdaSpace{}%
\AgdaKeyword{OPTIONS}\AgdaSpace{}%
\AgdaPragma{--rewriting}\AgdaSpace{}%
\AgdaPragma{--confluence-check}\AgdaSpace{}%
\AgdaPragma{--lossy-unification}\AgdaSpace{}%
\AgdaSymbol{\#-\}}\<%
\\
\>[0]\<%
\end{code}
\begin{code}%
\>[0]\AgdaKeyword{module}\AgdaSpace{}%
%
\AgdaModule{Notation.Sums}%
\AgdaSpace{}%
\AgdaKeyword{where}\<%
\end{code}
\begin{code}[hide]%
\>[0]\<%
\\
\>[0][@{}l@{\AgdaIndent{1}}]%
\>[2]\AgdaKeyword{open}\AgdaSpace{}%
\AgdaKeyword{import}\AgdaSpace{}%
%
\AgdaModule{Notation.Domains}%
\<%
\\
\>[2]\AgdaKeyword{open}\AgdaSpace{}%
\AgdaKeyword{import}\AgdaSpace{}%
%
\AgdaModule{Notation.Functions}%
\<%
\\
\>[2]\AgdaKeyword{open}\AgdaSpace{}%
\AgdaKeyword{import}\AgdaSpace{}%
%
\AgdaModule{Notation.Flat.Booleans}%
\<%
\\
\>[0]\<%
\end{code}
\begin{code}%
\>[0][@{}l@{\AgdaIndent{1}}]%
\>[2]\AgdaKeyword{postulate}\<%
\\
\>[2][@{}l@{\AgdaIndent{0}}]%
\>[4]\AgdaOperator{%
\AgdaPostulate{\AgdaUnderscore{}%
+\AgdaUnderscore{}}}%
\>[11]\AgdaSymbol{:}\AgdaSpace{}%
%
\AgdaPostulate{Domain}%
\AgdaSpace{}%
\AgdaSymbol{→}\AgdaSpace{}%
%
\AgdaPostulate{Domain}%
\AgdaSpace{}%
\AgdaSymbol{→}\AgdaSpace{}%
%
\AgdaPostulate{Domain}%
%
\>[40]\AgdaComment{--\ D\ +\ E\ is\ separated\ sum}\<%
\\
\>[4]%
\AgdaPostulate{inj₁}%
%
\>[11]\AgdaSymbol{:}\AgdaSpace{}%
\AgdaOperator{%
\AgdaPostulate{⟪}%
}\AgdaSpace{}%
\AgdaGeneralizable{D}\AgdaSpace{}%
\AgdaOperator{%
\AgdaPostulate{→ᶜ}%
}\AgdaSpace{}%
\AgdaSymbol{(}\AgdaGeneralizable{D}\AgdaSpace{}%
\AgdaOperator{%
\AgdaPostulate{+}%
}\AgdaSpace{}%
\AgdaGeneralizable{E}\AgdaSymbol{)}\AgdaSpace{}%
\AgdaOperator{%
\AgdaPostulate{⟫}%
}%
\>[40]\AgdaComment{--\ inj₁\ δ\ is\ injection\ from\ D}\<%
\\
\>[4]%
\AgdaPostulate{inj₂}%
%
\>[11]\AgdaSymbol{:}\AgdaSpace{}%
\AgdaOperator{%
\AgdaPostulate{⟪}%
}\AgdaSpace{}%
\AgdaGeneralizable{E}\AgdaSpace{}%
\AgdaOperator{%
\AgdaPostulate{→ᶜ}%
}\AgdaSpace{}%
\AgdaSymbol{(}\AgdaGeneralizable{D}\AgdaSpace{}%
\AgdaOperator{%
\AgdaPostulate{+}%
}\AgdaSpace{}%
\AgdaGeneralizable{E}\AgdaSymbol{)}\AgdaSpace{}%
\AgdaOperator{%
\AgdaPostulate{⟫}%
}%
\>[40]\AgdaComment{--\ inj₂\ ε\ is\ injection\ from\ E}\<%
\\
\>[4]\AgdaOperator{%
\AgdaPostulate{[\AgdaUnderscore{}%
,\AgdaUnderscore{}]}}%
\>[11]\AgdaSymbol{:}\AgdaSpace{}%
\AgdaOperator{%
\AgdaPostulate{⟪}%
}\AgdaSpace{}%
\AgdaSymbol{(}\AgdaGeneralizable{D}\AgdaSpace{}%
\AgdaOperator{%
\AgdaPostulate{→ᶜ}%
}\AgdaSpace{}%
\AgdaGeneralizable{F}\AgdaSymbol{)}\AgdaSpace{}%
\AgdaOperator{%
\AgdaPostulate{→ᶜ}%
}\AgdaSpace{}%
\AgdaSymbol{(}\AgdaGeneralizable{E}\AgdaSpace{}%
\AgdaOperator{%
\AgdaPostulate{→ᶜ}%
}\AgdaSpace{}%
\AgdaGeneralizable{F}\AgdaSymbol{)}\AgdaSpace{}%
\AgdaOperator{%
\AgdaPostulate{→ᶜ}%
}\AgdaSpace{}%
\AgdaSymbol{((}\AgdaGeneralizable{D}\AgdaSpace{}%
\AgdaOperator{%
\AgdaPostulate{+}%
}\AgdaSpace{}%
\AgdaGeneralizable{E}\AgdaSymbol{)}\AgdaSpace{}%
\AgdaOperator{%
\AgdaPostulate{→ᶜ}%
}\AgdaSpace{}%
\AgdaGeneralizable{F}\AgdaSymbol{)}\AgdaSpace{}%
\AgdaOperator{%
\AgdaPostulate{⟫}%
}\<%
\\
\>[4]\AgdaComment{--\ [\ φ\ ,\ ψ\ ]\ applies\ φ\ to\ arguments\ in\ D,\ and\ ψ\ to\ arguments\ in\ E}\<%
\end{code}
\end{AgdaSuppressSpace}
Conventional denotational definitions of programming languages (e.g.,
the \emph{Scheme} semantics
\cite{Scheme}) use domain names instead
of numerical indices in operations associated with separated sums. The
inherently \emph{dependent} types of the Agda embedding of these
operations are as follows.
\begin{AgdaSuppressSpace}
\begin{code}%
\>[2]\AgdaKeyword{open}\AgdaSpace{}%
\AgdaKeyword{import}\AgdaSpace{}%
%
\AgdaModule{Agda.Builtin.Nat}%
\AgdaSpace{}%
\AgdaKeyword{using}\AgdaSpace{}%
\AgdaSymbol{(}%
\AgdaDatatype{Nat}%
\AgdaSymbol{)}\<%
\\
\>[2]\AgdaKeyword{variable}\AgdaSpace{}%
\AgdaGeneralizable{n}\AgdaSpace{}%
\AgdaSymbol{:}\AgdaSpace{}%
%
\AgdaDatatype{Nat}%
\<%
\\
\>[2]\AgdaKeyword{postulate}\<%
\\
\>[2][@{}l@{\AgdaIndent{0}}]%
\>[4]\AgdaOperator{%
\AgdaPostulate{\AgdaUnderscore{}%
≳\AgdaUnderscore{}↦\AgdaUnderscore{}}}%
\>[11]\AgdaSymbol{:}\AgdaSpace{}%
%
\AgdaPostulate{Domain}%
\AgdaSpace{}%
\AgdaSymbol{→}\AgdaSpace{}%
%
\AgdaDatatype{Nat}%
\AgdaSpace{}%
\AgdaSymbol{→}\AgdaSpace{}%
%
\AgdaPostulate{Domain}%
\AgdaSpace{}%
\AgdaSymbol{→}\AgdaSpace{}%
\AgdaPrimitive{Set}\<%
\\
\>[4]\AgdaOperator{%
\AgdaPostulate{\AgdaUnderscore{}%
in⊥\AgdaUnderscore{}}}%
\>[11]\AgdaSymbol{:}\AgdaSpace{}%
\AgdaOperator{%
\AgdaPostulate{⟪}%
}\AgdaSpace{}%
\AgdaGeneralizable{D}\AgdaSpace{}%
\AgdaOperator{%
\AgdaPostulate{⟫}%
}\AgdaSpace{}%
\AgdaSymbol{→}\AgdaSpace{}%
\AgdaSymbol{(}\AgdaBound{E}\AgdaSpace{}%
\AgdaSymbol{:}\AgdaSpace{}%
%
\AgdaPostulate{Domain}%
\AgdaSymbol{)}\AgdaSpace{}%
\AgdaSymbol{→}\AgdaSpace{}%
\AgdaSymbol{\{\{}\AgdaBound{E}\AgdaSpace{}%
\AgdaOperator{%
\AgdaPostulate{≳}%
}\AgdaSpace{}%
\AgdaGeneralizable{n}\AgdaSpace{}%
\AgdaOperator{%
\AgdaPostulate{↦}%
}\AgdaSpace{}%
\AgdaGeneralizable{D}\AgdaSymbol{\}\}}\AgdaSpace{}%
\AgdaSymbol{→}\AgdaSpace{}%
\AgdaOperator{%
\AgdaPostulate{⟪}%
}\AgdaSpace{}%
\AgdaBound{E}\AgdaSpace{}%
\AgdaOperator{%
\AgdaPostulate{⟫}%
}%
\>[63]\AgdaComment{--\ δ\ in⊥\ E\ injection}\<%
\\
\>[4]\AgdaOperator{%
\AgdaPostulate{\AgdaUnderscore{}%
|⊥\AgdaUnderscore{}}}%
\>[11]\AgdaSymbol{:}\AgdaSpace{}%
\AgdaOperator{%
\AgdaPostulate{⟪}%
}\AgdaSpace{}%
\AgdaGeneralizable{E}\AgdaSpace{}%
\AgdaOperator{%
\AgdaPostulate{⟫}%
}\AgdaSpace{}%
\AgdaSymbol{→}\AgdaSpace{}%
\AgdaSymbol{(}\AgdaBound{D}\AgdaSpace{}%
\AgdaSymbol{:}\AgdaSpace{}%
%
\AgdaPostulate{Domain}%
\AgdaSymbol{)}\AgdaSpace{}%
\AgdaSymbol{→}\AgdaSpace{}%
\AgdaSymbol{\{\{}\AgdaGeneralizable{E}\AgdaSpace{}%
\AgdaOperator{%
\AgdaPostulate{≳}%
}\AgdaSpace{}%
\AgdaGeneralizable{n}\AgdaSpace{}%
\AgdaOperator{%
\AgdaPostulate{↦}%
}\AgdaSpace{}%
\AgdaBound{D}\AgdaSymbol{\}\}}\AgdaSpace{}%
\AgdaSymbol{→}\AgdaSpace{}%
\AgdaOperator{%
\AgdaPostulate{⟪}%
}\AgdaSpace{}%
\AgdaBound{D}\AgdaSpace{}%
\AgdaOperator{%
\AgdaPostulate{⟫}%
}%
\>[63]\AgdaComment{--\ ε\ |⊥\ D\ \ projection}\<%
\\
\>[4]\AgdaOperator{%
\AgdaPostulate{\AgdaUnderscore{}%
∈⊥\AgdaUnderscore{}}}%
\>[11]\AgdaSymbol{:}\AgdaSpace{}%
\AgdaOperator{%
\AgdaPostulate{⟪}%
}\AgdaSpace{}%
\AgdaGeneralizable{E}\AgdaSpace{}%
\AgdaOperator{%
\AgdaPostulate{⟫}%
}\AgdaSpace{}%
\AgdaSymbol{→}\AgdaSpace{}%
\AgdaSymbol{(}\AgdaBound{D}\AgdaSpace{}%
\AgdaSymbol{:}\AgdaSpace{}%
%
\AgdaPostulate{Domain}%
\AgdaSymbol{)}\AgdaSpace{}%
\AgdaSymbol{→}\AgdaSpace{}%
\AgdaSymbol{\{\{}\AgdaGeneralizable{E}\AgdaSpace{}%
\AgdaOperator{%
\AgdaPostulate{≳}%
}\AgdaSpace{}%
\AgdaGeneralizable{n}\AgdaSpace{}%
\AgdaOperator{%
\AgdaPostulate{↦}%
}\AgdaSpace{}%
\AgdaBound{D}\AgdaSymbol{\}\}}\AgdaSpace{}%
\AgdaSymbol{→}\AgdaSpace{}%
\AgdaOperator{%
\AgdaPostulate{⟪}%
}\AgdaSpace{}%
%
\AgdaFunction{Bool⊥}%
\AgdaSpace{}%
\AgdaOperator{%
\AgdaPostulate{⟫}%
}%
\>[63]\AgdaComment{--\ ε\ ∈⊥\ D\ \ inspection}\<%
\end{code}
\end{AgdaSuppressSpace}
The operations are defined only for \AgdaRef{D} and \AgdaRef{E} where an
instance of type \AgdaRef{E\ ≳\ n\ ↦\ D} is declared for some~\AgdaRef{n}.
Instead of defining the summands \AgdaRef{D} of a separated sum domain
\AgdaRef{E} by an equation \AgdaRef{E\ =\ ...\ +\ D\ +\ ...}, the domain
\AgdaRef{E} is merely \emph{postulated}, and each summand is declared
separately by \AgdaRef{instance\ \_\ :\ E\ ≳\ n\ ↦\ D} (where~\AgdaRef{n}
should be a different natural number for each summand, but is otherwise
arbitrary).

\subsection{Product Domains}\label{Notation/Products-product-domains}

The carrier of the binary product \AgdaRef{D\ ×\ E} of two domains
consists of all pairs \AgdaRef{(d\ ,\ e)} of elements of \AgdaRef{D} and
\AgdaRef{E} with the pair \AgdaRef{(⊥\{D\}\ ,\ ⊥\{E\})} as the bottom
element \AgdaRef{⊥\{D\ ×\ E\}}. The following operations can be used
directly for binary products, and iterated for products of more than two
domains.
\begin{AgdaSuppressSpace}
\begin{code}[hide]%
\>[0]\AgdaSymbol{\{-\#}\AgdaSpace{}%
\AgdaKeyword{OPTIONS}\AgdaSpace{}%
\AgdaPragma{--rewriting}\AgdaSpace{}%
\AgdaPragma{--confluence-check}\AgdaSpace{}%
\AgdaSymbol{\#-\}}\<%
\\
\>[0]\<%
\end{code}
\begin{code}%
\>[0]\AgdaKeyword{module}\AgdaSpace{}%
%
\AgdaModule{Notation.Products}%
\AgdaSpace{}%
\AgdaKeyword{where}\<%
\end{code}
\begin{code}[hide]%
\>[0]\<%
\\
\>[0][@{}l@{\AgdaIndent{1}}]%
\>[2]\AgdaKeyword{open}\AgdaSpace{}%
\AgdaKeyword{import}\AgdaSpace{}%
%
\AgdaModule{Notation.Domains}%
\<%
\\
\>[2]\AgdaKeyword{open}\AgdaSpace{}%
\AgdaKeyword{import}\AgdaSpace{}%
%
\AgdaModule{Notation.Functions}%
\<%
\\
\>[0]\<%
\end{code}
\begin{code}%
\>[0][@{}l@{\AgdaIndent{1}}]%
\>[2]\AgdaKeyword{postulate}\<%
\\
\>[2][@{}l@{\AgdaIndent{0}}]%
\>[4]\AgdaOperator{%
\AgdaPostulate{\AgdaUnderscore{}%
×\AgdaUnderscore{}}}%
\>[9]\AgdaSymbol{:}\AgdaSpace{}%
%
\AgdaPostulate{Domain}%
\AgdaSpace{}%
\AgdaSymbol{→}\AgdaSpace{}%
%
\AgdaPostulate{Domain}%
\AgdaSpace{}%
\AgdaSymbol{→}\AgdaSpace{}%
%
\AgdaPostulate{Domain}%
%
\>[37]\AgdaComment{--\ D\ ×\ E\ is\ the\ categorical\ product}\<%
\\
\>[4]\AgdaOperator{%
\AgdaPostulate{\AgdaUnderscore{}%
,\AgdaUnderscore{}}}%
\>[9]\AgdaSymbol{:}\AgdaSpace{}%
\AgdaOperator{%
\AgdaPostulate{⟪}%
}\AgdaSpace{}%
\AgdaGeneralizable{D}\AgdaSpace{}%
\AgdaOperator{%
\AgdaPostulate{→ᶜ}%
}\AgdaSpace{}%
\AgdaGeneralizable{E}\AgdaSpace{}%
\AgdaOperator{%
\AgdaPostulate{→ᶜ}%
}\AgdaSpace{}%
\AgdaSymbol{(}\AgdaGeneralizable{D}\AgdaSpace{}%
\AgdaOperator{%
\AgdaPostulate{×}%
}\AgdaSpace{}%
\AgdaGeneralizable{E}\AgdaSymbol{)}\AgdaSpace{}%
\AgdaOperator{%
\AgdaPostulate{⟫}%
}%
\>[37]\AgdaComment{--\ (δ\ ,\ ε)\ is\ a\ pair\ of\ elements}\<%
\\
\>[4]\AgdaOperator{%
\AgdaPostulate{\AgdaUnderscore{}%
↓₁}}%
\>[9]\AgdaSymbol{:}\AgdaSpace{}%
\AgdaOperator{%
\AgdaPostulate{⟪}%
}\AgdaSpace{}%
\AgdaSymbol{(}\AgdaGeneralizable{D}\AgdaSpace{}%
\AgdaOperator{%
\AgdaPostulate{×}%
}\AgdaSpace{}%
\AgdaGeneralizable{E}\AgdaSymbol{)}\AgdaSpace{}%
\AgdaOperator{%
\AgdaPostulate{→ᶜ}%
}\AgdaSpace{}%
\AgdaGeneralizable{D}\AgdaSpace{}%
\AgdaOperator{%
\AgdaPostulate{⟫}%
}%
\>[37]\AgdaComment{--\ (δ\ ,\ ε)↓₁\ is\ δ}\<%
\\
\>[4]\AgdaOperator{%
\AgdaPostulate{\AgdaUnderscore{}%
↓₂}}%
\>[9]\AgdaSymbol{:}\AgdaSpace{}%
\AgdaOperator{%
\AgdaPostulate{⟪}%
}\AgdaSpace{}%
\AgdaSymbol{(}\AgdaGeneralizable{D}\AgdaSpace{}%
\AgdaOperator{%
\AgdaPostulate{×}%
}\AgdaSpace{}%
\AgdaGeneralizable{E}\AgdaSymbol{)}\AgdaSpace{}%
\AgdaOperator{%
\AgdaPostulate{→ᶜ}%
}\AgdaSpace{}%
\AgdaGeneralizable{E}\AgdaSpace{}%
\AgdaOperator{%
\AgdaPostulate{⟫}%
}%
\>[37]\AgdaComment{--\ (δ\ ,\ ε)↓₂\ is\ ε}\<%
\end{code}
\begin{code}[hide]%
\>[2]\AgdaKeyword{infixr}\AgdaSpace{}%
\AgdaNumber{2}\AgdaSpace{}%
\AgdaOperator{%
\AgdaPostulate{\AgdaUnderscore{}%
×\AgdaUnderscore{}}}\<%
\\
\>[2]\AgdaKeyword{infixr}\AgdaSpace{}%
\AgdaNumber{4}\AgdaSpace{}%
\AgdaOperator{%
\AgdaPostulate{\AgdaUnderscore{}%
,\AgdaUnderscore{}}}\<%
\end{code}
\end{AgdaSuppressSpace}
Neither the product nor pairing is associative in the embedding -- they
are associative up to isomorphism, but the Agda type-checker does not
take account of that.

\subsubsection{Tuple Domains}\label{Notation/Products/Tuples-tuple-domains}

The domain \AgdaRef{D\ \^{}\ n} of \AgdaRef{n}-tuples of elements of a
domain \AgdaRef{D} is conventionally written \(D^n\), but Agda does not
support the use of variables as superscripts.
\begin{AgdaSuppressSpace}
\begin{code}[hide]%
\>[0]\AgdaSymbol{\{-\#}\AgdaSpace{}%
\AgdaKeyword{OPTIONS}\AgdaSpace{}%
\AgdaPragma{--rewriting}\AgdaSpace{}%
\AgdaPragma{--confluence-check}\AgdaSpace{}%
\AgdaSymbol{\#-\}}\<%
\\
\>[0]\<%
\end{code}
\begin{code}%
\>[0]\AgdaKeyword{module}\AgdaSpace{}%
%
\AgdaModule{Notation.Products.Tuples}%
\AgdaSpace{}%
\AgdaKeyword{where}\<%
\end{code}
\begin{code}[hide]%
\>[0]\<%
\\
\>[0][@{}l@{\AgdaIndent{1}}]%
\>[2]\AgdaKeyword{open}\AgdaSpace{}%
\AgdaKeyword{import}\AgdaSpace{}%
%
\AgdaModule{Notation.Domains}%
\<%
\\
\>[2]\AgdaKeyword{open}\AgdaSpace{}%
\AgdaKeyword{import}\AgdaSpace{}%
%
\AgdaModule{Notation.Products}%
\<%
\end{code}
\begin{code}%
\>[2]\AgdaKeyword{open}\AgdaSpace{}%
\AgdaKeyword{import}\AgdaSpace{}%
%
\AgdaModule{Agda.Builtin.Nat}%
\AgdaSpace{}%
\AgdaKeyword{public}\AgdaSpace{}%
\AgdaKeyword{using}\AgdaSpace{}%
\AgdaSymbol{(}%
\AgdaDatatype{Nat}%
\AgdaSymbol{;}\AgdaSpace{}%
%
\AgdaInductiveConstructor{suc}%
\AgdaSymbol{)}\<%
\\
\\[\AgdaEmptyExtraSkip]%
\>[2]\AgdaOperator{%
\AgdaFunction{\AgdaUnderscore{}%
\textasciicircum{}\AgdaUnderscore{}}}\AgdaSpace{}%
\AgdaSymbol{:}\AgdaSpace{}%
%
\AgdaPostulate{Domain}%
\AgdaSpace{}%
\AgdaSymbol{→}\AgdaSpace{}%
%
\AgdaDatatype{Nat}%
\AgdaSpace{}%
\AgdaSymbol{→}\AgdaSpace{}%
%
\AgdaPostulate{Domain}%
%
\>[31]\AgdaComment{--\ D\ \textasciicircum{}\ n\ is\ the\ domain\ of\ n-tuples\ (n\ ≥\ 0)}\<%
\\
\>[2]\AgdaBound{D}\AgdaSpace{}%
\AgdaOperator{%
\AgdaFunction{\textasciicircum{}%
}}\AgdaSpace{}%
\AgdaNumber{0}%
\>[19]\AgdaSymbol{=}\AgdaSpace{}%
%
\AgdaPostulate{𝟙}%
\<%
\\
\>[2]\AgdaBound{D}\AgdaSpace{}%
\AgdaOperator{%
\AgdaFunction{\textasciicircum{}%
}}\AgdaSpace{}%
\AgdaNumber{1}%
\>[19]\AgdaSymbol{=}\AgdaSpace{}%
\AgdaBound{D}\<%
\\
\>[2]\AgdaBound{D}\AgdaSpace{}%
\AgdaOperator{%
\AgdaFunction{\textasciicircum{}%
}}\AgdaSpace{}%
%
\AgdaInductiveConstructor{suc}%
\AgdaSpace{}%
\AgdaSymbol{(}%
\AgdaInductiveConstructor{suc}%
\AgdaSpace{}%
\AgdaBound{n}\AgdaSymbol{)}%
\>[19]\AgdaSymbol{=}\AgdaSpace{}%
\AgdaBound{D}\AgdaSpace{}%
\AgdaOperator{%
\AgdaPostulate{×}%
}\AgdaSpace{}%
\AgdaSymbol{(}\AgdaBound{D}\AgdaSpace{}%
\AgdaOperator{%
\AgdaFunction{\textasciicircum{}%
}}\AgdaSpace{}%
%
\AgdaInductiveConstructor{suc}%
\AgdaSpace{}%
\AgdaBound{n}\AgdaSymbol{)}\<%
\end{code}
\end{AgdaSuppressSpace}

\subsubsection{Sequence Domains}\label{Notation/Products/Sequences-sequence-domains}

The domain \AgdaRef{D\ ⋆} of finite sequences of elements of a domain
\AgdaRef{D} is conventionally written \(D^*\).

The following notation for the various operations on sequences was
introduced in the early 1970s, and is used in the \emph{Scheme}
semantics \cite{Scheme}. (The single
angle-brackets \AgdaRef{⟨...⟩} used to form sequences are unrelated to
the double angle-brackets \AgdaRef{⟪\ D\ ⟫} used for the carrier of
domain \AgdaRef{D}.)
\begin{AgdaSuppressSpace}
\begin{code}[hide]%
\>[0]\AgdaSymbol{\{-\#}\AgdaSpace{}%
\AgdaKeyword{OPTIONS}\AgdaSpace{}%
\AgdaPragma{--rewriting}\AgdaSpace{}%
\AgdaPragma{--confluence-check}\AgdaSpace{}%
\AgdaSymbol{\#-\}}\<%
\\
\>[0]\<%
\end{code}
\begin{code}%
\>[0]\AgdaKeyword{module}\AgdaSpace{}%
%
\AgdaModule{Notation.Products.Sequences}%
\AgdaSpace{}%
\AgdaKeyword{where}\<%
\end{code}
\begin{code}[hide]%
\>[0]\<%
\\
\>[0][@{}l@{\AgdaIndent{1}}]%
\>[2]\AgdaKeyword{open}\AgdaSpace{}%
\AgdaKeyword{import}\AgdaSpace{}%
%
\AgdaModule{Notation.Domains}%
\<%
\\
\>[2]\AgdaKeyword{open}\AgdaSpace{}%
\AgdaKeyword{import}\AgdaSpace{}%
%
\AgdaModule{Notation.Functions}%
\<%
\\
\>[2]\AgdaKeyword{open}\AgdaSpace{}%
\AgdaKeyword{import}\AgdaSpace{}%
%
\AgdaModule{Notation.Flat.Naturals}%
\<%
\\
\>[2]\AgdaKeyword{open}\AgdaSpace{}%
\AgdaKeyword{import}\AgdaSpace{}%
%
\AgdaModule{Notation.Products.Tuples}%
\<%
\end{code}
\begin{code}%
\>[2]\AgdaKeyword{variable}\AgdaSpace{}%
\AgdaGeneralizable{n}\AgdaSpace{}%
\AgdaSymbol{:}\AgdaSpace{}%
%
\AgdaDatatype{Nat}%
\<%
\\
\\[\AgdaEmptyExtraSkip]%
\>[2]\AgdaKeyword{postulate}\<%
\\
\>[2][@{}l@{\AgdaIndent{0}}]%
\>[4]\AgdaOperator{%
\AgdaPostulate{\AgdaUnderscore{}%
⋆}}%
\>[11]\AgdaSymbol{:}\AgdaSpace{}%
%
\AgdaPostulate{Domain}%
\AgdaSpace{}%
\AgdaSymbol{→}\AgdaSpace{}%
%
\AgdaPostulate{Domain}%
%
\>[37]\AgdaComment{--\ D\ ⋆\ is\ the\ finite\ sequence\ domain}\<%
\\
\>[4]%
\AgdaPostulate{⟨⟩}%
%
\>[11]\AgdaSymbol{:}\AgdaSpace{}%
\AgdaOperator{%
\AgdaPostulate{⟪}%
}\AgdaSpace{}%
\AgdaGeneralizable{D}\AgdaSpace{}%
\AgdaOperator{%
\AgdaPostulate{⋆}%
}\AgdaSpace{}%
\AgdaOperator{%
\AgdaPostulate{⟫}%
}%
\>[37]\AgdaComment{--\ ⟨⟩\ is\ the\ empty\ sequence}\<%
\\
\>[4]\AgdaOperator{%
\AgdaPostulate{⟨\AgdaUnderscore{}%
⟩}}%
\>[11]\AgdaSymbol{:}\AgdaSpace{}%
\AgdaOperator{%
\AgdaPostulate{⟪}%
}\AgdaSpace{}%
\AgdaSymbol{(}\AgdaGeneralizable{D}\AgdaSpace{}%
\AgdaOperator{%
\AgdaFunction{\textasciicircum{}%
}}\AgdaSpace{}%
%
\AgdaInductiveConstructor{suc}%
\AgdaSpace{}%
\AgdaGeneralizable{n}\AgdaSymbol{)}\AgdaSpace{}%
\AgdaOperator{%
\AgdaPostulate{→ᶜ}%
}\AgdaSpace{}%
\AgdaGeneralizable{D}\AgdaSpace{}%
\AgdaOperator{%
\AgdaPostulate{⋆}%
}\AgdaSpace{}%
\AgdaOperator{%
\AgdaPostulate{⟫}%
}%
\>[37]\AgdaComment{--\ ⟨\ δ₁\ ,\ ...\ ⟩\ is\ a\ non-empty\ sequence}\<%
\\
\>[4]%
\AgdaPostulate{\#}%
%
\>[11]\AgdaSymbol{:}\AgdaSpace{}%
\AgdaOperator{%
\AgdaPostulate{⟪}%
}\AgdaSpace{}%
\AgdaGeneralizable{D}\AgdaSpace{}%
\AgdaOperator{%
\AgdaPostulate{⋆}%
}\AgdaSpace{}%
\AgdaOperator{%
\AgdaPostulate{→ᶜ}%
}\AgdaSpace{}%
%
\AgdaFunction{Nat⊥}%
\AgdaSpace{}%
\AgdaOperator{%
\AgdaPostulate{⟫}%
}%
\>[37]\AgdaComment{--\ \#\ δ⋆\ is\ the\ length\ of\ sequence\ δ⋆}\<%
\\
\>[4]\AgdaOperator{%
\AgdaPostulate{\AgdaUnderscore{}%
§\AgdaUnderscore{}}}%
\>[11]\AgdaSymbol{:}\AgdaSpace{}%
\AgdaOperator{%
\AgdaPostulate{⟪}%
}\AgdaSpace{}%
\AgdaGeneralizable{D}\AgdaSpace{}%
\AgdaOperator{%
\AgdaPostulate{⋆}%
}\AgdaSpace{}%
\AgdaOperator{%
\AgdaPostulate{→ᶜ}%
}\AgdaSpace{}%
\AgdaGeneralizable{D}\AgdaSpace{}%
\AgdaOperator{%
\AgdaPostulate{⋆}%
}\AgdaSpace{}%
\AgdaOperator{%
\AgdaPostulate{→ᶜ}%
}\AgdaSpace{}%
\AgdaGeneralizable{D}\AgdaSpace{}%
\AgdaOperator{%
\AgdaPostulate{⋆}%
}\AgdaSpace{}%
\AgdaOperator{%
\AgdaPostulate{⟫}%
}%
\>[37]\AgdaComment{--\ δ⋆₁\ §\ δ⋆₂\ is\ sequence\ concatenation}\<%
\\
\>[4]\AgdaOperator{%
\AgdaPostulate{\AgdaUnderscore{}%
↓\AgdaUnderscore{}}}%
\>[11]\AgdaSymbol{:}\AgdaSpace{}%
\AgdaOperator{%
\AgdaPostulate{⟪}%
}\AgdaSpace{}%
\AgdaGeneralizable{D}\AgdaSpace{}%
\AgdaOperator{%
\AgdaPostulate{⋆}%
}\AgdaSpace{}%
\AgdaOperator{%
\AgdaPostulate{→ᶜ}%
}\AgdaSpace{}%
%
\AgdaDatatype{Nat}%
\AgdaSpace{}%
\AgdaOperator{%
\AgdaPostulate{→ˢ}%
}\AgdaSpace{}%
\AgdaGeneralizable{D}\AgdaSpace{}%
\AgdaOperator{%
\AgdaPostulate{⟫}%
}%
\>[37]\AgdaComment{--\ δ⋆\ ↓\ n\ is\ the\ nth\ element}\<%
\\
\>[4]\AgdaOperator{%
\AgdaPostulate{\AgdaUnderscore{}%
†\AgdaUnderscore{}}}%
\>[11]\AgdaSymbol{:}\AgdaSpace{}%
\AgdaOperator{%
\AgdaPostulate{⟪}%
}\AgdaSpace{}%
\AgdaGeneralizable{D}\AgdaSpace{}%
\AgdaOperator{%
\AgdaPostulate{⋆}%
}\AgdaSpace{}%
\AgdaOperator{%
\AgdaPostulate{→ᶜ}%
}\AgdaSpace{}%
%
\AgdaDatatype{Nat}%
\AgdaSpace{}%
\AgdaOperator{%
\AgdaPostulate{→ˢ}%
}\AgdaSpace{}%
\AgdaGeneralizable{D}\AgdaSpace{}%
\AgdaOperator{%
\AgdaPostulate{⋆}%
}\AgdaSpace{}%
\AgdaOperator{%
\AgdaPostulate{⟫}%
}%
\>[37]\AgdaComment{--\ δ⋆\ †\ n\ is\ the\ nth\ tail}\<%
\end{code}
\end{AgdaSuppressSpace}

\subsection{Recursive Domains}\label{Notation/Recursion-recursive-domains}

Conventional denotational definitions often involve groups of mutually
recursive domain equations. In Agda, recursive type definitions lead to
non-termination of the type-checker. To avoid non-termination, it is
sufficient to break the recursion by leaving (one or more) domains as
\emph{postulated}. The following operations can then be used to map
values from a postulated domain to its structure and \emph{vice versa}.
\begin{AgdaSuppressSpace}
\begin{code}[hide]%
\>[0]\AgdaSymbol{\{-\#}\AgdaSpace{}%
\AgdaKeyword{OPTIONS}\AgdaSpace{}%
\AgdaPragma{--rewriting}\AgdaSpace{}%
\AgdaPragma{--confluence-check}\AgdaSpace{}%
\AgdaSymbol{\#-\}}\<%
\\
\>[0]\<%
\end{code}
\begin{code}%
\>[0]\AgdaKeyword{module}\AgdaSpace{}%
%
\AgdaModule{Notation.Recursion}%
\AgdaSpace{}%
\AgdaKeyword{where}\<%
\end{code}
\begin{code}[hide]%
\>[0]\<%
\\
\>[0][@{}l@{\AgdaIndent{1}}]%
\>[2]\AgdaKeyword{open}\AgdaSpace{}%
\AgdaKeyword{import}\AgdaSpace{}%
%
\AgdaModule{Notation.Domains}%
\<%
\\
\>[2]\AgdaKeyword{open}\AgdaSpace{}%
\AgdaKeyword{import}\AgdaSpace{}%
%
\AgdaModule{Notation.Functions}%
\<%
\\
\>[0]\<%
\end{code}
\begin{code}%
\>[0][@{}l@{\AgdaIndent{1}}]%
\>[2]\AgdaKeyword{postulate}\<%
\\
\>[2][@{}l@{\AgdaIndent{0}}]%
\>[4]\AgdaOperator{%
\AgdaPostulate{\AgdaUnderscore{}%
≅\AgdaUnderscore{}}}\AgdaSpace{}%
\AgdaSymbol{:}\AgdaSpace{}%
%
\AgdaPostulate{Domain}%
\AgdaSpace{}%
\AgdaSymbol{→}\AgdaSpace{}%
%
\AgdaPostulate{Domain}%
\AgdaSpace{}%
\AgdaSymbol{→}\AgdaSpace{}%
\AgdaPrimitive{Set}\<%
\\
\>[4]\AgdaComment{--\ an\ instance\ of\ D\ ≅\ E\ declares\ that\ the\ structure\ of\ D\ is\ the\ same\ as\ E}\<%
\\
\>[4]%
\AgdaPostulate{unfold}%
%
\>[12]\AgdaSymbol{:}\AgdaSpace{}%
\AgdaSymbol{\{\{}\AgdaGeneralizable{D}\AgdaSpace{}%
\AgdaOperator{%
\AgdaPostulate{≅}%
}\AgdaSpace{}%
\AgdaGeneralizable{E}\AgdaSymbol{\}\}}\AgdaSpace{}%
\AgdaSymbol{→}\AgdaSpace{}%
\AgdaOperator{%
\AgdaPostulate{⟪}%
}\AgdaSpace{}%
\AgdaGeneralizable{D}\AgdaSpace{}%
\AgdaOperator{%
\AgdaPostulate{→ᶜ}%
}\AgdaSpace{}%
\AgdaGeneralizable{E}\AgdaSpace{}%
\AgdaOperator{%
\AgdaPostulate{⟫}%
}\<%
\\
\>[4]%
\AgdaPostulate{fold}%
%
\>[12]\AgdaSymbol{:}\AgdaSpace{}%
\AgdaSymbol{\{\{}\AgdaGeneralizable{D}\AgdaSpace{}%
\AgdaOperator{%
\AgdaPostulate{≅}%
}\AgdaSpace{}%
\AgdaGeneralizable{E}\AgdaSymbol{\}\}}\AgdaSpace{}%
\AgdaSymbol{→}\AgdaSpace{}%
\AgdaOperator{%
\AgdaPostulate{⟪}%
}\AgdaSpace{}%
\AgdaGeneralizable{E}\AgdaSpace{}%
\AgdaOperator{%
\AgdaPostulate{→ᶜ}%
}\AgdaSpace{}%
\AgdaGeneralizable{D}\AgdaSpace{}%
\AgdaOperator{%
\AgdaPostulate{⟫}%
}\<%
\end{code}
\end{AgdaSuppressSpace}
The \emph{instance parameter} \AgdaRef{\{\{D\ ≅\ E\}\}} of the above
operations restricts them to domains \AgdaRef{D} and \AgdaRef{E} for which
\AgdaRef{instance\ \_\ :\ D\ ≅\ E} has been declared. The
untyped
lambda-calculus (Section~\ref{Examples/LC-untyped-lambda-calculus}) example illustrates the use of this notation in an Agda
embedding.

\subsection{Updates}\label{Notation/Updates-updates}

When a type \AgdaRef{A} has an equality operation
\AgdaRef{\_==\_\ :\ A\ →\ A\ →\ Bool}, environments
\AgdaRef{ρ\ :\ ⟪\ A\ →ˢ\ D\ ⟫} can be `updated' (i.e., extended or
overridden) using the conventional notation
\AgdaRef{ρ\ {[}\ δ\ /\ a\ {]}}, defined as follows.
\begin{AgdaSuppressSpace}
\begin{code}[hide]%
\>[0]\AgdaSymbol{\{-\#}\AgdaSpace{}%
\AgdaKeyword{OPTIONS}\AgdaSpace{}%
\AgdaPragma{--rewriting}\AgdaSpace{}%
\AgdaPragma{--confluence-check}\AgdaSpace{}%
\AgdaPragma{--lossy-unification}\AgdaSpace{}%
\AgdaSymbol{\#-\}}\<%
\\
\>[0]\<%
\end{code}
\begin{code}%
\>[0]\AgdaKeyword{module}\AgdaSpace{}%
%
\AgdaModule{Notation.Updates}%
\AgdaSpace{}%
\AgdaKeyword{where}\<%
\end{code}
\begin{code}[hide]%
\>[0]\<%
\\
\>[0][@{}l@{\AgdaIndent{1}}]%
\>[2]\AgdaKeyword{open}\AgdaSpace{}%
\AgdaKeyword{import}\AgdaSpace{}%
%
\AgdaModule{Notation.Domains}%
\<%
\\
\>[2]\AgdaKeyword{open}\AgdaSpace{}%
\AgdaKeyword{import}\AgdaSpace{}%
%
\AgdaModule{Notation.Functions}%
\<%
\\
\>[2]\AgdaKeyword{open}\AgdaSpace{}%
\AgdaKeyword{import}\AgdaSpace{}%
%
\AgdaModule{Notation.Flat}%
\<%
\\
\>[2]\AgdaKeyword{open}\AgdaSpace{}%
\AgdaKeyword{import}\AgdaSpace{}%
%
\AgdaModule{Notation.Flat.Booleans}%
\<%
\\
\>[0]\<%
\end{code}
\begin{code}%
\>[0][@{}l@{\AgdaIndent{1}}]%
\>[2]\AgdaOperator{%
\AgdaFunction{\AgdaUnderscore{}%
[\AgdaUnderscore{}/\AgdaUnderscore{}]}}\AgdaSpace{}%
\AgdaSymbol{:}\AgdaSpace{}%
\AgdaSymbol{\{\{}%
\AgdaRecord{Eq}%
\AgdaSpace{}%
\AgdaGeneralizable{A}\AgdaSymbol{\}\}}\AgdaSpace{}%
\AgdaSymbol{→}\AgdaSpace{}%
\AgdaOperator{%
\AgdaPostulate{⟪}%
}\AgdaSpace{}%
\AgdaSymbol{(}\AgdaGeneralizable{A}\AgdaSpace{}%
\AgdaOperator{%
\AgdaPostulate{→ˢ}%
}\AgdaSpace{}%
\AgdaGeneralizable{D}\AgdaSymbol{)}\AgdaSpace{}%
\AgdaOperator{%
\AgdaPostulate{→ᶜ}%
}\AgdaSpace{}%
\AgdaGeneralizable{D}\AgdaSpace{}%
\AgdaOperator{%
\AgdaPostulate{→ᶜ}%
}\AgdaSpace{}%
\AgdaGeneralizable{A}\AgdaSpace{}%
\AgdaOperator{%
\AgdaPostulate{→ˢ}%
}\AgdaSpace{}%
\AgdaSymbol{(}\AgdaGeneralizable{A}\AgdaSpace{}%
\AgdaOperator{%
\AgdaPostulate{→ˢ}%
}\AgdaSpace{}%
\AgdaGeneralizable{D}\AgdaSymbol{)}\AgdaSpace{}%
\AgdaOperator{%
\AgdaPostulate{⟫}%
}\<%
\end{code}
\begin{code}[hide]%
\>[2]\AgdaComment{--\ ρ\ [\ δ\ /\ a\ ]\ maps\ a\ to\ δ,\ and\ other\ arguments\ a′\ to\ ρ\ a′}\<%
\end{code}
\begin{code}%
\>[2]\AgdaBound{ρ}\AgdaSpace{}%
\AgdaOperator{%
\AgdaFunction{[}%
}\AgdaSpace{}%
\AgdaBound{δ}\AgdaSpace{}%
\AgdaOperator{%
\AgdaFunction{/}%
}\AgdaSpace{}%
\AgdaBound{a}\AgdaSpace{}%
\AgdaOperator{%
\AgdaFunction{]}%
}\AgdaSpace{}%
\AgdaSymbol{=}\AgdaSpace{}%
\AgdaSymbol{λ}\AgdaSpace{}%
\AgdaBound{a′}\AgdaSpace{}%
\AgdaSymbol{→}\AgdaSpace{}%
\AgdaOperator{%
\AgdaFunction{if}%
}\AgdaSpace{}%
\AgdaBound{a}\AgdaSpace{}%
\AgdaOperator{%
\AgdaField{==}%
}\AgdaSpace{}%
\AgdaBound{a′}\AgdaSpace{}%
\AgdaOperator{%
\AgdaFunction{then}%
}\AgdaSpace{}%
\AgdaBound{δ}\AgdaSpace{}%
\AgdaOperator{%
\AgdaFunction{else}%
}\AgdaSpace{}%
\AgdaBound{ρ}\AgdaSpace{}%
\AgdaBound{a′}\<%
\end{code}
\end{AgdaSuppressSpace}
Similarly for stores \AgdaRef{σ\ :\ ⟪\ (A\ +⊥)\ →ᶜ\ D\ ⟫} (appending
\(\bot\) to the conventional notation):
\begin{AgdaSuppressSpace}
\begin{code}%
\>[2]\AgdaOperator{%
\AgdaFunction{\AgdaUnderscore{}%
[\AgdaUnderscore{}/\AgdaUnderscore{}]⊥}}\AgdaSpace{}%
\AgdaSymbol{:}\AgdaSpace{}%
\AgdaSymbol{\{\{}%
\AgdaRecord{Eq}%
\AgdaSpace{}%
\AgdaGeneralizable{A}\AgdaSymbol{\}\}}\AgdaSpace{}%
\AgdaSymbol{→}\AgdaSpace{}%
\AgdaOperator{%
\AgdaPostulate{⟪}%
}\AgdaSpace{}%
\AgdaSymbol{((}\AgdaGeneralizable{A}\AgdaSpace{}%
\AgdaOperator{%
\AgdaPostulate{+⊥}%
}\AgdaSymbol{)}\AgdaSpace{}%
\AgdaOperator{%
\AgdaPostulate{→ᶜ}%
}\AgdaSpace{}%
\AgdaGeneralizable{D}\AgdaSymbol{)}\AgdaSpace{}%
\AgdaOperator{%
\AgdaPostulate{→ᶜ}%
}\AgdaSpace{}%
\AgdaGeneralizable{D}\AgdaSpace{}%
\AgdaOperator{%
\AgdaPostulate{→ᶜ}%
}\AgdaSpace{}%
\AgdaSymbol{(}\AgdaGeneralizable{A}\AgdaSpace{}%
\AgdaOperator{%
\AgdaPostulate{+⊥}%
}\AgdaSymbol{)}\AgdaSpace{}%
\AgdaOperator{%
\AgdaPostulate{→ᶜ}%
}\AgdaSpace{}%
\AgdaSymbol{((}\AgdaGeneralizable{A}\AgdaSpace{}%
\AgdaOperator{%
\AgdaPostulate{+⊥}%
}\AgdaSymbol{)}\AgdaSpace{}%
\AgdaOperator{%
\AgdaPostulate{→ᶜ}%
}\AgdaSpace{}%
\AgdaGeneralizable{D}\AgdaSymbol{)}\AgdaSpace{}%
\AgdaOperator{%
\AgdaPostulate{⟫}%
}\<%
\end{code}
\begin{code}[hide]%
\>[2]\AgdaComment{--\ σ\ [\ δ\ /\ α\ ]⊥\ maps\ α\ to\ δ,\ and\ other\ arguments\ α′\ to\ σ\ α′}\<%
\end{code}
\begin{code}%
\>[2]\AgdaBound{σ}\AgdaSpace{}%
\AgdaOperator{%
\AgdaFunction{[}%
}\AgdaSpace{}%
\AgdaBound{δ}\AgdaSpace{}%
\AgdaOperator{%
\AgdaFunction{/}%
}\AgdaSpace{}%
\AgdaBound{α}\AgdaSpace{}%
\AgdaOperator{%
\AgdaFunction{]⊥}%
}\AgdaSpace{}%
\AgdaSymbol{=}\AgdaSpace{}%
\AgdaSymbol{λ}\AgdaSpace{}%
\AgdaBound{α′}\AgdaSpace{}%
\AgdaSymbol{→}\AgdaSpace{}%
\AgdaSymbol{(}\AgdaBound{α}\AgdaSpace{}%
\AgdaOperator{%
\AgdaFunction{==⊥}%
}\AgdaSpace{}%
\AgdaBound{α′}\AgdaSymbol{)}\AgdaSpace{}%
\AgdaOperator{%
\AgdaFunction{⟶}%
}\AgdaSpace{}%
\AgdaBound{δ}\AgdaSpace{}%
\AgdaOperator{%
\AgdaFunction{,}%
}\AgdaSpace{}%
\AgdaBound{σ}\AgdaSpace{}%
\AgdaBound{α′}\<%
\end{code}
\end{AgdaSuppressSpace}
Defining an operation \AgdaRef{m\ {[}\ x\ ←\ y\ {]}} for updating a
\emph{dependent} map \AgdaRef{m} to map \AgdaRef{x} to \AgdaRef{y} is
somewhat less straightforward, as it involves an equality test that may
return an \emph{equivalence proof}.
\begin{AgdaSuppressSpace}
\begin{code}%
\>[2]\AgdaKeyword{open}\AgdaSpace{}%
\AgdaKeyword{import}\AgdaSpace{}%
%
\AgdaModule{Data.Maybe.Base}%
\AgdaSpace{}%
\AgdaKeyword{public}\AgdaSpace{}%
\AgdaKeyword{using}\AgdaSpace{}%
\AgdaSymbol{(}%
\AgdaDatatype{Maybe}%
\AgdaSymbol{;}\AgdaSpace{}%
%
\AgdaInductiveConstructor{just}%
\AgdaSymbol{;}\AgdaSpace{}%
%
\AgdaInductiveConstructor{nothing}%
\AgdaSymbol{)}\<%
\\
\>[2]\AgdaKeyword{open}\AgdaSpace{}%
\AgdaKeyword{import}\AgdaSpace{}%
%
\AgdaModule{Relation.Binary.PropositionalEquality.Core}%
\AgdaSpace{}%
\AgdaKeyword{public}\AgdaSpace{}%
\AgdaKeyword{using}\AgdaSpace{}%
\AgdaSymbol{(}\AgdaOperator{%
\AgdaDatatype{\AgdaUnderscore{}%
≡\AgdaUnderscore{}}}\AgdaSymbol{;}\AgdaSpace{}%
%
\AgdaInductiveConstructor{refl}%
\AgdaSymbol{)}\<%
\\
\\[\AgdaEmptyExtraSkip]%
\>[2]\AgdaKeyword{record}\AgdaSpace{}%
%
\AgdaRecord{MaybeEq}%
\AgdaSpace{}%
\AgdaSymbol{(}\AgdaBound{A}\AgdaSpace{}%
\AgdaSymbol{:}\AgdaSpace{}%
\AgdaPrimitive{Set}\AgdaSymbol{)}\AgdaSpace{}%
\AgdaSymbol{:}\AgdaSpace{}%
\AgdaPrimitive{Set}\AgdaSpace{}%
\AgdaKeyword{where}\AgdaSpace{}%
\AgdaKeyword{field}\AgdaSpace{}%
\AgdaOperator{%
\AgdaField{\AgdaUnderscore{}%
==?\AgdaUnderscore{}}}\AgdaSpace{}%
\AgdaSymbol{:}\AgdaSpace{}%
\AgdaSymbol{(}\AgdaBound{a}\AgdaSpace{}%
\AgdaBound{a′}\AgdaSpace{}%
\AgdaSymbol{:}\AgdaSpace{}%
\AgdaBound{A}\AgdaSymbol{)}\AgdaSpace{}%
\AgdaSymbol{→}\AgdaSpace{}%
%
\AgdaDatatype{Maybe}%
\AgdaSpace{}%
\AgdaSymbol{(}\AgdaBound{a}\AgdaSpace{}%
\AgdaOperator{%
\AgdaDatatype{≡}%
}\AgdaSpace{}%
\AgdaBound{a′}\AgdaSymbol{)}\<%
\\
\>[2]\AgdaKeyword{open}\AgdaSpace{}%
%
\AgdaRecord{MaybeEq}%
\AgdaSpace{}%
\AgdaSymbol{\{\{...\}\}}\AgdaSpace{}%
\AgdaKeyword{public}\<%
\\
\\[\AgdaEmptyExtraSkip]%
\>[2]\AgdaKeyword{variable}\AgdaSpace{}%
\AgdaGeneralizable{X}\AgdaSpace{}%
\AgdaSymbol{:}\AgdaSpace{}%
\AgdaPrimitive{Set}\AgdaSymbol{;}\AgdaSpace{}%
\AgdaGeneralizable{Y}\AgdaSpace{}%
\AgdaSymbol{:}\AgdaSpace{}%
\AgdaGeneralizable{X}\AgdaSpace{}%
\AgdaSymbol{→}\AgdaSpace{}%
\AgdaPrimitive{Set}\<%
\\
\\[\AgdaEmptyExtraSkip]%
\>[2]\AgdaOperator{%
\AgdaFunction{\AgdaUnderscore{}%
[\AgdaUnderscore{}←\AgdaUnderscore{}]}}\AgdaSpace{}%
\AgdaSymbol{:}%
\>[12]\AgdaSymbol{\{\{}%
\AgdaRecord{MaybeEq}%
\AgdaSpace{}%
\AgdaGeneralizable{X}\AgdaSymbol{\}\}}\AgdaSpace{}%
\AgdaSymbol{→}\AgdaSpace{}%
\AgdaSymbol{(∀}\AgdaSpace{}%
\AgdaBound{x′}\AgdaSpace{}%
\AgdaSymbol{→}\AgdaSpace{}%
\AgdaGeneralizable{Y}\AgdaSpace{}%
\AgdaBound{x′}\AgdaSymbol{)}\AgdaSpace{}%
\AgdaSymbol{→}\AgdaSpace{}%
\AgdaSymbol{(}\AgdaBound{x}\AgdaSpace{}%
\AgdaSymbol{:}\AgdaSpace{}%
\AgdaGeneralizable{X}\AgdaSymbol{)}\AgdaSpace{}%
\AgdaSymbol{→}\AgdaSpace{}%
\AgdaGeneralizable{Y}\AgdaSpace{}%
\AgdaBound{x}\AgdaSpace{}%
\AgdaSymbol{→}\AgdaSpace{}%
\AgdaSymbol{(∀}\AgdaSpace{}%
\AgdaBound{x′}\AgdaSpace{}%
\AgdaSymbol{→}\AgdaSpace{}%
\AgdaGeneralizable{Y}\AgdaSpace{}%
\AgdaBound{x′}\AgdaSymbol{)}\<%
\\
\>[2]\AgdaOperator{%
\AgdaFunction{\AgdaUnderscore{}%
[\AgdaUnderscore{}←\AgdaUnderscore{}]}}\AgdaSpace{}%
\AgdaSymbol{\{}\AgdaBound{X}\AgdaSymbol{\}}\AgdaSpace{}%
\AgdaSymbol{\{}\AgdaBound{Y}\AgdaSymbol{\}}\AgdaSpace{}%
\AgdaBound{m}\AgdaSpace{}%
\AgdaBound{x}\AgdaSpace{}%
\AgdaBound{y}\AgdaSpace{}%
\AgdaSymbol{=}\AgdaSpace{}%
\AgdaSymbol{λ}\AgdaSpace{}%
\AgdaBound{x′}\AgdaSpace{}%
\AgdaSymbol{→}\AgdaSpace{}%
%
\AgdaFunction{h}%
\AgdaSpace{}%
\AgdaBound{x′}\AgdaSpace{}%
\AgdaSymbol{(}\AgdaBound{x}\AgdaSpace{}%
\AgdaOperator{%
\AgdaField{==?}%
}\AgdaSpace{}%
\AgdaBound{x′}\AgdaSymbol{)}\<%
\\
\>[2][@{}l@{\AgdaIndent{0}}]%
\>[4]\AgdaKeyword{where}\<%
\\
\>[4][@{}l@{\AgdaIndent{0}}]%
\>[6]%
\AgdaFunction{h}%
\AgdaSpace{}%
\AgdaSymbol{:}\AgdaSpace{}%
\AgdaSymbol{(}\AgdaBound{x′}\AgdaSpace{}%
\AgdaSymbol{:}\AgdaSpace{}%
\AgdaBound{X}\AgdaSymbol{)}\AgdaSpace{}%
\AgdaSymbol{→}\AgdaSpace{}%
%
\AgdaDatatype{Maybe}%
\AgdaSpace{}%
\AgdaSymbol{(}\AgdaBound{x}\AgdaSpace{}%
\AgdaOperator{%
\AgdaDatatype{≡}%
}\AgdaSpace{}%
\AgdaBound{x′}\AgdaSymbol{)}\AgdaSpace{}%
\AgdaSymbol{→}\AgdaSpace{}%
\AgdaBound{Y}\AgdaSpace{}%
\AgdaBound{x′}\<%
\\
\>[6]%
\AgdaFunction{h}%
\AgdaSpace{}%
\AgdaBound{x′}\AgdaSpace{}%
\AgdaSymbol{(}%
\AgdaInductiveConstructor{just}%
\AgdaSpace{}%
%
\AgdaInductiveConstructor{refl}%
\AgdaSymbol{)}\AgdaSpace{}%
\AgdaSymbol{=}\AgdaSpace{}%
\AgdaBound{y}\<%
\\
\>[6]%
\AgdaFunction{h}%
\AgdaSpace{}%
\AgdaBound{x′}\AgdaSpace{}%
%
\AgdaInductiveConstructor{nothing}%
\AgdaSpace{}%
\AgdaSymbol{=}\AgdaSpace{}%
\AgdaBound{m}\AgdaSpace{}%
\AgdaBound{x′}\<%
\end{code}
\end{AgdaSuppressSpace}
Note that the order of the arguments differs from that in the notation
for non-dependent updates, as the type of \AgdaRef{y} may depend on the
value of \AgdaRef{x}.

\end{AgdaAlign}

\section{Illustrative Examples}\label{Examples-illustrative-examples}

This section illustrates mechanisation of denotational semantics in Agda
with three examples, all using the
postulated
notation (Section~\ref{Notation-postulated-domain-notation}) for domains and their associated operations:
LC: An Untyped
Lambda-Calculus (Section~\ref{Examples/LC-untyped-lambda-calculus}),
PCF:
A Programming Language for Computable Functions (Section~\ref{Examples/PCF-pcf-a-programming-language-for-computable-functions}), and
Scm: A
Sublanguage of Scheme (Section~\ref{Examples/Scm-scm-a-sublanguage-of-scheme}).
\begin{AgdaSuppressSpace}
\begin{code}[hide]%
\>[0]\AgdaSymbol{\{-\#}\AgdaSpace{}%
\AgdaKeyword{OPTIONS}\AgdaSpace{}%
\AgdaPragma{--rewriting}\AgdaSpace{}%
\AgdaPragma{--confluence-check}\AgdaSpace{}%
\AgdaPragma{--lossy-unification}\AgdaSpace{}%
\AgdaSymbol{\#-\}}\<%
\\
\\[\AgdaEmptyExtraSkip]%
\>[0]\AgdaKeyword{module}\AgdaSpace{}%
%
\AgdaModule{Examples}%
\AgdaSpace{}%
\AgdaKeyword{where}\<%
\\
\\[\AgdaEmptyExtraSkip]%
\>[0][@{}l@{\AgdaIndent{0}}]%
\>[2]\AgdaKeyword{import}\AgdaSpace{}%
%
\AgdaModule{Examples.LC}%
\<%
\\
\>[2]\AgdaKeyword{import}\AgdaSpace{}%
%
\AgdaModule{Examples.PCF}%
\<%
\\
\>[2]\AgdaKeyword{import}\AgdaSpace{}%
%
\AgdaModule{Examples.Scm}%
\<%
\end{code}
\end{AgdaSuppressSpace}

\begin{AgdaAlign}
\subsection{Untyped Lambda-Calculus}\label{Examples/LC-untyped-lambda-calculus}

This section presents our Agda embedding of a denotational semantics of
an untyped \(\lambda\)-calculus.
\begin{AgdaSuppressSpace}
\begin{code}[hide]%
\>[0]\AgdaSymbol{\{-\#}\AgdaSpace{}%
\AgdaKeyword{OPTIONS}\AgdaSpace{}%
\AgdaPragma{--rewriting}\AgdaSpace{}%
\AgdaPragma{--confluence-check}\AgdaSpace{}%
\AgdaSymbol{\#-\}}\<%
\\
\\[\AgdaEmptyExtraSkip]%
\>[0]\AgdaKeyword{module}\AgdaSpace{}%
%
\AgdaModule{Examples.LC}%
\AgdaSpace{}%
\AgdaKeyword{where}\<%
\\
\\[\AgdaEmptyExtraSkip]%
\>[0][@{}l@{\AgdaIndent{0}}]%
\>[2]\AgdaKeyword{import}\AgdaSpace{}%
%
\AgdaModule{Examples.LC.Abstract-Syntax}%
\<%
\\
\>[2]\AgdaKeyword{import}\AgdaSpace{}%
%
\AgdaModule{Examples.LC.Domain-Equations}%
\<%
\\
\>[2]\AgdaKeyword{import}\AgdaSpace{}%
%
\AgdaModule{Examples.LC.Semantic-Functions}%
\<%
\end{code}
\end{AgdaSuppressSpace}

\subsubsection{Abstract Syntax}\label{Examples/LC/Abstract-Syntax-abstract-syntax}

Abstract syntax is \emph{not} regarded as a domain. In conventional
Scott--Strachey style denotational semantics, abstract syntax is
generally first-order: terms are finite, totally-defined elements.

A variable is written \AgdaRef{x\ n}. The argument \AgdaRef{n} merely
distinguishes between variables -- it is \emph{not} a De Bruijn index.
The term constructor \AgdaRef{var} below includes variables in terms.

In Agda, mixfix notation requires argument positions \AgdaRef{\_} to be
separated by characters other than spaces. The term constructors for
function abstraction and application use the Unicode character
\AgdaRef{␣} as a separator.
\begin{AgdaSuppressSpace}
\begin{code}[hide]%
\>[0]\AgdaSymbol{\{-\#}\AgdaSpace{}%
\AgdaKeyword{OPTIONS}\AgdaSpace{}%
\AgdaPragma{--rewriting}\AgdaSpace{}%
\AgdaPragma{--confluence-check}\AgdaSpace{}%
\AgdaSymbol{\#-\}}\<%
\\
\>[0]\<%
\end{code}
\begin{code}%
\>[0]\AgdaKeyword{module}\AgdaSpace{}%
%
\AgdaModule{Examples.LC.Abstract-Syntax}%
\AgdaSpace{}%
\AgdaKeyword{where}\<%
\end{code}
\begin{code}[hide]%
\>[0]\<%
\\
\>[0][@{}l@{\AgdaIndent{1}}]%
\>[2]\AgdaKeyword{open}\AgdaSpace{}%
\AgdaKeyword{import}\AgdaSpace{}%
%
\AgdaModule{Agda.Builtin.Nat}%
\AgdaSpace{}%
\AgdaKeyword{public}\AgdaSpace{}%
\AgdaKeyword{using}\AgdaSpace{}%
\AgdaSymbol{(}%
\AgdaDatatype{Nat}%
\AgdaSymbol{)}\<%
\\
\>[0]\<%
\end{code}
\begin{code}%
\>[0][@{}l@{\AgdaIndent{1}}]%
\>[2]\AgdaKeyword{data}\AgdaSpace{}%
%
\AgdaDatatype{Var}%
\AgdaSpace{}%
\AgdaSymbol{:}\AgdaSpace{}%
\AgdaPrimitive{Set}\AgdaSpace{}%
\AgdaKeyword{where}\<%
\\
\>[2][@{}l@{\AgdaIndent{0}}]%
\>[4]%
\AgdaInductiveConstructor{x}%
\AgdaSpace{}%
\AgdaSymbol{:}\AgdaSpace{}%
%
\AgdaDatatype{Nat}%
\AgdaSpace{}%
\AgdaSymbol{→}\AgdaSpace{}%
%
\AgdaDatatype{Var}%
\<%
\end{code}
\begin{code}[hide]%
\>[2]\AgdaKeyword{variable}\AgdaSpace{}%
\AgdaGeneralizable{v}\AgdaSpace{}%
\AgdaSymbol{:}\AgdaSpace{}%
%
\AgdaDatatype{Var}%
\<%
\\
\>[0]\<%
\end{code}
\begin{code}%
\>[0][@{}l@{\AgdaIndent{1}}]%
\>[2]\AgdaKeyword{data}\AgdaSpace{}%
%
\AgdaDatatype{Exp}%
\AgdaSpace{}%
\AgdaSymbol{:}\AgdaSpace{}%
\AgdaPrimitive{Set}\AgdaSpace{}%
\AgdaKeyword{where}\<%
\\
\>[2][@{}l@{\AgdaIndent{0}}]%
\>[4]\AgdaOperator{%
\AgdaInductiveConstructor{var\AgdaUnderscore{}%
}}%
\>[12]\AgdaSymbol{:}\AgdaSpace{}%
%
\AgdaDatatype{Var}%
\AgdaSpace{}%
\AgdaSymbol{→}\AgdaSpace{}%
%
\AgdaDatatype{Exp}%
%
\>[31]\AgdaComment{--\ variable\ reference}\<%
\\
\>[4]\AgdaOperator{%
\AgdaInductiveConstructor{⦅λ\AgdaUnderscore{}%
␣\AgdaUnderscore{}⦆}}%
\>[12]\AgdaSymbol{:}\AgdaSpace{}%
%
\AgdaDatatype{Var}%
\AgdaSpace{}%
\AgdaSymbol{→}\AgdaSpace{}%
%
\AgdaDatatype{Exp}%
\AgdaSpace{}%
\AgdaSymbol{→}\AgdaSpace{}%
%
\AgdaDatatype{Exp}%
%
\>[31]\AgdaComment{--\ function\ abstraction}\<%
\\
\>[4]\AgdaOperator{%
\AgdaInductiveConstructor{⦅\AgdaUnderscore{}%
␣\AgdaUnderscore{}⦆}}%
\>[12]\AgdaSymbol{:}\AgdaSpace{}%
%
\AgdaDatatype{Exp}%
\AgdaSpace{}%
\AgdaSymbol{→}\AgdaSpace{}%
%
\AgdaDatatype{Exp}%
\AgdaSpace{}%
\AgdaSymbol{→}\AgdaSpace{}%
%
\AgdaDatatype{Exp}%
%
\>[31]\AgdaComment{--\ function\ application}\<%
\end{code}
\begin{code}[hide]%
\>[2]\AgdaKeyword{variable}\AgdaSpace{}%
\AgdaGeneralizable{e}\AgdaSpace{}%
\AgdaGeneralizable{e₁}\AgdaSpace{}%
\AgdaGeneralizable{e₂}\AgdaSpace{}%
\AgdaSymbol{:}\AgdaSpace{}%
%
\AgdaDatatype{Exp}%
\<%
\end{code}
\end{AgdaSuppressSpace}

\subsubsection{Domain Equations}\label{Examples/LC/Domain-Equations-domain-equations}

Simply defining \AgdaRef{D∞\ =\ (D∞\ →ᶜ\ D∞)} would lead to
non-termination of the Agda type-checker. Instead, we postulate the
domain \AgdaRef{D∞}, together with a bijection
\AgdaRef{D∞\ ≅\ (D∞\ →ᶜ\ D∞)}. This declares the operations
\AgdaRef{unfold\ :\ ⟪\ D∞\ →ᶜ\ (D∞\ →ᶜ\ D∞)\ ⟫} and
\AgdaRef{fold\ :\ ⟪\ (D∞\ →ᶜ\ D∞)\ →ᶜ\ D∞\ ⟫} associated with
recursive
domains (Section~\ref{Notation/Recursion-recursive-domains}).
\begin{AgdaSuppressSpace}
\begin{code}[hide]%
\>[0]\AgdaSymbol{\{-\#}\AgdaSpace{}%
\AgdaKeyword{OPTIONS}\AgdaSpace{}%
\AgdaPragma{--rewriting}\AgdaSpace{}%
\AgdaPragma{--confluence-check}\AgdaSpace{}%
\AgdaSymbol{\#-\}}\<%
\\
\>[0]\<%
\end{code}
\begin{code}%
\>[0]\AgdaKeyword{module}\AgdaSpace{}%
%
\AgdaModule{Examples.LC.Domain-Equations}%
\AgdaSpace{}%
\AgdaKeyword{where}\<%
\end{code}
\begin{code}[hide]%
\>[0]\<%
\\
\>[0][@{}l@{\AgdaIndent{1}}]%
\>[2]\AgdaKeyword{open}\AgdaSpace{}%
\AgdaKeyword{import}\AgdaSpace{}%
%
\AgdaModule{Examples.LC.Abstract-Syntax}%
\<%
\\
\>[2]\AgdaKeyword{open}\AgdaSpace{}%
\AgdaKeyword{import}\AgdaSpace{}%
%
\AgdaModule{Notation.Domains}%
\<%
\\
\>[2]\AgdaKeyword{open}\AgdaSpace{}%
\AgdaKeyword{import}\AgdaSpace{}%
%
\AgdaModule{Notation.Functions}%
\<%
\\
\>[2]\AgdaKeyword{open}\AgdaSpace{}%
\AgdaKeyword{import}\AgdaSpace{}%
%
\AgdaModule{Notation.Recursion}%
\<%
\\
\>[2]\AgdaKeyword{open}\AgdaSpace{}%
\AgdaKeyword{import}\AgdaSpace{}%
%
\AgdaModule{Notation.Flat.Booleans}%
\<%
\\
\>[2]\AgdaKeyword{open}\AgdaSpace{}%
\AgdaKeyword{import}\AgdaSpace{}%
%
\AgdaModule{Notation.Flat.Naturals}%
\<%
\\
\>[2]\AgdaKeyword{open}\AgdaSpace{}%
\AgdaKeyword{import}\AgdaSpace{}%
%
\AgdaModule{Notation.Updates}%
\<%
\\
\>[2]\AgdaKeyword{open}\AgdaSpace{}%
\AgdaKeyword{import}\AgdaSpace{}%
%
\AgdaModule{Agda.Builtin.Nat}%
\AgdaSpace{}%
\AgdaKeyword{renaming}\AgdaSpace{}%
\AgdaSymbol{(}\AgdaOperator{\AgdaPrimitive{\AgdaUnderscore{}==\AgdaUnderscore{}}}\AgdaSpace{}%
\AgdaSymbol{to}\AgdaSpace{}%
\AgdaOperator{\AgdaPrimitive{\AgdaUnderscore{}==ᴺ\AgdaUnderscore{}}}\AgdaSymbol{)}\AgdaSpace{}%
\AgdaKeyword{public}\<%
\\
\>[0]\<%
\end{code}
\begin{code}%
\>[0][@{}l@{\AgdaIndent{1}}]%
\>[2]\AgdaKeyword{postulate}\<%
\\
\>[2][@{}l@{\AgdaIndent{0}}]%
\>[4]%
\AgdaPostulate{D∞}%
\AgdaSpace{}%
\AgdaSymbol{:}\AgdaSpace{}%
%
\AgdaPostulate{Domain}%
%
\>[37]\AgdaComment{--\ corresponds\ to\ Scott's\ domain\ }\<%
\\
\>[4]\AgdaKeyword{instance}\AgdaSpace{}%
%
\AgdaPostulate{eqD∞}%
\AgdaSpace{}%
\AgdaSymbol{:}\AgdaSpace{}%
%
\AgdaPostulate{D∞}%
\AgdaSpace{}%
\AgdaOperator{%
\AgdaPostulate{≅}%
}\AgdaSpace{}%
\AgdaSymbol{(}%
\AgdaPostulate{D∞}%
\AgdaSpace{}%
\AgdaOperator{%
\AgdaPostulate{→ᶜ}%
}\AgdaSpace{}%
%
\AgdaPostulate{D∞}%
\AgdaSymbol{)}%
\>[37]\AgdaComment{--\ bijection}\<%
\end{code}
\begin{code}[hide]%
\>[2]\AgdaKeyword{variable}\AgdaSpace{}%
\AgdaGeneralizable{δ}\AgdaSpace{}%
\AgdaSymbol{:}\AgdaSpace{}%
\AgdaOperator{%
\AgdaPostulate{⟪}%
}\AgdaSpace{}%
%
\AgdaPostulate{D∞}%
\AgdaSpace{}%
\AgdaOperator{%
\AgdaPostulate{⟫}%
}\<%
\\
\>[0]\<%
\end{code}
\begin{code}%
\>[0][@{}l@{\AgdaIndent{1}}]%
\>[2]%
\AgdaFunction{Env}%
\AgdaSpace{}%
\AgdaSymbol{=}\AgdaSpace{}%
%
\AgdaDatatype{Var}%
\AgdaSpace{}%
\AgdaOperator{%
\AgdaPostulate{→ˢ}%
}\AgdaSpace{}%
%
\AgdaPostulate{D∞}%
%
\>[37]\AgdaComment{--\ environments}\<%
\end{code}
\begin{code}[hide]%
\>[2]\AgdaKeyword{variable}\AgdaSpace{}%
\AgdaGeneralizable{ρ}\AgdaSpace{}%
\AgdaSymbol{:}\AgdaSpace{}%
\AgdaOperator{%
\AgdaPostulate{⟪}%
}\AgdaSpace{}%
%
\AgdaFunction{Env}%
\AgdaSpace{}%
\AgdaOperator{%
\AgdaPostulate{⟫}%
}\<%
\end{code}
\end{AgdaSuppressSpace}
Use of the conventional notation \AgdaRef{ρ\ {[}\ δ\ /\ v\ {]}} for
updating an environment \AgdaRef{ρ} to map \AgdaRef{v} to \AgdaRef{δ}
requires an equality test for variables; its definition is elided
here.
\begin{AgdaSuppressSpace}
\begin{code}[hide]%
\>[2]\AgdaOperator{%
\AgdaFunction{\AgdaUnderscore{}%
==ⱽ\AgdaUnderscore{}}}\AgdaSpace{}%
\AgdaSymbol{:}\AgdaSpace{}%
%
\AgdaDatatype{Var}%
\AgdaSpace{}%
\AgdaSymbol{→}\AgdaSpace{}%
%
\AgdaDatatype{Var}%
\AgdaSpace{}%
\AgdaSymbol{→}\AgdaSpace{}%
%
\AgdaDatatype{Bool}%
\<%
\\
\>[2]\AgdaSymbol{(}%
\AgdaInductiveConstructor{x}%
\AgdaSpace{}%
\AgdaBound{n}\AgdaSpace{}%
\AgdaOperator{%
\AgdaFunction{==ⱽ}%
}\AgdaSpace{}%
%
\AgdaInductiveConstructor{x}%
\AgdaSpace{}%
\AgdaBound{n′}\AgdaSymbol{)}\AgdaSpace{}%
\AgdaSymbol{=}\AgdaSpace{}%
\AgdaSymbol{(}\AgdaBound{n}\AgdaSpace{}%
\AgdaOperator{\AgdaPrimitive{==ᴺ}}\AgdaSpace{}%
\AgdaBound{n′}\AgdaSymbol{)}\<%
\\
\>[2]\AgdaKeyword{instance}\AgdaSpace{}%
%
\AgdaFunction{eqVar}%
\AgdaSpace{}%
\AgdaSymbol{:}\AgdaSpace{}%
%
\AgdaRecord{Eq}%
\AgdaSpace{}%
%
\AgdaDatatype{Var}%
\<%
\\
\>[2]\AgdaOperator{%
\AgdaField{\AgdaUnderscore{}%
==\AgdaUnderscore{}}}\AgdaSpace{}%
\AgdaSymbol{\{\{}%
\AgdaFunction{eqVar}%
\AgdaSymbol{\}\}}\AgdaSpace{}%
\AgdaSymbol{=}\AgdaSpace{}%
\AgdaOperator{%
\AgdaFunction{\AgdaUnderscore{}%
==ⱽ\AgdaUnderscore{}}}\<%
\end{code}
\end{AgdaSuppressSpace}

\subsubsection{Semantic Functions}\label{Examples/LC/Semantic-Functions-semantic-functions}

The semantic equations below correspond closely to those found in
textbooks on denotational semantics (e.g.,
\cite{Reynolds1998TPL}). In
larger conventional definitions, \AgdaRef{fold} and \AgdaRef{unfold} are
usually left implicit, but Agda does not support this.
\begin{AgdaSuppressSpace}
\begin{code}[hide]%
\>[0]\AgdaSymbol{\{-\#}\AgdaSpace{}%
\AgdaKeyword{OPTIONS}\AgdaSpace{}%
\AgdaPragma{--rewriting}\AgdaSpace{}%
\AgdaPragma{--confluence-check}\AgdaSpace{}%
\AgdaPragma{--lossy-unification}\AgdaSpace{}%
\AgdaSymbol{\#-\}}\<%
\\
\>[0]\<%
\end{code}
\begin{code}%
\>[0]\AgdaKeyword{module}\AgdaSpace{}%
%
\AgdaModule{Examples.LC.Semantic-Functions}%
\AgdaSpace{}%
\AgdaKeyword{where}\<%
\end{code}
\begin{code}[hide]%
\>[0]\<%
\\
\>[0][@{}l@{\AgdaIndent{1}}]%
\>[2]\AgdaKeyword{open}\AgdaSpace{}%
\AgdaKeyword{import}\AgdaSpace{}%
%
\AgdaModule{Examples.LC.Abstract-Syntax}%
\<%
\\
\>[2]\AgdaKeyword{open}\AgdaSpace{}%
\AgdaKeyword{import}\AgdaSpace{}%
%
\AgdaModule{Examples.LC.Domain-Equations}%
\<%
\\
\>[2]\AgdaKeyword{open}\AgdaSpace{}%
\AgdaKeyword{import}\AgdaSpace{}%
%
\AgdaModule{Notation.Domains}%
\<%
\\
\>[2]\AgdaKeyword{open}\AgdaSpace{}%
\AgdaKeyword{import}\AgdaSpace{}%
%
\AgdaModule{Notation.Functions}%
\<%
\\
\>[2]\AgdaKeyword{open}\AgdaSpace{}%
\AgdaKeyword{import}\AgdaSpace{}%
%
\AgdaModule{Notation.Recursion}%
\<%
\\
\>[2]\AgdaKeyword{open}\AgdaSpace{}%
\AgdaKeyword{import}\AgdaSpace{}%
%
\AgdaModule{Notation.Updates}%
\<%
\\
\>[0]\<%
\end{code}
\begin{code}%
\>[0][@{}l@{\AgdaIndent{1}}]%
\>[2]\AgdaOperator{%
\AgdaFunction{⟦\AgdaUnderscore{}%
⟧}}\AgdaSpace{}%
\AgdaSymbol{:}\AgdaSpace{}%
%
\AgdaDatatype{Exp}%
\AgdaSpace{}%
\AgdaSymbol{→}\AgdaSpace{}%
\AgdaOperator{%
\AgdaPostulate{⟪}%
}\AgdaSpace{}%
%
\AgdaFunction{Env}%
\AgdaSpace{}%
\AgdaOperator{%
\AgdaPostulate{→ᶜ}%
}\AgdaSpace{}%
%
\AgdaPostulate{D∞}%
\AgdaSpace{}%
\AgdaOperator{%
\AgdaPostulate{⟫}%
}\<%
\\
\>[2]\AgdaOperator{%
\AgdaFunction{⟦}%
}\AgdaSpace{}%
\AgdaOperator{%
\AgdaInductiveConstructor{var}%
}\AgdaSpace{}%
\AgdaBound{v}\AgdaSpace{}%
\AgdaOperator{%
\AgdaFunction{⟧}%
}\AgdaSpace{}%
\AgdaBound{ρ}%
\>[21]\AgdaSymbol{=}\AgdaSpace{}%
\AgdaBound{ρ}\AgdaSpace{}%
\AgdaBound{v}\<%
\\
\>[2]\AgdaOperator{%
\AgdaFunction{⟦}%
}\AgdaSpace{}%
\AgdaOperator{%
\AgdaInductiveConstructor{⦅λ}%
}\AgdaSpace{}%
\AgdaBound{v}\AgdaSpace{}%
\AgdaOperator{%
\AgdaInductiveConstructor{␣}%
}\AgdaSpace{}%
\AgdaBound{e}\AgdaSpace{}%
\AgdaOperator{%
\AgdaInductiveConstructor{⦆}%
}\AgdaSpace{}%
\AgdaOperator{%
\AgdaFunction{⟧}%
}\AgdaSpace{}%
\AgdaBound{ρ}%
\>[21]\AgdaSymbol{=}\AgdaSpace{}%
%
\AgdaPostulate{fold}%
\AgdaSpace{}%
\AgdaSymbol{(}\AgdaSpace{}%
\AgdaSymbol{λ}\AgdaSpace{}%
\AgdaBound{δ}\AgdaSpace{}%
\AgdaSymbol{→}\AgdaSpace{}%
\AgdaOperator{%
\AgdaFunction{⟦}%
}\AgdaSpace{}%
\AgdaBound{e}\AgdaSpace{}%
\AgdaOperator{%
\AgdaFunction{⟧}%
}\AgdaSpace{}%
\AgdaSymbol{(}\AgdaBound{ρ}\AgdaSpace{}%
\AgdaOperator{%
\AgdaFunction{[}%
}\AgdaSpace{}%
\AgdaBound{δ}\AgdaSpace{}%
\AgdaOperator{%
\AgdaFunction{/}%
}\AgdaSpace{}%
\AgdaBound{v}\AgdaSpace{}%
\AgdaOperator{%
\AgdaFunction{]}%
}\AgdaSymbol{)}\AgdaSpace{}%
\AgdaSymbol{)}\<%
\\
\>[2]\AgdaOperator{%
\AgdaFunction{⟦}%
}\AgdaSpace{}%
\AgdaOperator{%
\AgdaInductiveConstructor{⦅}%
}\AgdaSpace{}%
\AgdaBound{e₁}\AgdaSpace{}%
\AgdaOperator{%
\AgdaInductiveConstructor{␣}%
}\AgdaSpace{}%
\AgdaBound{e₂}\AgdaSpace{}%
\AgdaOperator{%
\AgdaInductiveConstructor{⦆}%
}\AgdaSpace{}%
\AgdaOperator{%
\AgdaFunction{⟧}%
}\AgdaSpace{}%
\AgdaBound{ρ}%
\>[21]\AgdaSymbol{=}\AgdaSpace{}%
%
\AgdaPostulate{unfold}%
\AgdaSpace{}%
\AgdaSymbol{(}\AgdaSpace{}%
\AgdaOperator{%
\AgdaFunction{⟦}%
}\AgdaSpace{}%
\AgdaBound{e₁}\AgdaSpace{}%
\AgdaOperator{%
\AgdaFunction{⟧}%
}\AgdaSpace{}%
\AgdaBound{ρ}\AgdaSpace{}%
\AgdaSymbol{)}\AgdaSpace{}%
\AgdaSymbol{(}\AgdaSpace{}%
\AgdaOperator{%
\AgdaFunction{⟦}%
}\AgdaSpace{}%
\AgdaBound{e₂}\AgdaSpace{}%
\AgdaOperator{%
\AgdaFunction{⟧}%
}\AgdaSpace{}%
\AgdaBound{ρ}\AgdaSpace{}%
\AgdaSymbol{)}\<%
\end{code}
\end{AgdaSuppressSpace}

\end{AgdaAlign}

\begin{AgdaAlign}
\subsection{PCF: A Programming Language for Computable
Functions}\label{Examples/PCF-pcf-a-programming-language-for-computable-functions}

PCF and its denotational semantics were orginally defined by Dana Scott
in 1969
\cite{Scott1993TTA} with
combinators (\AgdaRef{S}, \AgdaRef{K}) instead of \(\lambda\)-abstraction.
Gordon Plotkin subsequently defined a denotational semantics for PCF
including \(\lambda\)-abstraction
\cite{Plotkin1977LCP}.
The Agda modules presented below are an embedding of the latter
definition.

PCF is an intrinsically typed language: every well-formed term has a
unique type. The following grammar summarises the context-free abstract
syntax of types \(\sigma, \tau\) and terms \(M, N\) with variables
\(\alpha_i^\sigma\) (\(i \geq 0\)) and constants \(c\). In
Section~\ref{Examples/PCF/Abstract-Syntax-abstract-syntax} we reflect Plotkin's presentation of PCF more accurately by
exploiting Agda's support for dependent types.  
\begin{align}
  \sigma, \tau  & ::=
    \iota \mid o \mid (\sigma \to \tau)
  \\
  c & ::=
    \textit{tt} \mid \textit{ff} \mid {\supset} \mid \textbf{Y} \mid
    k_n \mid (+1) \mid (-1) \mid \textbf{Z}
  \\
  M, N & ::=
    \alpha_i^\sigma \mid c \mid (M \, N) \mid (\lambda \alpha_i^\sigma M)
\end{align}
\begin{AgdaSuppressSpace}
\begin{code}[hide]%
\>[0]\AgdaSymbol{\{-\#}\AgdaSpace{}%
\AgdaKeyword{OPTIONS}\AgdaSpace{}%
\AgdaPragma{--rewriting}\AgdaSpace{}%
\AgdaPragma{--confluence-check}\AgdaSpace{}%
\AgdaSymbol{\#-\}}\<%
\\
\\[\AgdaEmptyExtraSkip]%
\>[0]\AgdaKeyword{module}\AgdaSpace{}%
%
\AgdaModule{Examples.PCF}%
\AgdaSpace{}%
\AgdaKeyword{where}\<%
\\
\\[\AgdaEmptyExtraSkip]%
\>[0][@{}l@{\AgdaIndent{0}}]%
\>[2]\AgdaKeyword{import}\AgdaSpace{}%
%
\AgdaModule{Examples.PCF.Abstract-Syntax}%
\<%
\\
\>[2]\AgdaKeyword{import}\AgdaSpace{}%
%
\AgdaModule{Examples.PCF.Domain-Equations}%
\<%
\\
\>[2]\AgdaKeyword{import}\AgdaSpace{}%
%
\AgdaModule{Examples.PCF.Semantic-Functions}%
\<%
\end{code}
\end{AgdaSuppressSpace}

\subsubsection{Abstract Syntax}\label{Examples/PCF/Abstract-Syntax-abstract-syntax}

The following abstract syntax of well-formed PCF terms in Agda uses
indexed datatype definitions. PCF function types \(\sigma \to \tau\) are
written \AgdaRef{σ\ ⇒\ τ}, and variables \(\alpha_i^\sigma\) are written
\AgdaRef{α\ i\ σ} (where the argument \AgdaRef{i} merely distinguishes
between variables -- it is \emph{not} a De Bruijn index).
\begin{AgdaSuppressSpace}
\begin{code}[hide]%
\>[0]\AgdaSymbol{\{-\#}\AgdaSpace{}%
\AgdaKeyword{OPTIONS}\AgdaSpace{}%
\AgdaPragma{--rewriting}\AgdaSpace{}%
\AgdaPragma{--confluence-check}\AgdaSpace{}%
\AgdaSymbol{\#-\}}\<%
\\
\>[0]\<%
\end{code}
\begin{code}%
\>[0]\AgdaKeyword{module}\AgdaSpace{}%
%
\AgdaModule{Examples.PCF.Abstract-Syntax}%
\AgdaSpace{}%
\AgdaKeyword{where}\<%
\\
\>[0]\<%
\end{code}
\begin{code}[hide]%
\>[0][@{}l@{\AgdaIndent{1}}]%
\>[2]\AgdaKeyword{open}\AgdaSpace{}%
\AgdaKeyword{import}\AgdaSpace{}%
%
\AgdaModule{Agda.Builtin.Nat}%
\AgdaSpace{}%
\AgdaKeyword{public}\AgdaSpace{}%
\AgdaKeyword{using}\AgdaSpace{}%
\AgdaSymbol{(}%
\AgdaDatatype{Nat}%
\AgdaSymbol{)}\<%
\\
\>[0]\<%
\end{code}
\begin{code}%
\>[0][@{}l@{\AgdaIndent{1}}]%
\>[2]\AgdaKeyword{data}\AgdaSpace{}%
%
\AgdaDatatype{Types}%
%
\>[14]\AgdaSymbol{:}\AgdaSpace{}%
\AgdaPrimitive{Set}\AgdaSpace{}%
\AgdaKeyword{where}%
\>[51]\AgdaComment{--\ type\ terms}\<%
\\
\>[2][@{}l@{\AgdaIndent{0}}]%
\>[4]%
\AgdaInductiveConstructor{ι}%
%
\>[14]\AgdaSymbol{:}\AgdaSpace{}%
%
\AgdaDatatype{Types}%
%
\>[51]\AgdaComment{--\ individuals}\<%
\\
\>[4]%
\AgdaInductiveConstructor{o}%
%
\>[14]\AgdaSymbol{:}\AgdaSpace{}%
%
\AgdaDatatype{Types}%
%
\>[51]\AgdaComment{--\ truth-values}\<%
\\
\>[4]\AgdaOperator{%
\AgdaInductiveConstructor{\AgdaUnderscore{}%
⇒\AgdaUnderscore{}}}%
\>[14]\AgdaSymbol{:}\AgdaSpace{}%
%
\AgdaDatatype{Types}%
\AgdaSpace{}%
\AgdaSymbol{→}\AgdaSpace{}%
%
\AgdaDatatype{Types}%
\AgdaSpace{}%
\AgdaSymbol{→}\AgdaSpace{}%
%
\AgdaDatatype{Types}%
%
\>[51]\AgdaComment{--\ functions}\<%
\end{code}
\begin{code}[hide]%
\>[2]\AgdaKeyword{infixr}\AgdaSpace{}%
\AgdaNumber{1}\AgdaSpace{}%
\AgdaOperator{%
\AgdaInductiveConstructor{\AgdaUnderscore{}%
⇒\AgdaUnderscore{}}}\<%
\\
\>[2]\AgdaKeyword{variable}\AgdaSpace{}%
\AgdaGeneralizable{σ}\AgdaSpace{}%
\AgdaGeneralizable{τ}\AgdaSpace{}%
\AgdaSymbol{:}\AgdaSpace{}%
%
\AgdaDatatype{Types}%
\<%
\\
\>[0]\<%
\end{code}
\begin{code}%
\>[0]\<%
\\
\>[0][@{}l@{\AgdaIndent{1}}]%
\>[2]\AgdaKeyword{data}\AgdaSpace{}%
%
\AgdaDatatype{Vars}%
%
\>[14]\AgdaSymbol{:}\AgdaSpace{}%
%
\AgdaDatatype{Types}%
\AgdaSpace{}%
\AgdaSymbol{→}\AgdaSpace{}%
\AgdaPrimitive{Set}\AgdaSpace{}%
\AgdaKeyword{where}%
\>[51]\AgdaComment{--\ typed\ variables}\<%
\\
\>[2][@{}l@{\AgdaIndent{0}}]%
\>[4]%
\AgdaInductiveConstructor{α}%
%
\>[14]\AgdaSymbol{:}\AgdaSpace{}%
%
\AgdaDatatype{Nat}%
\AgdaSpace{}%
\AgdaSymbol{→}\AgdaSpace{}%
\AgdaSymbol{(}\AgdaBound{σ}\AgdaSpace{}%
\AgdaSymbol{:}\AgdaSpace{}%
%
\AgdaDatatype{Types}%
\AgdaSymbol{)}\AgdaSpace{}%
\AgdaSymbol{→}\AgdaSpace{}%
%
\AgdaDatatype{Vars}%
\AgdaSpace{}%
\AgdaBound{σ}%
\>[51]\AgdaComment{--\ α\ i\ σ\ is\ a\ variable\ of\ type\ σ}\<%
\end{code}
\begin{code}[hide]%
\>[2]\AgdaKeyword{variable}\AgdaSpace{}%
\AgdaGeneralizable{i}%
\>[14]\AgdaSymbol{:}\AgdaSpace{}%
%
\AgdaDatatype{Nat}%
\<%
\end{code}
\begin{code}%
\>[0]\<%
\\
\>[2]\AgdaKeyword{data}\AgdaSpace{}%
%
\AgdaDatatype{ℒᴬ}%
%
\>[14]\AgdaSymbol{:}\AgdaSpace{}%
%
\AgdaDatatype{Types}%
\AgdaSpace{}%
\AgdaSymbol{→}\AgdaSpace{}%
\AgdaPrimitive{Set}\AgdaSpace{}%
\AgdaKeyword{where}%
\>[51]\AgdaComment{--\ typed\ constants}\<%
\\
\>[2][@{}l@{\AgdaIndent{0}}]%
\>[4]%
\AgdaInductiveConstructor{tt}%
%
\>[14]\AgdaSymbol{:}\AgdaSpace{}%
%
\AgdaDatatype{ℒᴬ}%
\AgdaSpace{}%
%
\AgdaInductiveConstructor{o}%
%
\>[51]\AgdaComment{--\ true}\<%
\\
\>[4]%
\AgdaInductiveConstructor{ff}%
%
\>[14]\AgdaSymbol{:}\AgdaSpace{}%
%
\AgdaDatatype{ℒᴬ}%
\AgdaSpace{}%
%
\AgdaInductiveConstructor{o}%
%
\>[51]\AgdaComment{--\ false}\<%
\\
\>[4]%
\AgdaInductiveConstructor{⊃}%
%
\>[14]\AgdaSymbol{:}\AgdaSpace{}%
%
\AgdaDatatype{ℒᴬ}%
\AgdaSpace{}%
\AgdaSymbol{(}%
\AgdaInductiveConstructor{o}%
\AgdaSpace{}%
\AgdaOperator{%
\AgdaInductiveConstructor{⇒}%
}\AgdaSpace{}%
\AgdaGeneralizable{σ}\AgdaSpace{}%
\AgdaOperator{%
\AgdaInductiveConstructor{⇒}%
}\AgdaSpace{}%
\AgdaGeneralizable{σ}\AgdaSpace{}%
\AgdaOperator{%
\AgdaInductiveConstructor{⇒}%
}\AgdaSpace{}%
\AgdaGeneralizable{σ}\AgdaSymbol{)}%
\>[51]\AgdaComment{--\ conditional}\<%
\\
\>[4]%
\AgdaInductiveConstructor{Y}%
%
\>[14]\AgdaSymbol{:}\AgdaSpace{}%
%
\AgdaDatatype{ℒᴬ}%
\AgdaSpace{}%
\AgdaSymbol{((}\AgdaGeneralizable{σ}\AgdaSpace{}%
\AgdaOperator{%
\AgdaInductiveConstructor{⇒}%
}\AgdaSpace{}%
\AgdaGeneralizable{σ}\AgdaSymbol{)}\AgdaSpace{}%
\AgdaOperator{%
\AgdaInductiveConstructor{⇒}%
}\AgdaSpace{}%
\AgdaGeneralizable{σ}\AgdaSymbol{)}%
\>[51]\AgdaComment{--\ fixed\ point}\<%
\\
\>[4]%
\AgdaInductiveConstructor{k}%
%
\>[14]\AgdaSymbol{:}\AgdaSpace{}%
%
\AgdaDatatype{Nat}%
\AgdaSpace{}%
\AgdaSymbol{→}\AgdaSpace{}%
%
\AgdaDatatype{ℒᴬ}%
\AgdaSpace{}%
%
\AgdaInductiveConstructor{ι}%
%
\>[51]\AgdaComment{--\ numerals}\<%
\\
\>[4]%
\AgdaInductiveConstructor{⦅+1⦆}%
%
\>[14]\AgdaSymbol{:}\AgdaSpace{}%
%
\AgdaDatatype{ℒᴬ}%
\AgdaSpace{}%
\AgdaSymbol{(}%
\AgdaInductiveConstructor{ι}%
\AgdaSpace{}%
\AgdaOperator{%
\AgdaInductiveConstructor{⇒}%
}\AgdaSpace{}%
%
\AgdaInductiveConstructor{ι}%
\AgdaSymbol{)}%
\>[51]\AgdaComment{--\ successor}\<%
\\
\>[4]%
\AgdaInductiveConstructor{⦅−1⦆}%
%
\>[14]\AgdaSymbol{:}\AgdaSpace{}%
%
\AgdaDatatype{ℒᴬ}%
\AgdaSpace{}%
\AgdaSymbol{(}%
\AgdaInductiveConstructor{ι}%
\AgdaSpace{}%
\AgdaOperator{%
\AgdaInductiveConstructor{⇒}%
}\AgdaSpace{}%
%
\AgdaInductiveConstructor{ι}%
\AgdaSymbol{)}%
\>[51]\AgdaComment{--\ predecessor}\<%
\\
\>[4]%
\AgdaInductiveConstructor{Z}%
%
\>[14]\AgdaSymbol{:}\AgdaSpace{}%
%
\AgdaDatatype{ℒᴬ}%
\AgdaSpace{}%
\AgdaSymbol{(}%
\AgdaInductiveConstructor{ι}%
\AgdaSpace{}%
\AgdaOperator{%
\AgdaInductiveConstructor{⇒}%
}\AgdaSpace{}%
%
\AgdaInductiveConstructor{o}%
\AgdaSymbol{)}%
\>[51]\AgdaComment{--\ zero\ test}\<%
\end{code}
\begin{code}[hide]%
\>[2]\AgdaKeyword{variable}\AgdaSpace{}%
\AgdaGeneralizable{c}%
\>[14]\AgdaSymbol{:}\AgdaSpace{}%
%
\AgdaDatatype{ℒᴬ}%
\AgdaSpace{}%
\AgdaGeneralizable{σ}\<%
\end{code}
\begin{code}%
\>[0]\<%
\\
\>[2]\AgdaKeyword{data}\AgdaSpace{}%
%
\AgdaDatatype{Terms}%
%
\>[14]\AgdaSymbol{:}\AgdaSpace{}%
%
\AgdaDatatype{Types}%
\AgdaSpace{}%
\AgdaSymbol{→}\AgdaSpace{}%
\AgdaPrimitive{Set}\AgdaSpace{}%
\AgdaKeyword{where}%
\>[51]\AgdaComment{--\ typed\ terms}\<%
\\
\>[2][@{}l@{\AgdaIndent{0}}]%
\>[4]\AgdaOperator{%
\AgdaInductiveConstructor{𝑉\AgdaUnderscore{}%
}}%
\>[14]\AgdaSymbol{:}\AgdaSpace{}%
%
\AgdaDatatype{Vars}%
\AgdaSpace{}%
\AgdaGeneralizable{σ}\AgdaSpace{}%
\AgdaSymbol{→}\AgdaSpace{}%
%
\AgdaDatatype{Terms}%
\AgdaSpace{}%
\AgdaGeneralizable{σ}%
\>[51]\AgdaComment{--\ variable}\<%
\\
\>[4]\AgdaOperator{%
\AgdaInductiveConstructor{𝐿\AgdaUnderscore{}%
}}%
\>[14]\AgdaSymbol{:}\AgdaSpace{}%
%
\AgdaDatatype{ℒᴬ}%
\AgdaSpace{}%
\AgdaGeneralizable{σ}\AgdaSpace{}%
\AgdaSymbol{→}\AgdaSpace{}%
%
\AgdaDatatype{Terms}%
\AgdaSpace{}%
\AgdaGeneralizable{σ}%
\>[51]\AgdaComment{--\ constant}\<%
\\
\>[4]\AgdaOperator{%
\AgdaInductiveConstructor{⦅\AgdaUnderscore{}%
␣\AgdaUnderscore{}⦆}}%
\>[14]\AgdaSymbol{:}\AgdaSpace{}%
%
\AgdaDatatype{Terms}%
\AgdaSpace{}%
\AgdaSymbol{(}\AgdaGeneralizable{σ}\AgdaSpace{}%
\AgdaOperator{%
\AgdaInductiveConstructor{⇒}%
}\AgdaSpace{}%
\AgdaGeneralizable{τ}\AgdaSymbol{)}\AgdaSpace{}%
\AgdaSymbol{→}\AgdaSpace{}%
%
\AgdaDatatype{Terms}%
\AgdaSpace{}%
\AgdaGeneralizable{σ}\AgdaSpace{}%
\AgdaSymbol{→}\AgdaSpace{}%
%
\AgdaDatatype{Terms}%
\AgdaSpace{}%
\AgdaGeneralizable{τ}%
\>[51]\AgdaComment{--\ function\ application}\<%
\\
\>[4]\AgdaOperator{%
\AgdaInductiveConstructor{⦅λ\AgdaUnderscore{}%
␣\AgdaUnderscore{}⦆}}%
\>[14]\AgdaSymbol{:}\AgdaSpace{}%
%
\AgdaDatatype{Vars}%
\AgdaSpace{}%
\AgdaGeneralizable{σ}\AgdaSpace{}%
\AgdaSymbol{→}\AgdaSpace{}%
%
\AgdaDatatype{Terms}%
\AgdaSpace{}%
\AgdaGeneralizable{τ}\AgdaSpace{}%
\AgdaSymbol{→}\AgdaSpace{}%
%
\AgdaDatatype{Terms}%
\AgdaSpace{}%
\AgdaSymbol{(}\AgdaGeneralizable{σ}\AgdaSpace{}%
\AgdaOperator{%
\AgdaInductiveConstructor{⇒}%
}\AgdaSpace{}%
\AgdaGeneralizable{τ}\AgdaSymbol{)}%
\>[51]\AgdaComment{--\ function\ abstraction}\<%
\end{code}
\begin{code}[hide]%
\>[2]\AgdaKeyword{variable}\AgdaSpace{}%
\AgdaGeneralizable{M}\AgdaSpace{}%
\AgdaGeneralizable{N}\AgdaSpace{}%
\AgdaSymbol{:}\AgdaSpace{}%
%
\AgdaDatatype{Terms}%
\AgdaSpace{}%
\AgdaGeneralizable{σ}\<%
\end{code}
\end{AgdaSuppressSpace}

\subsubsection{Domain Equations}\label{Examples/PCF/Domain-Equations-domain-equations}

The domains \AgdaRef{𝒟\ σ} form a ``\emph{standard collection of domains
for arithmetic}'' in PCF, written \(\mathcal D_\sigma\) in
\cite{Plotkin1977LCP}.
As PCF is a simply-typed language, the domains \AgdaRef{𝒟\ σ} are not
reflexive, so their embedding in Agda can use ordinary type definitions,
not involving bijections.
\begin{AgdaSuppressSpace}
\begin{code}[hide]%
\>[0]\AgdaSymbol{\{-\#}\AgdaSpace{}%
\AgdaKeyword{OPTIONS}\AgdaSpace{}%
\AgdaPragma{--rewriting}\AgdaSpace{}%
\AgdaPragma{--confluence-check}\AgdaSpace{}%
\AgdaPragma{--lossy-unification}\AgdaSpace{}%
\AgdaSymbol{\#-\}}\<%
\\
\>[0]\<%
\end{code}
\begin{code}%
\>[0]\AgdaKeyword{module}\AgdaSpace{}%
%
\AgdaModule{Examples.PCF.Domain-Equations}%
\AgdaSpace{}%
\AgdaKeyword{where}\<%
\end{code}
\begin{code}[hide]%
\>[0]\<%
\\
\>[0][@{}l@{\AgdaIndent{1}}]%
\>[2]\AgdaKeyword{open}\AgdaSpace{}%
\AgdaKeyword{import}\AgdaSpace{}%
%
\AgdaModule{Examples.PCF.Abstract-Syntax}%
\<%
\\
\>[2]\AgdaKeyword{open}\AgdaSpace{}%
\AgdaKeyword{import}\AgdaSpace{}%
%
\AgdaModule{Notation.Domains}%
\<%
\\
\>[2]\AgdaKeyword{open}\AgdaSpace{}%
\AgdaKeyword{import}\AgdaSpace{}%
%
\AgdaModule{Notation.Functions}%
\<%
\\
\>[2]\AgdaKeyword{open}\AgdaSpace{}%
\AgdaKeyword{import}\AgdaSpace{}%
%
\AgdaModule{Notation.Flat.Booleans}%
\<%
\\
\>[2]\AgdaKeyword{open}\AgdaSpace{}%
\AgdaKeyword{import}\AgdaSpace{}%
%
\AgdaModule{Notation.Flat.Naturals}%
\<%
\\
\>[2]\AgdaKeyword{open}\AgdaSpace{}%
\AgdaKeyword{import}\AgdaSpace{}%
%
\AgdaModule{Notation.Updates}%
\<%
\\
\>[2]\AgdaKeyword{open}\AgdaSpace{}%
\AgdaKeyword{import}\AgdaSpace{}%
%
\AgdaModule{Agda.Builtin.Nat}%
\AgdaSpace{}%
\AgdaKeyword{renaming}\AgdaSpace{}%
\AgdaSymbol{(}\AgdaOperator{\AgdaPrimitive{\AgdaUnderscore{}==\AgdaUnderscore{}}}\AgdaSpace{}%
\AgdaSymbol{to}\AgdaSpace{}%
\AgdaOperator{\AgdaPrimitive{\AgdaUnderscore{}==ᴺ\AgdaUnderscore{}}}\AgdaSymbol{)}\AgdaSpace{}%
\AgdaKeyword{public}\<%
\end{code}
\begin{code}%
\>[0]\<%
\\
\>[2]%
\AgdaFunction{𝒟}%
\AgdaSpace{}%
\AgdaSymbol{:}\AgdaSpace{}%
%
\AgdaDatatype{Types}%
\AgdaSpace{}%
\AgdaSymbol{→}\AgdaSpace{}%
%
\AgdaPostulate{Domain}%
%
\>[27]\AgdaComment{--\ standard\ domains}\<%
\\
\>[2]%
\AgdaFunction{𝒟}%
\AgdaSpace{}%
%
\AgdaInductiveConstructor{ι}%
%
\>[13]\AgdaSymbol{=}\AgdaSpace{}%
%
\AgdaFunction{Nat⊥}%
%
\>[27]\AgdaComment{--\ natural\ numbers}\<%
\\
\>[2]%
\AgdaFunction{𝒟}%
\AgdaSpace{}%
%
\AgdaInductiveConstructor{o}%
%
\>[13]\AgdaSymbol{=}\AgdaSpace{}%
%
\AgdaFunction{Bool⊥}%
%
\>[27]\AgdaComment{--\ truth-values}\<%
\\
\>[2]%
\AgdaFunction{𝒟}%
\AgdaSpace{}%
\AgdaSymbol{(}\AgdaBound{σ}\AgdaSpace{}%
\AgdaOperator{%
\AgdaInductiveConstructor{⇒}%
}\AgdaSpace{}%
\AgdaBound{τ}\AgdaSymbol{)}%
\>[13]\AgdaSymbol{=}\AgdaSpace{}%
%
\AgdaFunction{𝒟}%
\AgdaSpace{}%
\AgdaBound{σ}\AgdaSpace{}%
\AgdaOperator{%
\AgdaPostulate{→ᶜ}%
}\AgdaSpace{}%
%
\AgdaFunction{𝒟}%
\AgdaSpace{}%
\AgdaBound{τ}%
\>[27]\AgdaComment{--\ functions}\<%
\end{code}
\begin{code}[hide]%
\>[2]\AgdaKeyword{variable}\AgdaSpace{}%
\AgdaGeneralizable{x}\AgdaSpace{}%
\AgdaGeneralizable{y}\AgdaSpace{}%
\AgdaGeneralizable{z}\AgdaSpace{}%
\AgdaSymbol{:}\AgdaSpace{}%
\AgdaOperator{%
\AgdaPostulate{⟪}%
}\AgdaSpace{}%
%
\AgdaFunction{𝒟}%
\AgdaSpace{}%
\AgdaGeneralizable{σ}\AgdaSpace{}%
\AgdaOperator{%
\AgdaPostulate{⟫}%
}\<%
\end{code}
\end{AgdaSuppressSpace}
Environments \AgdaRef{ρ} are type-preserving maps from variables to
values. They are naturally modeled by a dependent type: \AgdaRef{Env\ σ}
consists of type-preserving maps from variables in \AgdaRef{Vars\ σ} to
their values in the domain \AgdaRef{𝒟\ σ}. The environment \AgdaRef{ρ⊥}
maps all variables to \AgdaRef{⊥}.
\begin{AgdaSuppressSpace}
\begin{code}%
\>[2]%
\AgdaFunction{Env}%
\AgdaSpace{}%
\AgdaSymbol{=}\AgdaSpace{}%
\AgdaSymbol{(}\AgdaBound{σ}\AgdaSpace{}%
\AgdaSymbol{:}\AgdaSpace{}%
%
\AgdaDatatype{Types}%
\AgdaSymbol{)}\AgdaSpace{}%
\AgdaSymbol{→}\AgdaSpace{}%
\AgdaOperator{%
\AgdaPostulate{⟪}%
}\AgdaSpace{}%
%
\AgdaDatatype{Vars}%
\AgdaSpace{}%
\AgdaBound{σ}\AgdaSpace{}%
\AgdaOperator{%
\AgdaPostulate{→ˢ}%
}\AgdaSpace{}%
%
\AgdaFunction{𝒟}%
\AgdaSpace{}%
\AgdaBound{σ}\AgdaSpace{}%
\AgdaOperator{%
\AgdaPostulate{⟫}%
}%
\>[41]\AgdaComment{--\ typed\ environments}\<%
\end{code}
\begin{code}[hide]%
\>[2]\AgdaKeyword{variable}\AgdaSpace{}%
\AgdaGeneralizable{ρ}\AgdaSpace{}%
\AgdaSymbol{:}\AgdaSpace{}%
%
\AgdaFunction{Env}%
\<%
\end{code}
\begin{code}%
\>[2]%
\AgdaFunction{ρ⊥}%
\AgdaSpace{}%
\AgdaSymbol{:}\AgdaSpace{}%
%
\AgdaFunction{Env}%
%
\>[41]\AgdaComment{--\ initial\ environment}\<%
\\
\>[2]%
\AgdaFunction{ρ⊥}%
\AgdaSpace{}%
\AgdaSymbol{\AgdaUnderscore{}}\AgdaSpace{}%
\AgdaSymbol{\AgdaUnderscore{}}\AgdaSpace{}%
\AgdaSymbol{=}\AgdaSpace{}%
%
\AgdaPostulate{⊥}%
\<%
\end{code}
\end{AgdaSuppressSpace}
Extension or overriding typed environments, written
\AgdaRef{ρ\ {[}\ v\ /\ x\ {]}′} here, requires instances of the equality
tests for both variables and types. The definition of the latter is
somewhat tedious:  it is omitted here, but can found in the Agda
code repository and website
\cite{MFPS2026-Agda}.

\begin{AgdaSuppressSpace}
\begin{code}[hide]%
\>[2]\AgdaOperator{%
\AgdaFunction{\AgdaUnderscore{}%
==ⱽ\AgdaUnderscore{}}}\AgdaSpace{}%
\AgdaSymbol{:}\AgdaSpace{}%
%
\AgdaDatatype{Vars}%
\AgdaSpace{}%
\AgdaGeneralizable{σ}\AgdaSpace{}%
\AgdaSymbol{→}\AgdaSpace{}%
%
\AgdaDatatype{Vars}%
\AgdaSpace{}%
\AgdaGeneralizable{σ}\AgdaSpace{}%
\AgdaSymbol{→}\AgdaSpace{}%
%
\AgdaDatatype{Bool}%
\<%
\\
\>[2]\AgdaSymbol{(}%
\AgdaInductiveConstructor{α}%
\AgdaSpace{}%
\AgdaBound{i}\AgdaSpace{}%
\AgdaBound{σ}\AgdaSpace{}%
\AgdaOperator{%
\AgdaFunction{==ⱽ}%
}\AgdaSpace{}%
%
\AgdaInductiveConstructor{α}%
\AgdaSpace{}%
\AgdaBound{i′}\AgdaSpace{}%
\AgdaBound{σ}\AgdaSymbol{)}%
\>[22]\AgdaSymbol{=}%
\>[25]\AgdaSymbol{(}\AgdaBound{i}\AgdaSpace{}%
\AgdaOperator{\AgdaPrimitive{==ᴺ}}\AgdaSpace{}%
\AgdaBound{i′}\AgdaSymbol{)}\<%
\\
\>[2]\AgdaKeyword{instance}\<%
\\
\>[2][@{}l@{\AgdaIndent{0}}]%
\>[4]%
\AgdaFunction{eqV}%
\AgdaSpace{}%
\AgdaSymbol{:}\AgdaSpace{}%
%
\AgdaRecord{Eq}%
\AgdaSpace{}%
\AgdaSymbol{(}%
\AgdaDatatype{Vars}%
\AgdaSpace{}%
\AgdaGeneralizable{σ}\AgdaSymbol{)}\<%
\\
\>[4]\AgdaOperator{%
\AgdaField{\AgdaUnderscore{}%
==\AgdaUnderscore{}}}\AgdaSpace{}%
\AgdaSymbol{\{\{}%
\AgdaFunction{eqV}%
\AgdaSymbol{\}\}}\AgdaSpace{}%
\AgdaSymbol{=}\AgdaSpace{}%
\AgdaOperator{%
\AgdaFunction{\AgdaUnderscore{}%
==ⱽ\AgdaUnderscore{}}}\<%
\\
\>[2]\AgdaKeyword{instance}\<%
\\
\>[2][@{}l@{\AgdaIndent{0}}]%
\>[4]%
\AgdaFunction{eqT}%
\AgdaSpace{}%
\AgdaSymbol{:}\AgdaSpace{}%
%
\AgdaRecord{MaybeEq}%
\AgdaSpace{}%
%
\AgdaDatatype{Types}%
\<%
\\
\>[4]%
\AgdaFunction{eqT}%
\AgdaSpace{}%
\AgdaSymbol{.}\AgdaOperator{%
\AgdaField{\AgdaUnderscore{}%
==?\AgdaUnderscore{}}}\AgdaSpace{}%
%
\AgdaInductiveConstructor{ι}%
\AgdaSpace{}%
%
\AgdaInductiveConstructor{ι}%
\AgdaSpace{}%
\AgdaSymbol{=}\AgdaSpace{}%
%
\AgdaInductiveConstructor{just}%
\AgdaSpace{}%
%
\AgdaInductiveConstructor{refl}%
\<%
\\
\>[4]%
\AgdaFunction{eqT}%
\AgdaSpace{}%
\AgdaSymbol{.}\AgdaOperator{%
\AgdaField{\AgdaUnderscore{}%
==?\AgdaUnderscore{}}}\AgdaSpace{}%
%
\AgdaInductiveConstructor{o}%
\AgdaSpace{}%
%
\AgdaInductiveConstructor{o}%
\AgdaSpace{}%
\AgdaSymbol{=}\AgdaSpace{}%
%
\AgdaInductiveConstructor{just}%
\AgdaSpace{}%
%
\AgdaInductiveConstructor{refl}%
\<%
\\
\>[4]%
\AgdaFunction{eqT}%
\AgdaSpace{}%
\AgdaSymbol{.}\AgdaOperator{%
\AgdaField{\AgdaUnderscore{}%
==?\AgdaUnderscore{}}}\AgdaSpace{}%
\AgdaSymbol{(}\AgdaBound{σ}\AgdaSpace{}%
\AgdaOperator{%
\AgdaInductiveConstructor{⇒}%
}\AgdaSpace{}%
\AgdaBound{τ}\AgdaSymbol{)}\AgdaSpace{}%
\AgdaSymbol{(}\AgdaBound{σ₁}\AgdaSpace{}%
\AgdaOperator{%
\AgdaInductiveConstructor{⇒}%
}\AgdaSpace{}%
\AgdaBound{τ₁}\AgdaSymbol{)}\AgdaSpace{}%
\AgdaKeyword{with}\AgdaSpace{}%
\AgdaBound{σ}\AgdaSpace{}%
\AgdaOperator{%
\AgdaField{==?}%
}\AgdaSpace{}%
\AgdaBound{σ₁}%
\>[48]\AgdaSymbol{|}\AgdaSpace{}%
\AgdaBound{τ}\AgdaSpace{}%
\AgdaOperator{%
\AgdaField{==?}%
}\AgdaSpace{}%
\AgdaBound{τ₁}\<%
\\
\>[4]%
\AgdaFunction{eqT}%
\AgdaSpace{}%
\AgdaSymbol{.}\AgdaOperator{%
\AgdaField{\AgdaUnderscore{}%
==?\AgdaUnderscore{}}}\AgdaSpace{}%
\AgdaSymbol{(}\AgdaBound{σ}\AgdaSpace{}%
\AgdaOperator{%
\AgdaInductiveConstructor{⇒}%
}\AgdaSpace{}%
\AgdaBound{τ}\AgdaSymbol{)}\AgdaSpace{}%
\AgdaSymbol{(}\AgdaBound{σ₁}\AgdaSpace{}%
\AgdaOperator{%
\AgdaInductiveConstructor{⇒}%
}\AgdaSpace{}%
\AgdaBound{τ₁}\AgdaSymbol{)}%
\>[36]\AgdaSymbol{|}\AgdaSpace{}%
%
\AgdaInductiveConstructor{just}%
\AgdaSpace{}%
%
\AgdaInductiveConstructor{refl}%
\AgdaSpace{}%
\AgdaSymbol{|}\AgdaSpace{}%
%
\AgdaInductiveConstructor{just}%
\AgdaSpace{}%
%
\AgdaInductiveConstructor{refl}%
\AgdaSpace{}%
\AgdaSymbol{=}\AgdaSpace{}%
%
\AgdaInductiveConstructor{just}%
\AgdaSpace{}%
%
\AgdaInductiveConstructor{refl}%
\<%
\\
\>[4]\AgdaCatchallClause{%
\AgdaFunction{eqT}%
}\AgdaSpace{}%
\AgdaCatchallClause{\AgdaSymbol{.}}\AgdaCatchallClause{\AgdaOperator{%
\AgdaField{\AgdaUnderscore{}%
==?\AgdaUnderscore{}}}}\AgdaSpace{}%
\AgdaCatchallClause{\AgdaSymbol{(}}\AgdaCatchallClause{\AgdaBound{σ}}\AgdaSpace{}%
\AgdaCatchallClause{\AgdaOperator{%
\AgdaInductiveConstructor{⇒}%
}}\AgdaSpace{}%
\AgdaCatchallClause{\AgdaBound{τ}}\AgdaCatchallClause{\AgdaSymbol{)}}\AgdaSpace{}%
\AgdaCatchallClause{\AgdaSymbol{(}}\AgdaCatchallClause{\AgdaBound{σ₁}}\AgdaSpace{}%
\AgdaCatchallClause{\AgdaOperator{%
\AgdaInductiveConstructor{⇒}%
}}\AgdaSpace{}%
\AgdaCatchallClause{\AgdaBound{τ₁}}\AgdaCatchallClause{\AgdaSymbol{)}}%
\>[36]\AgdaCatchallClause{\AgdaSymbol{|}}\AgdaSpace{}%
\AgdaCatchallClause{\AgdaSymbol{\AgdaUnderscore{}}}%
\>[48]\AgdaCatchallClause{\AgdaSymbol{|}}\AgdaSpace{}%
\AgdaCatchallClause{\AgdaSymbol{\AgdaUnderscore{}}}%
\>[60]\AgdaSymbol{=}\AgdaSpace{}%
%
\AgdaInductiveConstructor{nothing}%
\<%
\\
\>[4]\AgdaCatchallClause{%
\AgdaFunction{eqT}%
}\AgdaSpace{}%
\AgdaCatchallClause{\AgdaSymbol{.}}\AgdaCatchallClause{\AgdaOperator{%
\AgdaField{\AgdaUnderscore{}%
==?\AgdaUnderscore{}}}}\AgdaSpace{}%
\AgdaCatchallClause{\AgdaSymbol{\AgdaUnderscore{}}}\AgdaSpace{}%
\AgdaCatchallClause{\AgdaSymbol{\AgdaUnderscore{}}}\AgdaSpace{}%
\AgdaSymbol{=}\AgdaSpace{}%
%
\AgdaInductiveConstructor{nothing}%
\<%
\\
\\[\AgdaEmptyExtraSkip]%
\>[2]\AgdaOperator{%
\AgdaFunction{\AgdaUnderscore{}%
[\AgdaUnderscore{}/\AgdaUnderscore{}]′}}\AgdaSpace{}%
\AgdaSymbol{:}\AgdaSpace{}%
%
\AgdaFunction{Env}%
\AgdaSpace{}%
\AgdaSymbol{→}\AgdaSpace{}%
\AgdaOperator{%
\AgdaPostulate{⟪}%
}\AgdaSpace{}%
%
\AgdaFunction{𝒟}%
\AgdaSpace{}%
\AgdaGeneralizable{σ}\AgdaSpace{}%
\AgdaOperator{%
\AgdaPostulate{⟫}%
}\AgdaSpace{}%
\AgdaSymbol{→}\AgdaSpace{}%
%
\AgdaDatatype{Vars}%
\AgdaSpace{}%
\AgdaGeneralizable{σ}\AgdaSpace{}%
\AgdaSymbol{→}\AgdaSpace{}%
%
\AgdaFunction{Env}%
\<%
\\
\>[2]\AgdaComment{--\ ρ\ [\ v\ /\ x\ ]′\ maps\ x\ to\ v,\ and\ other\ x′\ to\ ρ\ x′}\<%
\\
\>[2]\AgdaOperator{%
\AgdaFunction{\AgdaUnderscore{}%
[\AgdaUnderscore{}/\AgdaUnderscore{}]′}}\AgdaSpace{}%
\AgdaSymbol{\{}\AgdaBound{σ}\AgdaSymbol{\}}\AgdaSpace{}%
\AgdaBound{ρ}\AgdaSpace{}%
\AgdaBound{x}\AgdaSpace{}%
\AgdaBound{v}\AgdaSpace{}%
\AgdaSymbol{=}\AgdaSpace{}%
\AgdaBound{ρ}\AgdaSpace{}%
\AgdaOperator{%
\AgdaFunction{[}%
}\AgdaSpace{}%
\AgdaBound{σ}\AgdaSpace{}%
\AgdaOperator{%
\AgdaFunction{←}%
}\AgdaSpace{}%
\AgdaBound{ρ}\AgdaSpace{}%
\AgdaBound{σ}\AgdaSpace{}%
\AgdaOperator{%
\AgdaFunction{[}%
}\AgdaSpace{}%
\AgdaBound{x}\AgdaSpace{}%
\AgdaOperator{%
\AgdaFunction{/}%
}\AgdaSpace{}%
\AgdaBound{v}\AgdaSpace{}%
\AgdaOperator{%
\AgdaFunction{]}%
}\AgdaSpace{}%
\AgdaOperator{%
\AgdaFunction{]}%
}\<%
\end{code}
\end{AgdaSuppressSpace}

\subsubsection{Semantic functions}\label{Examples/PCF/Semantic-Functions-semantic-functions}

The notation \AgdaRef{ρ\ ⟦\ α\ i\ σ\ ⟧} gives the value of the variable
\AgdaRef{α\ i\ σ} in \AgdaRef{ρ} by applying \AgdaRef{ρ\ σ} to the
variable.
\begin{AgdaSuppressSpace}
\begin{code}[hide]%
\>[0]\AgdaSymbol{\{-\#}\AgdaSpace{}%
\AgdaKeyword{OPTIONS}\AgdaSpace{}%
\AgdaPragma{--rewriting}\AgdaSpace{}%
\AgdaPragma{--confluence-check}\AgdaSpace{}%
\AgdaPragma{--lossy-unification}\AgdaSpace{}%
\AgdaSymbol{\#-\}}\<%
\\
\>[0]\<%
\end{code}
\begin{code}%
\>[0]\AgdaKeyword{module}\AgdaSpace{}%
%
\AgdaModule{Examples.PCF.Semantic-Functions}%
\AgdaSpace{}%
\AgdaKeyword{where}\<%
\end{code}
\begin{code}[hide]%
\>[0]\<%
\\
\>[0][@{}l@{\AgdaIndent{1}}]%
\>[2]\AgdaKeyword{open}\AgdaSpace{}%
\AgdaKeyword{import}\AgdaSpace{}%
%
\AgdaModule{Examples.PCF.Abstract-Syntax}%
\<%
\\
\>[2]\AgdaKeyword{open}\AgdaSpace{}%
\AgdaKeyword{import}\AgdaSpace{}%
%
\AgdaModule{Examples.PCF.Domain-Equations}%
\<%
\\
\>[2]\AgdaKeyword{open}\AgdaSpace{}%
\AgdaKeyword{import}\AgdaSpace{}%
%
\AgdaModule{Notation.Domains}%
\<%
\\
\>[2]\AgdaKeyword{open}\AgdaSpace{}%
\AgdaKeyword{import}\AgdaSpace{}%
%
\AgdaModule{Notation.Functions}%
\<%
\\
\>[2]\AgdaKeyword{open}\AgdaSpace{}%
\AgdaKeyword{import}\AgdaSpace{}%
%
\AgdaModule{Notation.Flat}%
\<%
\\
\>[2]\AgdaKeyword{open}\AgdaSpace{}%
\AgdaKeyword{import}\AgdaSpace{}%
%
\AgdaModule{Notation.Flat.Booleans}%
\<%
\\
\>[2]\AgdaKeyword{open}\AgdaSpace{}%
\AgdaKeyword{import}\AgdaSpace{}%
%
\AgdaModule{Notation.Flat.Naturals}%
\<%
\\
\>[0]\<%
\end{code}
\begin{code}%
\>[0]\<%
\\
\>[0][@{}l@{\AgdaIndent{1}}]%
\>[2]\AgdaOperator{%
\AgdaFunction{\AgdaUnderscore{}%
⟦\AgdaUnderscore{}⟧}}\AgdaSpace{}%
\AgdaSymbol{:}\AgdaSpace{}%
%
\AgdaFunction{Env}%
\AgdaSpace{}%
\AgdaSymbol{→}\AgdaSpace{}%
%
\AgdaDatatype{Vars}%
\AgdaSpace{}%
\AgdaGeneralizable{σ}\AgdaSpace{}%
\AgdaSymbol{→}\AgdaSpace{}%
\AgdaOperator{%
\AgdaPostulate{⟪}%
}\AgdaSpace{}%
%
\AgdaFunction{𝒟}%
\AgdaSpace{}%
\AgdaGeneralizable{σ}\AgdaSpace{}%
\AgdaOperator{%
\AgdaPostulate{⟫}%
}%
\>[33]\AgdaComment{--\ typed\ variable\ denotations}\<%
\\
\>[2]\AgdaBound{ρ}\AgdaSpace{}%
\AgdaOperator{%
\AgdaFunction{⟦}%
}\AgdaSpace{}%
%
\AgdaInductiveConstructor{α}%
\AgdaSpace{}%
\AgdaBound{i}\AgdaSpace{}%
\AgdaBound{σ}\AgdaSpace{}%
\AgdaOperator{%
\AgdaFunction{⟧}%
}\AgdaSpace{}%
\AgdaSymbol{=}\AgdaSpace{}%
\AgdaBound{ρ}\AgdaSpace{}%
\AgdaBound{σ}\AgdaSpace{}%
\AgdaSymbol{(}%
\AgdaInductiveConstructor{α}%
\AgdaSpace{}%
\AgdaBound{i}\AgdaSpace{}%
\AgdaBound{σ}\AgdaSymbol{)}\<%
\end{code}
\end{AgdaSuppressSpace}
The semantic function \AgdaRef{𝒜⟦\ c\ ⟧} gives the standard
interpretation of the constant \AgdaRef{c} (making the injections and
lifting for flat domains explicit). The corresponding definitions in
\cite{Plotkin1977LCP}
use case analysis on the domain \AgdaRef{𝒟\ ι}, which our Agda embedding
does not support (partly because it can express non-continuous
functions).
\begin{AgdaSuppressSpace}
\begin{code}%
\>[2]\AgdaOperator{%
\AgdaFunction{𝒜⟦\AgdaUnderscore{}%
⟧}}\AgdaSpace{}%
\AgdaSymbol{:}\AgdaSpace{}%
%
\AgdaDatatype{ℒᴬ}%
\AgdaSpace{}%
\AgdaGeneralizable{σ}\AgdaSpace{}%
\AgdaSymbol{→}\AgdaSpace{}%
\AgdaOperator{%
\AgdaPostulate{⟪}%
}\AgdaSpace{}%
%
\AgdaFunction{𝒟}%
\AgdaSpace{}%
\AgdaGeneralizable{σ}\AgdaSpace{}%
\AgdaOperator{%
\AgdaPostulate{⟫}%
}%
\>[25]\AgdaComment{--\ typed\ constant\ denotations}\<%
\\
\>[2]\AgdaOperator{%
\AgdaFunction{𝒜⟦}%
}\AgdaSpace{}%
%
\AgdaInductiveConstructor{tt}%
\AgdaSpace{}%
\AgdaOperator{%
\AgdaFunction{⟧}%
}%
\>[13]\AgdaSymbol{=}%
\>[16]%
\AgdaPostulate{↑}%
\AgdaSpace{}%
%
\AgdaInductiveConstructor{true}%
\<%
\\
\>[2]\AgdaOperator{%
\AgdaFunction{𝒜⟦}%
}\AgdaSpace{}%
%
\AgdaInductiveConstructor{ff}%
\AgdaSpace{}%
\AgdaOperator{%
\AgdaFunction{⟧}%
}%
\>[13]\AgdaSymbol{=}%
\>[16]%
\AgdaPostulate{↑}%
\AgdaSpace{}%
%
\AgdaInductiveConstructor{false}%
\<%
\\
\>[2]\AgdaOperator{%
\AgdaFunction{𝒜⟦}%
}\AgdaSpace{}%
%
\AgdaInductiveConstructor{⊃}%
\AgdaSpace{}%
\AgdaOperator{%
\AgdaFunction{⟧}%
}%
\>[13]\AgdaSymbol{=}%
\>[16]\AgdaSymbol{λ}\AgdaSpace{}%
\AgdaBound{β}\AgdaSpace{}%
\AgdaBound{δ₁}\AgdaSpace{}%
\AgdaBound{δ₂}\AgdaSpace{}%
\AgdaSymbol{→}\AgdaSpace{}%
\AgdaSymbol{(}\AgdaBound{β}\AgdaSpace{}%
\AgdaOperator{%
\AgdaFunction{⟶}%
}\AgdaSpace{}%
\AgdaBound{δ₁}\AgdaSpace{}%
\AgdaOperator{%
\AgdaFunction{,}%
}\AgdaSpace{}%
\AgdaBound{δ₂}\AgdaSymbol{)}\<%
\\
\>[2]\AgdaOperator{%
\AgdaFunction{𝒜⟦}%
}\AgdaSpace{}%
%
\AgdaInductiveConstructor{Y}%
\AgdaSpace{}%
\AgdaOperator{%
\AgdaFunction{⟧}%
}%
\>[13]\AgdaSymbol{=}%
\>[16]%
\AgdaPostulate{fix}%
\<%
\\
\>[2]\AgdaOperator{%
\AgdaFunction{𝒜⟦}%
}\AgdaSpace{}%
%
\AgdaInductiveConstructor{k}%
\AgdaSpace{}%
\AgdaBound{n}\AgdaSpace{}%
\AgdaOperator{%
\AgdaFunction{⟧}%
}%
\>[13]\AgdaSymbol{=}%
\>[16]%
\AgdaPostulate{↑}%
\AgdaSpace{}%
\AgdaBound{n}\<%
\\
\>[2]\AgdaOperator{%
\AgdaFunction{𝒜⟦}%
}\AgdaSpace{}%
%
\AgdaInductiveConstructor{⦅+1⦆}%
\AgdaSpace{}%
\AgdaOperator{%
\AgdaFunction{⟧}%
}%
\>[13]\AgdaSymbol{=}%
\>[16]\AgdaSymbol{(λ}\AgdaSpace{}%
\AgdaBound{n}\AgdaSpace{}%
\AgdaSymbol{→}\AgdaSpace{}%
%
\AgdaPostulate{↑}%
\AgdaSpace{}%
\AgdaSymbol{(}\AgdaBound{n}\AgdaSpace{}%
\AgdaOperator{\AgdaPrimitive{+}}\AgdaSpace{}%
\AgdaNumber{1}\AgdaSymbol{))}\AgdaSpace{}%
\AgdaOperator{%
\AgdaPostulate{♯}%
}\<%
\\
\>[2]\AgdaOperator{%
\AgdaFunction{𝒜⟦}%
}\AgdaSpace{}%
%
\AgdaInductiveConstructor{⦅−1⦆}%
\AgdaSpace{}%
\AgdaOperator{%
\AgdaFunction{⟧}%
}%
\>[13]\AgdaSymbol{=}%
\>[16]\AgdaSymbol{(λ}\AgdaSpace{}%
\AgdaBound{n}\AgdaSpace{}%
\AgdaSymbol{→}\AgdaSpace{}%
%
\AgdaPostulate{↑}%
\AgdaSpace{}%
\AgdaSymbol{(}\AgdaBound{n}\AgdaSpace{}%
\AgdaOperator{\AgdaPrimitive{==ᴺ}}\AgdaSpace{}%
\AgdaNumber{0}\AgdaSymbol{)}\AgdaSpace{}%
\AgdaOperator{%
\AgdaFunction{⟶}%
}\AgdaSpace{}%
%
\AgdaPostulate{⊥}%
\AgdaSpace{}%
\AgdaOperator{%
\AgdaFunction{,}%
}\AgdaSpace{}%
%
\AgdaPostulate{↑}%
\AgdaSpace{}%
\AgdaSymbol{(}\AgdaBound{n}\AgdaSpace{}%
\AgdaOperator{\AgdaPrimitive{-}}\AgdaSpace{}%
\AgdaNumber{1}\AgdaSymbol{))}\AgdaSpace{}%
\AgdaOperator{%
\AgdaPostulate{♯}%
}\<%
\\
\>[2]\AgdaOperator{%
\AgdaFunction{𝒜⟦}%
}\AgdaSpace{}%
%
\AgdaInductiveConstructor{Z}%
\AgdaSpace{}%
\AgdaOperator{%
\AgdaFunction{⟧}%
}%
\>[13]\AgdaSymbol{=}%
\>[16]\AgdaSymbol{(λ}\AgdaSpace{}%
\AgdaBound{n}\AgdaSpace{}%
\AgdaSymbol{→}\AgdaSpace{}%
%
\AgdaPostulate{↑}%
\AgdaSpace{}%
\AgdaSymbol{(}\AgdaBound{n}\AgdaSpace{}%
\AgdaOperator{\AgdaPrimitive{==ᴺ}}\AgdaSpace{}%
\AgdaNumber{0}\AgdaSymbol{))}\AgdaSpace{}%
\AgdaOperator{%
\AgdaPostulate{♯}%
}\<%
\end{code}
\end{AgdaSuppressSpace}
The semantic function \AgdaRef{𝒜′⟦\ M\ ⟧} is written
\(\hat{\mathcal A} \llbracket M \rrbracket\) in
\cite{Plotkin1977LCP}.
It gives the denotation of the term \AgdaRef{M} as a function of the
environment \AgdaRef{ρ}.
\begin{AgdaSuppressSpace}
\begin{code}%
\>[2]\AgdaOperator{%
\AgdaFunction{𝒜′⟦\AgdaUnderscore{}%
⟧}}\AgdaSpace{}%
\AgdaSymbol{:}\AgdaSpace{}%
%
\AgdaDatatype{Terms}%
\AgdaSpace{}%
\AgdaGeneralizable{σ}\AgdaSpace{}%
\AgdaSymbol{→}\AgdaSpace{}%
\AgdaOperator{%
\AgdaPostulate{⟪}%
}\AgdaSpace{}%
%
\AgdaFunction{Env}%
\AgdaSpace{}%
\AgdaOperator{%
\AgdaPostulate{→ˢ}%
}\AgdaSpace{}%
%
\AgdaFunction{𝒟}%
\AgdaSpace{}%
\AgdaGeneralizable{σ}\AgdaSpace{}%
\AgdaOperator{%
\AgdaPostulate{⟫}%
}%
\>[36]\AgdaComment{--\ typed\ term\ denotations}\<%
\\
\>[2]\AgdaOperator{%
\AgdaFunction{𝒜′⟦}%
}\AgdaSpace{}%
\AgdaOperator{%
\AgdaInductiveConstructor{𝑉}%
}\AgdaSpace{}%
%
\AgdaInductiveConstructor{α}%
\AgdaSpace{}%
\AgdaBound{i}\AgdaSpace{}%
\AgdaBound{σ}\AgdaSpace{}%
\AgdaOperator{%
\AgdaFunction{⟧}%
}\AgdaSpace{}%
\AgdaBound{ρ}%
\>[28]\AgdaSymbol{=}%
\>[31]\AgdaBound{ρ}\AgdaSpace{}%
\AgdaOperator{%
\AgdaFunction{⟦}%
}\AgdaSpace{}%
%
\AgdaInductiveConstructor{α}%
\AgdaSpace{}%
\AgdaBound{i}\AgdaSpace{}%
\AgdaBound{σ}\AgdaSpace{}%
\AgdaOperator{%
\AgdaFunction{⟧}%
}\<%
\\
\>[2]\AgdaOperator{%
\AgdaFunction{𝒜′⟦}%
}\AgdaSpace{}%
\AgdaOperator{%
\AgdaInductiveConstructor{𝐿}%
}\AgdaSpace{}%
\AgdaBound{c}\AgdaSpace{}%
\AgdaOperator{%
\AgdaFunction{⟧}%
}\AgdaSpace{}%
\AgdaBound{ρ}%
\>[28]\AgdaSymbol{=}%
\>[31]\AgdaOperator{%
\AgdaFunction{𝒜⟦}%
}\AgdaSpace{}%
\AgdaBound{c}\AgdaSpace{}%
\AgdaOperator{%
\AgdaFunction{⟧}%
}\<%
\\
\>[2]\AgdaOperator{%
\AgdaFunction{𝒜′⟦}%
}\AgdaSpace{}%
\AgdaOperator{%
\AgdaInductiveConstructor{⦅}%
}\AgdaSpace{}%
\AgdaBound{M}\AgdaSpace{}%
\AgdaOperator{%
\AgdaInductiveConstructor{␣}%
}\AgdaSpace{}%
\AgdaBound{N}\AgdaSpace{}%
\AgdaOperator{%
\AgdaInductiveConstructor{⦆}%
}\AgdaSpace{}%
\AgdaOperator{%
\AgdaFunction{⟧}%
}\AgdaSpace{}%
\AgdaBound{ρ}%
\>[28]\AgdaSymbol{=}%
\>[31]\AgdaOperator{%
\AgdaFunction{𝒜′⟦}%
}\AgdaSpace{}%
\AgdaBound{M}\AgdaSpace{}%
\AgdaOperator{%
\AgdaFunction{⟧}%
}\AgdaSpace{}%
\AgdaBound{ρ}\AgdaSpace{}%
\AgdaSymbol{(}\AgdaOperator{%
\AgdaFunction{𝒜′⟦}%
}\AgdaSpace{}%
\AgdaBound{N}\AgdaSpace{}%
\AgdaOperator{%
\AgdaFunction{⟧}%
}\AgdaSpace{}%
\AgdaBound{ρ}\AgdaSymbol{)}\<%
\\
\>[2]\AgdaOperator{%
\AgdaFunction{𝒜′⟦}%
}\AgdaSpace{}%
\AgdaOperator{%
\AgdaInductiveConstructor{⦅λ}%
}\AgdaSpace{}%
%
\AgdaInductiveConstructor{α}%
\AgdaSpace{}%
\AgdaBound{i}\AgdaSpace{}%
\AgdaBound{σ}\AgdaSpace{}%
\AgdaOperator{%
\AgdaInductiveConstructor{␣}%
}\AgdaSpace{}%
\AgdaBound{M}\AgdaSpace{}%
\AgdaOperator{%
\AgdaInductiveConstructor{⦆}%
}\AgdaSpace{}%
\AgdaOperator{%
\AgdaFunction{⟧}%
}\AgdaSpace{}%
\AgdaBound{ρ}\AgdaSpace{}%
\AgdaBound{x}%
\>[28]\AgdaSymbol{=}%
\>[31]\AgdaOperator{%
\AgdaFunction{𝒜′⟦}%
}\AgdaSpace{}%
\AgdaBound{M}\AgdaSpace{}%
\AgdaOperator{%
\AgdaFunction{⟧}%
}\AgdaSpace{}%
\AgdaSymbol{(}\AgdaBound{ρ}\AgdaSpace{}%
\AgdaOperator{%
\AgdaFunction{[}%
}\AgdaSpace{}%
\AgdaBound{x}\AgdaSpace{}%
\AgdaOperator{%
\AgdaFunction{/}%
}\AgdaSpace{}%
%
\AgdaInductiveConstructor{α}%
\AgdaSpace{}%
\AgdaBound{i}\AgdaSpace{}%
\AgdaBound{σ}\AgdaSpace{}%
\AgdaOperator{%
\AgdaFunction{]′}%
}\AgdaSymbol{)}\<%
\end{code}
\end{AgdaSuppressSpace}
Comparison with Plotkin's original definition of PCF
\cite{Plotkin1977LCP}
confirms the directness of our Agda embedding of the semantic equations
for typed terms.

\end{AgdaAlign}

\begin{AgdaAlign}
\subsection{\texorpdfstring{\emph{Scm}: A Sublanguage of
\emph{Scheme}}{Scm: A Sublanguage of Scheme}}\label{Examples/Scm-scm-a-sublanguage-of-scheme}

\emph{Scm} is a particularly basic sublanguage of the core \emph{Scheme}
expression language whose denotational semantics is defined in the
\emph{Scheme} reports \cite{Scheme}. The
domains and auxiliary functions declared in this section are explained
in the presentation of the conventional denotational semantics of
\emph{Scm}
\cite{Mosses2025CSE}; they
involve the notation for
sequence
domains (Section~\ref{Notation/Products/Sequences-sequence-domains}).
\begin{AgdaSuppressSpace}
\begin{code}[hide]%
\>[0]\AgdaSymbol{\{-\#}\AgdaSpace{}%
\AgdaKeyword{OPTIONS}\AgdaSpace{}%
\AgdaPragma{--rewriting}\AgdaSpace{}%
\AgdaPragma{--confluence-check}\AgdaSpace{}%
\AgdaSymbol{\#-\}}\<%
\\
\\[\AgdaEmptyExtraSkip]%
\>[0]\AgdaKeyword{module}\AgdaSpace{}%
%
\AgdaModule{Examples.Scm}%
\AgdaSpace{}%
\AgdaKeyword{where}\<%
\\
\\[\AgdaEmptyExtraSkip]%
\>[0][@{}l@{\AgdaIndent{0}}]%
\>[2]\AgdaKeyword{import}\AgdaSpace{}%
%
\AgdaModule{Examples.Scm.Abstract-Syntax}%
\<%
\\
\>[2]\AgdaKeyword{import}\AgdaSpace{}%
%
\AgdaModule{Examples.Scm.Domain-Equations}%
\<%
\\
\>[2]\AgdaKeyword{import}\AgdaSpace{}%
%
\AgdaModule{Examples.Scm.Semantic-Functions}%
\<%
\\
\>[2]\AgdaKeyword{import}\AgdaSpace{}%
%
\AgdaModule{Examples.Scm.Auxiliary-Functions}%
\<%
\end{code}
\end{AgdaSuppressSpace}

\subsubsection{Abstract Syntax}\label{Examples/Scm/Abstract-Syntax-abstract-syntax}

The following grammar
\cite{Mosses2025CSE}
summarises the abstract syntax of \emph{Scm} expressions
\(\text{E} : \text{Exp}\) with integers \(\text{Z} : \text{Int}\),
constants \(\text{K} : \text{Con}\) and identifiers
\(\text{I} : \text{Ide}\). The meta-variable
\(\text{E}^* : \text{Exp}^*\) implicitly ranges over arbitrary sequences
of expressions.   \begin{align}
  \text{K} & ::=
    \text{Z} \mid \text{\tt \#t} \mid \text{\tt \#f}
  \\
  \text{E} & ::=
    \text{K} \mid \text{I} \mid \texttt ( \text{E}_0~\text{E}^* \texttt ) \mid \text{\tt (lambda} ~ \text{I} ~ \text{E} \texttt ) \mid
    \text{\tt (if} ~ \text{E}_0 ~ \text{E}_1 ~ \text{E}_2 \texttt ) \mid \text{\tt (set!} ~ \text{I} ~ \text{E} \texttt )
\end{align}   In the following Agda specifcation of
\emph{Scm} ASTs, the abstract syntax of sequences \AgdaRef{E⋆\ :\ Exp⋆}
is made explicit: the empty sequence is represented by \AgdaRef{␣␣␣} ,
and sequence prefixing by \AgdaRef{E\ ␣␣\ E⋆}.
\begin{AgdaSuppressSpace}
\begin{code}[hide]%
\>[0]\AgdaSymbol{\{-\#}\AgdaSpace{}%
\AgdaKeyword{OPTIONS}\AgdaSpace{}%
\AgdaPragma{--rewriting}\AgdaSpace{}%
\AgdaPragma{--confluence-check}\AgdaSpace{}%
\AgdaSymbol{\#-\}}\<%
\\
\>[0]\<%
\end{code}
\begin{code}%
\>[0]\AgdaKeyword{module}\AgdaSpace{}%
%
\AgdaModule{Examples.Scm.Abstract-Syntax}%
\AgdaSpace{}%
\AgdaKeyword{where}\<%
\end{code}
\begin{code}[hide]%
\>[0][@{}l@{\AgdaIndent{1}}]%
\>[2]\AgdaKeyword{open}\AgdaSpace{}%
\AgdaKeyword{import}\AgdaSpace{}%
%
\AgdaModule{Notation.Flat.Booleans}%
\<%
\\
\>[2]\AgdaKeyword{open}\AgdaSpace{}%
\AgdaKeyword{import}\AgdaSpace{}%
%
\AgdaModule{Data.Integer.Base}%
\AgdaSpace{}%
\AgdaKeyword{public}\AgdaSpace{}%
\AgdaKeyword{renaming}\AgdaSpace{}%
\AgdaSymbol{(}%
\AgdaDatatype{ℤ}%
\AgdaSpace{}%
\AgdaSymbol{to}\AgdaSpace{}%
%
\AgdaDatatype{Int}%
\AgdaSymbol{)}\AgdaSpace{}%
\AgdaKeyword{using}\AgdaSpace{}%
\AgdaSymbol{(}%
\AgdaInductiveConstructor{+\AgdaUnderscore{}%
}\AgdaSymbol{)}\<%
\\
\>[2]\AgdaKeyword{open}\AgdaSpace{}%
\AgdaKeyword{import}\AgdaSpace{}%
%
\AgdaModule{Data.String.Base}%
\AgdaSpace{}%
\AgdaKeyword{public}\AgdaSpace{}%
\AgdaKeyword{using}\AgdaSpace{}%
\AgdaSymbol{(}%
\AgdaPostulate{String}%
\AgdaSymbol{)}\<%
\end{code}
\begin{code}%
\>[0]\<%
\\
\>[2]%
\AgdaFunction{Ide}%
\AgdaSpace{}%
\AgdaSymbol{=}\AgdaSpace{}%
%
\AgdaPostulate{String}%
%
\>[16]\AgdaComment{--\ identifiers}\<%
\end{code}
\begin{code}[hide]%
\>[2]\AgdaKeyword{variable}\AgdaSpace{}%
\AgdaGeneralizable{I}\AgdaSpace{}%
\AgdaSymbol{:}\AgdaSpace{}%
%
\AgdaFunction{Ide}%
\<%
\\
\>[0]\<%
\end{code}
\begin{code}%
\>[0]\<%
\\
\>[0][@{}l@{\AgdaIndent{1}}]%
\>[2]\AgdaKeyword{data}\AgdaSpace{}%
%
\AgdaDatatype{Con}%
%
\>[12]\AgdaSymbol{:}\AgdaSpace{}%
\AgdaPrimitive{Set}\AgdaSpace{}%
\AgdaKeyword{where}%
\>[25]\AgdaComment{--\ constants}\<%
\\
\>[2][@{}l@{\AgdaIndent{0}}]%
\>[4]%
\AgdaInductiveConstructor{int}%
%
\>[12]\AgdaSymbol{:}\AgdaSpace{}%
%
\AgdaDatatype{Int}%
\AgdaSpace{}%
\AgdaSymbol{→}\AgdaSpace{}%
%
\AgdaDatatype{Con}%
%
\>[25]\AgdaComment{--\ integer\ numerals}\<%
\\
\>[4]%
\AgdaInductiveConstructor{\#t}%
%
\>[12]\AgdaSymbol{:}\AgdaSpace{}%
%
\AgdaDatatype{Con}%
%
\>[25]\AgdaComment{--\ true}\<%
\\
\>[4]%
\AgdaInductiveConstructor{\#f}%
%
\>[12]\AgdaSymbol{:}\AgdaSpace{}%
%
\AgdaDatatype{Con}%
%
\>[25]\AgdaComment{--\ false}\<%
\end{code}
\begin{code}[hide]%
\>[2]\AgdaKeyword{variable}\AgdaSpace{}%
\AgdaGeneralizable{K}\AgdaSpace{}%
\AgdaSymbol{:}\AgdaSpace{}%
%
\AgdaDatatype{Con}%
\<%
\end{code}
\begin{code}%
\>[0]\<%
\\
\>[2]\AgdaKeyword{mutual}\<%
\\
\>[2][@{}l@{\AgdaIndent{0}}]%
\>[4]\AgdaKeyword{data}\AgdaSpace{}%
%
\AgdaDatatype{Exp}%
%
\>[19]\AgdaSymbol{:}\AgdaSpace{}%
\AgdaPrimitive{Set}\AgdaSpace{}%
\AgdaKeyword{where}%
\>[44]\AgdaComment{--\ expressions}\<%
\\
\>[4][@{}l@{\AgdaIndent{0}}]%
\>[6]%
\AgdaInductiveConstructor{con}%
%
\>[19]\AgdaSymbol{:}\AgdaSpace{}%
%
\AgdaDatatype{Con}%
\AgdaSpace{}%
\AgdaSymbol{→}\AgdaSpace{}%
%
\AgdaDatatype{Exp}%
%
\>[44]\AgdaComment{--\ constants}\<%
\\
\>[6]%
\AgdaInductiveConstructor{ide}%
%
\>[19]\AgdaSymbol{:}\AgdaSpace{}%
%
\AgdaFunction{Ide}%
\AgdaSpace{}%
\AgdaSymbol{→}\AgdaSpace{}%
%
\AgdaDatatype{Exp}%
%
\>[44]\AgdaComment{--\ identifiers}\<%
\\
\>[6]\AgdaOperator{%
\AgdaInductiveConstructor{⦅\AgdaUnderscore{}%
␣\AgdaUnderscore{}⦆}}%
\>[19]\AgdaSymbol{:}\AgdaSpace{}%
%
\AgdaDatatype{Exp}%
\AgdaSpace{}%
\AgdaSymbol{→}\AgdaSpace{}%
%
\AgdaDatatype{Exp⋆}%
\AgdaSpace{}%
\AgdaSymbol{→}\AgdaSpace{}%
%
\AgdaDatatype{Exp}%
%
\>[44]\AgdaComment{--\ procedure\ application\ }\<%
\\
\>[6]\AgdaOperator{%
\AgdaInductiveConstructor{⦅lambda\AgdaUnderscore{}%
␣\AgdaUnderscore{}⦆}}%
\>[19]\AgdaSymbol{:}\AgdaSpace{}%
%
\AgdaFunction{Ide}%
\AgdaSpace{}%
\AgdaSymbol{→}\AgdaSpace{}%
%
\AgdaDatatype{Exp}%
\AgdaSpace{}%
\AgdaSymbol{→}\AgdaSpace{}%
%
\AgdaDatatype{Exp}%
%
\>[44]\AgdaComment{--\ procedure\ abstraction}\<%
\\
\>[6]\AgdaOperator{%
\AgdaInductiveConstructor{⦅if\AgdaUnderscore{}%
␣\AgdaUnderscore{}␣\AgdaUnderscore{}⦆}}%
\>[19]\AgdaSymbol{:}\AgdaSpace{}%
%
\AgdaDatatype{Exp}%
\AgdaSpace{}%
\AgdaSymbol{→}\AgdaSpace{}%
%
\AgdaDatatype{Exp}%
\AgdaSpace{}%
\AgdaSymbol{→}\AgdaSpace{}%
%
\AgdaDatatype{Exp}%
\AgdaSpace{}%
\AgdaSymbol{→}\AgdaSpace{}%
%
\AgdaDatatype{Exp}%
%
\>[44]\AgdaComment{--\ conditional\ choice}\<%
\\
\>[6]\AgdaOperator{%
\AgdaInductiveConstructor{⦅set!\AgdaUnderscore{}%
␣\AgdaUnderscore{}⦆}}%
\>[19]\AgdaSymbol{:}\AgdaSpace{}%
%
\AgdaFunction{Ide}%
\AgdaSpace{}%
\AgdaSymbol{→}\AgdaSpace{}%
%
\AgdaDatatype{Exp}%
\AgdaSpace{}%
\AgdaSymbol{→}\AgdaSpace{}%
%
\AgdaDatatype{Exp}%
%
\>[44]\AgdaComment{--\ assignment}\<%
\\
\>[4]\AgdaKeyword{data}\AgdaSpace{}%
%
\AgdaDatatype{Exp⋆}%
%
\>[19]\AgdaSymbol{:}\AgdaSpace{}%
\AgdaPrimitive{Set}\AgdaSpace{}%
\AgdaKeyword{where}%
\>[44]\AgdaComment{--\ expression\ sequences}\<%
\\
\>[4][@{}l@{\AgdaIndent{0}}]%
\>[6]%
\AgdaInductiveConstructor{␣␣␣}%
%
\>[19]\AgdaSymbol{:}\AgdaSpace{}%
%
\AgdaDatatype{Exp⋆}%
%
\>[44]\AgdaComment{--\ empty\ sequence}\<%
\\
\>[6]\AgdaOperator{%
\AgdaInductiveConstructor{\AgdaUnderscore{}%
␣␣\AgdaUnderscore{}}}%
\>[19]\AgdaSymbol{:}\AgdaSpace{}%
%
\AgdaDatatype{Exp}%
\AgdaSpace{}%
\AgdaSymbol{→}\AgdaSpace{}%
%
\AgdaDatatype{Exp⋆}%
\AgdaSpace{}%
\AgdaSymbol{→}\AgdaSpace{}%
%
\AgdaDatatype{Exp⋆}%
%
\>[44]\AgdaComment{--\ sequence\ prefix}\<%
\end{code}
\begin{code}[hide]%
\>[2]\AgdaKeyword{variable}\AgdaSpace{}%
\AgdaGeneralizable{E}\AgdaSpace{}%
\AgdaSymbol{:}\AgdaSpace{}%
%
\AgdaDatatype{Exp}%
\AgdaSymbol{;}\AgdaSpace{}%
\AgdaGeneralizable{E⋆}\AgdaSpace{}%
\AgdaSymbol{:}\AgdaSpace{}%
%
\AgdaDatatype{Exp⋆}%
\<%
\\
\\[\AgdaEmptyExtraSkip]%
\>[2]\AgdaKeyword{mutual}\<%
\\
\>[2][@{}l@{\AgdaIndent{0}}]%
\>[4]\AgdaKeyword{data}\AgdaSpace{}%
%
\AgdaDatatype{Body}%
%
\>[19]\AgdaSymbol{:}\AgdaSpace{}%
\AgdaPrimitive{Set}\AgdaSpace{}%
\AgdaKeyword{where}%
\>[44]\AgdaComment{--\ bodies}\<%
\\
\>[4][@{}l@{\AgdaIndent{0}}]%
\>[6]\AgdaOperator{%
\AgdaInductiveConstructor{␣␣\AgdaUnderscore{}%
}}%
\>[19]\AgdaSymbol{:}\AgdaSpace{}%
%
\AgdaDatatype{Exp}%
\AgdaSpace{}%
\AgdaSymbol{→}\AgdaSpace{}%
%
\AgdaDatatype{Body}%
%
\>[44]\AgdaComment{--\ expression\ body}\<%
\\
\>[6]\AgdaOperator{%
\AgdaInductiveConstructor{⦅define\AgdaUnderscore{}%
␣\AgdaUnderscore{}⦆}}%
\>[19]\AgdaSymbol{:}\AgdaSpace{}%
%
\AgdaFunction{Ide}%
\AgdaSpace{}%
\AgdaSymbol{→}\AgdaSpace{}%
%
\AgdaDatatype{Exp}%
\AgdaSpace{}%
\AgdaSymbol{→}\AgdaSpace{}%
%
\AgdaDatatype{Body}%
%
\>[44]\AgdaComment{--\ definition\ body}\<%
\\
\>[6]\AgdaOperator{%
\AgdaInductiveConstructor{⦅begin\AgdaUnderscore{}%
⦆}}%
\>[19]\AgdaSymbol{:}\AgdaSpace{}%
%
\AgdaDatatype{Body⁺}%
\AgdaSpace{}%
\AgdaSymbol{→}\AgdaSpace{}%
%
\AgdaDatatype{Body}%
%
\>[44]\AgdaComment{--\ body\ sequence}\<%
\\
\>[4]\AgdaKeyword{data}\AgdaSpace{}%
%
\AgdaDatatype{Body⁺}%
%
\>[19]\AgdaSymbol{:}\AgdaSpace{}%
\AgdaPrimitive{Set}\AgdaSpace{}%
\AgdaKeyword{where}%
\>[44]\AgdaComment{--\ body\ sequences}\<%
\\
\>[4][@{}l@{\AgdaIndent{0}}]%
\>[6]\AgdaOperator{%
\AgdaInductiveConstructor{␣␣\AgdaUnderscore{}%
}}%
\>[19]\AgdaSymbol{:}\AgdaSpace{}%
%
\AgdaDatatype{Body}%
\AgdaSpace{}%
\AgdaSymbol{→}\AgdaSpace{}%
%
\AgdaDatatype{Body⁺}%
%
\>[44]\AgdaComment{--\ empty\ sequence}\<%
\\
\>[6]\AgdaOperator{%
\AgdaInductiveConstructor{\AgdaUnderscore{}%
␣␣\AgdaUnderscore{}}}%
\>[19]\AgdaSymbol{:}\AgdaSpace{}%
%
\AgdaDatatype{Body}%
\AgdaSpace{}%
\AgdaSymbol{→}\AgdaSpace{}%
%
\AgdaDatatype{Body⁺}%
\AgdaSpace{}%
\AgdaSymbol{→}\AgdaSpace{}%
%
\AgdaDatatype{Body⁺}%
%
\>[44]\AgdaComment{--\ sequence\ prefix}\<%
\\
\>[2]\AgdaKeyword{data}\AgdaSpace{}%
%
\AgdaDatatype{Prog}%
%
\>[19]\AgdaSymbol{:}\AgdaSpace{}%
\AgdaPrimitive{Set}\AgdaSpace{}%
\AgdaKeyword{where}%
\>[44]\AgdaComment{--\ programs}\<%
\\
\>[2][@{}l@{\AgdaIndent{0}}]%
\>[4]%
\AgdaInductiveConstructor{␣␣␣}%
%
\>[19]\AgdaSymbol{:}\AgdaSpace{}%
%
\AgdaDatatype{Prog}%
%
\>[44]\AgdaComment{--\ empty\ program}\<%
\\
\>[4]\AgdaOperator{%
\AgdaInductiveConstructor{␣␣\AgdaUnderscore{}%
}}%
\>[19]\AgdaSymbol{:}\AgdaSpace{}%
%
\AgdaDatatype{Body⁺}%
\AgdaSpace{}%
\AgdaSymbol{→}\AgdaSpace{}%
%
\AgdaDatatype{Prog}%
%
\>[44]\AgdaComment{--\ body\ sequence\ program}\<%
\\
\>[2]\AgdaKeyword{variable}\AgdaSpace{}%
\AgdaGeneralizable{B}\AgdaSpace{}%
\AgdaSymbol{:}\AgdaSpace{}%
%
\AgdaDatatype{Body}%
\AgdaSymbol{;}\AgdaSpace{}%
\AgdaGeneralizable{B⁺}\AgdaSpace{}%
\AgdaSymbol{:}\AgdaSpace{}%
%
\AgdaDatatype{Body⁺}%
\AgdaSymbol{;}\AgdaSpace{}%
\AgdaGeneralizable{Π}\AgdaSpace{}%
\AgdaSymbol{:}\AgdaSpace{}%
%
\AgdaDatatype{Prog}%
\<%
\end{code}
\end{AgdaSuppressSpace}

\subsubsection{Domain Equations}\label{Examples/Scm/Domain-Equations-domain-equations}

The domains for \emph{Scm} are somewhat simpler than for the
denotational semantics in the Scheme standards
\cite{Scheme}, but still involve all our
postulated domain constructors. Using definitional equations
\AgdaRef{D\ =\ E} instead of postulated bijections \AgdaRef{D\ ≅\ E} for
non-recursive collections of domains avoids the need for the functions
\AgdaRef{fold} and \AgdaRef{unfold} in definitions of denotations.
\begin{AgdaSuppressSpace}
\begin{code}[hide]%
\>[0]\AgdaSymbol{\{-\#}\AgdaSpace{}%
\AgdaKeyword{OPTIONS}\AgdaSpace{}%
\AgdaPragma{--rewriting}\AgdaSpace{}%
\AgdaPragma{--confluence-check}\AgdaSpace{}%
\AgdaSymbol{\#-\}}\<%
\\
\>[0]\<%
\end{code}
\begin{code}%
\>[0]\AgdaKeyword{module}\AgdaSpace{}%
%
\AgdaModule{Examples.Scm.Domain-Equations}%
\AgdaSpace{}%
\AgdaKeyword{where}\<%
\end{code}
\begin{code}[hide]%
\>[0]\<%
\\
\>[0][@{}l@{\AgdaIndent{1}}]%
\>[2]\AgdaKeyword{open}\AgdaSpace{}%
\AgdaKeyword{import}\AgdaSpace{}%
%
\AgdaModule{Examples.Scm.Abstract-Syntax}%
\<%
\\
\>[2]\AgdaKeyword{open}\AgdaSpace{}%
\AgdaKeyword{import}\AgdaSpace{}%
%
\AgdaModule{Notation.Domains}%
\<%
\\
\>[2]\AgdaKeyword{open}\AgdaSpace{}%
\AgdaKeyword{import}\AgdaSpace{}%
%
\AgdaModule{Notation.Functions}%
\<%
\\
\>[2]\AgdaKeyword{open}\AgdaSpace{}%
\AgdaKeyword{import}\AgdaSpace{}%
%
\AgdaModule{Notation.Flat}%
\<%
\\
\>[2]\AgdaKeyword{open}\AgdaSpace{}%
\AgdaKeyword{import}\AgdaSpace{}%
%
\AgdaModule{Notation.Flat.Booleans}%
\<%
\\
\>[2]\AgdaKeyword{open}\AgdaSpace{}%
\AgdaKeyword{import}\AgdaSpace{}%
%
\AgdaModule{Notation.Flat.Naturals}%
\<%
\\
\>[2]\AgdaKeyword{open}\AgdaSpace{}%
\AgdaKeyword{import}\AgdaSpace{}%
%
\AgdaModule{Notation.Sums}%
\<%
\\
\>[2]\AgdaKeyword{open}\AgdaSpace{}%
\AgdaKeyword{import}\AgdaSpace{}%
%
\AgdaModule{Notation.Products}%
\<%
\\
\>[2]\AgdaKeyword{open}\AgdaSpace{}%
\AgdaKeyword{import}\AgdaSpace{}%
%
\AgdaModule{Notation.Products.Sequences}%
\<%
\end{code}
\begin{code}%
\>[0]\<%
\\
\>[2]\AgdaKeyword{postulate}\AgdaSpace{}%
%
\AgdaPostulate{Loc}%
\AgdaSpace{}%
\AgdaSymbol{:}\AgdaSpace{}%
\AgdaPrimitive{Set}\<%
\\
\>[2]%
\AgdaFunction{𝐋}%
%
\>[5]\AgdaSymbol{=}%
\>[8]%
\AgdaPostulate{Loc}%
\AgdaSpace{}%
\AgdaOperator{%
\AgdaPostulate{+⊥}%
}%
\>[30]\AgdaComment{--\ locations}\<%
\\
\>[2]%
\AgdaFunction{𝐍}%
%
\>[5]\AgdaSymbol{=}%
\>[8]%
\AgdaFunction{Nat⊥}%
%
\>[30]\AgdaComment{--\ natural\ numbers}\<%
\\
\>[2]%
\AgdaFunction{𝐓}%
%
\>[5]\AgdaSymbol{=}%
\>[8]%
\AgdaFunction{Bool⊥}%
%
\>[30]\AgdaComment{--\ booleans}\<%
\\
\>[2]%
\AgdaFunction{𝐑}%
%
\>[5]\AgdaSymbol{=}%
\>[8]%
\AgdaDatatype{Int}%
\AgdaSpace{}%
\AgdaOperator{%
\AgdaPostulate{+⊥}%
}%
\>[30]\AgdaComment{--\ numbers}\<%
\\
\>[2]%
\AgdaFunction{𝐏}%
%
\>[5]\AgdaSymbol{=}%
\>[8]%
\AgdaFunction{𝐋}%
\AgdaSpace{}%
\AgdaOperator{%
\AgdaPostulate{×}%
}\AgdaSpace{}%
%
\AgdaFunction{𝐋}%
%
\>[30]\AgdaComment{--\ pairs}\<%
\\
\>[2]%
\AgdaFunction{𝐔}%
%
\>[5]\AgdaSymbol{=}%
\>[8]%
\AgdaFunction{Ide}%
\AgdaSpace{}%
\AgdaOperator{%
\AgdaPostulate{→ˢ}%
}\AgdaSpace{}%
%
\AgdaFunction{𝐋}%
%
\>[30]\AgdaComment{--\ environments}\<%
\\
\>[2]\AgdaKeyword{data}\AgdaSpace{}%
%
\AgdaDatatype{Misc}%
\AgdaSpace{}%
\AgdaSymbol{:}\AgdaSpace{}%
\AgdaPrimitive{Set}\AgdaSpace{}%
\AgdaKeyword{where}\AgdaSpace{}%
%
\AgdaInductiveConstructor{null}%
\AgdaSpace{}%
%
\AgdaInductiveConstructor{unallocated}%
\AgdaSpace{}%
%
\AgdaInductiveConstructor{undefined}%
\AgdaSpace{}%
%
\AgdaInductiveConstructor{unspecified}%
\AgdaSpace{}%
\AgdaSymbol{:}\AgdaSpace{}%
%
\AgdaDatatype{Misc}%
\<%
\\
\>[2]%
\AgdaFunction{𝐌}%
%
\>[5]\AgdaSymbol{=}%
\>[8]%
\AgdaDatatype{Misc}%
\AgdaSpace{}%
\AgdaOperator{%
\AgdaPostulate{+⊥}%
}%
\>[30]\AgdaComment{--\ miscellaneous}\<%
\\
\>[2]\AgdaKeyword{postulate}\AgdaSpace{}%
%
\AgdaPostulate{𝐄}%
\AgdaSpace{}%
\AgdaSymbol{:}\AgdaSpace{}%
%
\AgdaPostulate{Domain}%
%
\>[30]\AgdaComment{--\ expressed\ values}\<%
\\
\>[2]%
\AgdaFunction{𝐒}%
%
\>[5]\AgdaSymbol{=}%
\>[8]%
\AgdaFunction{𝐋}%
\AgdaSpace{}%
\AgdaOperator{%
\AgdaPostulate{→ᶜ}%
}\AgdaSpace{}%
%
\AgdaPostulate{𝐄}%
%
\>[30]\AgdaComment{--\ stores}\<%
\\
\>[2]\AgdaKeyword{postulate}\AgdaSpace{}%
%
\AgdaPostulate{𝐀}%
\AgdaSpace{}%
\AgdaSymbol{:}\AgdaSpace{}%
%
\AgdaPostulate{Domain}%
%
\>[30]\AgdaComment{--\ answers}\<%
\\
\>[2]%
\AgdaFunction{𝐂}%
%
\>[5]\AgdaSymbol{=}%
\>[8]%
\AgdaFunction{𝐒}%
\AgdaSpace{}%
\AgdaOperator{%
\AgdaPostulate{→ᶜ}%
}\AgdaSpace{}%
%
\AgdaPostulate{𝐀}%
%
\>[30]\AgdaComment{--\ command\ continuations}\<%
\\
\>[2]%
\AgdaFunction{𝐅}%
%
\>[5]\AgdaSymbol{=}%
\>[8]%
\AgdaPostulate{𝐄}%
\AgdaSpace{}%
\AgdaOperator{%
\AgdaPostulate{⋆}%
}\AgdaSpace{}%
\AgdaOperator{%
\AgdaPostulate{→ᶜ}%
}\AgdaSpace{}%
\AgdaSymbol{(}%
\AgdaPostulate{𝐄}%
\AgdaSpace{}%
\AgdaOperator{%
\AgdaPostulate{→ᶜ}%
}\AgdaSpace{}%
%
\AgdaFunction{𝐂}%
\AgdaSymbol{)}\AgdaSpace{}%
\AgdaOperator{%
\AgdaPostulate{→ᶜ}%
}\AgdaSpace{}%
%
\AgdaFunction{𝐂}%
%
\>[30]\AgdaComment{--\ procedure\ values}\<%
\end{code}
\end{AgdaSuppressSpace}
The published denotational semantics of \emph{Scm}
\cite{Mosses2025CSE} defines
the domain \(𝐄\) of expression values by the equation
\(𝐄 = 𝐓 + 𝐑 + 𝐏 + 𝐌 + 𝐅\), and the domain \(𝐅\) of procedure values by
\(𝐅 = 𝐄^* \to (𝐄 \to 𝐂) \to 𝐂\). Mutually recursive groups of domain
equations have well-defined solutions, but in Agda, defining both the
corresponding domains \AgdaRef{𝐄} and \AgdaRef{𝐅} by equations would cause
the type checker to diverge. Postulating one (or both) of these domains
avoids divergence; postulating \AgdaRef{𝐄} also has the benefit that the
embeddings and projections for its summands subsume the bijection
between \AgdaRef{𝐄} and its intended structure.

The following postulates instantiate the notation used in denotational
semantics for sum domains (Section~\ref{Notation/Sums-sum-domains}):
injection (\AgdaRef{δ\ in⊥\ 𝐄}), inspection (\AgdaRef{ε\ ∈⊥\ D}), and
projection (\AgdaRef{ε\ \textbar{}⊥\ D}) for each summand \AgdaRef{D} of
\AgdaRef{𝐄}.
\begin{AgdaSuppressSpace}
\begin{code}%
\>[2]\AgdaKeyword{postulate}\AgdaSpace{}%
\AgdaKeyword{instance}\<%
\\
\>[2][@{}l@{\AgdaIndent{0}}]%
\>[4]%
\AgdaPostulate{E+=T}%
%
\>[10]\AgdaSymbol{:}\AgdaSpace{}%
%
\AgdaPostulate{𝐄}%
\AgdaSpace{}%
\AgdaOperator{%
\AgdaPostulate{≳}%
}\AgdaSpace{}%
\AgdaNumber{1}\AgdaSpace{}%
\AgdaOperator{%
\AgdaPostulate{↦}%
}\AgdaSpace{}%
%
\AgdaFunction{𝐓}%
\<%
\\
\>[4]%
\AgdaPostulate{E+=R}%
%
\>[10]\AgdaSymbol{:}\AgdaSpace{}%
%
\AgdaPostulate{𝐄}%
\AgdaSpace{}%
\AgdaOperator{%
\AgdaPostulate{≳}%
}\AgdaSpace{}%
\AgdaNumber{2}\AgdaSpace{}%
\AgdaOperator{%
\AgdaPostulate{↦}%
}\AgdaSpace{}%
%
\AgdaFunction{𝐑}%
\<%
\\
\>[4]%
\AgdaPostulate{E+=P}%
%
\>[10]\AgdaSymbol{:}\AgdaSpace{}%
%
\AgdaPostulate{𝐄}%
\AgdaSpace{}%
\AgdaOperator{%
\AgdaPostulate{≳}%
}\AgdaSpace{}%
\AgdaNumber{3}\AgdaSpace{}%
\AgdaOperator{%
\AgdaPostulate{↦}%
}\AgdaSpace{}%
%
\AgdaFunction{𝐏}%
\<%
\\
\>[4]%
\AgdaPostulate{E+=M}%
%
\>[10]\AgdaSymbol{:}\AgdaSpace{}%
%
\AgdaPostulate{𝐄}%
\AgdaSpace{}%
\AgdaOperator{%
\AgdaPostulate{≳}%
}\AgdaSpace{}%
\AgdaNumber{4}\AgdaSpace{}%
\AgdaOperator{%
\AgdaPostulate{↦}%
}\AgdaSpace{}%
%
\AgdaFunction{𝐌}%
\<%
\\
\>[4]%
\AgdaPostulate{E+=F}%
%
\>[10]\AgdaSymbol{:}\AgdaSpace{}%
%
\AgdaPostulate{𝐄}%
\AgdaSpace{}%
\AgdaOperator{%
\AgdaPostulate{≳}%
}\AgdaSpace{}%
\AgdaNumber{5}\AgdaSpace{}%
\AgdaOperator{%
\AgdaPostulate{↦}%
}\AgdaSpace{}%
%
\AgdaFunction{𝐅}%
\<%
\end{code}
\end{AgdaSuppressSpace}
(where the numerical indices and order of the summand declarations are
arbitrary).
\begin{AgdaSuppressSpace}
\begin{code}[hide]%
\>[0]\<%
\\
\>[2]\AgdaKeyword{postulate}\AgdaSpace{}%
\AgdaKeyword{instance}\<%
\\
\>[2][@{}l@{\AgdaIndent{0}}]%
\>[4]%
\AgdaPostulate{eqL}%
%
\>[10]\AgdaSymbol{:}\AgdaSpace{}%
%
\AgdaRecord{Eq}%
\AgdaSpace{}%
%
\AgdaPostulate{Loc}%
\<%
\\
\>[4]%
\AgdaPostulate{eqM}%
%
\>[10]\AgdaSymbol{:}\AgdaSpace{}%
%
\AgdaRecord{Eq}%
\AgdaSpace{}%
%
\AgdaDatatype{Misc}%
\<%
\\
\>[4]%
\AgdaPostulate{eqR}%
%
\>[10]\AgdaSymbol{:}\AgdaSpace{}%
%
\AgdaRecord{Eq}%
\AgdaSpace{}%
%
\AgdaDatatype{Int}%
\<%
\\
\\[\AgdaEmptyExtraSkip]%
\>[2]\AgdaKeyword{variable}\<%
\\
\>[2][@{}l@{\AgdaIndent{0}}]%
\>[4]\AgdaGeneralizable{α}\AgdaSpace{}%
\AgdaSymbol{:}\AgdaSpace{}%
\AgdaOperator{%
\AgdaPostulate{⟪}%
}\AgdaSpace{}%
%
\AgdaFunction{𝐋}%
\AgdaSpace{}%
\AgdaOperator{%
\AgdaPostulate{⟫}%
}\AgdaSymbol{;}%
\>[16]\AgdaGeneralizable{ρ}\AgdaSpace{}%
\AgdaSymbol{:}\AgdaSpace{}%
\AgdaOperator{%
\AgdaPostulate{⟪}%
}\AgdaSpace{}%
%
\AgdaFunction{𝐔}%
\AgdaSpace{}%
\AgdaOperator{%
\AgdaPostulate{⟫}%
}\AgdaSymbol{;}%
\>[28]\AgdaGeneralizable{μ}%
\>[31]\AgdaSymbol{:}\AgdaSpace{}%
\AgdaOperator{%
\AgdaPostulate{⟪}%
}\AgdaSpace{}%
%
\AgdaFunction{𝐌}%
\AgdaSpace{}%
\AgdaOperator{%
\AgdaPostulate{⟫}%
}\AgdaSymbol{;}%
\>[43]\AgdaGeneralizable{ϵ}\AgdaSpace{}%
\AgdaSymbol{:}\AgdaSpace{}%
\AgdaOperator{%
\AgdaPostulate{⟪}%
}\AgdaSpace{}%
%
\AgdaPostulate{𝐄}%
\AgdaSpace{}%
\AgdaOperator{%
\AgdaPostulate{⟫}%
}\<%
\\
\>[4]\AgdaGeneralizable{σ}\AgdaSpace{}%
\AgdaSymbol{:}\AgdaSpace{}%
\AgdaOperator{%
\AgdaPostulate{⟪}%
}\AgdaSpace{}%
%
\AgdaFunction{𝐒}%
\AgdaSpace{}%
\AgdaOperator{%
\AgdaPostulate{⟫}%
}\AgdaSymbol{;}%
\>[16]\AgdaGeneralizable{θ}\AgdaSpace{}%
\AgdaSymbol{:}\AgdaSpace{}%
\AgdaOperator{%
\AgdaPostulate{⟪}%
}\AgdaSpace{}%
%
\AgdaFunction{𝐂}%
\AgdaSpace{}%
\AgdaOperator{%
\AgdaPostulate{⟫}%
}\AgdaSymbol{;}%
\>[28]\AgdaGeneralizable{ϵ⋆}\AgdaSpace{}%
\AgdaSymbol{:}\AgdaSpace{}%
\AgdaOperator{%
\AgdaPostulate{⟪}%
}\AgdaSpace{}%
%
\AgdaPostulate{𝐄}%
\AgdaSpace{}%
\AgdaOperator{%
\AgdaPostulate{⋆}%
}\AgdaSpace{}%
\AgdaOperator{%
\AgdaPostulate{⟫}%
}\AgdaSymbol{;}%
\>[44]\AgdaGeneralizable{φ}\AgdaSpace{}%
\AgdaSymbol{:}\AgdaSpace{}%
\AgdaOperator{%
\AgdaPostulate{⟪}%
}\AgdaSpace{}%
%
\AgdaFunction{𝐅}%
\AgdaSpace{}%
\AgdaOperator{%
\AgdaPostulate{⟫}%
}\AgdaSymbol{;}%
\>[56]\AgdaGeneralizable{κ}\AgdaSpace{}%
\AgdaSymbol{:}\AgdaSpace{}%
\AgdaOperator{%
\AgdaPostulate{⟪}%
}\AgdaSpace{}%
%
\AgdaPostulate{𝐄}%
\AgdaSpace{}%
\AgdaOperator{%
\AgdaPostulate{→ᶜ}%
}\AgdaSpace{}%
%
\AgdaFunction{𝐂}%
\AgdaSpace{}%
\AgdaOperator{%
\AgdaPostulate{⟫}%
}\<%
\end{code}
\end{AgdaSuppressSpace}

\subsubsection{Auxiliary Functions}\label{Examples/Scm/Auxiliary-Functions-auxiliary-functions}

The \(\lambda\)-notation in the Agda definitions of auxiliary functions
for \emph{Scm} corresponds closely to that in its published denotational
semantics
\cite{Mosses2025CSE}.
\begin{AgdaSuppressSpace}
\begin{code}[hide]%
\>[0]\AgdaSymbol{\{-\#}\AgdaSpace{}%
\AgdaKeyword{OPTIONS}\AgdaSpace{}%
\AgdaPragma{--rewriting}\AgdaSpace{}%
\AgdaPragma{--confluence-check}\AgdaSpace{}%
\AgdaPragma{--lossy-unification}\AgdaSpace{}%
\AgdaSymbol{\#-\}}\<%
\\
\>[0]\<%
\end{code}
\begin{code}%
\>[0]\AgdaKeyword{module}\AgdaSpace{}%
%
\AgdaModule{Examples.Scm.Auxiliary-Functions}%
\AgdaSpace{}%
\AgdaKeyword{where}\<%
\end{code}
\begin{code}[hide]%
\>[0]\<%
\\
\>[0][@{}l@{\AgdaIndent{1}}]%
\>[2]\AgdaKeyword{open}\AgdaSpace{}%
\AgdaKeyword{import}\AgdaSpace{}%
%
\AgdaModule{Examples.Scm.Abstract-Syntax}%
\<%
\\
\>[2]\AgdaKeyword{open}\AgdaSpace{}%
\AgdaKeyword{import}\AgdaSpace{}%
%
\AgdaModule{Examples.Scm.Domain-Equations}%
\<%
\\
\>[2]\AgdaKeyword{open}\AgdaSpace{}%
\AgdaKeyword{import}\AgdaSpace{}%
%
\AgdaModule{Notation.Domains}%
\<%
\\
\>[2]\AgdaKeyword{open}\AgdaSpace{}%
\AgdaKeyword{import}\AgdaSpace{}%
%
\AgdaModule{Notation.Functions}%
\<%
\\
\>[2]\AgdaKeyword{open}\AgdaSpace{}%
\AgdaKeyword{import}\AgdaSpace{}%
%
\AgdaModule{Notation.Flat}%
\<%
\\
\>[2]\AgdaKeyword{open}\AgdaSpace{}%
\AgdaKeyword{import}\AgdaSpace{}%
%
\AgdaModule{Notation.Flat.Booleans}%
\<%
\\
\>[2]\AgdaKeyword{open}\AgdaSpace{}%
\AgdaKeyword{import}\AgdaSpace{}%
%
\AgdaModule{Notation.Flat.Naturals}%
\<%
\\
\>[2]\AgdaKeyword{open}\AgdaSpace{}%
\AgdaKeyword{import}\AgdaSpace{}%
%
\AgdaModule{Notation.Sums}%
\<%
\\
\>[2]\AgdaKeyword{open}\AgdaSpace{}%
\AgdaKeyword{import}\AgdaSpace{}%
%
\AgdaModule{Notation.Products}%
\<%
\\
\>[2]\AgdaKeyword{open}\AgdaSpace{}%
\AgdaKeyword{import}\AgdaSpace{}%
%
\AgdaModule{Notation.Products.Sequences}%
\<%
\\
\>[2]\AgdaKeyword{open}\AgdaSpace{}%
\AgdaKeyword{import}\AgdaSpace{}%
%
\AgdaModule{Notation.Updates}%
\<%
\\
\>[2]\AgdaKeyword{open}\AgdaSpace{}%
\AgdaKeyword{import}\AgdaSpace{}%
%
\AgdaModule{Agda.Builtin.Nat}%
\AgdaSpace{}%
\AgdaKeyword{using}\AgdaSpace{}%
\AgdaSymbol{(}%
\AgdaDatatype{Nat}%
\AgdaSymbol{)}\<%
\\
\>[2]\AgdaKeyword{open}\AgdaSpace{}%
\AgdaKeyword{import}\AgdaSpace{}%
%
\AgdaModule{Data.String.Properties}%
\AgdaSpace{}%
\AgdaKeyword{renaming}\AgdaSpace{}%
\AgdaSymbol{(}\AgdaOperator{%
\AgdaFunction{\AgdaUnderscore{}%
==\AgdaUnderscore{}}}\AgdaSpace{}%
\AgdaSymbol{to}\AgdaSpace{}%
\AgdaOperator{%
\AgdaFunction{\AgdaUnderscore{}%
==ˢ\AgdaUnderscore{}}}\AgdaSymbol{)}\AgdaSpace{}%
\AgdaKeyword{using}\AgdaSpace{}%
\AgdaSymbol{()}\<%
\\
\\[\AgdaEmptyExtraSkip]%
\>[2]\AgdaKeyword{postulate}\<%
\\
\>[2][@{}l@{\AgdaIndent{0}}]%
\>[4]\AgdaOperator{%
\AgdaPostulate{\AgdaUnderscore{}%
<ᴿ\AgdaUnderscore{}}}%
\>[10]\AgdaSymbol{:}\AgdaSpace{}%
\AgdaOperator{%
\AgdaPostulate{⟪}%
}\AgdaSpace{}%
%
\AgdaFunction{𝐑}%
\AgdaSpace{}%
\AgdaOperator{%
\AgdaPostulate{→ᶜ}%
}\AgdaSpace{}%
%
\AgdaFunction{𝐑}%
\AgdaSpace{}%
\AgdaOperator{%
\AgdaPostulate{→ᶜ}%
}\AgdaSpace{}%
%
\AgdaFunction{𝐓}%
\AgdaSpace{}%
\AgdaOperator{%
\AgdaPostulate{⟫}%
}\<%
\\
\>[4]\AgdaOperator{%
\AgdaPostulate{\AgdaUnderscore{}%
+ᴿ\AgdaUnderscore{}}}%
\>[10]\AgdaSymbol{:}\AgdaSpace{}%
\AgdaOperator{%
\AgdaPostulate{⟪}%
}\AgdaSpace{}%
%
\AgdaFunction{𝐑}%
\AgdaSpace{}%
\AgdaOperator{%
\AgdaPostulate{→ᶜ}%
}\AgdaSpace{}%
%
\AgdaFunction{𝐑}%
\AgdaSpace{}%
\AgdaOperator{%
\AgdaPostulate{→ᶜ}%
}\AgdaSpace{}%
%
\AgdaFunction{𝐑}%
\AgdaSpace{}%
\AgdaOperator{%
\AgdaPostulate{⟫}%
}\<%
\\
\>[4]\AgdaOperator{%
\AgdaPostulate{\AgdaUnderscore{}%
∧ᵀ\AgdaUnderscore{}}}%
\>[10]\AgdaSymbol{:}\AgdaSpace{}%
\AgdaOperator{%
\AgdaPostulate{⟪}%
}\AgdaSpace{}%
%
\AgdaFunction{𝐓}%
\AgdaSpace{}%
\AgdaOperator{%
\AgdaPostulate{→ᶜ}%
}\AgdaSpace{}%
%
\AgdaFunction{𝐓}%
\AgdaSpace{}%
\AgdaOperator{%
\AgdaPostulate{→ᶜ}%
}\AgdaSpace{}%
%
\AgdaFunction{𝐓}%
\AgdaSpace{}%
\AgdaOperator{%
\AgdaPostulate{⟫}%
}\<%
\\
\\[\AgdaEmptyExtraSkip]%
\>[2]\AgdaKeyword{instance}\<%
\\
\>[2][@{}l@{\AgdaIndent{0}}]%
\>[4]%
\AgdaFunction{eqIde}%
\AgdaSpace{}%
\AgdaSymbol{:}\AgdaSpace{}%
%
\AgdaRecord{Eq}%
\AgdaSpace{}%
%
\AgdaFunction{Ide}%
\<%
\\
\>[4]\AgdaOperator{%
\AgdaField{\AgdaUnderscore{}%
==\AgdaUnderscore{}}}\AgdaSpace{}%
\AgdaSymbol{\{\{}%
\AgdaFunction{eqIde}%
\AgdaSymbol{\}\}}\AgdaSpace{}%
\AgdaSymbol{=}\AgdaSpace{}%
\AgdaOperator{%
\AgdaFunction{\AgdaUnderscore{}%
==ˢ\AgdaUnderscore{}}}\<%
\\
\\[\AgdaEmptyExtraSkip]%
\>[2]\AgdaKeyword{postulate}\AgdaSpace{}%
%
\AgdaPostulate{unknown}%
\AgdaSpace{}%
\AgdaSymbol{:}\AgdaSpace{}%
%
\AgdaPostulate{Loc}%
\<%
\\
\\[\AgdaEmptyExtraSkip]%
\>[2]\AgdaKeyword{postulate}\AgdaSpace{}%
%
\AgdaPostulate{initial-env}%
\AgdaSpace{}%
\AgdaSymbol{:}\AgdaSpace{}%
\AgdaOperator{%
\AgdaPostulate{⟪}%
}\AgdaSpace{}%
%
\AgdaFunction{𝐔}%
\AgdaSpace{}%
\AgdaOperator{%
\AgdaPostulate{⟫}%
}\<%
\end{code}
\begin{code}%
\>[0]\<%
\\
\>[2]%
\AgdaFunction{assign}%
\AgdaSpace{}%
\AgdaSymbol{:}\AgdaSpace{}%
\AgdaOperator{%
\AgdaPostulate{⟪}%
}\AgdaSpace{}%
%
\AgdaFunction{𝐋}%
\AgdaSpace{}%
\AgdaOperator{%
\AgdaPostulate{→ᶜ}%
}\AgdaSpace{}%
%
\AgdaPostulate{𝐄}%
\AgdaSpace{}%
\AgdaOperator{%
\AgdaPostulate{→ᶜ}%
}\AgdaSpace{}%
%
\AgdaFunction{𝐂}%
\AgdaSpace{}%
\AgdaOperator{%
\AgdaPostulate{→ᶜ}%
}\AgdaSpace{}%
%
\AgdaFunction{𝐂}%
\AgdaSpace{}%
\AgdaOperator{%
\AgdaPostulate{⟫}%
}%
\>[37]\AgdaComment{--\ assign\ α\ ϵ\ stores\ ϵ\ at\ location\ α}\<%
\\
\>[2]%
\AgdaFunction{assign}%
\AgdaSpace{}%
\AgdaBound{α}\AgdaSpace{}%
\AgdaBound{ϵ}\AgdaSpace{}%
\AgdaBound{θ}\AgdaSpace{}%
\AgdaBound{σ}\AgdaSpace{}%
\AgdaSymbol{=}\AgdaSpace{}%
\AgdaBound{θ}\AgdaSpace{}%
\AgdaSymbol{(}\AgdaBound{σ}\AgdaSpace{}%
\AgdaOperator{%
\AgdaFunction{[}%
}\AgdaSpace{}%
\AgdaBound{ϵ}\AgdaSpace{}%
\AgdaOperator{%
\AgdaFunction{/}%
}\AgdaSpace{}%
\AgdaBound{α}\AgdaSpace{}%
\AgdaOperator{%
\AgdaFunction{]⊥}%
}\AgdaSymbol{)}\<%
\\
\\[\AgdaEmptyExtraSkip]%
\>[2]%
\AgdaFunction{hold}%
\AgdaSpace{}%
\AgdaSymbol{:}\AgdaSpace{}%
\AgdaOperator{%
\AgdaPostulate{⟪}%
}\AgdaSpace{}%
%
\AgdaFunction{𝐋}%
\AgdaSpace{}%
\AgdaOperator{%
\AgdaPostulate{→ᶜ}%
}\AgdaSpace{}%
\AgdaSymbol{(}%
\AgdaPostulate{𝐄}%
\AgdaSpace{}%
\AgdaOperator{%
\AgdaPostulate{→ᶜ}%
}\AgdaSpace{}%
%
\AgdaFunction{𝐂}%
\AgdaSymbol{)}\AgdaSpace{}%
\AgdaOperator{%
\AgdaPostulate{→ᶜ}%
}\AgdaSpace{}%
%
\AgdaFunction{𝐂}%
\AgdaSpace{}%
\AgdaOperator{%
\AgdaPostulate{⟫}%
}%
\>[37]\AgdaComment{--\ hold\ α\ gives\ the\ value\ stored\ at\ α}\<%
\\
\>[2]%
\AgdaFunction{hold}%
\AgdaSpace{}%
\AgdaBound{α}\AgdaSpace{}%
\AgdaBound{κ}\AgdaSpace{}%
\AgdaBound{σ}\AgdaSpace{}%
\AgdaSymbol{=}\AgdaSpace{}%
\AgdaBound{κ}\AgdaSpace{}%
\AgdaSymbol{(}\AgdaBound{σ}\AgdaSpace{}%
\AgdaBound{α}\AgdaSymbol{)}\AgdaSpace{}%
\AgdaBound{σ}\<%
\end{code}
\end{AgdaSuppressSpace}
In the continuation-passing style used to define auxilary functions for
\emph{Scm}, giving explicit continuity proofs would be particulary
tedious. For example, the function \AgdaRef{hold} is simply a combination
of \(\lambda\)-abstraction and application, which is well-known to
ensure continuity.
\begin{AgdaSuppressSpace}
\begin{code}%
\>[2]\AgdaKeyword{postulate}\AgdaSpace{}%
%
\AgdaPostulate{new}%
\AgdaSpace{}%
\AgdaSymbol{:}\AgdaSpace{}%
\AgdaOperator{%
\AgdaPostulate{⟪}%
}\AgdaSpace{}%
\AgdaSymbol{(}%
\AgdaFunction{𝐋}%
\AgdaSpace{}%
\AgdaOperator{%
\AgdaPostulate{→ᶜ}%
}\AgdaSpace{}%
%
\AgdaFunction{𝐂}%
\AgdaSymbol{)}\AgdaSpace{}%
\AgdaOperator{%
\AgdaPostulate{→ᶜ}%
}\AgdaSpace{}%
%
\AgdaFunction{𝐂}%
\AgdaSpace{}%
\AgdaOperator{%
\AgdaPostulate{⟫}%
}%
\>[37]\AgdaComment{--\ new\ gives\ an\ unallocated\ location}\<%
\\
\\[\AgdaEmptyExtraSkip]%
\>[2]%
\AgdaFunction{alloc}%
\AgdaSpace{}%
\AgdaSymbol{:}\AgdaSpace{}%
\AgdaOperator{%
\AgdaPostulate{⟪}%
}\AgdaSpace{}%
%
\AgdaPostulate{𝐄}%
\AgdaSpace{}%
\AgdaOperator{%
\AgdaPostulate{→ᶜ}%
}\AgdaSpace{}%
\AgdaSymbol{(}%
\AgdaFunction{𝐋}%
\AgdaSpace{}%
\AgdaOperator{%
\AgdaPostulate{→ᶜ}%
}\AgdaSpace{}%
%
\AgdaFunction{𝐂}%
\AgdaSymbol{)}\AgdaSpace{}%
\AgdaOperator{%
\AgdaPostulate{→ᶜ}%
}\AgdaSpace{}%
%
\AgdaFunction{𝐂}%
\AgdaSpace{}%
\AgdaOperator{%
\AgdaPostulate{⟫}%
}%
\>[37]\AgdaComment{--\ alloc\ ϵ\ allocates\ a\ location\ for\ ϵ}\<%
\\
\>[2]%
\AgdaFunction{alloc}%
\AgdaSpace{}%
\AgdaBound{ϵ}\AgdaSpace{}%
\AgdaBound{κ}\AgdaSpace{}%
\AgdaSymbol{=}\AgdaSpace{}%
%
\AgdaPostulate{new}%
\AgdaSpace{}%
\AgdaSymbol{(λ}\AgdaSpace{}%
\AgdaBound{α}\AgdaSpace{}%
\AgdaSymbol{→}\AgdaSpace{}%
%
\AgdaFunction{assign}%
\AgdaSpace{}%
\AgdaBound{α}\AgdaSpace{}%
\AgdaBound{ϵ}\AgdaSpace{}%
\AgdaSymbol{(}\AgdaBound{κ}\AgdaSpace{}%
\AgdaBound{α}\AgdaSymbol{))}\<%
\\
\\[\AgdaEmptyExtraSkip]%
\>[2]\AgdaKeyword{postulate}\AgdaSpace{}%
%
\AgdaPostulate{initial-store}%
\AgdaSpace{}%
\AgdaSymbol{:}\AgdaSpace{}%
\AgdaOperator{%
\AgdaPostulate{⟪}%
}\AgdaSpace{}%
%
\AgdaFunction{𝐒}%
\AgdaSpace{}%
\AgdaOperator{%
\AgdaPostulate{⟫}%
}%
\>[37]\AgdaComment{--\ may\ have\ initialised\ locations}\<%
\\
\>[0]\<%
\end{code}
\begin{code}[hide]%
\>[0][@{}l@{\AgdaIndent{1}}]%
\>[2]\AgdaKeyword{postulate}\AgdaSpace{}%
%
\AgdaPostulate{finished}%
\AgdaSpace{}%
\AgdaSymbol{:}\AgdaSpace{}%
\AgdaOperator{%
\AgdaPostulate{⟪}%
}\AgdaSpace{}%
%
\AgdaFunction{𝐒}%
\AgdaSpace{}%
\AgdaOperator{%
\AgdaPostulate{→ᶜ}%
}\AgdaSpace{}%
%
\AgdaPostulate{𝐀}%
\AgdaSpace{}%
\AgdaOperator{%
\AgdaPostulate{⟫}%
}%
\>[37]\AgdaComment{--\ obtain\ answer\ from\ the\ final\ store}\<%
\end{code}
\end{AgdaSuppressSpace}
Conventional denotational definitions usually leave the injection
function \AgdaRef{↑} from sets into
flat domains (Section~\ref{Notation/Flat-flat-domains})
implicit, in contrast to the following embedding of the definition of
\AgdaRef{truish}:
\begin{AgdaSuppressSpace}
\begin{code}%
\>[2]%
\AgdaFunction{truish}%
\AgdaSpace{}%
\AgdaSymbol{:}\AgdaSpace{}%
\AgdaOperator{%
\AgdaPostulate{⟪}%
}\AgdaSpace{}%
%
\AgdaPostulate{𝐄}%
\AgdaSpace{}%
\AgdaOperator{%
\AgdaPostulate{→ᶜ}%
}\AgdaSpace{}%
%
\AgdaFunction{𝐓}%
\AgdaSpace{}%
\AgdaOperator{%
\AgdaPostulate{⟫}%
}%
\>[37]\AgdaComment{--\ truish\ ε\ is\ true\ for\ all\ ε\ except\ false}\<%
\\
\>[2]%
\AgdaFunction{truish}%
\AgdaSpace{}%
\AgdaBound{ϵ}\AgdaSpace{}%
\AgdaSymbol{=}%
\>[14]\AgdaSymbol{(}\AgdaBound{ϵ}\AgdaSpace{}%
\AgdaOperator{%
\AgdaPostulate{∈⊥}%
}\AgdaSpace{}%
%
\AgdaFunction{𝐓}%
\AgdaSymbol{)}\AgdaSpace{}%
\AgdaOperator{%
\AgdaFunction{⟶}%
}\AgdaSpace{}%
\AgdaSymbol{(((}\AgdaBound{ϵ}\AgdaSpace{}%
\AgdaOperator{%
\AgdaPostulate{|⊥}%
}\AgdaSpace{}%
%
\AgdaFunction{𝐓}%
\AgdaSymbol{)}\AgdaSpace{}%
\AgdaOperator{%
\AgdaFunction{==⊥}%
}\AgdaSpace{}%
%
\AgdaPostulate{↑}%
\AgdaSpace{}%
%
\AgdaInductiveConstructor{false}%
\AgdaSymbol{)}\AgdaSpace{}%
\AgdaOperator{%
\AgdaFunction{⟶}%
}\AgdaSpace{}%
%
\AgdaPostulate{↑}%
\AgdaSpace{}%
%
\AgdaInductiveConstructor{false}%
\AgdaSpace{}%
\AgdaOperator{%
\AgdaFunction{,}%
}\AgdaSpace{}%
%
\AgdaPostulate{↑}%
\AgdaSpace{}%
%
\AgdaInductiveConstructor{true}%
\AgdaSymbol{)}\AgdaSpace{}%
\AgdaOperator{%
\AgdaFunction{,}%
}\<%
\\
\>[14]%
\AgdaPostulate{↑}%
\AgdaSpace{}%
%
\AgdaInductiveConstructor{true}%
\<%
\end{code}
\end{AgdaSuppressSpace}
The remaining auxiliary function definitions shown here involve the
operations for (finite) sequences \AgdaRef{ϵ⋆} declared in the module for
sequence
domains (Section~\ref{Notation/Products/Sequences-sequence-domains}).
\begin{AgdaSuppressSpace}
\begin{code}%
\>[2]%
\AgdaFunction{cons}%
\AgdaSpace{}%
\AgdaSymbol{:}\AgdaSpace{}%
\AgdaOperator{%
\AgdaPostulate{⟪}%
}\AgdaSpace{}%
%
\AgdaFunction{𝐅}%
\AgdaSpace{}%
\AgdaOperator{%
\AgdaPostulate{⟫}%
}%
\>[37]\AgdaComment{--\ cons\ ⟨\ ϵ₁\ ,\ ϵ₂\ ⟩\ allocates\ and\ initialises\ a\ pair}\<%
\\
\>[2]%
\AgdaFunction{cons}%
\AgdaSpace{}%
\AgdaBound{ϵ⋆}\AgdaSpace{}%
\AgdaBound{κ}\AgdaSpace{}%
\AgdaSymbol{=}%
\>[15]\AgdaSymbol{(}%
\AgdaPostulate{\#}%
%
\>[191I]\AgdaBound{ϵ⋆}\AgdaSpace{}%
\AgdaOperator{%
\AgdaFunction{==⊥}%
}\AgdaSpace{}%
%
\AgdaPostulate{↑}%
\AgdaSpace{}%
\AgdaNumber{2}\AgdaSymbol{)}\AgdaSpace{}%
\AgdaOperator{%
\AgdaFunction{⟶}%
}\<%
\\
\>[.][@{}l@{}]\<[191I]%
\>[18]%
\AgdaFunction{alloc}%
\AgdaSpace{}%
\AgdaSymbol{(}\AgdaBound{ϵ⋆}\AgdaSpace{}%
\AgdaOperator{%
\AgdaPostulate{↓}%
}\AgdaSpace{}%
\AgdaNumber{1}\AgdaSymbol{)}\AgdaSpace{}%
\AgdaSymbol{(λ}\AgdaSpace{}%
\AgdaBound{α₁}\AgdaSpace{}%
\AgdaSymbol{→}\<%
\\
\>[18][@{}l@{\AgdaIndent{0}}]%
\>[20]%
\AgdaFunction{alloc}%
\AgdaSpace{}%
\AgdaSymbol{(}\AgdaBound{ϵ⋆}\AgdaSpace{}%
\AgdaOperator{%
\AgdaPostulate{↓}%
}\AgdaSpace{}%
\AgdaNumber{2}\AgdaSymbol{)}\AgdaSpace{}%
\AgdaSymbol{(λ}\AgdaSpace{}%
\AgdaBound{α₂}\AgdaSpace{}%
\AgdaSymbol{→}\AgdaSpace{}%
\AgdaBound{κ}\AgdaSpace{}%
\AgdaSymbol{((}\AgdaBound{α₁}\AgdaSpace{}%
\AgdaOperator{%
\AgdaPostulate{,}%
}\AgdaSpace{}%
\AgdaBound{α₂}\AgdaSymbol{)}\AgdaSpace{}%
\AgdaOperator{%
\AgdaPostulate{in⊥}%
}\AgdaSpace{}%
%
\AgdaPostulate{𝐄}%
\AgdaSymbol{)))}\AgdaSpace{}%
\AgdaOperator{%
\AgdaFunction{,}%
}\<%
\\
\>[15][@{}l@{\AgdaIndent{0}}]%
\>[16]%
\AgdaPostulate{⊥}%
\<%
\end{code}
\end{AgdaSuppressSpace}
In \cite{Mosses2025CSE} the
auxiliary function \(\textit{list}\) is defined by recursion on
\AgdaRef{ϵ⋆}. Agda accepts recursive definitions only when it can
mechanically prove that the recursion terminates, which is not the case
for arguments in postulated types. The following definition uses the
postulated operation \AgdaRef{fix} to avoid recursion.
\begin{AgdaSuppressSpace}
\begin{code}%
\>[2]%
\AgdaFunction{list}%
\AgdaSpace{}%
\AgdaSymbol{:}\AgdaSpace{}%
\AgdaOperator{%
\AgdaPostulate{⟪}%
}\AgdaSpace{}%
%
\AgdaFunction{𝐅}%
\AgdaSpace{}%
\AgdaOperator{%
\AgdaPostulate{⟫}%
}%
\>[37]\AgdaComment{--\ list\ ϵ⋆\ allocates\ and\ initialises\ a\ list}\<%
\\
\>[2]%
\AgdaFunction{list}%
\AgdaSpace{}%
\AgdaSymbol{=}%
\>[11]%
\AgdaPostulate{fix}%
\AgdaSpace{}%
\AgdaSymbol{λ}\AgdaSpace{}%
\AgdaSymbol{(}\AgdaBound{list′}\AgdaSpace{}%
\AgdaSymbol{:}\AgdaSpace{}%
\AgdaOperator{%
\AgdaPostulate{⟪}%
}\AgdaSpace{}%
%
\AgdaFunction{𝐅}%
\AgdaSpace{}%
\AgdaOperator{%
\AgdaPostulate{⟫}%
}\AgdaSymbol{)}\AgdaSpace{}%
\AgdaSymbol{→}\AgdaSpace{}%
\AgdaSymbol{λ}\AgdaSpace{}%
\AgdaBound{ϵ⋆}\AgdaSpace{}%
\AgdaBound{κ}\AgdaSpace{}%
\AgdaSymbol{→}\<%
\\
\>[11][@{}l@{\AgdaIndent{0}}]%
\>[12]\AgdaSymbol{(}%
\AgdaPostulate{\#}%
\AgdaSpace{}%
\AgdaBound{ϵ⋆}\AgdaSpace{}%
\AgdaOperator{%
\AgdaFunction{==⊥}%
}\AgdaSpace{}%
%
\AgdaPostulate{↑}%
\AgdaSpace{}%
\AgdaNumber{0}\AgdaSymbol{)}\AgdaSpace{}%
\AgdaOperator{%
\AgdaFunction{⟶}%
}\AgdaSpace{}%
\AgdaBound{κ}\AgdaSpace{}%
\AgdaSymbol{(}%
\AgdaPostulate{↑}%
\AgdaSpace{}%
%
\AgdaInductiveConstructor{null}%
\AgdaSpace{}%
\AgdaOperator{%
\AgdaPostulate{in⊥}%
}\AgdaSpace{}%
%
\AgdaPostulate{𝐄}%
\AgdaSymbol{)}\AgdaSpace{}%
\AgdaOperator{%
\AgdaFunction{,}%
}\<%
\\
\>[12]\AgdaBound{list′}\AgdaSpace{}%
\AgdaSymbol{(}\AgdaBound{ϵ⋆}\AgdaSpace{}%
\AgdaOperator{%
\AgdaPostulate{†}%
}\AgdaSpace{}%
\AgdaNumber{1}\AgdaSymbol{)}\AgdaSpace{}%
\AgdaSymbol{(λ}\AgdaSpace{}%
\AgdaBound{ϵ}\AgdaSpace{}%
\AgdaSymbol{→}\AgdaSpace{}%
%
\AgdaFunction{cons}%
\AgdaSpace{}%
\AgdaOperator{%
\AgdaPostulate{⟨}%
}\AgdaSpace{}%
\AgdaSymbol{(}\AgdaBound{ϵ⋆}\AgdaSpace{}%
\AgdaOperator{%
\AgdaPostulate{↓}%
}\AgdaSpace{}%
\AgdaNumber{1}\AgdaSymbol{)}\AgdaSpace{}%
\AgdaOperator{%
\AgdaPostulate{,}%
}\AgdaSpace{}%
\AgdaBound{ϵ}\AgdaSpace{}%
\AgdaOperator{%
\AgdaPostulate{⟩}%
}\AgdaSpace{}%
\AgdaBound{κ}\AgdaSymbol{)}\<%
\end{code}
\begin{code}[hide]%
\>[0]\<%
\\
\>[2]%
\AgdaFunction{car}%
\AgdaSpace{}%
\AgdaSymbol{:}\AgdaSpace{}%
\AgdaOperator{%
\AgdaPostulate{⟪}%
}\AgdaSpace{}%
%
\AgdaFunction{𝐅}%
\AgdaSpace{}%
\AgdaOperator{%
\AgdaPostulate{⟫}%
}%
\>[37]\AgdaComment{--\ car\ ⟨\ ϵ\ ⟩\ gives\ the\ head\ of\ the\ list\ ϵ}\<%
\\
\>[2]%
\AgdaFunction{car}%
\AgdaSpace{}%
\AgdaBound{ϵ⋆}\AgdaSpace{}%
\AgdaBound{κ}\AgdaSpace{}%
\AgdaSymbol{=}\AgdaSpace{}%
\AgdaSymbol{(}%
\AgdaPostulate{\#}%
\AgdaSpace{}%
\AgdaBound{ϵ⋆}\AgdaSpace{}%
\AgdaOperator{%
\AgdaFunction{==⊥}%
}\AgdaSpace{}%
%
\AgdaPostulate{↑}%
\AgdaSpace{}%
\AgdaNumber{1}\AgdaSymbol{)}\AgdaSpace{}%
\AgdaOperator{%
\AgdaFunction{⟶}%
}\AgdaSpace{}%
%
\AgdaFunction{hold}%
\AgdaSpace{}%
\AgdaSymbol{(((}\AgdaBound{ϵ⋆}\AgdaSpace{}%
\AgdaOperator{%
\AgdaPostulate{↓}%
}\AgdaSpace{}%
\AgdaNumber{1}\AgdaSymbol{)}\AgdaSpace{}%
\AgdaOperator{%
\AgdaPostulate{|⊥}%
}\AgdaSpace{}%
%
\AgdaFunction{𝐏}%
\AgdaSymbol{)}\AgdaSpace{}%
\AgdaOperator{%
\AgdaPostulate{↓₁}%
}\AgdaSymbol{)}\AgdaSpace{}%
\AgdaBound{κ}\AgdaSpace{}%
\AgdaOperator{%
\AgdaFunction{,}%
}\AgdaSpace{}%
%
\AgdaPostulate{⊥}%
\<%
\\
\\[\AgdaEmptyExtraSkip]%
\>[2]%
\AgdaFunction{cdr}%
\AgdaSpace{}%
\AgdaSymbol{:}\AgdaSpace{}%
\AgdaOperator{%
\AgdaPostulate{⟪}%
}\AgdaSpace{}%
%
\AgdaFunction{𝐅}%
\AgdaSpace{}%
\AgdaOperator{%
\AgdaPostulate{⟫}%
}%
\>[37]\AgdaComment{--\ cdr\ ⟨\ ϵ\ ⟩\ gives\ the\ tail\ of\ the\ list\ ϵ}\<%
\\
\>[2]%
\AgdaFunction{cdr}%
\AgdaSpace{}%
\AgdaBound{ϵ⋆}\AgdaSpace{}%
\AgdaBound{κ}\AgdaSpace{}%
\AgdaSymbol{=}\AgdaSpace{}%
\AgdaSymbol{(}%
\AgdaPostulate{\#}%
\AgdaSpace{}%
\AgdaBound{ϵ⋆}\AgdaSpace{}%
\AgdaOperator{%
\AgdaFunction{==⊥}%
}\AgdaSpace{}%
%
\AgdaPostulate{↑}%
\AgdaSpace{}%
\AgdaNumber{1}\AgdaSymbol{)}\AgdaSpace{}%
\AgdaOperator{%
\AgdaFunction{⟶}%
}\AgdaSpace{}%
%
\AgdaFunction{hold}%
\AgdaSpace{}%
\AgdaSymbol{(((}\AgdaBound{ϵ⋆}\AgdaSpace{}%
\AgdaOperator{%
\AgdaPostulate{↓}%
}\AgdaSpace{}%
\AgdaNumber{1}\AgdaSymbol{)}\AgdaSpace{}%
\AgdaOperator{%
\AgdaPostulate{|⊥}%
}\AgdaSpace{}%
%
\AgdaFunction{𝐏}%
\AgdaSymbol{)}\AgdaSpace{}%
\AgdaOperator{%
\AgdaPostulate{↓₂}%
}\AgdaSymbol{)}\AgdaSpace{}%
\AgdaBound{κ}\AgdaSpace{}%
\AgdaOperator{%
\AgdaFunction{,}%
}\AgdaSpace{}%
%
\AgdaPostulate{⊥}%
\<%
\\
\\[\AgdaEmptyExtraSkip]%
\>[2]%
\AgdaFunction{setcar}%
\AgdaSpace{}%
\AgdaSymbol{:}\AgdaSpace{}%
\AgdaOperator{%
\AgdaPostulate{⟪}%
}\AgdaSpace{}%
%
\AgdaFunction{𝐅}%
\AgdaSpace{}%
\AgdaOperator{%
\AgdaPostulate{⟫}%
}%
\>[37]\AgdaComment{--\ setcar\ ⟨\ ϵ₁\ ,\ ϵ₂\ ⟩\ stores\ ϵ₂\ in\ the\ head\ of\ list\ ϵ₁}\<%
\\
\>[2]%
\AgdaFunction{setcar}%
\AgdaSpace{}%
\AgdaBound{ϵ⋆}\AgdaSpace{}%
\AgdaBound{κ}\AgdaSpace{}%
\AgdaSymbol{=}\<%
\\
\>[2][@{}l@{\AgdaIndent{0}}]%
\>[4]\AgdaSymbol{(}%
\AgdaPostulate{\#}%
\AgdaSpace{}%
\AgdaBound{ϵ⋆}\AgdaSpace{}%
\AgdaOperator{%
\AgdaFunction{==⊥}%
}\AgdaSpace{}%
%
\AgdaPostulate{↑}%
\AgdaSpace{}%
\AgdaNumber{2}\AgdaSymbol{)}\AgdaSpace{}%
\AgdaOperator{%
\AgdaFunction{⟶}%
}\<%
\\
\>[4][@{}l@{\AgdaIndent{0}}]%
\>[6]%
\AgdaFunction{assign}%
\AgdaSpace{}%
\AgdaSymbol{(((}\AgdaBound{ϵ⋆}\AgdaSpace{}%
\AgdaOperator{%
\AgdaPostulate{↓}%
}\AgdaSpace{}%
\AgdaNumber{1}\AgdaSymbol{)}\AgdaSpace{}%
\AgdaOperator{%
\AgdaPostulate{|⊥}%
}\AgdaSpace{}%
%
\AgdaFunction{𝐏}%
\AgdaSymbol{)}\AgdaSpace{}%
\AgdaOperator{%
\AgdaPostulate{↓₁}%
}\AgdaSymbol{)}\AgdaSpace{}%
\AgdaSymbol{(}\AgdaBound{ϵ⋆}\AgdaSpace{}%
\AgdaOperator{%
\AgdaPostulate{↓}%
}\AgdaSpace{}%
\AgdaNumber{2}\AgdaSymbol{)}\AgdaSpace{}%
\AgdaSymbol{(}\AgdaBound{κ}\AgdaSpace{}%
\AgdaSymbol{(}%
\AgdaPostulate{↑}%
\AgdaSpace{}%
%
\AgdaInductiveConstructor{unspecified}%
\AgdaSpace{}%
\AgdaOperator{%
\AgdaPostulate{in⊥}%
}\AgdaSpace{}%
%
\AgdaPostulate{𝐄}%
\AgdaSymbol{))}\AgdaSpace{}%
\AgdaOperator{%
\AgdaFunction{,}%
}\<%
\\
\>[4]%
\AgdaPostulate{⊥}%
\<%
\\
\\[\AgdaEmptyExtraSkip]%
\>[2]%
\AgdaFunction{setcdr}%
\AgdaSpace{}%
\AgdaSymbol{:}\AgdaSpace{}%
\AgdaOperator{%
\AgdaPostulate{⟪}%
}\AgdaSpace{}%
%
\AgdaFunction{𝐅}%
\AgdaSpace{}%
\AgdaOperator{%
\AgdaPostulate{⟫}%
}%
\>[37]\AgdaComment{--\ setcdr\ ⟨\ ϵ₁\ ,\ ϵ₂\ ⟩\ stores\ ϵ₂\ in\ the\ tail\ of\ list\ ϵ₁}\<%
\\
\>[2]%
\AgdaFunction{setcdr}%
\AgdaSpace{}%
\AgdaBound{ϵ⋆}\AgdaSpace{}%
\AgdaBound{κ}\AgdaSpace{}%
\AgdaSymbol{=}\<%
\\
\>[2][@{}l@{\AgdaIndent{0}}]%
\>[4]\AgdaSymbol{(}%
\AgdaPostulate{\#}%
\AgdaSpace{}%
\AgdaBound{ϵ⋆}\AgdaSpace{}%
\AgdaOperator{%
\AgdaFunction{==⊥}%
}\AgdaSpace{}%
%
\AgdaPostulate{↑}%
\AgdaSpace{}%
\AgdaNumber{2}\AgdaSymbol{)}\AgdaSpace{}%
\AgdaOperator{%
\AgdaFunction{⟶}%
}\<%
\\
\>[4][@{}l@{\AgdaIndent{0}}]%
\>[6]%
\AgdaFunction{assign}%
\AgdaSpace{}%
\AgdaSymbol{(((}\AgdaBound{ϵ⋆}\AgdaSpace{}%
\AgdaOperator{%
\AgdaPostulate{↓}%
}\AgdaSpace{}%
\AgdaNumber{1}\AgdaSymbol{)}\AgdaSpace{}%
\AgdaOperator{%
\AgdaPostulate{|⊥}%
}\AgdaSpace{}%
%
\AgdaFunction{𝐏}%
\AgdaSymbol{)}\AgdaSpace{}%
\AgdaOperator{%
\AgdaPostulate{↓₂}%
}\AgdaSymbol{)}\AgdaSpace{}%
\AgdaSymbol{(}\AgdaBound{ϵ⋆}\AgdaSpace{}%
\AgdaOperator{%
\AgdaPostulate{↓}%
}\AgdaSpace{}%
\AgdaNumber{2}\AgdaSymbol{)}\AgdaSpace{}%
\AgdaSymbol{(}\AgdaBound{κ}\AgdaSpace{}%
\AgdaSymbol{(}%
\AgdaPostulate{↑}%
\AgdaSpace{}%
%
\AgdaInductiveConstructor{unspecified}%
\AgdaSpace{}%
\AgdaOperator{%
\AgdaPostulate{in⊥}%
}\AgdaSpace{}%
%
\AgdaPostulate{𝐄}%
\AgdaSymbol{))}\AgdaSpace{}%
\AgdaOperator{%
\AgdaFunction{,}%
}\<%
\\
\>[4]%
\AgdaPostulate{⊥}%
\<%
\end{code}
\end{AgdaSuppressSpace}

\subsubsection{Semantic Functions}\label{Examples/Scm/Semantic-Functions-semantic-functions}

The \(\lambda\)-notation in the Agda definitions of semantic functions
for \emph{Scm} corresponds closely to that in its published denotational
semantics
\cite{Mosses2025CSE}.
\begin{AgdaSuppressSpace}
\begin{code}[hide]%
\>[0]\<%
\\
\>[0]\AgdaSymbol{\{-\#}\AgdaSpace{}%
\AgdaKeyword{OPTIONS}\AgdaSpace{}%
\AgdaPragma{--rewriting}\AgdaSpace{}%
\AgdaPragma{--confluence-check}\AgdaSpace{}%
\AgdaPragma{--lossy-unification}\AgdaSpace{}%
\AgdaSymbol{\#-\}}\<%
\\
\>[0]\<%
\end{code}
\begin{code}%
\>[0]\AgdaKeyword{module}\AgdaSpace{}%
%
\AgdaModule{Examples.Scm.Semantic-Functions}%
\AgdaSpace{}%
\AgdaKeyword{where}\<%
\end{code}
\begin{code}[hide]%
\>[0]\<%
\\
\>[0][@{}l@{\AgdaIndent{1}}]%
\>[2]\AgdaKeyword{open}\AgdaSpace{}%
\AgdaKeyword{import}\AgdaSpace{}%
%
\AgdaModule{Examples.Scm.Abstract-Syntax}%
\<%
\\
\>[2]\AgdaKeyword{open}\AgdaSpace{}%
\AgdaKeyword{import}\AgdaSpace{}%
%
\AgdaModule{Examples.Scm.Domain-Equations}%
\<%
\\
\>[2]\AgdaKeyword{open}\AgdaSpace{}%
\AgdaKeyword{import}\AgdaSpace{}%
%
\AgdaModule{Examples.Scm.Auxiliary-Functions}%
\<%
\\
\>[2]\AgdaKeyword{open}\AgdaSpace{}%
\AgdaKeyword{import}\AgdaSpace{}%
%
\AgdaModule{Notation.Domains}%
\<%
\\
\>[2]\AgdaKeyword{open}\AgdaSpace{}%
\AgdaKeyword{import}\AgdaSpace{}%
%
\AgdaModule{Notation.Functions}%
\<%
\\
\>[2]\AgdaKeyword{open}\AgdaSpace{}%
\AgdaKeyword{import}\AgdaSpace{}%
%
\AgdaModule{Notation.Flat}%
\<%
\\
\>[2]\AgdaKeyword{open}\AgdaSpace{}%
\AgdaKeyword{import}\AgdaSpace{}%
%
\AgdaModule{Notation.Flat.Booleans}%
\<%
\\
\>[2]\AgdaKeyword{open}\AgdaSpace{}%
\AgdaKeyword{import}\AgdaSpace{}%
%
\AgdaModule{Notation.Sums}%
\<%
\\
\>[2]\AgdaKeyword{open}\AgdaSpace{}%
\AgdaKeyword{import}\AgdaSpace{}%
%
\AgdaModule{Notation.Products.Sequences}%
\<%
\\
\>[2]\AgdaKeyword{open}\AgdaSpace{}%
\AgdaKeyword{import}\AgdaSpace{}%
%
\AgdaModule{Notation.Updates}%
\<%
\end{code}
\begin{code}%
\>[0]\<%
\\
\>[2]\AgdaOperator{%
\AgdaFunction{𝒦⟦\AgdaUnderscore{}%
⟧}}%
\>[9]\AgdaSymbol{:}%
\>[12]\AgdaOperator{%
\AgdaPostulate{⟪}%
}\AgdaSpace{}%
%
\AgdaDatatype{Con}%
\AgdaSpace{}%
\AgdaOperator{%
\AgdaPostulate{→ˢ}%
}\AgdaSpace{}%
%
\AgdaPostulate{𝐄}%
\AgdaSpace{}%
\AgdaOperator{%
\AgdaPostulate{⟫}%
}%
\>[46]\AgdaComment{--\ constant\ denotations}\<%
\\
\>[2]\AgdaOperator{%
\AgdaFunction{ℰ⟦\AgdaUnderscore{}%
⟧}}%
\>[9]\AgdaSymbol{:}%
\>[12]\AgdaOperator{%
\AgdaPostulate{⟪}%
}\AgdaSpace{}%
%
\AgdaDatatype{Exp}%
\AgdaSpace{}%
\AgdaOperator{%
\AgdaPostulate{→ˢ}%
}\AgdaSpace{}%
%
\AgdaFunction{𝐔}%
\AgdaSpace{}%
\AgdaOperator{%
\AgdaPostulate{→ᶜ}%
}\AgdaSpace{}%
\AgdaSymbol{(}%
\AgdaPostulate{𝐄}%
\AgdaSpace{}%
\AgdaOperator{%
\AgdaPostulate{→ᶜ}%
}\AgdaSpace{}%
%
\AgdaFunction{𝐂}%
\AgdaSymbol{)}\AgdaSpace{}%
\AgdaOperator{%
\AgdaPostulate{→ᶜ}%
}\AgdaSpace{}%
%
\AgdaFunction{𝐂}%
\AgdaSpace{}%
\AgdaOperator{%
\AgdaPostulate{⟫}%
}%
\>[46]\AgdaComment{--\ expression\ denotations}\<%
\\
\>[2]\AgdaOperator{%
\AgdaFunction{ℰ⋆⟦\AgdaUnderscore{}%
⟧}}%
\>[9]\AgdaSymbol{:}%
\>[12]\AgdaOperator{%
\AgdaPostulate{⟪}%
}\AgdaSpace{}%
%
\AgdaDatatype{Exp⋆}%
\AgdaSpace{}%
\AgdaOperator{%
\AgdaPostulate{→ˢ}%
}\AgdaSpace{}%
%
\AgdaFunction{𝐔}%
\AgdaSpace{}%
\AgdaOperator{%
\AgdaPostulate{→ᶜ}%
}\AgdaSpace{}%
\AgdaSymbol{(}%
\AgdaPostulate{𝐄}%
\AgdaSpace{}%
\AgdaOperator{%
\AgdaPostulate{⋆}%
}\AgdaSpace{}%
\AgdaOperator{%
\AgdaPostulate{→ᶜ}%
}\AgdaSpace{}%
%
\AgdaFunction{𝐂}%
\AgdaSymbol{)}\AgdaSpace{}%
\AgdaOperator{%
\AgdaPostulate{→ᶜ}%
}\AgdaSpace{}%
%
\AgdaFunction{𝐂}%
\AgdaSpace{}%
\AgdaOperator{%
\AgdaPostulate{⟫}%
}%
\>[46]\AgdaComment{--\ sequence\ denotations}\<%
\\
\\[\AgdaEmptyExtraSkip]%
\>[2]\AgdaOperator{%
\AgdaFunction{𝒦⟦}%
}\AgdaSpace{}%
%
\AgdaInductiveConstructor{int}%
\AgdaSpace{}%
\AgdaBound{Z}\AgdaSpace{}%
\AgdaOperator{%
\AgdaFunction{⟧}%
}%
\>[14]\AgdaSymbol{=}\AgdaSpace{}%
%
\AgdaPostulate{↑}%
\AgdaSpace{}%
\AgdaBound{Z}\AgdaSpace{}%
\AgdaOperator{%
\AgdaPostulate{in⊥}%
}\AgdaSpace{}%
%
\AgdaPostulate{𝐄}%
\<%
\\
\>[2]\AgdaOperator{%
\AgdaFunction{𝒦⟦}%
}\AgdaSpace{}%
%
\AgdaInductiveConstructor{\#t}%
\AgdaSpace{}%
\AgdaOperator{%
\AgdaFunction{⟧}%
}%
\>[14]\AgdaSymbol{=}\AgdaSpace{}%
%
\AgdaPostulate{↑}%
\AgdaSpace{}%
%
\AgdaInductiveConstructor{true}%
\AgdaSpace{}%
\AgdaOperator{%
\AgdaPostulate{in⊥}%
}\AgdaSpace{}%
%
\AgdaPostulate{𝐄}%
\<%
\\
\>[2]\AgdaOperator{%
\AgdaFunction{𝒦⟦}%
}\AgdaSpace{}%
%
\AgdaInductiveConstructor{\#f}%
\AgdaSpace{}%
\AgdaOperator{%
\AgdaFunction{⟧}%
}%
\>[14]\AgdaSymbol{=}\AgdaSpace{}%
%
\AgdaPostulate{↑}%
\AgdaSpace{}%
%
\AgdaInductiveConstructor{false}%
\AgdaSpace{}%
\AgdaOperator{%
\AgdaPostulate{in⊥}%
}\AgdaSpace{}%
%
\AgdaPostulate{𝐄}%
\<%
\\
\\[\AgdaEmptyExtraSkip]%
\>[2]\AgdaOperator{%
\AgdaFunction{ℰ⟦}%
}\AgdaSpace{}%
%
\AgdaInductiveConstructor{con}%
\AgdaSpace{}%
\AgdaBound{K}\AgdaSpace{}%
\AgdaOperator{%
\AgdaFunction{⟧}%
}\AgdaSpace{}%
\AgdaBound{ρ}\AgdaSpace{}%
\AgdaBound{κ}%
\>[31]\AgdaSymbol{=}\AgdaSpace{}%
\AgdaBound{κ}\AgdaSpace{}%
\AgdaSymbol{(}\AgdaOperator{%
\AgdaFunction{𝒦⟦}%
}\AgdaSpace{}%
\AgdaBound{K}\AgdaSpace{}%
\AgdaOperator{%
\AgdaFunction{⟧}%
}\AgdaSymbol{)}\<%
\\
\>[2]\AgdaOperator{%
\AgdaFunction{ℰ⟦}%
}\AgdaSpace{}%
%
\AgdaInductiveConstructor{ide}%
\AgdaSpace{}%
\AgdaBound{I}\AgdaSpace{}%
\AgdaOperator{%
\AgdaFunction{⟧}%
}\AgdaSpace{}%
\AgdaBound{ρ}\AgdaSpace{}%
\AgdaBound{κ}%
\>[31]\AgdaSymbol{=}\AgdaSpace{}%
%
\AgdaFunction{hold}%
\AgdaSpace{}%
\AgdaSymbol{(}\AgdaBound{ρ}\AgdaSpace{}%
\AgdaBound{I}\AgdaSymbol{)}\AgdaSpace{}%
\AgdaBound{κ}\<%
\\
\>[2]\AgdaOperator{%
\AgdaFunction{ℰ⟦}%
}\AgdaSpace{}%
\AgdaOperator{%
\AgdaInductiveConstructor{⦅}%
}\AgdaSpace{}%
\AgdaBound{E}\AgdaSpace{}%
\AgdaOperator{%
\AgdaInductiveConstructor{␣}%
}\AgdaSpace{}%
\AgdaBound{E⋆}\AgdaSpace{}%
\AgdaOperator{%
\AgdaInductiveConstructor{⦆}%
}\AgdaSpace{}%
\AgdaOperator{%
\AgdaFunction{⟧}%
}\AgdaSpace{}%
\AgdaBound{ρ}\AgdaSpace{}%
\AgdaBound{κ}%
\>[31]\AgdaSymbol{=}\AgdaSpace{}%
\AgdaOperator{%
\AgdaFunction{ℰ⟦}%
}\AgdaSpace{}%
\AgdaBound{E}\AgdaSpace{}%
\AgdaOperator{%
\AgdaFunction{⟧}%
}\AgdaSpace{}%
\AgdaBound{ρ}\AgdaSpace{}%
\AgdaSymbol{(λ}\AgdaSpace{}%
\AgdaBound{ϵ}\AgdaSpace{}%
\AgdaSymbol{→}\AgdaSpace{}%
\AgdaOperator{%
\AgdaFunction{ℰ⋆⟦}%
}\AgdaSpace{}%
\AgdaBound{E⋆}\AgdaSpace{}%
\AgdaOperator{%
\AgdaFunction{⟧}%
}\AgdaSpace{}%
\AgdaBound{ρ}\AgdaSpace{}%
\AgdaSymbol{(λ}\AgdaSpace{}%
\AgdaBound{ϵ⋆}\AgdaSpace{}%
\AgdaSymbol{→}\AgdaSpace{}%
\AgdaSymbol{(}\AgdaBound{ϵ}\AgdaSpace{}%
\AgdaOperator{%
\AgdaPostulate{|⊥}%
}\AgdaSpace{}%
%
\AgdaFunction{𝐅}%
\AgdaSymbol{)}\AgdaSpace{}%
\AgdaBound{ϵ⋆}\AgdaSpace{}%
\AgdaBound{κ}\AgdaSymbol{))}\<%
\\
\>[2]\AgdaOperator{%
\AgdaFunction{ℰ⟦}%
}\AgdaSpace{}%
\AgdaOperator{%
\AgdaInductiveConstructor{⦅lambda}%
}\AgdaSpace{}%
\AgdaBound{I}\AgdaSpace{}%
\AgdaOperator{%
\AgdaInductiveConstructor{␣}%
}\AgdaSpace{}%
\AgdaBound{E}\AgdaSpace{}%
\AgdaOperator{%
\AgdaInductiveConstructor{⦆}%
}\AgdaSpace{}%
\AgdaOperator{%
\AgdaFunction{⟧}%
}\AgdaSpace{}%
\AgdaBound{ρ}\AgdaSpace{}%
\AgdaBound{κ}%
\>[31]\AgdaSymbol{=}\AgdaSpace{}%
\AgdaBound{κ}\AgdaSpace{}%
\AgdaSymbol{(}%
\>[39]\AgdaSymbol{(λ}\AgdaSpace{}%
\AgdaBound{ϵ⋆}\AgdaSpace{}%
\AgdaBound{κ′}\AgdaSpace{}%
\AgdaSymbol{→}\<%
\\
\>[39][@{}l@{\AgdaIndent{0}}]%
\>[40]%
\AgdaFunction{list}%
\AgdaSpace{}%
\AgdaBound{ϵ⋆}\AgdaSpace{}%
\AgdaSymbol{(λ}\AgdaSpace{}%
\AgdaBound{ϵ}\AgdaSpace{}%
\AgdaSymbol{→}\AgdaSpace{}%
%
\AgdaFunction{alloc}%
\AgdaSpace{}%
\AgdaBound{ϵ}\AgdaSpace{}%
\AgdaSymbol{(λ}\AgdaSpace{}%
\AgdaBound{α}\AgdaSpace{}%
\AgdaSymbol{→}\AgdaSpace{}%
\AgdaOperator{%
\AgdaFunction{ℰ⟦}%
}\AgdaSpace{}%
\AgdaBound{E}\AgdaSpace{}%
\AgdaOperator{%
\AgdaFunction{⟧}%
}\AgdaSpace{}%
\AgdaSymbol{(}\AgdaBound{ρ}\AgdaSpace{}%
\AgdaOperator{%
\AgdaFunction{[}%
}\AgdaSpace{}%
\AgdaBound{α}\AgdaSpace{}%
\AgdaOperator{%
\AgdaFunction{/}%
}\AgdaSpace{}%
\AgdaBound{I}\AgdaSpace{}%
\AgdaOperator{%
\AgdaFunction{]}%
}\AgdaSymbol{)}\AgdaSpace{}%
\AgdaBound{κ′}\AgdaSymbol{))}\<%
\\
\>[39]\AgdaSymbol{)}\AgdaSpace{}%
\AgdaOperator{%
\AgdaPostulate{in⊥}%
}\AgdaSpace{}%
%
\AgdaPostulate{𝐄}%
\AgdaSpace{}%
\AgdaSymbol{)}\<%
\\
\>[2]\AgdaOperator{%
\AgdaFunction{ℰ⟦}%
}\AgdaSpace{}%
\AgdaOperator{%
\AgdaInductiveConstructor{⦅if}%
}\AgdaSpace{}%
\AgdaBound{E}\AgdaSpace{}%
\AgdaOperator{%
\AgdaInductiveConstructor{␣}%
}\AgdaSpace{}%
\AgdaBound{E₁}\AgdaSpace{}%
\AgdaOperator{%
\AgdaInductiveConstructor{␣}%
}\AgdaSpace{}%
\AgdaBound{E₂}\AgdaSpace{}%
\AgdaOperator{%
\AgdaInductiveConstructor{⦆}%
}\AgdaSpace{}%
\AgdaOperator{%
\AgdaFunction{⟧}%
}\AgdaSpace{}%
\AgdaBound{ρ}\AgdaSpace{}%
\AgdaBound{κ}%
\>[31]\AgdaSymbol{=}\AgdaSpace{}%
\AgdaOperator{%
\AgdaFunction{ℰ⟦}%
}\AgdaSpace{}%
\AgdaBound{E}\AgdaSpace{}%
\AgdaOperator{%
\AgdaFunction{⟧}%
}\AgdaSpace{}%
\AgdaBound{ρ}\AgdaSpace{}%
\AgdaSymbol{(λ}\AgdaSpace{}%
\AgdaBound{ϵ}\AgdaSpace{}%
\AgdaSymbol{→}\AgdaSpace{}%
%
\AgdaFunction{truish}%
\AgdaSpace{}%
\AgdaBound{ϵ}\AgdaSpace{}%
\AgdaOperator{%
\AgdaFunction{⟶}%
}\AgdaSpace{}%
\AgdaOperator{%
\AgdaFunction{ℰ⟦}%
}\AgdaSpace{}%
\AgdaBound{E₁}\AgdaSpace{}%
\AgdaOperator{%
\AgdaFunction{⟧}%
}\AgdaSpace{}%
\AgdaBound{ρ}\AgdaSpace{}%
\AgdaBound{κ}\AgdaSpace{}%
\AgdaOperator{%
\AgdaFunction{,}%
}\AgdaSpace{}%
\AgdaOperator{%
\AgdaFunction{ℰ⟦}%
}\AgdaSpace{}%
\AgdaBound{E₂}\AgdaSpace{}%
\AgdaOperator{%
\AgdaFunction{⟧}%
}\AgdaSpace{}%
\AgdaBound{ρ}\AgdaSpace{}%
\AgdaBound{κ}\AgdaSymbol{)}\<%
\\
\>[2]\AgdaOperator{%
\AgdaFunction{ℰ⟦}%
}\AgdaSpace{}%
\AgdaOperator{%
\AgdaInductiveConstructor{⦅set!}%
}\AgdaSpace{}%
\AgdaBound{I}\AgdaSpace{}%
\AgdaOperator{%
\AgdaInductiveConstructor{␣}%
}\AgdaSpace{}%
\AgdaBound{E}\AgdaSpace{}%
\AgdaOperator{%
\AgdaInductiveConstructor{⦆}%
}\AgdaSpace{}%
\AgdaOperator{%
\AgdaFunction{⟧}%
}\AgdaSpace{}%
\AgdaBound{ρ}\AgdaSpace{}%
\AgdaBound{κ}%
\>[31]\AgdaSymbol{=}\AgdaSpace{}%
\AgdaOperator{%
\AgdaFunction{ℰ⟦}%
}\AgdaSpace{}%
\AgdaBound{E}\AgdaSpace{}%
\AgdaOperator{%
\AgdaFunction{⟧}%
}\AgdaSpace{}%
\AgdaBound{ρ}\AgdaSpace{}%
\AgdaSymbol{(λ}\AgdaSpace{}%
\AgdaBound{ϵ}\AgdaSpace{}%
\AgdaSymbol{→}\AgdaSpace{}%
%
\AgdaFunction{assign}%
\AgdaSpace{}%
\AgdaSymbol{(}\AgdaBound{ρ}\AgdaSpace{}%
\AgdaBound{I}\AgdaSymbol{)}\AgdaSpace{}%
\AgdaBound{ϵ}\AgdaSpace{}%
\AgdaSymbol{(}\AgdaBound{κ}\AgdaSpace{}%
\AgdaSymbol{(}%
\AgdaPostulate{↑}%
\AgdaSpace{}%
%
\AgdaInductiveConstructor{unspecified}%
\AgdaSpace{}%
\AgdaOperator{%
\AgdaPostulate{in⊥}%
}\AgdaSpace{}%
%
\AgdaPostulate{𝐄}%
\AgdaSymbol{)))}\<%
\\
\\[\AgdaEmptyExtraSkip]%
\>[2]\AgdaOperator{%
\AgdaFunction{ℰ⋆⟦}%
}\AgdaSpace{}%
%
\AgdaInductiveConstructor{␣␣␣}%
\AgdaSpace{}%
\AgdaOperator{%
\AgdaFunction{⟧}%
}\AgdaSpace{}%
\AgdaBound{ρ}\AgdaSpace{}%
\AgdaBound{κ⋆}%
\>[22]\AgdaSymbol{=}\AgdaSpace{}%
\AgdaBound{κ⋆}\AgdaSpace{}%
%
\AgdaPostulate{⟨⟩}%
\<%
\\
\>[2]\AgdaOperator{%
\AgdaFunction{ℰ⋆⟦}%
}\AgdaSpace{}%
\AgdaBound{E}\AgdaSpace{}%
\AgdaOperator{%
\AgdaInductiveConstructor{␣␣}%
}\AgdaSpace{}%
\AgdaBound{E⋆}\AgdaSpace{}%
\AgdaOperator{%
\AgdaFunction{⟧}%
}\AgdaSpace{}%
\AgdaBound{ρ}\AgdaSpace{}%
\AgdaBound{κ⋆}%
\>[22]\AgdaSymbol{=}\AgdaSpace{}%
\AgdaOperator{%
\AgdaFunction{ℰ⟦}%
}\AgdaSpace{}%
\AgdaBound{E}\AgdaSpace{}%
\AgdaOperator{%
\AgdaFunction{⟧}%
}\AgdaSpace{}%
\AgdaBound{ρ}\AgdaSpace{}%
\AgdaSymbol{(λ}\AgdaSpace{}%
\AgdaBound{ϵ}\AgdaSpace{}%
\AgdaSymbol{→}\AgdaSpace{}%
\AgdaOperator{%
\AgdaFunction{ℰ⋆⟦}%
}\AgdaSpace{}%
\AgdaBound{E⋆}\AgdaSpace{}%
\AgdaOperator{%
\AgdaFunction{⟧}%
}\AgdaSpace{}%
\AgdaBound{ρ}\AgdaSpace{}%
\AgdaSymbol{(λ}\AgdaSpace{}%
\AgdaBound{ϵ⋆}\AgdaSpace{}%
\AgdaSymbol{→}\AgdaSpace{}%
\AgdaBound{κ⋆}\AgdaSpace{}%
\AgdaSymbol{(}\AgdaOperator{%
\AgdaPostulate{⟨}%
}\AgdaSpace{}%
\AgdaBound{ϵ}\AgdaSpace{}%
\AgdaOperator{%
\AgdaPostulate{⟩}%
}\AgdaSpace{}%
\AgdaOperator{%
\AgdaPostulate{§}%
}\AgdaSpace{}%
\AgdaBound{ϵ⋆}\AgdaSymbol{)))}\<%
\end{code}
\begin{code}[hide]%
\>[0]\<%
\\
\>[2]\AgdaOperator{%
\AgdaFunction{ℬ⟦\AgdaUnderscore{}%
⟧}}%
\>[9]\AgdaSymbol{:}%
\>[12]\AgdaOperator{%
\AgdaPostulate{⟪}%
}\AgdaSpace{}%
%
\AgdaDatatype{Body}%
\AgdaSpace{}%
\AgdaOperator{%
\AgdaPostulate{→ˢ}%
}\AgdaSpace{}%
%
\AgdaFunction{𝐔}%
\AgdaSpace{}%
\AgdaOperator{%
\AgdaPostulate{→ᶜ}%
}\AgdaSpace{}%
\AgdaSymbol{(}%
\AgdaFunction{𝐔}%
\AgdaSpace{}%
\AgdaOperator{%
\AgdaPostulate{→ᶜ}%
}\AgdaSpace{}%
%
\AgdaFunction{𝐂}%
\AgdaSymbol{)}\AgdaSpace{}%
\AgdaOperator{%
\AgdaPostulate{→ᶜ}%
}\AgdaSpace{}%
%
\AgdaFunction{𝐂}%
\AgdaSpace{}%
\AgdaOperator{%
\AgdaPostulate{⟫}%
}%
\>[46]\AgdaComment{--\ body\ denotations}\<%
\\
\>[2]\AgdaOperator{%
\AgdaFunction{ℬ⁺⟦\AgdaUnderscore{}%
⟧}}%
\>[9]\AgdaSymbol{:}%
\>[12]\AgdaOperator{%
\AgdaPostulate{⟪}%
}\AgdaSpace{}%
%
\AgdaDatatype{Body⁺}%
\AgdaSpace{}%
\AgdaOperator{%
\AgdaPostulate{→ˢ}%
}\AgdaSpace{}%
%
\AgdaFunction{𝐔}%
\AgdaSpace{}%
\AgdaOperator{%
\AgdaPostulate{→ᶜ}%
}\AgdaSpace{}%
\AgdaSymbol{(}%
\AgdaFunction{𝐔}%
\AgdaSpace{}%
\AgdaOperator{%
\AgdaPostulate{→ᶜ}%
}\AgdaSpace{}%
%
\AgdaFunction{𝐂}%
\AgdaSymbol{)}\AgdaSpace{}%
\AgdaOperator{%
\AgdaPostulate{→ᶜ}%
}\AgdaSpace{}%
%
\AgdaFunction{𝐂}%
\AgdaSpace{}%
\AgdaOperator{%
\AgdaPostulate{⟫}%
}%
\>[46]\AgdaComment{--\ sequence\ denotations}\<%
\\
\>[2]\AgdaOperator{%
\AgdaFunction{𝒫⟦\AgdaUnderscore{}%
⟧}}%
\>[9]\AgdaSymbol{:}%
\>[12]\AgdaOperator{%
\AgdaPostulate{⟪}%
}\AgdaSpace{}%
%
\AgdaDatatype{Prog}%
\AgdaSpace{}%
\AgdaOperator{%
\AgdaPostulate{→ˢ}%
}\AgdaSpace{}%
%
\AgdaPostulate{𝐀}%
\AgdaSpace{}%
\AgdaOperator{%
\AgdaPostulate{⟫}%
}%
\>[46]\AgdaComment{--\ program\ execution\ answers}\<%
\\
\\[\AgdaEmptyExtraSkip]%
\>[2]\AgdaOperator{%
\AgdaFunction{ℬ⟦}%
}\AgdaSpace{}%
\AgdaOperator{%
\AgdaInductiveConstructor{␣␣}%
}\AgdaSpace{}%
\AgdaBound{E}\AgdaSpace{}%
\AgdaOperator{%
\AgdaFunction{⟧}%
}\AgdaSpace{}%
\AgdaBound{ρ}\AgdaSpace{}%
\AgdaBound{κ}\AgdaSpace{}%
\AgdaSymbol{=}\AgdaSpace{}%
\AgdaOperator{%
\AgdaFunction{ℰ⟦}%
}\AgdaSpace{}%
\AgdaBound{E}\AgdaSpace{}%
\AgdaOperator{%
\AgdaFunction{⟧}%
}\AgdaSpace{}%
\AgdaBound{ρ}\AgdaSpace{}%
\AgdaSymbol{(λ}\AgdaSpace{}%
\AgdaBound{ϵ}\AgdaSpace{}%
\AgdaSymbol{→}\AgdaSpace{}%
\AgdaBound{κ}\AgdaSpace{}%
\AgdaBound{ρ}\AgdaSymbol{)}\<%
\\
\>[2]\AgdaOperator{%
\AgdaFunction{ℬ⟦}%
}\AgdaSpace{}%
\AgdaOperator{%
\AgdaInductiveConstructor{⦅define}%
}\AgdaSpace{}%
\AgdaBound{I}\AgdaSpace{}%
\AgdaOperator{%
\AgdaInductiveConstructor{␣}%
}\AgdaSpace{}%
\AgdaBound{E}\AgdaSpace{}%
\AgdaOperator{%
\AgdaInductiveConstructor{⦆}%
}\AgdaSpace{}%
\AgdaOperator{%
\AgdaFunction{⟧}%
}\AgdaSpace{}%
\AgdaBound{ρ}\AgdaSpace{}%
\AgdaBound{κ}\AgdaSpace{}%
\AgdaSymbol{=}\<%
\\
\>[2][@{}l@{\AgdaIndent{0}}]%
\>[4]\AgdaOperator{%
\AgdaFunction{ℰ⟦}%
}\AgdaSpace{}%
\AgdaBound{E}\AgdaSpace{}%
\AgdaOperator{%
\AgdaFunction{⟧}%
}\AgdaSpace{}%
\AgdaBound{ρ}\AgdaSpace{}%
\AgdaSymbol{(λ}\AgdaSpace{}%
\AgdaBound{ϵ}\AgdaSpace{}%
\AgdaSymbol{→}\AgdaSpace{}%
\AgdaSymbol{(}\AgdaBound{ρ}\AgdaSpace{}%
\AgdaBound{I}\AgdaSpace{}%
\AgdaOperator{%
\AgdaFunction{==⊥}%
}\AgdaSpace{}%
%
\AgdaPostulate{↑}%
\AgdaSpace{}%
%
\AgdaPostulate{unknown}%
\AgdaSymbol{)}\AgdaSpace{}%
\AgdaOperator{%
\AgdaFunction{⟶}%
}\AgdaSpace{}%
%
\AgdaFunction{alloc}%
\AgdaSpace{}%
\AgdaBound{ϵ}\AgdaSpace{}%
\AgdaSymbol{(λ}\AgdaSpace{}%
\AgdaBound{α}\AgdaSpace{}%
\AgdaSymbol{→}\AgdaSpace{}%
\AgdaBound{κ}\AgdaSpace{}%
\AgdaSymbol{(}\AgdaBound{ρ}\AgdaSpace{}%
\AgdaOperator{%
\AgdaFunction{[}%
}\AgdaSpace{}%
\AgdaBound{α}\AgdaSpace{}%
\AgdaOperator{%
\AgdaFunction{/}%
}\AgdaSpace{}%
\AgdaBound{I}\AgdaSpace{}%
\AgdaOperator{%
\AgdaFunction{]}%
}\AgdaSymbol{))}\AgdaSpace{}%
\AgdaOperator{%
\AgdaFunction{,}%
}\<%
\\
\>[4]%
\AgdaFunction{assign}%
\AgdaSpace{}%
\AgdaSymbol{(}\AgdaBound{ρ}\AgdaSpace{}%
\AgdaBound{I}\AgdaSymbol{)}\AgdaSpace{}%
\AgdaBound{ϵ}\AgdaSpace{}%
\AgdaSymbol{(}\AgdaBound{κ}\AgdaSpace{}%
\AgdaBound{ρ}\AgdaSymbol{))}\<%
\\
\>[2]\AgdaOperator{%
\AgdaFunction{ℬ⟦}%
}\AgdaSpace{}%
\AgdaOperator{%
\AgdaInductiveConstructor{⦅begin}%
}\AgdaSpace{}%
\AgdaBound{B⁺}\AgdaSpace{}%
\AgdaOperator{%
\AgdaInductiveConstructor{⦆}%
}\AgdaSpace{}%
\AgdaOperator{%
\AgdaFunction{⟧}%
}\AgdaSpace{}%
\AgdaBound{ρ}\AgdaSpace{}%
\AgdaBound{κ}\AgdaSpace{}%
\AgdaSymbol{=}\<%
\\
\>[2][@{}l@{\AgdaIndent{0}}]%
\>[4]\AgdaOperator{%
\AgdaFunction{ℬ⁺⟦}%
}\AgdaSpace{}%
\AgdaBound{B⁺}\AgdaSpace{}%
\AgdaOperator{%
\AgdaFunction{⟧}%
}\AgdaSpace{}%
\AgdaBound{ρ}\AgdaSpace{}%
\AgdaBound{κ}\<%
\\
\\[\AgdaEmptyExtraSkip]%
\>[2]\AgdaOperator{%
\AgdaFunction{ℬ⁺⟦}%
}\AgdaSpace{}%
\AgdaOperator{%
\AgdaInductiveConstructor{␣␣}%
}\AgdaSpace{}%
\AgdaBound{B}\AgdaSpace{}%
\AgdaOperator{%
\AgdaFunction{⟧}%
}\AgdaSpace{}%
\AgdaBound{ρ}\AgdaSpace{}%
\AgdaBound{κ}\AgdaSpace{}%
\AgdaSymbol{=}\<%
\\
\>[2][@{}l@{\AgdaIndent{0}}]%
\>[4]\AgdaOperator{%
\AgdaFunction{ℬ⟦}%
}\AgdaSpace{}%
\AgdaBound{B}\AgdaSpace{}%
\AgdaOperator{%
\AgdaFunction{⟧}%
}\AgdaSpace{}%
\AgdaBound{ρ}\AgdaSpace{}%
\AgdaBound{κ}\<%
\\
\>[2]\AgdaOperator{%
\AgdaFunction{ℬ⁺⟦}%
}\AgdaSpace{}%
\AgdaBound{B}\AgdaSpace{}%
\AgdaOperator{%
\AgdaInductiveConstructor{␣␣}%
}\AgdaSpace{}%
\AgdaBound{B⁺}\AgdaSpace{}%
\AgdaOperator{%
\AgdaFunction{⟧}%
}\AgdaSpace{}%
\AgdaBound{ρ}\AgdaSpace{}%
\AgdaBound{κ}\AgdaSpace{}%
\AgdaSymbol{=}\<%
\\
\>[2][@{}l@{\AgdaIndent{0}}]%
\>[4]\AgdaOperator{%
\AgdaFunction{ℬ⟦}%
}\AgdaSpace{}%
\AgdaBound{B}\AgdaSpace{}%
\AgdaOperator{%
\AgdaFunction{⟧}%
}\AgdaSpace{}%
\AgdaBound{ρ}\AgdaSpace{}%
\AgdaSymbol{(λ}\AgdaSpace{}%
\AgdaBound{ρ′}\AgdaSpace{}%
\AgdaSymbol{→}\AgdaSpace{}%
\AgdaOperator{%
\AgdaFunction{ℬ⁺⟦}%
}\AgdaSpace{}%
\AgdaBound{B⁺}\AgdaSpace{}%
\AgdaOperator{%
\AgdaFunction{⟧}%
}\AgdaSpace{}%
\AgdaBound{ρ′}\AgdaSpace{}%
\AgdaBound{κ}\AgdaSymbol{)}\<%
\\
\\[\AgdaEmptyExtraSkip]%
\>[2]\AgdaOperator{%
\AgdaFunction{𝒫⟦}%
}\AgdaSpace{}%
%
\AgdaInductiveConstructor{␣␣␣}%
\AgdaSpace{}%
\AgdaOperator{%
\AgdaFunction{⟧}%
}%
\>[14]\AgdaSymbol{=}\AgdaSpace{}%
%
\AgdaPostulate{finished}%
\AgdaSpace{}%
%
\AgdaPostulate{initial-store}%
\<%
\\
\>[2]\AgdaOperator{%
\AgdaFunction{𝒫⟦}%
}\AgdaSpace{}%
\AgdaOperator{%
\AgdaInductiveConstructor{␣␣}%
}\AgdaSpace{}%
\AgdaBound{B⁺}\AgdaSpace{}%
\AgdaOperator{%
\AgdaFunction{⟧}%
}%
\>[14]\AgdaSymbol{=}\AgdaSpace{}%
\AgdaOperator{%
\AgdaFunction{ℬ⁺⟦}%
}\AgdaSpace{}%
\AgdaBound{B⁺}\AgdaSpace{}%
\AgdaOperator{%
\AgdaFunction{⟧}%
}\AgdaSpace{}%
%
\AgdaPostulate{initial-env}%
\AgdaSpace{}%
\AgdaSymbol{(λ}\AgdaSpace{}%
\AgdaBound{ρ}\AgdaSpace{}%
\AgdaSymbol{→}\AgdaSpace{}%
%
\AgdaPostulate{finished}%
\AgdaSymbol{)}\AgdaSpace{}%
%
\AgdaPostulate{initial-store}%
\<%
\end{code}
\end{AgdaSuppressSpace}

\end{AgdaAlign}

\begin{AgdaAlign}
\section{Postulated Properties}\label{Properties-postulated-properties}

The postulated
domain notation (Section~\ref{Notation-postulated-domain-notation}) declares domain constructors and their associated
operations, which we use in our embedding of conventional denotational
semantic definitions in Agda. The corresponding \AgdaRef{Properties}
modules postulate basic properties of some of the associated operations.

The postulated properties presented below support proofs that
denotations compute the expected values These properties are expected to
hold in various categories of domains, but they do \emph{not} define the
\emph{mathematical structure} of domains.

When postulated properties are declared as \emph{rewrite rules}, Agda
can use them \emph{automatically} in proofs. Agda also has an option to
check that the declared rewrite rules form a confluent system. Rewrite
rules are safe to use with \AgdaRef{Agda.Builtin.Equality} when that
option is enabled. Confluent but non-terminating rewrite rules cannot
break consistency
\cite{Cockx2021TRT}. The rewrite
rules declared in the modules imported below support \emph{automatic}
proof of all the illustrative tests: the proof terms are simply
\AgdaRef{refl} (i.e., reflexivity).
\begin{AgdaSuppressSpace}
\begin{code}[hide]%
\>[0]\AgdaSymbol{\{-\#}\AgdaSpace{}%
\AgdaKeyword{OPTIONS}\AgdaSpace{}%
\AgdaPragma{--rewriting}\AgdaSpace{}%
\AgdaPragma{--confluence-check}\AgdaSpace{}%
\AgdaSymbol{\#-\}}\<%
\\
\>[0]\<%
\end{code}
\begin{code}%
\>[0]\AgdaKeyword{module}\AgdaSpace{}%
%
\AgdaModule{Properties}%
\AgdaSpace{}%
\AgdaKeyword{where}\<%
\\
\>[0][@{}l@{\AgdaIndent{0}}]%
\>[2]\AgdaKeyword{import}\AgdaSpace{}%
%
\AgdaModule{Properties.Functions}%
\<%
\\
\>[2]\AgdaKeyword{import}\AgdaSpace{}%
%
\AgdaModule{Properties.Flat}%
\<%
\\
\>[2]\AgdaKeyword{import}\AgdaSpace{}%
%
\AgdaModule{Properties.Recursion}%
\<%
\end{code}
\end{AgdaSuppressSpace}
Removing any of the rewrite rules in the above modules would break the
proof for at least one of our tests of the \AgdaRef{LC} and \AgdaRef{PCF}
examples. Further modules with postulated properties will be needed when
tests for equivalence of denotations of \AgdaRef{Scm} expressions are
added.

Rewrite rules are safe to use with \AgdaRef{Agda.Builtin.Equality} when
that option is enabled. In principle, all \AgdaRef{refl} proof terms that
rely on rewrite rules could be replaced by proofs that apply the
postulated properties to specified subterms. However, we expect that it
would be quite tedious to develop such proofs, and reading them is
unlikely to provide new insights.
\begin{AgdaSuppressSpace}
\begin{code}[hide]%
\>[2]\AgdaKeyword{import}\AgdaSpace{}%
%
\AgdaModule{Properties.Flat.Booleans}%
\<%
\\
\>[2]\AgdaKeyword{import}\AgdaSpace{}%
%
\AgdaModule{Properties.Flat.Naturals}%
\<%
\\
\>[2]\AgdaKeyword{import}\AgdaSpace{}%
%
\AgdaModule{Properties.Sums}%
\<%
\\
\>[2]\AgdaKeyword{import}\AgdaSpace{}%
%
\AgdaModule{Properties.Products}%
\<%
\\
\>[2]\AgdaKeyword{import}\AgdaSpace{}%
%
\AgdaModule{Properties.Products.Tuples}%
\<%
\\
\>[2]\AgdaKeyword{import}\AgdaSpace{}%
%
\AgdaModule{Properties.Products.Sequences}%
\<%
\\
\>[2]\AgdaKeyword{import}\AgdaSpace{}%
%
\AgdaModule{Properties.Updates}%
\<%
\end{code}
\end{AgdaSuppressSpace}

\subsection{Function Domains}\label{Properties/Functions-function-domains}
\begin{AgdaSuppressSpace}
\begin{code}[hide]%
\>[0]\AgdaSymbol{\{-\#}\AgdaSpace{}%
\AgdaKeyword{OPTIONS}\AgdaSpace{}%
\AgdaPragma{--rewriting}\AgdaSpace{}%
\AgdaPragma{--confluence-check}\AgdaSpace{}%
\AgdaPragma{--lossy-unification}\AgdaSpace{}%
\AgdaSymbol{\#-\}}\<%
\\
\>[0]\<%
\end{code}
\begin{code}%
\>[0]\AgdaKeyword{module}\AgdaSpace{}%
%
\AgdaModule{Properties.Functions}%
\AgdaSpace{}%
\AgdaKeyword{where}\<%
\end{code}
\begin{code}[hide]%
\>[0][@{}l@{\AgdaIndent{1}}]%
\>[2]\AgdaKeyword{open}\AgdaSpace{}%
\AgdaKeyword{import}\AgdaSpace{}%
%
\AgdaModule{Notation.Domains}%
\<%
\\
\>[2]\AgdaKeyword{open}\AgdaSpace{}%
\AgdaKeyword{import}\AgdaSpace{}%
%
\AgdaModule{Notation.Functions}%
\AgdaSpace{}%
\AgdaKeyword{public}\<%
\\
\>[2]\AgdaKeyword{open}\AgdaSpace{}%
\AgdaKeyword{import}\AgdaSpace{}%
%
\AgdaModule{Agda.Builtin.Equality}%
\AgdaSpace{}%
\AgdaKeyword{public}\AgdaSpace{}%
\AgdaKeyword{using}\AgdaSpace{}%
\AgdaSymbol{(}\AgdaOperator{%
\AgdaDatatype{\AgdaUnderscore{}%
≡\AgdaUnderscore{}}}\AgdaSymbol{;}\AgdaSpace{}%
%
\AgdaInductiveConstructor{refl}%
\AgdaSymbol{)}\<%
\\
\>[2]\AgdaKeyword{open}\AgdaSpace{}%
\AgdaKeyword{import}\AgdaSpace{}%
%
\AgdaModule{Agda.Builtin.Equality.Rewrite}%
\AgdaSpace{}%
\AgdaKeyword{using}\AgdaSpace{}%
\AgdaSymbol{()}\<%
\\
\>[0]\<%
\end{code}
\begin{code}%
\>[0][@{}l@{\AgdaIndent{1}}]%
\>[2]\AgdaKeyword{postulate}\<%
\\
\>[2][@{}l@{\AgdaIndent{0}}]%
\>[4]%
\AgdaPostulate{apply-fix}%
\AgdaSpace{}%
\AgdaSymbol{:}\AgdaSpace{}%
\AgdaSymbol{\{}\AgdaBound{φ}\AgdaSpace{}%
\AgdaSymbol{:}\AgdaSpace{}%
\AgdaOperator{%
\AgdaPostulate{⟪}%
}\AgdaSpace{}%
\AgdaGeneralizable{D}\AgdaSpace{}%
\AgdaOperator{%
\AgdaPostulate{→ᶜ}%
}\AgdaSpace{}%
\AgdaGeneralizable{D}\AgdaSpace{}%
\AgdaOperator{%
\AgdaPostulate{⟫}%
}\AgdaSymbol{\}}\AgdaSpace{}%
\AgdaSymbol{→}\AgdaSpace{}%
%
\AgdaPostulate{fix}%
\AgdaSpace{}%
\AgdaBound{φ}\AgdaSpace{}%
\AgdaOperator{%
\AgdaDatatype{≡}%
}\AgdaSpace{}%
\AgdaBound{φ}\AgdaSpace{}%
\AgdaSymbol{(}%
\AgdaPostulate{fix}%
\AgdaSpace{}%
\AgdaBound{φ}\AgdaSymbol{)}\AgdaSpace{}%
\AgdaComment{--\ apply-fix\{φ\}\ unfolds\ fix\ φ\ once}\<%
\\
\>[2]\AgdaSymbol{\{-\#}\AgdaSpace{}%
\AgdaKeyword{REWRITE}\AgdaSpace{}%
%
\AgdaPostulate{apply-fix}%
\AgdaSpace{}%
\AgdaSymbol{\#-\}}\<%
\end{code}
\end{AgdaSuppressSpace}
The rewrite rule \AgdaRef{apply-fix} does not cause the type-checker to
diverge, despite the obvious non-termination. Agda's type checker uses
\emph{weak head evaluation}: it only unfolds expressions to the point
where the top-level constructor becomes visible. In particular, it will
not evaluate under a λ-abstraction unless it is being compared to
another λ-abstraction and the bodies are not syntactically equal.

\subsection{Flat Domains}\label{Properties/Flat-flat-domains}
\begin{AgdaSuppressSpace}
\begin{code}[hide]%
\>[0]\AgdaSymbol{\{-\#}\AgdaSpace{}%
\AgdaKeyword{OPTIONS}\AgdaSpace{}%
\AgdaPragma{--rewriting}\AgdaSpace{}%
\AgdaPragma{--confluence-check}\AgdaSpace{}%
\AgdaPragma{--lossy-unification}\AgdaSpace{}%
\AgdaSymbol{\#-\}}\<%
\\
\>[0]\<%
\end{code}
\begin{code}%
\>[0]\AgdaKeyword{module}\AgdaSpace{}%
%
\AgdaModule{Properties.Flat}%
\AgdaSpace{}%
\AgdaKeyword{where}\<%
\end{code}
\begin{code}[hide]%
\>[0][@{}l@{\AgdaIndent{1}}]%
\>[2]\AgdaKeyword{open}\AgdaSpace{}%
\AgdaKeyword{import}\AgdaSpace{}%
%
\AgdaModule{Notation.Domains}%
\<%
\\
\>[2]\AgdaKeyword{open}\AgdaSpace{}%
\AgdaKeyword{import}\AgdaSpace{}%
%
\AgdaModule{Notation.Flat}%
\AgdaSpace{}%
\AgdaKeyword{public}\<%
\\
\>[2]\AgdaKeyword{open}\AgdaSpace{}%
\AgdaKeyword{import}\AgdaSpace{}%
%
\AgdaModule{Agda.Builtin.Equality}%
\AgdaSpace{}%
\AgdaKeyword{public}\AgdaSpace{}%
\AgdaKeyword{using}\AgdaSpace{}%
\AgdaSymbol{(}\AgdaOperator{%
\AgdaDatatype{\AgdaUnderscore{}%
≡\AgdaUnderscore{}}}\AgdaSymbol{;}\AgdaSpace{}%
%
\AgdaInductiveConstructor{refl}%
\AgdaSymbol{)}\<%
\\
\>[2]\AgdaKeyword{open}\AgdaSpace{}%
\AgdaKeyword{import}\AgdaSpace{}%
%
\AgdaModule{Agda.Builtin.Equality.Rewrite}%
\AgdaSpace{}%
\AgdaKeyword{using}\AgdaSpace{}%
\AgdaSymbol{()}\<%
\\
\>[0]\<%
\end{code}
\begin{code}%
\>[0][@{}l@{\AgdaIndent{1}}]%
\>[2]\AgdaKeyword{variable}\AgdaSpace{}%
\AgdaGeneralizable{f}\AgdaSpace{}%
\AgdaSymbol{:}\AgdaSpace{}%
\AgdaGeneralizable{A}\AgdaSpace{}%
\AgdaSymbol{→}\AgdaSpace{}%
\AgdaOperator{%
\AgdaPostulate{⟪}%
}\AgdaSpace{}%
\AgdaGeneralizable{D}\AgdaSpace{}%
\AgdaOperator{%
\AgdaPostulate{⟫}%
}\AgdaSymbol{;}\AgdaSpace{}%
\AgdaGeneralizable{a′}\AgdaSpace{}%
\AgdaSymbol{:}\AgdaSpace{}%
\AgdaGeneralizable{A}\<%
\\
\>[2]\AgdaKeyword{postulate}\<%
\\
\>[2][@{}l@{\AgdaIndent{0}}]%
\>[4]%
\AgdaPostulate{elim-♯-↑}%
%
\>[14]\AgdaSymbol{:}\AgdaSpace{}%
\AgdaSymbol{(}\AgdaGeneralizable{f}\AgdaSpace{}%
\AgdaOperator{%
\AgdaPostulate{♯}%
}\AgdaSymbol{)}\AgdaSpace{}%
\AgdaSymbol{(}%
\AgdaPostulate{↑}%
\AgdaSpace{}%
\AgdaGeneralizable{a′}\AgdaSymbol{)}%
\>[30]\AgdaOperator{%
\AgdaDatatype{≡}%
}\AgdaSpace{}%
\AgdaGeneralizable{f}\AgdaSpace{}%
\AgdaGeneralizable{a′}\<%
\\
\>[4]%
\AgdaPostulate{elim-♯-⊥}%
%
\>[14]\AgdaSymbol{:}\AgdaSpace{}%
\AgdaSymbol{(}\AgdaGeneralizable{f}\AgdaSpace{}%
\AgdaOperator{%
\AgdaPostulate{♯}%
}\AgdaSymbol{)}\AgdaSpace{}%
%
\AgdaPostulate{⊥}%
%
\>[29]\AgdaOperator{%
\AgdaDatatype{≡}%
}\AgdaSpace{}%
%
\AgdaPostulate{⊥}%
\<%
\\
\>[2]\AgdaSymbol{\{-\#}\AgdaSpace{}%
\AgdaKeyword{REWRITE}\AgdaSpace{}%
%
\AgdaPostulate{elim-♯-↑}%
\AgdaSpace{}%
\AgdaSymbol{\#-\}}\<%
\end{code}
\end{AgdaSuppressSpace}
The \AgdaRef{elim-♯-⊥} property could be added as a rewrite rule for
automatic use in proofs about an embedding of denotations that represent
errors or failures by \AgdaRef{⊥}. When \AgdaRef{⊥} only represents
nontermination, rewrite rules that propagate it are irrelevant.

\subsection{Recursive Domains}\label{Properties/Recursion-recursive-domains}
\begin{AgdaSuppressSpace}
\begin{code}[hide]%
\>[0]\AgdaSymbol{\{-\#}\AgdaSpace{}%
\AgdaKeyword{OPTIONS}\AgdaSpace{}%
\AgdaPragma{--rewriting}\AgdaSpace{}%
\AgdaPragma{--confluence-check}\AgdaSpace{}%
\AgdaSymbol{\#-\}}\<%
\\
\>[0]\<%
\end{code}
\begin{code}%
\>[0]\AgdaKeyword{module}\AgdaSpace{}%
%
\AgdaModule{Properties.Recursion}%
\AgdaSpace{}%
\AgdaKeyword{where}\<%
\end{code}
\begin{code}[hide]%
\>[0][@{}l@{\AgdaIndent{1}}]%
\>[2]\AgdaKeyword{open}\AgdaSpace{}%
\AgdaKeyword{import}\AgdaSpace{}%
%
\AgdaModule{Notation.Domains}%
\<%
\\
\>[2]\AgdaKeyword{open}\AgdaSpace{}%
\AgdaKeyword{import}\AgdaSpace{}%
%
\AgdaModule{Notation.Recursion}%
\AgdaSpace{}%
\AgdaKeyword{public}\<%
\\
\>[2]\AgdaKeyword{open}\AgdaSpace{}%
\AgdaKeyword{import}\AgdaSpace{}%
%
\AgdaModule{Agda.Builtin.Equality}%
\AgdaSpace{}%
\AgdaKeyword{public}\AgdaSpace{}%
\AgdaKeyword{using}\AgdaSpace{}%
\AgdaSymbol{(}\AgdaOperator{%
\AgdaDatatype{\AgdaUnderscore{}%
≡\AgdaUnderscore{}}}\AgdaSymbol{;}\AgdaSpace{}%
%
\AgdaInductiveConstructor{refl}%
\AgdaSymbol{)}\<%
\\
\>[2]\AgdaKeyword{open}\AgdaSpace{}%
\AgdaKeyword{import}\AgdaSpace{}%
%
\AgdaModule{Agda.Builtin.Equality.Rewrite}%
\AgdaSpace{}%
\AgdaKeyword{using}\AgdaSpace{}%
\AgdaSymbol{()}\<%
\\
\>[0]\<%
\end{code}
\begin{code}%
\>[0][@{}l@{\AgdaIndent{1}}]%
\>[2]\AgdaKeyword{postulate}\<%
\\
\>[2][@{}l@{\AgdaIndent{0}}]%
\>[4]%
\AgdaPostulate{elim-unfold-fold}%
\AgdaSpace{}%
\AgdaSymbol{:}\AgdaSpace{}%
\AgdaSymbol{\{\{}\AgdaBound{\AgdaUnderscore{}}\AgdaSpace{}%
\AgdaSymbol{:}\AgdaSpace{}%
\AgdaGeneralizable{D}\AgdaSpace{}%
\AgdaOperator{%
\AgdaPostulate{≅}%
}\AgdaSpace{}%
\AgdaGeneralizable{E}\AgdaSymbol{\}\}}\AgdaSpace{}%
\AgdaSymbol{→}\AgdaSpace{}%
\AgdaSymbol{\{}\AgdaBound{e}\AgdaSpace{}%
\AgdaSymbol{:}\AgdaSpace{}%
\AgdaOperator{%
\AgdaPostulate{⟪}%
}\AgdaSpace{}%
\AgdaGeneralizable{E}\AgdaSpace{}%
\AgdaOperator{%
\AgdaPostulate{⟫}%
}\AgdaSymbol{\}}\AgdaSpace{}%
\AgdaSymbol{→}\AgdaSpace{}%
%
\AgdaPostulate{unfold}%
\AgdaSpace{}%
\AgdaSymbol{(}%
\AgdaPostulate{fold}%
\AgdaSpace{}%
\AgdaBound{e}\AgdaSymbol{)}\AgdaSpace{}%
\AgdaOperator{%
\AgdaDatatype{≡}%
}\AgdaSpace{}%
\AgdaBound{e}\<%
\\
\>[2]\AgdaSymbol{\{-\#}\AgdaSpace{}%
\AgdaKeyword{REWRITE}\AgdaSpace{}%
%
\AgdaPostulate{elim-unfold-fold}%
\AgdaSpace{}%
\AgdaSymbol{\#-\}}\<%
\end{code}
\end{AgdaSuppressSpace}
The property \AgdaRef{fold\ (unfold\ d)\ ≡\ d} could be added, but it is
not needed for the current illustrative tests.

\end{AgdaAlign}

\section{Mechanising Domain Theory}\label{mechanising}


We have shown above how to use a theorem prover (Agda) to mechanically check the well-definedness of a programming language semantics,
and showed how some properties of semantic definitions can be checked automatically. 
The denotational semantics requires \emph{domains} to model recursive functions and recursive types. 
The prover of choice must therefore provide libraries that offer a sufficiently rich domain theory.
Our approach was to axiomatise (postulate) the relevant part of domain theory rather than develop it fully as a mathematical theory. 
This was motivated by the fact that not many provers (including Agda) come with libraries for domain theory. 
And those libraries that do exist are usually difficult to use, as complex proofs are sometimes required to express even simple operations on domains. 
In the literature the following  mechanisations of implementations of domain theory are known:


\subsection{Overview of Mechanised Standard Domain Theory}

There have been previous successful attempts to mechanise classical domain theory
and Milner’s Logic for Computable Functions \cite{MilnerLCF}.
Franz Regensburger \cite{Regensburger95} developed a corresponding mechanisation
in Isabelle/HOL \cite{Nipkow-Paulson-Wenzel:2002} which was further extended
as described in \cite{Mueller_1999}. The authors of the latter use a special tactic
that ``\emph{reduces the continuity of an LCF term to the continuity of its
basic components by structural rules}'' \cite{Mueller_1999}. In Agda, there exists the possibility for users
to define tactics through reflection \cite{ReflectionAgda}. 
It might be possible to define tactics for automatically proving continuity for functions between domains, but it is unclear how well that would work in practice.

There have also been attempts to mechanise domain theory in a constructive setting
using intuitionistic logic. There are, for instance, developments of constructive
domain theory in Coq/Rocq \cite{BentonCoq,DockinsCoq,Paulin-Mohring2009}. 
A version of $\omega$-cpos has been developed in
\cite{BentonCoq} where the lifting is defined via a coinductive definition of
infinite streams. It is based on Paulin-Mohring's mechanisation \cite{Paulin-Mohring2009} 
of the semantics of Kahn networks where she developed a general library for cpos. 
Dockins  developed effective algebraic domains (more precisely: constructive models of profinite domains) in Coq/Rocq \cite{DockinsCoq}. 
Lennon-Bertrand \cite{lennonbertrand2026} ported Dockins' library to Rocq 9.1, rewriting it substantially using observational equality instead of setoids.

Another mechanisation of domain theory has been carried out in Agda based on
constructive and predicative univalent foundations \cite{dejong_escardo_CSL}.
In this work,  Scott's model of PCF is developed in  Agda  with the aim to 
``\emph{explore the development of domain theory from the univalent point of view}'' \cite{dejong_escardo_CSL}.
The authors use universes
to allow for algebraic directed complete partial orders to be large but locally small.
Their axiomatisation is based on propositional and functional extensionality.  
This is an excellent example of mechanising mathematics and semantics in univalent foundations, but it 
requires a rather well trained person. One must keep track of universe levels, and manage
the required proofs that types are propositions and that $\lambda$-abstractions are continuous.
The proof of continuity of combinator $S$ in
\texttt{PCF.Combinatory.PCFCombinators.html} \cite{Jong-TypeTopology/DomainTheory}
is an example of how detailed such continuity proofs must be.

\subsection{Synthetic Domain Theory}\label{synthetic}

All of the mechanisations considered above have in common that the continuity of functions
between domains needs to be proven for each function involved
(including anonymous functions expressed via $\lambda$).
This is extremely tedious, and discourages mechanisation of denotational semantics.
Developments like the one presented in this paper would be much easier to carry out
 if domains were sets and functions between domains
automatically continuous -- as originally suggested by Dana Scott  \cite{Scott1980RTL} -- such that
the category of domains becomes a full subcategory of the category of sets.
Based on this idea, several different flavours of domains have been defined over time
in specific realisability/topos models or axiomatically: replete objects \cite{Sreplete},
well-complete objects \cite{Longley1997UAD}, extensional $\Sigma$-spaces
\cite{Phoa1990EDI,Reus:SigSpaces}. As domains are ``synthesized'' as sets with properties,
rather than defined as sets with structure, i.e.\ partial order, the corresponding theory was called
\emph{Synthetic Domain Theory} (SDT) in analogy with Synthetic Differential Geometry.
Since ``domains as sets'' is only consistent with intuitionistic logic, any such approach
needs to be developed in a version of intuitionistic set theory. Domain theory requires
the notion of partiality which in SDT is expressed via $\Sigma$, the subset of those
propositions that express termination of programs. The set $\Sigma$ contains the
proposition $\top$ and is closed under composition (i.e., it is a dominance \cite{Fiore2001DH}). 
Based on $\Sigma$ one can define a lifting functor with final coalgebra $\mathbf{F}$ and initial algebra $\mathbf{I}$.
This $\mathbf{I}$ is the generic $\omega$-chain such that elements in $X^\mathbf{I}$ can be regarded as  $\omega$-chains in $X$.
And $\mathbf{F}$ can be viewed as $\mathbf{I}$'s chain completion.

\subsection{Mechanisations of SDT}
There are only very few mechanisations of SDT in theorem provers.

The mechanisation of SDT in \cite{Reus1999FSD}, based on
\cite{ReusStreicher99MSCS}, 
was carried out in the \emph{Lego} proof assistant developed in Edinburgh and used in the 1990s. It implements
the Extended Calculus of Constructions \cite{Luo_ECC} and uses the impredicative universe of
propositions. For its mechanisation an additional impredicative universe of sets had been
built into the Lego system. Complete extensional $\Sigma$-spaces serve as notion of domains
which can be represented as double-negation closed subsets of powers of $\Sigma$.
Domain constructors are defined as endofunctors on this category of domains and existence of
fixpoints for those functors was proven using the standard inverse limit construction.
A correctness proof of the Sieve of Eratosthenes on the domain of streams of natural numbers
was used to test the theory. Due to its use of two impredicative universes (Set and Prop), this mechanisation appears to be impossible to translate into Agda. The justification for this axiomatisation is the extensional $\Sigma$-per model \cite{Phoa1990EDI} which is impredicative. A more general approach that does not require impredicativity is Alex Simpson's wonderfully unifying account of SDT \cite{Simpson2004CAR}.

Simpson's purely axiomatic model appears to be the only
one that ``\emph{applies uniformly across the different types of model}'' \cite[p.~209]{Simpson2004CAR}.
``\emph{This is not trivial, however, \ldots . The problem is that elementary toposes,
although models of intuitionistic higher-order logic, are not, in general,
models of a sufficiently powerful set theory. Thus, \ldots we shall require that
\textbf{S} having enough structure to model full intuitionstic Zermelo-Fraenkel (IZF)
set theory asking for \textbf{S} to be given as the full subcategory of small objects
in a category \textbf{C} with class(ic) structure and universal object.}'' \cite[p.~208]{Simpson2004CAR} 
Within  \textbf{C} the category \textbf{P} of predomains is the full subcategory
of $\mathbf{S}$ with  small well-complete  sets as objects. 
An object is said to be well-complete if its lifting is complete 
w.r.t\ the embedding $\iota: \mathbf{I} \rightarrow \mathbf{F}$ or, in other words,
$X$ is complete if $X^\iota$ is an isomorphism. Here, $\mathbf{I}$ represents a generic $\omega$ chain and $\mathbf{F}$ its completion via $\iota$. An $\omega$-chain in $X$ is simply represented then as an element of $X^\mathbf{I}$.

If one stipulates that the single-element type $1$ is well-complete then many useful results hold, among them the following: 
the category of predomains  is cartesian closed,
 well-complete objects are automatically complete,
lifting preserves well-complete objects, 
and products of families of well-complete objects
indexed by a small set are well-complete. 
Simpson managed to prove that the category of $\Sigma$-partial maps is algebraically compact
(for internal endofunctors). This is shown by an adaptation of the  classic limit-colimit coincidence theorem \cite{SmythPlotkin}.
Instead of $\mathbf{N}$-indexed families of embedding-projection pairs, one uses general $\mathbf{I}$-indexed
  diagrams that meet some equational properties. Actually, Simpson shows more generally that certain kinds of suitable categories are algebraically compact and that this property is closed under products and duality, which is important to define fixpoints of mixed-variant internal bifunctors, and thus to model recursive type in FPC \cite{DBLP:conf/lics/FioreP94}.

Note that  the algebraic compactness result requires that \textbf{S}, \textbf{P} and
the category of partial maps between objects in \textbf{P} must be \emph{internal} subcategories of
\textbf{C}.
  The axiom that the type $2$ of Booleans is well-complete is required if one wishes to model  sum types,
and also implies that the empty type  $0$ is well-complete and thus that $\bot\in \Sigma$.

It remains future work to attempt to mechanise Simpson's setup of constructive set theory
with a few axioms in a prover like Agda. As Simpson himself, we speculate that such a mechanisation should
be possible. Unlike the one presented in this paper, this would prove
that all major notions and results of domain theory, like fixpoints of functions and functors between domains, 
can be developed in Agda from a very small number of axioms and postulates.  
However, this would not require any change to the Agda embedding of the definitions of denotational semantics presented in this paper.

 \subsection{Other Flavours of SDT}
There are also some mechanisations of a different kind of Synthetic Domain Theory.
Those would require some rewriting of the ways domains are defined.

\emph{Synthetic Guarded Domain Theory} (SGDT) \cite{BirkedalSGDT} is a variant
where Nakano's later modality is used to type the fixpoint operator. Fixpoints exist
for functions that factor through the unit of the later monad.
These principles can be applied to define (productive) recursive types. The underlying model
is the category of presheaves over $\omega$ which contains the category of complete
bisected ultrametric spaces as coreflective subcategory. The latter provides the
semantic justification for the later monad and existence of fixpoints for contractive maps.

\emph{Guarded Dependent Type Theory} (GDTT) \cite{BizjakGCMB16} is an extensional type theory
based on the later monad as type constructor. Paviotti et al.\ \cite{Paviotti} have developed
PCF in GDTT but no Agda implementation is available.

\emph{Clocked Cubical Type Theory} \cite{Kristensen2021AMO} is an intensional type theory
extending cubical type theory and thus permits developments in homotopy type theory.
Agda includes an implementation of some of clocked cubical type theory under the name
Guarded Cubical Agda. CCS and the $\pi$-calculus have been implemented in this version
of Agda \cite{VEZZOSI_2021} as well as gradual typing \cite{Giovannini_2025}. Those
mechanisations clearly involve some denotational semantics but a dedicated user friendly
library for domain theory or denotational semantics appears to be lacking.

\section{Conclusion}\label{conclusion}
We have shown how to formalise conventional Scott–Strachey denotational definitions
by embedding them in Agda with almost no changes to $\lambda$-notation
(apart from adding explicit injection and lifting from sets to flat domains,
and bijections between recursive domains and their structure)
or domain equations (apart from separated sums).
This supports mechanisation of semantic definitions of programming languages
with relatively little extra effort.
Mechanisation is essential for checking the wellformedness of larger definitions.

We have presented three illustrative examples of our shallow embedding of denotational semantics in Agda:
a model of the untyped $\lambda$-calculus,
Plotkin’s denotational semantics of PCF, and a semantics of a sublanguage \emph{Scm} of \emph{Scheme}.
Each example exhibited different benefits of our approach:
\begin{itemize}
  \item
  The  $\lambda$-calculus semantics was particularly simple to mechanise in Agda,
  thanks to the ease of postulating recursive domains as bijections, and to our support for
  the conventional notation for updating environments.
  \item
  The PCF semantics demonstrated how the mathematical meta-notation in Plotkin's article
  could be represented in Agda by exploiting its support for Unicode symbols.
  It also showed how easily dependent types are able to formalise Plotkin's verbal explanation
  of the syntax of an intrinsically typed language.
  \item
  The semantics of \emph{Scm} was developed by one of the authors from the denotational semantics
  of the full \emph{Scheme} expression language, which has been included in the \emph{Scheme} standards
  since 1986 \cite{Scheme}. The Agda formalisation of the \emph{Scheme} semantics \cite{Mosses2025CDS}
  was produced from a plain (Unicode)
  text preview of the PDF of a published \emph{Scheme} report, followed by some simple transformations
  in an interactive editor, and finally insertion of layout to recover the original alignment.
  The Agda type-checker reported several wellformedness issues; some of them stemmed from
  informally elided injections and projections, but some were trivial type errors
  that had not previously been reported. The presentation of the results at a \emph{Scheme} workshop
  led to interest in Agda mechanisation of future denotational definitions of \emph{Scheme}.
\end{itemize}

Although the main aim of our mechanisation of denotational semantics in Agda is to check wellformedness,
we have discovered that we can also use it to test whether denotations of terms compute the expected values,
as demonstrated in Appendix~\ref{Tests-illustrative-tests}.
The proof terms are trivial (\AgdaFontStyle{refl}) thanks to the automatic use of our
specified rewrite rules.

Previous versions of the Agda mechanisation of the $\lambda$-calculus model and the \emph{Scm} semantics
have been published \cite{Mosses2025CSE,Mosses2025LAF}. They both treated ordinary Agda types
as domains, which is inconsistent with the existence of the empty type in Agda.
Moreover, the previous version of the \emph{Scm} semantics did not make use of instance parameters,
so the postulated operations on separated sums had to be defined separately for each summand.
Those drawbacks have been eliminated in the current mechanisation.
Our mechanisation of the PCF semantics has not previously been published.

A remaining awkwardness with our approach is the lack of Agda support for the conventional
concise definition of abstract syntax by context-free grammars in formal semantics of programming languages.
The restrictions on mixfix notation for function symbols in Agda require the use of extra characters
as separators, and datatype definitions do not seem to be as perspicuous as grammars.

We have chosen to use a \emph{shallow} embedding of denotational semantics in Agda.
This has several disadvantages compared to a deep embedding. For example,
the Agda type-checker is far more general than required for checking wellformedness
of a denotational semantics, but does not allow domain definitions to be recursive;
the embedding of abstract syntax grammars as datatype definitions is undesirable;
and there are minor differences at the lexical level,
such as the need to replace conventional $\lambda$-abstraction notation $\lambda x. f x$ by
\AgdaFontStyle{λ~x~→~f~x} (with symbols also separated by layout).
However, implementing a deep embedding of denotational semantics in Agda can take a significantly greater effort,
as for example with the development of SIS (Semantics Implementation System) in the 1970s
\cite{DonzeauGouge1978CMD,Mosses1974SSE,Mosses1975PHD,Mosses1979SIS};
use of a modern language workbench would reduce the effort, but also introduce an extra level of meta-notation.

The plethora of postulates in our Agda code is unconventional: module parameters are generally
preferred to postulates, as they document dependence of proofs on undischarged assumptions.
However, adding rewrite rules on module parameters requires \emph{local}
rewriting~\cite{leray:hal-05160846}, which is only available in the (as of this
writing unreleased) Agda version 2.9.0~\cite{burke-local-rewrite}.
Postulates also have the advantage of being notationally simple, while the
requirement to declare module parameters for each module that uses them adds
significant notational overhead.
As discussed in detail in Section~\ref{synthetic}, an axiomatisation of synthetic domain theory in Agda
could lead not only to a much smaller number of postulates, but also to proof of the existence
of the various domain constructors declared in Section~\ref{Notation-postulated-domain-notation}.
This should establish solid mathematical foundations for our mechanisation of denotational semantics
in Agda, without undermining the directness of the shallow embedding of domain equations and $\lambda$-notation.

\paragraph*{Acknowledgements.}

The inspiration for the Agda development reported in this paper stems from Alex Simpson's
helpful response to our enquiry about the possibility of formalising his development of SDT
\cite{Simpson2004CAR} in Agda: he wrote
``\emph{All published theoretical work on SDT I'm aware of (including my own)
concentrated mainly on sorting out the underlying mathematical framework [\ldots]
It would be really nice to have a more black-box, modular and applicable approach.
For example, one might simply assume in Agda that there is a `universe' of (pre)domains
closed under many type constructions, and with a suitably well-behaved fixed-point operator,
and use that to do semantics}''.

We have also benefited from the advice of participants at an IFIP WG 2.11 meeting (Edinburgh, December 2024) and at
an Agda Implementors' Meeting (Angers, November 2025).

We are very grateful to the anonymous reviewers of our submitted paper for their
perceptive comments and suggestions for its improvement.

\appendix

\begin{AgdaAlign}
\section{Illustrative Tests}\label{Tests-illustrative-tests}

The tests below illustrate \emph{automatic} proof by the Agda
type-checker that the denotations of some terms compute the expected
values. The type-correctness of the \AgdaRef{refl} definitions of the
declared equivalences confirms that they hold.

Apart from confirming that a denotational semantics defines denotations
which compute the expected values at least for some terms, the tests
also depend on the imported rewrite rules for
postulated
properties (Section~\ref{Properties-postulated-properties}). The success of those tests indirectly checks that the
rewrite rules preserve denotations. (A more systematic approach would be
to develop a suite of unit tests for consequences of postulated
properties.)
\begin{AgdaSuppressSpace}
\begin{code}[hide]%
\>[0]\AgdaSymbol{\{-\#}\AgdaSpace{}%
\AgdaKeyword{OPTIONS}\AgdaSpace{}%
\AgdaPragma{--rewriting}\AgdaSpace{}%
\AgdaPragma{--confluence-check}\AgdaSpace{}%
\AgdaSymbol{\#-\}}\<%
\\
\>[0]\<%
\end{code}
\begin{code}%
\>[0]\AgdaKeyword{module}\AgdaSpace{}%
%
\AgdaModule{Tests}%
\AgdaSpace{}%
\AgdaKeyword{where}\<%
\\
\>[0][@{}l@{\AgdaIndent{0}}]%
\>[2]\AgdaKeyword{import}\AgdaSpace{}%
%
\AgdaModule{Tests.LC}%
\<%
\\
\>[2]\AgdaKeyword{import}\AgdaSpace{}%
%
\AgdaModule{Tests.PCF}%
\<%
\end{code}
\end{AgdaSuppressSpace}
We leave development of significant tests for the \AgdaRef{Scm} language
to future work. This is partly because the published denotational
definition of \emph{Scm}
\cite{Mosses2025CDS} leaves
various domains and operations completely unspecified, and our Agda
embedding merely postulates them, without specifying their properties.
Reduction of embedded denotations involving such operations by the Agda
type-checker to (weak) head form would not (in general) lead to
particular elements of domains.

As a workaround, we could replace the embedded declarations of the
postulated domains and operations by definitions; e.g., the postulated
type \AgdaRef{Loc} of locations could be defined by \AgdaRef{Loc\ =\ Nat},
and the postulated operation \AgdaRef{new} could then be defined to give
the first location mapped to \AgdaRef{unallocated} by the current store.
But that would undermine the correspondence between the published
denotational semantics and its Agda embedding.

\subsection{Untyped Lambda-Calculus
Tests}\label{Tests/LC-untyped-lambda-calculus-tests}

The following tests check that the denotations of some simple untyped
\(\lambda\)-expressions in the abstract syntax of the
LC language (Section~\ref{Examples/LC-untyped-lambda-calculus})
compute the expected values. In the absence of atomic values in this
language, we regard free variables as values, and apply denotations to
an arbitrary environment \AgdaRef{ρ}.

All the \AgdaRef{refl} proofs of the tests implicitly use the rewrite
rule postulated as a
property of
recursive domains (Section~\ref{Properties/Recursion-recursive-domains}) to eliminate compositions of \AgdaRef{unfold} and
\AgdaRef{fold}.
\begin{AgdaSuppressSpace}
\begin{code}[hide]%
\>[0]\AgdaSymbol{\{-\#}\AgdaSpace{}%
\AgdaKeyword{OPTIONS}\AgdaSpace{}%
\AgdaPragma{--rewriting}\AgdaSpace{}%
\AgdaPragma{--confluence-check}\AgdaSpace{}%
\AgdaSymbol{\#-\}}\<%
\\
\>[0]\<%
\end{code}
\begin{code}%
\>[0]\AgdaKeyword{module}\AgdaSpace{}%
%
\AgdaModule{Tests.LC}%
\AgdaSpace{}%
\AgdaKeyword{where}\<%
\end{code}
\begin{code}[hide]%
\>[0][@{}l@{\AgdaIndent{1}}]%
\>[2]\AgdaKeyword{open}\AgdaSpace{}%
\AgdaKeyword{import}\AgdaSpace{}%
%
\AgdaModule{Notation.Domains}%
\<%
\\
\>[2]\AgdaKeyword{open}\AgdaSpace{}%
\AgdaKeyword{import}\AgdaSpace{}%
%
\AgdaModule{Notation.Functions}%
\<%
\\
\\[\AgdaEmptyExtraSkip]%
\>[2]\AgdaKeyword{open}\AgdaSpace{}%
\AgdaKeyword{import}\AgdaSpace{}%
%
\AgdaModule{Examples.LC.Abstract-Syntax}%
\<%
\\
\>[2]\AgdaKeyword{open}\AgdaSpace{}%
\AgdaKeyword{import}\AgdaSpace{}%
%
\AgdaModule{Examples.LC.Domain-Equations}%
\<%
\\
\>[2]\AgdaKeyword{open}\AgdaSpace{}%
\AgdaKeyword{import}\AgdaSpace{}%
%
\AgdaModule{Examples.LC.Semantic-Functions}%
\<%
\\
\\[\AgdaEmptyExtraSkip]%
\>[2]\AgdaKeyword{open}\AgdaSpace{}%
\AgdaKeyword{import}\AgdaSpace{}%
%
\AgdaModule{Properties.Flat.Booleans}%
\<%
\\
\>[2]\AgdaKeyword{open}\AgdaSpace{}%
\AgdaKeyword{import}\AgdaSpace{}%
%
\AgdaModule{Properties.Updates}%
\<%
\\
\>[2]\AgdaKeyword{open}\AgdaSpace{}%
\AgdaKeyword{import}\AgdaSpace{}%
%
\AgdaModule{Properties.Recursion}%
\<%
\end{code}
\end{AgdaSuppressSpace}
The following trivial abbreviations improve the readbility of AST terms
in the tests:
\begin{AgdaSuppressSpace}
\begin{code}%
\>[2]%
\AgdaFunction{a}%
\AgdaSpace{}%
\AgdaSymbol{=}\AgdaSpace{}%
%
\AgdaInductiveConstructor{x}%
\AgdaSpace{}%
\AgdaNumber{0}\<%
\\
\>[2]%
\AgdaFunction{b}%
\AgdaSpace{}%
\AgdaSymbol{=}\AgdaSpace{}%
%
\AgdaInductiveConstructor{x}%
\AgdaSpace{}%
\AgdaNumber{1}\<%
\\
\>[2]%
\AgdaFunction{c}%
\AgdaSpace{}%
\AgdaSymbol{=}\AgdaSpace{}%
%
\AgdaInductiveConstructor{x}%
\AgdaSpace{}%
\AgdaNumber{2}\<%
\end{code}
\end{AgdaSuppressSpace}
Due to potential non-injectivity of the postulated operation
\AgdaRef{⟪\_⟫}, the definition of \AgdaRef{\_{[}\_/\_{]}} needs to be
instantiated at \AgdaRef{D∞} for the type-checker to resolve the
\AgdaRef{lookup-id} test:
\begin{AgdaSuppressSpace}
\begin{code}%
\>[2]\AgdaOperator{%
\AgdaFunction{\AgdaUnderscore{}%
[\AgdaUnderscore{}/\AgdaUnderscore{}]D∞}}\AgdaSpace{}%
\AgdaSymbol{:}\AgdaSpace{}%
\AgdaSymbol{\{\{}%
\AgdaRecord{Eq}%
\AgdaSpace{}%
\AgdaGeneralizable{A}\AgdaSymbol{\}\}}\AgdaSpace{}%
\AgdaSymbol{→}\AgdaSpace{}%
\AgdaOperator{%
\AgdaPostulate{⟪}%
}\AgdaSpace{}%
\AgdaSymbol{(}\AgdaGeneralizable{A}\AgdaSpace{}%
\AgdaOperator{%
\AgdaPostulate{→ˢ}%
}\AgdaSpace{}%
%
\AgdaPostulate{D∞}%
\AgdaSymbol{)}\AgdaSpace{}%
\AgdaOperator{%
\AgdaPostulate{→ᶜ}%
}\AgdaSpace{}%
%
\AgdaPostulate{D∞}%
\AgdaSpace{}%
\AgdaOperator{%
\AgdaPostulate{→ᶜ}%
}\AgdaSpace{}%
\AgdaGeneralizable{A}\AgdaSpace{}%
\AgdaOperator{%
\AgdaPostulate{→ˢ}%
}\AgdaSpace{}%
\AgdaSymbol{(}\AgdaGeneralizable{A}\AgdaSpace{}%
\AgdaOperator{%
\AgdaPostulate{→ˢ}%
}\AgdaSpace{}%
%
\AgdaPostulate{D∞}%
\AgdaSymbol{)}\AgdaSpace{}%
\AgdaOperator{%
\AgdaPostulate{⟫}%
}\<%
\\
\>[2]\AgdaBound{ρ}\AgdaSpace{}%
\AgdaOperator{%
\AgdaFunction{[}%
}\AgdaSpace{}%
\AgdaBound{δ}\AgdaSpace{}%
\AgdaOperator{%
\AgdaFunction{/}%
}\AgdaSpace{}%
\AgdaBound{a}\AgdaSpace{}%
\AgdaOperator{%
\AgdaFunction{]D∞}%
}\AgdaSpace{}%
\AgdaSymbol{=}\AgdaSpace{}%
\AgdaSymbol{λ}\AgdaSpace{}%
\AgdaBound{a′}\AgdaSpace{}%
\AgdaSymbol{→}\AgdaSpace{}%
\AgdaOperator{%
\AgdaFunction{if}%
}\AgdaSpace{}%
\AgdaBound{a}\AgdaSpace{}%
\AgdaOperator{%
\AgdaField{==}%
}\AgdaSpace{}%
\AgdaBound{a′}\AgdaSpace{}%
\AgdaOperator{%
\AgdaFunction{then}%
}\AgdaSpace{}%
\AgdaBound{δ}\AgdaSpace{}%
\AgdaOperator{%
\AgdaFunction{else}%
}\AgdaSpace{}%
\AgdaBound{ρ}\AgdaSpace{}%
\AgdaBound{a′}\<%
\\
\\[\AgdaEmptyExtraSkip]%
\>[2]%
\AgdaFunction{lookup-id}%
\AgdaSpace{}%
\AgdaSymbol{:}\<%
\\
\>[2][@{}l@{\AgdaIndent{0}}]%
\>[4]\AgdaOperator{%
\AgdaFunction{⟦}%
}\AgdaSpace{}%
\AgdaOperator{%
\AgdaInductiveConstructor{var}%
}\AgdaSpace{}%
%
\AgdaFunction{a}%
\AgdaSpace{}%
\AgdaOperator{%
\AgdaFunction{⟧}%
}\AgdaSpace{}%
\AgdaSymbol{(}\AgdaGeneralizable{ρ}\AgdaSpace{}%
\AgdaOperator{%
\AgdaFunction{[}%
}\AgdaSpace{}%
\AgdaGeneralizable{δ}\AgdaSpace{}%
\AgdaOperator{%
\AgdaFunction{/}%
}\AgdaSpace{}%
%
\AgdaFunction{a}%
\AgdaSpace{}%
\AgdaOperator{%
\AgdaFunction{]D∞}%
}\AgdaSymbol{)}\AgdaSpace{}%
\AgdaOperator{%
\AgdaDatatype{≡}%
}\AgdaSpace{}%
\AgdaGeneralizable{δ}\<%
\\
\>[2]%
\AgdaFunction{lookup-id}%
\AgdaSpace{}%
\AgdaSymbol{=}\AgdaSpace{}%
%
\AgdaInductiveConstructor{refl}%
\<%
\\
\\[\AgdaEmptyExtraSkip]%
\>[2]%
\AgdaFunction{app-id}%
\AgdaSpace{}%
\AgdaSymbol{:}\<%
\\
\>[2][@{}l@{\AgdaIndent{0}}]%
\>[4]\AgdaOperator{%
\AgdaFunction{⟦}%
}\AgdaSpace{}%
\AgdaOperator{%
\AgdaInductiveConstructor{⦅}%
}\AgdaSpace{}%
\AgdaOperator{%
\AgdaInductiveConstructor{⦅λ}%
}\AgdaSpace{}%
%
\AgdaFunction{a}%
\AgdaSpace{}%
\AgdaOperator{%
\AgdaInductiveConstructor{␣}%
}\AgdaSpace{}%
\AgdaOperator{%
\AgdaInductiveConstructor{var}%
}\AgdaSpace{}%
%
\AgdaFunction{a}%
\AgdaSpace{}%
\AgdaOperator{%
\AgdaInductiveConstructor{⦆}%
}\AgdaSpace{}%
\AgdaOperator{%
\AgdaInductiveConstructor{␣}%
}\AgdaSpace{}%
\AgdaOperator{%
\AgdaInductiveConstructor{var}%
}\AgdaSpace{}%
%
\AgdaFunction{b}%
\AgdaSpace{}%
\AgdaOperator{%
\AgdaInductiveConstructor{⦆}%
}\AgdaSpace{}%
\AgdaOperator{%
\AgdaFunction{⟧}%
}\AgdaSpace{}%
\AgdaGeneralizable{ρ}\AgdaSpace{}%
\AgdaOperator{%
\AgdaDatatype{≡}%
}\AgdaSpace{}%
\AgdaGeneralizable{ρ}\AgdaSpace{}%
%
\AgdaFunction{b}%
\<%
\\
\>[2]%
\AgdaFunction{app-id}%
\AgdaSpace{}%
\AgdaSymbol{=}\AgdaSpace{}%
%
\AgdaInductiveConstructor{refl}%
\<%
\\
\\[\AgdaEmptyExtraSkip]%
\>[2]%
\AgdaFunction{app-k}%
\AgdaSpace{}%
\AgdaSymbol{:}\<%
\\
\>[2][@{}l@{\AgdaIndent{0}}]%
\>[4]\AgdaOperator{%
\AgdaFunction{⟦}%
}\AgdaSpace{}%
\AgdaOperator{%
\AgdaInductiveConstructor{⦅}%
}\AgdaSpace{}%
\AgdaOperator{%
\AgdaInductiveConstructor{⦅λ}%
}\AgdaSpace{}%
%
\AgdaFunction{a}%
\AgdaSpace{}%
\AgdaOperator{%
\AgdaInductiveConstructor{␣}%
}\AgdaSpace{}%
\AgdaOperator{%
\AgdaInductiveConstructor{var}%
}\AgdaSpace{}%
%
\AgdaFunction{b}%
\AgdaSpace{}%
\AgdaOperator{%
\AgdaInductiveConstructor{⦆}%
}\AgdaSpace{}%
\AgdaOperator{%
\AgdaInductiveConstructor{␣}%
}\AgdaSpace{}%
\AgdaOperator{%
\AgdaInductiveConstructor{var}%
}\AgdaSpace{}%
%
\AgdaFunction{c}%
\AgdaSpace{}%
\AgdaOperator{%
\AgdaInductiveConstructor{⦆}%
}\AgdaSpace{}%
\AgdaOperator{%
\AgdaFunction{⟧}%
}\AgdaSpace{}%
\AgdaGeneralizable{ρ}\AgdaSpace{}%
\AgdaOperator{%
\AgdaDatatype{≡}%
}\AgdaSpace{}%
\AgdaGeneralizable{ρ}\AgdaSpace{}%
%
\AgdaFunction{b}%
\<%
\\
\>[2]%
\AgdaFunction{app-k}%
\AgdaSpace{}%
\AgdaSymbol{=}\AgdaSpace{}%
%
\AgdaInductiveConstructor{refl}%
\<%
\end{code}
\end{AgdaSuppressSpace}
The following test involves the diverging evaluation of a λ-abstraction
to itself. It is commented-out, to avoid nontermination of the Agda
type-checker.
\begin{AgdaSuppressSpace}
\begin{code}%
\>[2]\AgdaComment{--\ app-id-to-divergence\ :}\<%
\\
\>[2]\AgdaComment{--\ \ \ ⟦\ \ ⦅\ \ ⦅λ\ a\ ␣\ var\ a\ ⦆\ ␣}\<%
\\
\>[2]\AgdaComment{--\ \ \ \ \ \ \ \ \ ⦅\ ⦅λ\ c\ ␣\ ⦅\ var\ c\ ␣\ var\ c\ ⦆\ ⦆\ ␣\ ⦅λ\ c\ ␣\ ⦅\ var\ c\ ␣\ var\ c\ ⦆\ ⦆\ ⦆\ ⦆\ ⟧\ ρ\ ≡\ ρ\ b}\<%
\\
\>[2]\AgdaComment{--\ app-id-to-divergence\ =\ refl}\<%
\end{code}
\end{AgdaSuppressSpace}
The following test illustrates that application of a λ-abstraction can
terminate when evaluation of its argument would diverge.
\begin{AgdaSuppressSpace}
\begin{code}%
\>[2]%
\AgdaFunction{app-k-to-divergence}%
\AgdaSpace{}%
\AgdaSymbol{:}\<%
\\
\>[2][@{}l@{\AgdaIndent{0}}]%
\>[4]\AgdaOperator{%
\AgdaFunction{⟦}%
}%
\>[7]\AgdaOperator{%
\AgdaInductiveConstructor{⦅}%
}%
\>[10]\AgdaOperator{%
\AgdaInductiveConstructor{⦅λ}%
}\AgdaSpace{}%
%
\AgdaFunction{a}%
\AgdaSpace{}%
\AgdaOperator{%
\AgdaInductiveConstructor{␣}%
}\AgdaSpace{}%
\AgdaOperator{%
\AgdaInductiveConstructor{var}%
}\AgdaSpace{}%
%
\AgdaFunction{b}%
\AgdaSpace{}%
\AgdaOperator{%
\AgdaInductiveConstructor{⦆}%
}\AgdaSpace{}%
\AgdaOperator{%
\AgdaInductiveConstructor{␣}%
}\<%
\\
\>[10]\AgdaOperator{%
\AgdaInductiveConstructor{⦅}%
}\AgdaSpace{}%
\AgdaOperator{%
\AgdaInductiveConstructor{⦅λ}%
}\AgdaSpace{}%
%
\AgdaFunction{c}%
\AgdaSpace{}%
\AgdaOperator{%
\AgdaInductiveConstructor{␣}%
}\AgdaSpace{}%
\AgdaOperator{%
\AgdaInductiveConstructor{⦅}%
}\AgdaSpace{}%
\AgdaOperator{%
\AgdaInductiveConstructor{var}%
}\AgdaSpace{}%
%
\AgdaFunction{c}%
\AgdaSpace{}%
\AgdaOperator{%
\AgdaInductiveConstructor{␣}%
}\AgdaSpace{}%
\AgdaOperator{%
\AgdaInductiveConstructor{var}%
}\AgdaSpace{}%
%
\AgdaFunction{c}%
\AgdaSpace{}%
\AgdaOperator{%
\AgdaInductiveConstructor{⦆}%
}\AgdaSpace{}%
\AgdaOperator{%
\AgdaInductiveConstructor{⦆}%
}\AgdaSpace{}%
\AgdaOperator{%
\AgdaInductiveConstructor{␣}%
}\AgdaSpace{}%
\AgdaOperator{%
\AgdaInductiveConstructor{⦅λ}%
}\AgdaSpace{}%
%
\AgdaFunction{c}%
\AgdaSpace{}%
\AgdaOperator{%
\AgdaInductiveConstructor{␣}%
}\AgdaSpace{}%
\AgdaOperator{%
\AgdaInductiveConstructor{⦅}%
}\AgdaSpace{}%
\AgdaOperator{%
\AgdaInductiveConstructor{var}%
}\AgdaSpace{}%
%
\AgdaFunction{c}%
\AgdaSpace{}%
\AgdaOperator{%
\AgdaInductiveConstructor{␣}%
}\AgdaSpace{}%
\AgdaOperator{%
\AgdaInductiveConstructor{var}%
}\AgdaSpace{}%
%
\AgdaFunction{c}%
\AgdaSpace{}%
\AgdaOperator{%
\AgdaInductiveConstructor{⦆}%
}\AgdaSpace{}%
\AgdaOperator{%
\AgdaInductiveConstructor{⦆}%
}\AgdaSpace{}%
\AgdaOperator{%
\AgdaInductiveConstructor{⦆}%
}\AgdaSpace{}%
\AgdaOperator{%
\AgdaInductiveConstructor{⦆}%
}\AgdaSpace{}%
\AgdaOperator{%
\AgdaFunction{⟧}%
}\AgdaSpace{}%
\AgdaGeneralizable{ρ}\AgdaSpace{}%
\AgdaOperator{%
\AgdaDatatype{≡}%
}\AgdaSpace{}%
\AgdaGeneralizable{ρ}\AgdaSpace{}%
%
\AgdaFunction{b}%
\<%
\\
\>[2]%
\AgdaFunction{app-k-to-divergence}%
\AgdaSpace{}%
\AgdaSymbol{=}\AgdaSpace{}%
%
\AgdaInductiveConstructor{refl}%
\<%
\end{code}
\end{AgdaSuppressSpace}
This final illustrative test shows that the free variable \AgdaRef{a} is
not captured by the λ-abstraction on \AgdaRef{a}.
\begin{AgdaSuppressSpace}
\begin{code}%
\>[2]%
\AgdaFunction{app-k-abs}%
\AgdaSpace{}%
\AgdaSymbol{:}\<%
\\
\>[2][@{}l@{\AgdaIndent{0}}]%
\>[4]\AgdaOperator{%
\AgdaFunction{⟦}%
}\AgdaSpace{}%
\AgdaOperator{%
\AgdaInductiveConstructor{⦅}%
}\AgdaSpace{}%
\AgdaOperator{%
\AgdaInductiveConstructor{⦅λ}%
}\AgdaSpace{}%
%
\AgdaFunction{b}%
\AgdaSpace{}%
\AgdaOperator{%
\AgdaInductiveConstructor{␣}%
}\AgdaSpace{}%
\AgdaOperator{%
\AgdaInductiveConstructor{⦅}%
}\AgdaSpace{}%
\AgdaOperator{%
\AgdaInductiveConstructor{⦅λ}%
}\AgdaSpace{}%
%
\AgdaFunction{a}%
\AgdaSpace{}%
\AgdaOperator{%
\AgdaInductiveConstructor{␣}%
}\AgdaSpace{}%
\AgdaOperator{%
\AgdaInductiveConstructor{var}%
}\AgdaSpace{}%
%
\AgdaFunction{b}%
\AgdaSpace{}%
\AgdaOperator{%
\AgdaInductiveConstructor{⦆}%
}\AgdaSpace{}%
\AgdaOperator{%
\AgdaInductiveConstructor{␣}%
}\AgdaSpace{}%
\AgdaOperator{%
\AgdaInductiveConstructor{var}%
}\AgdaSpace{}%
%
\AgdaFunction{c}%
\AgdaSpace{}%
\AgdaOperator{%
\AgdaInductiveConstructor{⦆}%
}\AgdaSpace{}%
\AgdaOperator{%
\AgdaInductiveConstructor{⦆}%
}\AgdaSpace{}%
\AgdaOperator{%
\AgdaInductiveConstructor{␣}%
}\AgdaSpace{}%
\AgdaOperator{%
\AgdaInductiveConstructor{var}%
}\AgdaSpace{}%
%
\AgdaFunction{a}%
\AgdaSpace{}%
\AgdaOperator{%
\AgdaInductiveConstructor{⦆}%
}\AgdaSpace{}%
\AgdaOperator{%
\AgdaFunction{⟧}%
}\AgdaSpace{}%
\AgdaGeneralizable{ρ}\AgdaSpace{}%
\AgdaOperator{%
\AgdaDatatype{≡}%
}\AgdaSpace{}%
\AgdaGeneralizable{ρ}\AgdaSpace{}%
%
\AgdaFunction{a}%
\<%
\\
\>[2]%
\AgdaFunction{app-k-abs}%
\AgdaSpace{}%
\AgdaSymbol{=}\AgdaSpace{}%
%
\AgdaInductiveConstructor{refl}%
\<%
\end{code}
\end{AgdaSuppressSpace}
A reviewer pointed out that the only interpretation of \AgdaRef{⟪\ D∞\ ⟫}
in Agda could be a singleton type, in which case one should expect many
equalities to hold. However, when the Agda proof of an equality is
simply by \AgdaRef{refl}, it cannot use such reasoning about cardinality.

\subsection{PCF Tests}\label{Tests/PCF-pcf-tests}

The following tests check that the denotations of some expressions in
the abstract syntax of the PCF language compute the expected values when
applied to an arbitrary environment \AgdaRef{ρ}.

All the \AgdaRef{refl} proofs of the tests below implicitly use the
rewrite rule postulated as a
property of the
fixed-point operator (Section~\ref{Properties/Functions-function-domains}), together with the rule for
eliminating injections (Section~\ref{Properties/Flat-flat-domains})
in arguments of extensions of functions to flat domains.
\begin{AgdaSuppressSpace}
\begin{code}[hide]%
\>[0]\AgdaSymbol{\{-\#}\AgdaSpace{}%
\AgdaKeyword{OPTIONS}\AgdaSpace{}%
\AgdaPragma{--rewriting}\AgdaSpace{}%
\AgdaPragma{--confluence-check}\AgdaSpace{}%
\AgdaSymbol{\#-\}}\<%
\\
\>[0]\<%
\end{code}
\begin{code}%
\>[0]\AgdaKeyword{module}\AgdaSpace{}%
%
\AgdaModule{Tests.PCF}%
\AgdaSpace{}%
\AgdaKeyword{where}\<%
\end{code}
\begin{code}[hide]%
\>[0][@{}l@{\AgdaIndent{1}}]%
\>[2]\AgdaKeyword{open}\AgdaSpace{}%
\AgdaKeyword{import}\AgdaSpace{}%
%
\AgdaModule{Notation.Domains}%
\<%
\\
\\[\AgdaEmptyExtraSkip]%
\>[2]\AgdaKeyword{open}\AgdaSpace{}%
\AgdaKeyword{import}\AgdaSpace{}%
%
\AgdaModule{Examples.PCF.Abstract-Syntax}%
\<%
\\
\>[2]\AgdaKeyword{open}\AgdaSpace{}%
\AgdaKeyword{import}\AgdaSpace{}%
%
\AgdaModule{Examples.PCF.Domain-Equations}%
\<%
\\
\>[2]\AgdaKeyword{open}\AgdaSpace{}%
\AgdaKeyword{import}\AgdaSpace{}%
%
\AgdaModule{Examples.PCF.Semantic-Functions}%
\<%
\\
\\[\AgdaEmptyExtraSkip]%
\>[2]\AgdaKeyword{open}\AgdaSpace{}%
\AgdaKeyword{import}\AgdaSpace{}%
%
\AgdaModule{Properties.Functions}%
\<%
\\
\>[2]\AgdaKeyword{open}\AgdaSpace{}%
\AgdaKeyword{import}\AgdaSpace{}%
%
\AgdaModule{Properties.Flat}%
\<%
\end{code}
\end{AgdaSuppressSpace}
The following trivial abbreviations improve the readbility of AST terms
in the tests:
\begin{AgdaSuppressSpace}
\begin{code}%
\>[2]%
\AgdaFunction{a}%
%
\>[5]\AgdaSymbol{=}\AgdaSpace{}%
%
\AgdaInductiveConstructor{α}%
\AgdaSpace{}%
\AgdaNumber{0}\AgdaSpace{}%
%
\AgdaInductiveConstructor{ι}%
\<%
\\
\>[2]%
\AgdaFunction{b}%
%
\>[5]\AgdaSymbol{=}\AgdaSpace{}%
%
\AgdaInductiveConstructor{α}%
\AgdaSpace{}%
\AgdaNumber{1}\AgdaSpace{}%
%
\AgdaInductiveConstructor{ι}%
\<%
\\
\>[2]%
\AgdaFunction{e}%
%
\>[5]\AgdaSymbol{=}\AgdaSpace{}%
%
\AgdaInductiveConstructor{α}%
\AgdaSpace{}%
\AgdaNumber{2}\AgdaSpace{}%
%
\AgdaInductiveConstructor{ι}%
\<%
\\
\>[0]\<%
\\
\>[2]%
\AgdaFunction{g}%
%
\>[5]\AgdaSymbol{=}\AgdaSpace{}%
%
\AgdaInductiveConstructor{α}%
\AgdaSpace{}%
\AgdaNumber{0}\AgdaSpace{}%
\AgdaSymbol{(}%
\AgdaInductiveConstructor{ι}%
\AgdaSpace{}%
\AgdaOperator{%
\AgdaInductiveConstructor{⇒}%
}\AgdaSpace{}%
%
\AgdaInductiveConstructor{ι}%
\AgdaSymbol{)}\<%
\\
\>[2]%
\AgdaFunction{h}%
%
\>[5]\AgdaSymbol{=}\AgdaSpace{}%
%
\AgdaInductiveConstructor{α}%
\AgdaSpace{}%
\AgdaNumber{0}\AgdaSpace{}%
\AgdaSymbol{(}%
\AgdaInductiveConstructor{ι}%
\AgdaSpace{}%
\AgdaOperator{%
\AgdaInductiveConstructor{⇒}%
}\AgdaSpace{}%
%
\AgdaInductiveConstructor{ι}%
\AgdaSpace{}%
\AgdaOperator{%
\AgdaInductiveConstructor{⇒}%
}\AgdaSpace{}%
%
\AgdaInductiveConstructor{ι}%
\AgdaSymbol{)}\<%
\end{code}
\end{AgdaSuppressSpace}
Checking applications of successor and predecessor:
\begin{AgdaSuppressSpace}
\begin{code}%
\>[2]%
\AgdaFunction{check-41+1}%
\AgdaSpace{}%
\AgdaSymbol{:}\<%
\\
\>[2][@{}l@{\AgdaIndent{0}}]%
\>[4]\AgdaOperator{%
\AgdaFunction{𝒜′⟦}%
}\AgdaSpace{}%
\AgdaOperator{%
\AgdaInductiveConstructor{⦅}%
}\AgdaSpace{}%
\AgdaOperator{%
\AgdaInductiveConstructor{𝐿}%
}\AgdaSpace{}%
%
\AgdaInductiveConstructor{⦅+1⦆}%
\AgdaSpace{}%
\AgdaOperator{%
\AgdaInductiveConstructor{␣}%
}\AgdaSpace{}%
\AgdaOperator{%
\AgdaInductiveConstructor{𝐿}%
}\AgdaSpace{}%
%
\AgdaInductiveConstructor{k}%
\AgdaSpace{}%
\AgdaNumber{41}\AgdaSpace{}%
\AgdaOperator{%
\AgdaInductiveConstructor{⦆}%
}\AgdaSpace{}%
\AgdaOperator{%
\AgdaFunction{⟧}%
}\AgdaSpace{}%
\AgdaGeneralizable{ρ}\AgdaSpace{}%
\AgdaOperator{%
\AgdaDatatype{≡}%
}\AgdaSpace{}%
%
\AgdaPostulate{↑}%
\AgdaSpace{}%
\AgdaNumber{42}\<%
\\
\>[2]%
\AgdaFunction{check-41+1}%
\AgdaSpace{}%
\AgdaSymbol{=}\AgdaSpace{}%
%
\AgdaInductiveConstructor{refl}%
\<%
\\
\\[\AgdaEmptyExtraSkip]%
\>[2]%
\AgdaFunction{check-43-1}%
\AgdaSpace{}%
\AgdaSymbol{:}\<%
\\
\>[2][@{}l@{\AgdaIndent{0}}]%
\>[4]\AgdaOperator{%
\AgdaFunction{𝒜′⟦}%
}\AgdaSpace{}%
\AgdaOperator{%
\AgdaInductiveConstructor{⦅}%
}\AgdaSpace{}%
\AgdaOperator{%
\AgdaInductiveConstructor{𝐿}%
}\AgdaSpace{}%
%
\AgdaInductiveConstructor{⦅−1⦆}%
\AgdaSpace{}%
\AgdaOperator{%
\AgdaInductiveConstructor{␣}%
}\AgdaSpace{}%
\AgdaOperator{%
\AgdaInductiveConstructor{𝐿}%
}\AgdaSpace{}%
%
\AgdaInductiveConstructor{k}%
\AgdaSpace{}%
\AgdaNumber{43}\AgdaSpace{}%
\AgdaOperator{%
\AgdaInductiveConstructor{⦆}%
}\AgdaSpace{}%
\AgdaOperator{%
\AgdaFunction{⟧}%
}\AgdaSpace{}%
\AgdaGeneralizable{ρ}\AgdaSpace{}%
\AgdaOperator{%
\AgdaDatatype{≡}%
}\AgdaSpace{}%
%
\AgdaPostulate{↑}%
\AgdaSpace{}%
\AgdaNumber{42}\<%
\\
\>[2]%
\AgdaFunction{check-43-1}%
\AgdaSpace{}%
\AgdaSymbol{=}\AgdaSpace{}%
%
\AgdaInductiveConstructor{refl}%
\<%
\end{code}
\end{AgdaSuppressSpace}
Checking that an environment binds a variable:
\begin{AgdaSuppressSpace}
\begin{code}%
\>[2]%
\AgdaFunction{lookup-var}%
\AgdaSpace{}%
\AgdaSymbol{:}\<%
\\
\>[2][@{}l@{\AgdaIndent{0}}]%
\>[4]\AgdaOperator{%
\AgdaFunction{𝒜′⟦}%
}\AgdaSpace{}%
\AgdaOperator{%
\AgdaInductiveConstructor{𝑉}%
}\AgdaSpace{}%
%
\AgdaFunction{a}%
\AgdaSpace{}%
\AgdaOperator{%
\AgdaFunction{⟧}%
}\AgdaSpace{}%
\AgdaSymbol{(}\AgdaGeneralizable{ρ}\AgdaSpace{}%
\AgdaOperator{%
\AgdaFunction{[}%
}\AgdaSpace{}%
%
\AgdaPostulate{↑}%
\AgdaSpace{}%
\AgdaNumber{42}\AgdaSpace{}%
\AgdaOperator{%
\AgdaFunction{/}%
}\AgdaSpace{}%
%
\AgdaFunction{a}%
\AgdaSpace{}%
\AgdaOperator{%
\AgdaFunction{]′}%
}\AgdaSymbol{)}\AgdaSpace{}%
\AgdaOperator{%
\AgdaDatatype{≡}%
}\AgdaSpace{}%
%
\AgdaPostulate{↑}%
\AgdaSpace{}%
\AgdaNumber{42}\<%
\\
\>[2]%
\AgdaFunction{lookup-var}%
\AgdaSpace{}%
\AgdaSymbol{=}\AgdaSpace{}%
%
\AgdaInductiveConstructor{refl}%
\<%
\end{code}
\end{AgdaSuppressSpace}
Checking applications of some combinators (\(I\), \(K\), \(K\,I\)):
\begin{AgdaSuppressSpace}
\begin{code}%
\>[2]%
\AgdaFunction{check-id}%
\AgdaSpace{}%
\AgdaSymbol{:}\<%
\\
\>[2][@{}l@{\AgdaIndent{0}}]%
\>[4]\AgdaOperator{%
\AgdaFunction{𝒜′⟦}%
}\AgdaSpace{}%
\AgdaOperator{%
\AgdaInductiveConstructor{⦅}%
}\AgdaSpace{}%
\AgdaOperator{%
\AgdaInductiveConstructor{⦅λ}%
}\AgdaSpace{}%
%
\AgdaFunction{a}%
\AgdaSpace{}%
\AgdaOperator{%
\AgdaInductiveConstructor{␣}%
}\AgdaSpace{}%
\AgdaOperator{%
\AgdaInductiveConstructor{𝑉}%
}\AgdaSpace{}%
%
\AgdaFunction{a}%
\AgdaSpace{}%
\AgdaOperator{%
\AgdaInductiveConstructor{⦆}%
}\AgdaSpace{}%
\AgdaOperator{%
\AgdaInductiveConstructor{␣}%
}\AgdaSpace{}%
\AgdaOperator{%
\AgdaInductiveConstructor{𝐿}%
}\AgdaSpace{}%
%
\AgdaInductiveConstructor{k}%
\AgdaSpace{}%
\AgdaNumber{42}\AgdaSpace{}%
\AgdaOperator{%
\AgdaInductiveConstructor{⦆}%
}\AgdaSpace{}%
\AgdaOperator{%
\AgdaFunction{⟧}%
}\AgdaSpace{}%
\AgdaGeneralizable{ρ}\AgdaSpace{}%
\AgdaOperator{%
\AgdaDatatype{≡}%
}\AgdaSpace{}%
%
\AgdaPostulate{↑}%
\AgdaSpace{}%
\AgdaNumber{42}\<%
\\
\>[2]%
\AgdaFunction{check-id}%
\AgdaSpace{}%
\AgdaSymbol{=}\AgdaSpace{}%
%
\AgdaInductiveConstructor{refl}%
\<%
\\
\\[\AgdaEmptyExtraSkip]%
\>[2]%
\AgdaFunction{check-k}%
\AgdaSpace{}%
\AgdaSymbol{:}\<%
\\
\>[2][@{}l@{\AgdaIndent{0}}]%
\>[4]\AgdaOperator{%
\AgdaFunction{𝒜′⟦}%
}\AgdaSpace{}%
\AgdaOperator{%
\AgdaInductiveConstructor{⦅}%
}\AgdaSpace{}%
\AgdaOperator{%
\AgdaInductiveConstructor{⦅}%
}\AgdaSpace{}%
\AgdaOperator{%
\AgdaInductiveConstructor{⦅λ}%
}\AgdaSpace{}%
%
\AgdaFunction{a}%
\AgdaSpace{}%
\AgdaOperator{%
\AgdaInductiveConstructor{␣}%
}\AgdaSpace{}%
\AgdaOperator{%
\AgdaInductiveConstructor{⦅λ}%
}\AgdaSpace{}%
%
\AgdaFunction{b}%
\AgdaSpace{}%
\AgdaOperator{%
\AgdaInductiveConstructor{␣}%
}\AgdaSpace{}%
\AgdaOperator{%
\AgdaInductiveConstructor{𝑉}%
}\AgdaSpace{}%
%
\AgdaFunction{a}%
\AgdaSpace{}%
\AgdaOperator{%
\AgdaInductiveConstructor{⦆}%
}\AgdaSpace{}%
\AgdaOperator{%
\AgdaInductiveConstructor{⦆}%
}\AgdaSpace{}%
\AgdaOperator{%
\AgdaInductiveConstructor{␣}%
}\AgdaSpace{}%
\AgdaOperator{%
\AgdaInductiveConstructor{𝐿}%
}\AgdaSpace{}%
%
\AgdaInductiveConstructor{k}%
\AgdaSpace{}%
\AgdaNumber{42}\AgdaSpace{}%
\AgdaOperator{%
\AgdaInductiveConstructor{⦆}%
}\AgdaSpace{}%
\AgdaOperator{%
\AgdaInductiveConstructor{␣}%
}\AgdaSpace{}%
\AgdaOperator{%
\AgdaInductiveConstructor{𝐿}%
}\AgdaSpace{}%
%
\AgdaInductiveConstructor{k}%
\AgdaSpace{}%
\AgdaNumber{41}\AgdaSpace{}%
\AgdaOperator{%
\AgdaInductiveConstructor{⦆}%
}\AgdaSpace{}%
\AgdaOperator{%
\AgdaFunction{⟧}%
}\AgdaSpace{}%
\AgdaGeneralizable{ρ}\AgdaSpace{}%
\AgdaOperator{%
\AgdaDatatype{≡}%
}\AgdaSpace{}%
%
\AgdaPostulate{↑}%
\AgdaSpace{}%
\AgdaNumber{42}\<%
\\
\>[2]%
\AgdaFunction{check-k}%
\AgdaSpace{}%
\AgdaSymbol{=}\AgdaSpace{}%
%
\AgdaInductiveConstructor{refl}%
\<%
\\
\\[\AgdaEmptyExtraSkip]%
\>[2]%
\AgdaFunction{check-ki}%
\AgdaSpace{}%
\AgdaSymbol{:}\<%
\\
\>[2][@{}l@{\AgdaIndent{0}}]%
\>[4]\AgdaOperator{%
\AgdaFunction{𝒜′⟦}%
}\AgdaSpace{}%
\AgdaOperator{%
\AgdaInductiveConstructor{⦅}%
}\AgdaSpace{}%
\AgdaOperator{%
\AgdaInductiveConstructor{⦅}%
}\AgdaSpace{}%
\AgdaOperator{%
\AgdaInductiveConstructor{⦅λ}%
}\AgdaSpace{}%
%
\AgdaFunction{a}%
\AgdaSpace{}%
\AgdaOperator{%
\AgdaInductiveConstructor{␣}%
}\AgdaSpace{}%
\AgdaOperator{%
\AgdaInductiveConstructor{⦅λ}%
}\AgdaSpace{}%
%
\AgdaFunction{b}%
\AgdaSpace{}%
\AgdaOperator{%
\AgdaInductiveConstructor{␣}%
}\AgdaSpace{}%
\AgdaOperator{%
\AgdaInductiveConstructor{𝑉}%
}\AgdaSpace{}%
%
\AgdaFunction{b}%
\AgdaSpace{}%
\AgdaOperator{%
\AgdaInductiveConstructor{⦆}%
}\AgdaSpace{}%
\AgdaOperator{%
\AgdaInductiveConstructor{⦆}%
}\AgdaSpace{}%
\AgdaOperator{%
\AgdaInductiveConstructor{␣}%
}\AgdaSpace{}%
\AgdaOperator{%
\AgdaInductiveConstructor{𝐿}%
}\AgdaSpace{}%
%
\AgdaInductiveConstructor{k}%
\AgdaSpace{}%
\AgdaNumber{41}\AgdaSpace{}%
\AgdaOperator{%
\AgdaInductiveConstructor{⦆}%
}\AgdaSpace{}%
\AgdaOperator{%
\AgdaInductiveConstructor{␣}%
}\AgdaSpace{}%
\AgdaOperator{%
\AgdaInductiveConstructor{𝐿}%
}\AgdaSpace{}%
%
\AgdaInductiveConstructor{k}%
\AgdaSpace{}%
\AgdaNumber{42}\AgdaSpace{}%
\AgdaOperator{%
\AgdaInductiveConstructor{⦆}%
}\AgdaSpace{}%
\AgdaOperator{%
\AgdaFunction{⟧}%
}\AgdaSpace{}%
\AgdaGeneralizable{ρ}\AgdaSpace{}%
\AgdaOperator{%
\AgdaDatatype{≡}%
}\AgdaSpace{}%
%
\AgdaPostulate{↑}%
\AgdaSpace{}%
\AgdaNumber{42}\<%
\\
\>[2]%
\AgdaFunction{check-ki}%
\AgdaSpace{}%
\AgdaSymbol{=}\AgdaSpace{}%
%
\AgdaInductiveConstructor{refl}%
\<%
\end{code}
\end{AgdaSuppressSpace}
Checking applications of conditional choice:
\begin{AgdaSuppressSpace}
\begin{code}%
\>[2]%
\AgdaFunction{check-if-zero}%
\AgdaSpace{}%
\AgdaSymbol{:}\<%
\\
\>[2][@{}l@{\AgdaIndent{0}}]%
\>[4]\AgdaOperator{%
\AgdaFunction{𝒜′⟦}%
}\AgdaSpace{}%
\AgdaOperator{%
\AgdaInductiveConstructor{⦅}%
}\AgdaSpace{}%
\AgdaOperator{%
\AgdaInductiveConstructor{⦅}%
}\AgdaSpace{}%
\AgdaOperator{%
\AgdaInductiveConstructor{⦅}%
}\AgdaSpace{}%
\AgdaOperator{%
\AgdaInductiveConstructor{𝐿}%
}\AgdaSpace{}%
%
\AgdaInductiveConstructor{⊃}%
\AgdaSpace{}%
\AgdaOperator{%
\AgdaInductiveConstructor{␣}%
}\AgdaSpace{}%
\AgdaOperator{%
\AgdaInductiveConstructor{⦅}%
}\AgdaSpace{}%
\AgdaOperator{%
\AgdaInductiveConstructor{𝐿}%
}\AgdaSpace{}%
%
\AgdaInductiveConstructor{Z}%
%
\>[27]\AgdaOperator{%
\AgdaInductiveConstructor{␣}%
}\AgdaSpace{}%
\AgdaOperator{%
\AgdaInductiveConstructor{𝐿}%
}\AgdaSpace{}%
%
\AgdaInductiveConstructor{k}%
\AgdaSpace{}%
\AgdaNumber{0}\AgdaSpace{}%
\AgdaOperator{%
\AgdaInductiveConstructor{⦆}%
}\AgdaSpace{}%
\AgdaOperator{%
\AgdaInductiveConstructor{⦆}%
}\AgdaSpace{}%
\AgdaOperator{%
\AgdaInductiveConstructor{␣}%
}\AgdaSpace{}%
\AgdaOperator{%
\AgdaInductiveConstructor{𝐿}%
}\AgdaSpace{}%
%
\AgdaInductiveConstructor{k}%
\AgdaSpace{}%
\AgdaNumber{42}\AgdaSpace{}%
\AgdaOperator{%
\AgdaInductiveConstructor{⦆}%
}\AgdaSpace{}%
\AgdaOperator{%
\AgdaInductiveConstructor{␣}%
}\AgdaSpace{}%
\AgdaOperator{%
\AgdaInductiveConstructor{𝐿}%
}\AgdaSpace{}%
%
\AgdaInductiveConstructor{k}%
\AgdaSpace{}%
\AgdaNumber{0}\AgdaSpace{}%
\AgdaOperator{%
\AgdaInductiveConstructor{⦆}%
}\AgdaSpace{}%
\AgdaOperator{%
\AgdaFunction{⟧}%
}\AgdaSpace{}%
\AgdaGeneralizable{ρ}\AgdaSpace{}%
\AgdaOperator{%
\AgdaDatatype{≡}%
}\AgdaSpace{}%
%
\AgdaPostulate{↑}%
\AgdaSpace{}%
\AgdaNumber{42}\<%
\\
\>[2]%
\AgdaFunction{check-if-zero}%
\AgdaSpace{}%
\AgdaSymbol{=}\AgdaSpace{}%
%
\AgdaInductiveConstructor{refl}%
\<%
\\
\\[\AgdaEmptyExtraSkip]%
\>[2]%
\AgdaFunction{check-if-nonzero}%
\AgdaSpace{}%
\AgdaSymbol{:}\<%
\\
\>[2][@{}l@{\AgdaIndent{0}}]%
\>[4]\AgdaOperator{%
\AgdaFunction{𝒜′⟦}%
}\AgdaSpace{}%
\AgdaOperator{%
\AgdaInductiveConstructor{⦅}%
}\AgdaSpace{}%
\AgdaOperator{%
\AgdaInductiveConstructor{⦅}%
}\AgdaSpace{}%
\AgdaOperator{%
\AgdaInductiveConstructor{⦅}%
}\AgdaSpace{}%
\AgdaOperator{%
\AgdaInductiveConstructor{𝐿}%
}\AgdaSpace{}%
%
\AgdaInductiveConstructor{⊃}%
\AgdaSpace{}%
\AgdaOperator{%
\AgdaInductiveConstructor{␣}%
}\AgdaSpace{}%
\AgdaOperator{%
\AgdaInductiveConstructor{⦅}%
}\AgdaSpace{}%
\AgdaOperator{%
\AgdaInductiveConstructor{𝐿}%
}\AgdaSpace{}%
%
\AgdaInductiveConstructor{Z}%
%
\>[27]\AgdaOperator{%
\AgdaInductiveConstructor{␣}%
}\AgdaSpace{}%
\AgdaOperator{%
\AgdaInductiveConstructor{𝐿}%
}\AgdaSpace{}%
%
\AgdaInductiveConstructor{k}%
\AgdaSpace{}%
\AgdaNumber{42}\AgdaSpace{}%
\AgdaOperator{%
\AgdaInductiveConstructor{⦆}%
}\AgdaSpace{}%
\AgdaOperator{%
\AgdaInductiveConstructor{⦆}%
}\AgdaSpace{}%
\AgdaOperator{%
\AgdaInductiveConstructor{␣}%
}\AgdaSpace{}%
\AgdaOperator{%
\AgdaInductiveConstructor{𝐿}%
}\AgdaSpace{}%
%
\AgdaInductiveConstructor{k}%
\AgdaSpace{}%
\AgdaNumber{0}\AgdaSpace{}%
\AgdaOperator{%
\AgdaInductiveConstructor{⦆}%
}\AgdaSpace{}%
\AgdaOperator{%
\AgdaInductiveConstructor{␣}%
}\AgdaSpace{}%
\AgdaOperator{%
\AgdaInductiveConstructor{𝐿}%
}\AgdaSpace{}%
%
\AgdaInductiveConstructor{k}%
\AgdaSpace{}%
\AgdaNumber{42}\AgdaSpace{}%
\AgdaOperator{%
\AgdaInductiveConstructor{⦆}%
}\AgdaSpace{}%
\AgdaOperator{%
\AgdaFunction{⟧}%
}\AgdaSpace{}%
\AgdaGeneralizable{ρ}\AgdaSpace{}%
\AgdaOperator{%
\AgdaDatatype{≡}%
}\AgdaSpace{}%
%
\AgdaPostulate{↑}%
\AgdaSpace{}%
\AgdaNumber{42}\<%
\\
\>[2]%
\AgdaFunction{check-if-nonzero}%
\AgdaSpace{}%
\AgdaSymbol{=}\AgdaSpace{}%
%
\AgdaInductiveConstructor{refl}%
\<%
\end{code}
\end{AgdaSuppressSpace}
Checking applications of the fixed-point operator:
\begin{AgdaSuppressSpace}
\begin{code}%
\>[2]%
\AgdaFunction{check-fix-const}%
\AgdaSpace{}%
\AgdaSymbol{:}\AgdaSpace{}%
\AgdaComment{--\ Y\ (λe.\ 42)\ ≡\ 42}\<%
\\
\>[2][@{}l@{\AgdaIndent{0}}]%
\>[4]\AgdaOperator{%
\AgdaFunction{𝒜′⟦}%
}\AgdaSpace{}%
\AgdaOperator{%
\AgdaInductiveConstructor{⦅}%
}\AgdaSpace{}%
\AgdaOperator{%
\AgdaInductiveConstructor{𝐿}%
}\AgdaSpace{}%
%
\AgdaInductiveConstructor{Y}%
\AgdaSpace{}%
\AgdaOperator{%
\AgdaInductiveConstructor{␣}%
}\AgdaSpace{}%
\AgdaOperator{%
\AgdaInductiveConstructor{⦅λ}%
}\AgdaSpace{}%
%
\AgdaFunction{e}%
\AgdaSpace{}%
\AgdaOperator{%
\AgdaInductiveConstructor{␣}%
}\AgdaSpace{}%
\AgdaOperator{%
\AgdaInductiveConstructor{𝐿}%
}\AgdaSpace{}%
%
\AgdaInductiveConstructor{k}%
\AgdaSpace{}%
\AgdaNumber{42}\AgdaSpace{}%
\AgdaOperator{%
\AgdaInductiveConstructor{⦆}%
}\AgdaSpace{}%
\AgdaOperator{%
\AgdaInductiveConstructor{⦆}%
}\AgdaSpace{}%
\AgdaOperator{%
\AgdaFunction{⟧}%
}\AgdaSpace{}%
\AgdaGeneralizable{ρ}\AgdaSpace{}%
\AgdaOperator{%
\AgdaDatatype{≡}%
}\AgdaSpace{}%
%
\AgdaPostulate{↑}%
\AgdaSpace{}%
\AgdaNumber{42}\<%
\\
\>[2]%
\AgdaFunction{check-fix-const}%
\AgdaSpace{}%
\AgdaSymbol{=}\AgdaSpace{}%
%
\AgdaInductiveConstructor{refl}%
\<%
\\
\\[\AgdaEmptyExtraSkip]%
\>[2]%
\AgdaFunction{check-fix-lambda}%
\AgdaSpace{}%
\AgdaSymbol{:}\AgdaSpace{}%
\AgdaComment{--\ Y\ (λg.\ λa.\ 42)\ 2\ ≡\ 42}\<%
\\
\>[2][@{}l@{\AgdaIndent{0}}]%
\>[4]\AgdaOperator{%
\AgdaFunction{𝒜′⟦}%
}\AgdaSpace{}%
\AgdaOperator{%
\AgdaInductiveConstructor{⦅}%
}\AgdaSpace{}%
\AgdaOperator{%
\AgdaInductiveConstructor{⦅}%
}\AgdaSpace{}%
\AgdaOperator{%
\AgdaInductiveConstructor{𝐿}%
}\AgdaSpace{}%
%
\AgdaInductiveConstructor{Y}%
\AgdaSpace{}%
\AgdaOperator{%
\AgdaInductiveConstructor{␣}%
}\AgdaSpace{}%
\AgdaOperator{%
\AgdaInductiveConstructor{⦅λ}%
}\AgdaSpace{}%
%
\AgdaFunction{g}%
\AgdaSpace{}%
\AgdaOperator{%
\AgdaInductiveConstructor{␣}%
}\AgdaSpace{}%
\AgdaOperator{%
\AgdaInductiveConstructor{⦅λ}%
}\AgdaSpace{}%
%
\AgdaFunction{a}%
\AgdaSpace{}%
\AgdaOperator{%
\AgdaInductiveConstructor{␣}%
}\AgdaSpace{}%
\AgdaOperator{%
\AgdaInductiveConstructor{𝐿}%
}\AgdaSpace{}%
%
\AgdaInductiveConstructor{k}%
\AgdaSpace{}%
\AgdaNumber{42}\AgdaSpace{}%
\AgdaOperator{%
\AgdaInductiveConstructor{⦆}%
}\AgdaSpace{}%
\AgdaOperator{%
\AgdaInductiveConstructor{⦆}%
}\AgdaSpace{}%
\AgdaOperator{%
\AgdaInductiveConstructor{⦆}%
}\AgdaSpace{}%
\AgdaOperator{%
\AgdaInductiveConstructor{␣}%
}\AgdaSpace{}%
\AgdaOperator{%
\AgdaInductiveConstructor{𝐿}%
}\AgdaSpace{}%
%
\AgdaInductiveConstructor{k}%
\AgdaSpace{}%
\AgdaNumber{2}\AgdaSpace{}%
\AgdaOperator{%
\AgdaInductiveConstructor{⦆}%
}\AgdaSpace{}%
\AgdaOperator{%
\AgdaFunction{⟧}%
}\AgdaSpace{}%
\AgdaGeneralizable{ρ}\AgdaSpace{}%
\AgdaOperator{%
\AgdaDatatype{≡}%
}\AgdaSpace{}%
%
\AgdaPostulate{↑}%
\AgdaSpace{}%
\AgdaNumber{42}\<%
\\
\>[2]%
\AgdaFunction{check-fix-lambda}%
\AgdaSpace{}%
\AgdaSymbol{=}\AgdaSpace{}%
%
\AgdaInductiveConstructor{refl}%
\<%
\end{code}
\end{AgdaSuppressSpace}
Checking counting down from 5 to zero:
\begin{AgdaSuppressSpace}
\begin{code}%
\>[2]%
\AgdaFunction{check-countdown}%
\AgdaSpace{}%
\AgdaSymbol{:}\AgdaSpace{}%
\AgdaComment{--\ Y\ (λg.\ λa.\ ifz\ a\ then\ 42\ else\ g\ (pred\ a))\ 5\ ≡\ 42}\<%
\\
\>[2][@{}l@{\AgdaIndent{0}}]%
\>[4]\AgdaOperator{%
\AgdaFunction{𝒜′⟦}%
}\AgdaSpace{}%
\AgdaOperator{%
\AgdaInductiveConstructor{⦅}%
}\AgdaSpace{}%
\AgdaOperator{%
\AgdaInductiveConstructor{⦅}%
}\AgdaSpace{}%
\AgdaOperator{%
\AgdaInductiveConstructor{𝐿}%
}\AgdaSpace{}%
%
\AgdaInductiveConstructor{Y}%
%
\>[289I]\AgdaOperator{%
\AgdaInductiveConstructor{␣}%
}\AgdaSpace{}%
\AgdaOperator{%
\AgdaInductiveConstructor{⦅λ}%
}\AgdaSpace{}%
%
\AgdaFunction{g}%
\AgdaSpace{}%
\AgdaOperator{%
\AgdaInductiveConstructor{␣}%
}\AgdaSpace{}%
\AgdaOperator{%
\AgdaInductiveConstructor{⦅λ}%
}\AgdaSpace{}%
%
\AgdaFunction{a}%
\AgdaSpace{}%
\AgdaOperator{%
\AgdaInductiveConstructor{␣}%
}\<%
\\
\>[.][@{}l@{}]\<[289I]%
\>[16]\AgdaOperator{%
\AgdaInductiveConstructor{⦅}%
}\AgdaSpace{}%
\AgdaOperator{%
\AgdaInductiveConstructor{⦅}%
}\AgdaSpace{}%
\AgdaOperator{%
\AgdaInductiveConstructor{⦅}%
}%
\>[298I]\AgdaOperator{%
\AgdaInductiveConstructor{𝐿}%
}\AgdaSpace{}%
%
\AgdaInductiveConstructor{⊃}%
\AgdaSpace{}%
\AgdaOperator{%
\AgdaInductiveConstructor{␣}%
}\AgdaSpace{}%
\AgdaOperator{%
\AgdaInductiveConstructor{⦅}%
}\AgdaSpace{}%
\AgdaOperator{%
\AgdaInductiveConstructor{𝐿}%
}\AgdaSpace{}%
%
\AgdaInductiveConstructor{Z}%
%
\>[35]\AgdaOperator{%
\AgdaInductiveConstructor{␣}%
}\AgdaSpace{}%
\AgdaOperator{%
\AgdaInductiveConstructor{𝑉}%
}\AgdaSpace{}%
%
\AgdaFunction{a}%
\AgdaSpace{}%
\AgdaOperator{%
\AgdaInductiveConstructor{⦆}%
}\AgdaSpace{}%
\AgdaOperator{%
\AgdaInductiveConstructor{⦆}%
}\AgdaSpace{}%
\AgdaOperator{%
\AgdaInductiveConstructor{␣}%
}\AgdaSpace{}%
\AgdaOperator{%
\AgdaInductiveConstructor{𝐿}%
}\AgdaSpace{}%
%
\AgdaInductiveConstructor{k}%
\AgdaSpace{}%
\AgdaNumber{42}\AgdaSpace{}%
\AgdaOperator{%
\AgdaInductiveConstructor{⦆}%
}\AgdaSpace{}%
\AgdaOperator{%
\AgdaInductiveConstructor{␣}%
}\<%
\\
\>[.][@{}l@{}]\<[298I]%
\>[22]\AgdaOperator{%
\AgdaInductiveConstructor{⦅}%
}\AgdaSpace{}%
\AgdaOperator{%
\AgdaInductiveConstructor{𝑉}%
}\AgdaSpace{}%
%
\AgdaFunction{g}%
\AgdaSpace{}%
\AgdaOperator{%
\AgdaInductiveConstructor{␣}%
}\AgdaSpace{}%
\AgdaOperator{%
\AgdaInductiveConstructor{⦅}%
}\AgdaSpace{}%
\AgdaOperator{%
\AgdaInductiveConstructor{𝐿}%
}\AgdaSpace{}%
%
\AgdaInductiveConstructor{⦅−1⦆}%
\AgdaSpace{}%
\AgdaOperator{%
\AgdaInductiveConstructor{␣}%
}\AgdaSpace{}%
\AgdaOperator{%
\AgdaInductiveConstructor{𝑉}%
}\AgdaSpace{}%
%
\AgdaFunction{a}%
\AgdaSpace{}%
\AgdaOperator{%
\AgdaInductiveConstructor{⦆}%
}\AgdaSpace{}%
\AgdaOperator{%
\AgdaInductiveConstructor{⦆}%
}\AgdaSpace{}%
\AgdaOperator{%
\AgdaInductiveConstructor{⦆}%
}\AgdaSpace{}%
\AgdaOperator{%
\AgdaInductiveConstructor{⦆}%
}\AgdaSpace{}%
\AgdaOperator{%
\AgdaInductiveConstructor{⦆}%
}\AgdaSpace{}%
\AgdaOperator{%
\AgdaInductiveConstructor{⦆}%
}\AgdaSpace{}%
\AgdaOperator{%
\AgdaInductiveConstructor{␣}%
}\AgdaSpace{}%
\AgdaOperator{%
\AgdaInductiveConstructor{𝐿}%
}\AgdaSpace{}%
%
\AgdaInductiveConstructor{k}%
\AgdaSpace{}%
\AgdaNumber{5}\AgdaSpace{}%
\AgdaOperator{%
\AgdaInductiveConstructor{⦆}%
}\AgdaSpace{}%
\AgdaOperator{%
\AgdaFunction{⟧}%
}\AgdaSpace{}%
\AgdaGeneralizable{ρ}\AgdaSpace{}%
\AgdaOperator{%
\AgdaDatatype{≡}%
}\AgdaSpace{}%
%
\AgdaPostulate{↑}%
\AgdaSpace{}%
\AgdaNumber{42}\<%
\\
\>[2]%
\AgdaFunction{check-countdown}%
\AgdaSpace{}%
\AgdaSymbol{=}\AgdaSpace{}%
%
\AgdaInductiveConstructor{refl}%
\<%
\end{code}
\end{AgdaSuppressSpace}
Checking addition of 4 to 38:
\begin{AgdaSuppressSpace}
\begin{code}%
\>[2]%
\AgdaFunction{check-sum-42}%
\AgdaSpace{}%
\AgdaSymbol{:}\AgdaSpace{}%
\AgdaComment{--\ Y\ (λh.λa.λb.\ ifz\ a\ then\ b\ else\ h(pred\ a)(𝐿\ ⦅+1⦆\ b))\ 4\ 38\ ≡\ 42}\<%
\\
\>[2][@{}l@{\AgdaIndent{0}}]%
\>[4]\AgdaOperator{%
\AgdaFunction{𝒜′⟦}%
}%
\>[343I]\AgdaOperator{%
\AgdaInductiveConstructor{⦅}%
}\AgdaSpace{}%
\AgdaOperator{%
\AgdaInductiveConstructor{⦅}%
}\AgdaSpace{}%
\AgdaOperator{%
\AgdaInductiveConstructor{⦅}%
}\AgdaSpace{}%
\AgdaOperator{%
\AgdaInductiveConstructor{𝐿}%
}\AgdaSpace{}%
%
\AgdaInductiveConstructor{Y}%
\AgdaSpace{}%
\AgdaOperator{%
\AgdaInductiveConstructor{␣}%
}%
\>[349I]\AgdaOperator{%
\AgdaInductiveConstructor{⦅λ}%
}\AgdaSpace{}%
%
\AgdaFunction{h}%
\AgdaSpace{}%
\AgdaOperator{%
\AgdaInductiveConstructor{␣}%
}\AgdaSpace{}%
\AgdaOperator{%
\AgdaInductiveConstructor{⦅λ}%
}\AgdaSpace{}%
%
\AgdaFunction{a}%
\AgdaSpace{}%
\AgdaOperator{%
\AgdaInductiveConstructor{␣}%
}\AgdaSpace{}%
\AgdaOperator{%
\AgdaInductiveConstructor{⦅λ}%
}\AgdaSpace{}%
%
\AgdaFunction{b}%
\AgdaSpace{}%
\AgdaOperator{%
\AgdaInductiveConstructor{␣}%
}\<%
\\
\>[.][@{}l@{}]\<[349I]%
\>[20]\AgdaOperator{%
\AgdaInductiveConstructor{⦅}%
}%
\>[358I]\AgdaOperator{%
\AgdaInductiveConstructor{⦅}%
}\AgdaSpace{}%
\AgdaOperator{%
\AgdaInductiveConstructor{⦅}%
}\AgdaSpace{}%
\AgdaOperator{%
\AgdaInductiveConstructor{𝐿}%
}\AgdaSpace{}%
%
\AgdaInductiveConstructor{⊃}%
\AgdaSpace{}%
\AgdaOperator{%
\AgdaInductiveConstructor{␣}%
}\AgdaSpace{}%
\AgdaOperator{%
\AgdaInductiveConstructor{⦅}%
}\AgdaSpace{}%
\AgdaOperator{%
\AgdaInductiveConstructor{𝐿}%
}\AgdaSpace{}%
%
\AgdaInductiveConstructor{Z}%
%
\>[39]\AgdaOperator{%
\AgdaInductiveConstructor{␣}%
}\AgdaSpace{}%
\AgdaOperator{%
\AgdaInductiveConstructor{𝑉}%
}\AgdaSpace{}%
%
\AgdaFunction{a}%
\AgdaSpace{}%
\AgdaOperator{%
\AgdaInductiveConstructor{⦆}%
}\AgdaSpace{}%
\AgdaOperator{%
\AgdaInductiveConstructor{⦆}%
}\AgdaSpace{}%
\AgdaOperator{%
\AgdaInductiveConstructor{␣}%
}\AgdaSpace{}%
\AgdaOperator{%
\AgdaInductiveConstructor{𝑉}%
}\AgdaSpace{}%
%
\AgdaFunction{b}%
\AgdaSpace{}%
\AgdaOperator{%
\AgdaInductiveConstructor{⦆}%
}\AgdaSpace{}%
\AgdaOperator{%
\AgdaInductiveConstructor{␣}%
}\<%
\\
\>[.][@{}l@{}]\<[358I]%
\>[22]\AgdaOperator{%
\AgdaInductiveConstructor{⦅}%
}\AgdaSpace{}%
\AgdaOperator{%
\AgdaInductiveConstructor{⦅}%
}\AgdaSpace{}%
\AgdaOperator{%
\AgdaInductiveConstructor{𝑉}%
}\AgdaSpace{}%
%
\AgdaFunction{h}%
\AgdaSpace{}%
\AgdaOperator{%
\AgdaInductiveConstructor{␣}%
}\AgdaSpace{}%
\AgdaOperator{%
\AgdaInductiveConstructor{⦅}%
}\AgdaSpace{}%
\AgdaOperator{%
\AgdaInductiveConstructor{𝐿}%
}\AgdaSpace{}%
%
\AgdaInductiveConstructor{⦅−1⦆}%
\AgdaSpace{}%
\AgdaOperator{%
\AgdaInductiveConstructor{␣}%
}\AgdaSpace{}%
\AgdaOperator{%
\AgdaInductiveConstructor{𝑉}%
}\AgdaSpace{}%
%
\AgdaFunction{a}%
\AgdaSpace{}%
\AgdaOperator{%
\AgdaInductiveConstructor{⦆}%
}\AgdaSpace{}%
\AgdaOperator{%
\AgdaInductiveConstructor{⦆}%
}\AgdaSpace{}%
\AgdaOperator{%
\AgdaInductiveConstructor{␣}%
}\AgdaSpace{}%
\AgdaOperator{%
\AgdaInductiveConstructor{⦅}%
}\AgdaSpace{}%
\AgdaOperator{%
\AgdaInductiveConstructor{𝐿}%
}\AgdaSpace{}%
%
\AgdaInductiveConstructor{⦅+1⦆}%
\AgdaSpace{}%
\AgdaOperator{%
\AgdaInductiveConstructor{␣}%
}\AgdaSpace{}%
\AgdaOperator{%
\AgdaInductiveConstructor{𝑉}%
}\AgdaSpace{}%
%
\AgdaFunction{b}%
\AgdaSpace{}%
\AgdaOperator{%
\AgdaInductiveConstructor{⦆}%
}\AgdaSpace{}%
\AgdaOperator{%
\AgdaInductiveConstructor{⦆}%
}\AgdaSpace{}%
\AgdaOperator{%
\AgdaInductiveConstructor{⦆}%
}\AgdaSpace{}%
\AgdaOperator{%
\AgdaInductiveConstructor{⦆}%
}\AgdaSpace{}%
\AgdaOperator{%
\AgdaInductiveConstructor{⦆}%
}\AgdaSpace{}%
\AgdaOperator{%
\AgdaInductiveConstructor{⦆}%
}\AgdaSpace{}%
\AgdaOperator{%
\AgdaInductiveConstructor{⦆}%
}\<%
\\
\>[.][@{}l@{}]\<[343I]%
\>[8]\AgdaOperator{%
\AgdaInductiveConstructor{␣}%
}\AgdaSpace{}%
\AgdaOperator{%
\AgdaInductiveConstructor{𝐿}%
}\AgdaSpace{}%
%
\AgdaInductiveConstructor{k}%
\AgdaSpace{}%
\AgdaNumber{4}\AgdaSpace{}%
\AgdaOperator{%
\AgdaInductiveConstructor{⦆}%
}\AgdaSpace{}%
\AgdaOperator{%
\AgdaInductiveConstructor{␣}%
}\AgdaSpace{}%
\AgdaOperator{%
\AgdaInductiveConstructor{𝐿}%
}\AgdaSpace{}%
%
\AgdaInductiveConstructor{k}%
\AgdaSpace{}%
\AgdaNumber{38}\AgdaSpace{}%
\AgdaOperator{%
\AgdaInductiveConstructor{⦆}%
}\AgdaSpace{}%
\AgdaOperator{%
\AgdaFunction{⟧}%
}\AgdaSpace{}%
\AgdaGeneralizable{ρ}\AgdaSpace{}%
\AgdaOperator{%
\AgdaDatatype{≡}%
}\AgdaSpace{}%
%
\AgdaPostulate{↑}%
\AgdaSpace{}%
\AgdaNumber{42}\<%
\\
\>[2]%
\AgdaFunction{check-sum-42}%
\AgdaSpace{}%
\AgdaSymbol{=}\AgdaSpace{}%
%
\AgdaInductiveConstructor{refl}%
\<%
\end{code}
\end{AgdaSuppressSpace}

\end{AgdaAlign}


\bibliographystyle{./entics}
\bibliography{Peter,Jesper,Bernhard}

\end{document}